\documentclass[12pt]{report}
\usepackage{yalethesis1}
\usepackage{graphicx}
\usepackage{subfigure}
\usepackage{multirow}
\usepackage{hhline}
\usepackage{epsf}
\usepackage{amsmath}

\begin{document}

\input epsf
\title{Pion, Kaon, Proton and Antiproton Spectra in d+Au and p+p Collisions\\
at $\sqrt{s_{NN}}$= 200GeV at the Relativistic Heavy Ion Collider}
    \author{Lijuan Ruan}
    \awardate{Jan. 2005}
    \advisor{Prof. Hongfang Chen\\Off-campus Co-adviser: Dr. Zhangbu Xu}
{

\begin{abstract}
  Identified mid-rapidity particle spectra of $\pi^{\pm}$, $K^{\pm}$,
and $p(\bar{p})$ from 200 GeV p+p and d+Au collisions are
reported. The d+Au collisions were divided into 3 centralities.
This data were taken in 2003 run from the Solenoidal Tracker at
RHIC (STAR) experiment. A time-of-flight detector based on
multi-gap resistive plate chamber (MRPC) technology is used for
particle identification. This is the first time that MRPC detector
was installed to take data as a time-of-flight detector in the
collider experiment.

The calibration method was set up in the STAR experiment for the
first time, which will be described in the analysis chapter in
detail and has been applied to the experimental data successfully.
The intrinsic timing resolution 85 ps was achieved after the
calibration. In 2003 run, the pion/kaon can be separated up to
transverse momentum ($p_T$) 1.6 GeV/c while proton can be
identified up to 3.0 GeV/c. Thus the identified particle spectra
can be extended to intermediate $p_T$ in STAR. This proved that
MRPC as a time-of-flight detector works in the heavy ion collider
experiment.

We observe that the spectra of $\pi^{\pm}$, $K^{\pm}$, $p$ and
$\bar{p}$ are considerably harder in d+Au than those in p+p
collisions. In $\sqrt{s_{_{NN}}} = 200$ GeV d+Au collisions, the
nuclear modification factor $R_{dAu}$ of protons rise faster than
those of pions and kaons. The $R_{dAu}$ of proton is larger than 1
at intermediate $p_T$ while the proton production follows binary
scaling at the same $p_T$ range in 200 GeV Au+Au collisions. These
results further prove that the suppression observed in Au+Au
collisions at intermediate and high $p_T$ is due to final state
interactions in a dense and dissipative medium produced during the
collision and not due to the initial state wave function of the Au
nucleus. Since the initial state in d+Au collisions is similar to
that in Au+Au collisions, and, it's believed that the quark-gluon
plasma doesn't exist in d+Au collisions, these results from d+Au
collisions are very important for us to judge whether the
quark-gluon plasma exists in Au+Au collisions or not and to
understand the property of the dense matter created in Au+Au
collisions. Besides, the particle-species dependence of the Cronin
effect is observed to be significantly smaller than that at lower
energies. The ratio of the nuclear modification factor ($R_{dAu}$)
between $(p+\bar{p})$ and charged hadrons ($h$) in the transverse
momentum range $1.2<p_{T}<3.0$ GeV/c is measured to be
$1.19\pm0.05$(stat)$\pm0.03$(syst) in minimum-bias collisions and
shows little centrality dependence. The yield ratio of
$(p+\bar{p})/h$ in minimum-bias d+Au collisions is found to be a
factor of 2 lower than that in Au+Au collisions, indicating that
the relative baryon enhancement observed in heavy ion collisions
at RHIC is due to the final state effects in Au+Au collisions.

The mechanism for Cronin effect is also discussed in this thesis
by comparison with the recombination model~\cite{hwayang} and with
the initial multiple parton scattering model~\cite{accardi}, which
will be described in the discussion chapter in detail. Usually the
Cronin effect has been explained to be the initial state effect
only~\cite{accardi}. However, from the comparisons, we conclude
that the Cronin effect in $\sqrt{s_{_{NN}}} = 200$ GeV d+Au
collisions is not initial state effect only, and that final state
effect plays an important role.

These physics results has been at e-Print Archives (nu-ex/0309012)
and submitted for publication. The excellent particle
identification from the prototype MRPC tray and the important
physics from it have provided a solid basis for the STAR
full-time-of-flight-system proposal.

\end{abstract}

\beforepreface
\prefacesection{Acknowledgments}

I would like to thank my adviser, Prof. Hongfang Chen. She has
been my adviser since I was an undergraduate student. Her enormous
physics knowledge and serious attitude on science give me a deep
impression. Her enthusiasm on the research sets an example for me.
I also thank her for giving me a chance to stay in BNL to do my
research on STAR physics. I would like to thank Dr. Zhangbu Xu,
the co-adviser of my PhD research. He guided me through all the
detailed analysis in my research. He also gave me a lot of help on
life when I stayed at BNL.

Special thanks to Xin Dong, my classmate and my friend, a great
partner of the research. He gave me a lot of help and
encouragement. I would like to thank Dr. Nu Xu and Prof. Huanzhong
Huang for many helpful and inspiring discussion on my research.
I'd like to thank Prof. Jian Wu for taking care of me and
encouraging me when he stayed in BNL. I thank Dr. Haibin Zhang to
give me a lot of help on life when I just arrived at BNL. He also
guided me through the $K^*$ analysis.

I'd like to thank the high energy group in USTC for their hardwork
on the MRPC detector construction. Spectra thanks to Prof. Cheng
Li, and Dr. Ming Shao.

I'd like to thank my parents for their sacrifice and support.
Without the sacrifice, I will be only the body without spirit. I'd
like to thank my husband, Dr. Shengli Huang for his encouragement
and support.

I'd like to thank all my friends for their support. Without you,
the life is meaningless.

I thank every member of  the STAR collaboration for their hard
work to make the experiment run smoothly and successfully, to
construct the detector and develop the software.

\pagenumbering{roman} \tableofcontents \figurespagetrue
\tablespagetrue \listoffigures \listoftables } \pagebreak
\pagenumbering{arabic}
\chapter{Physics}
\label{chp:physics}

\section{Deconfinement and phase diagram}
The theory which describes the interaction of the color charges of
quarks and gluons is called Quantum Chromodynamics (QCD). In
phenomenological quark models, mesons can be described as
quark-antiquark bound states, while baryons can be considered as
three quark bound states. Up to now, it's found that all the
hadron states which can be observed in isolation is colorless
singlet states. Experimentally, no single quark, which is
described by a color-triplet state, has ever been isolated. The
absence of the observation of a single quark in isolation suggests
that the interaction between quarks and gluons must be strong on
large distance scale. In the other extreme, much insight into the
nature of the interaction between quarks and gluons on short
distance scales was provides by deep inelastic scattering
experiments. In these experiments, the incident electron interacts
with a quark within a hadron and is accompanied by the momentum
transfer from the electron to the quark. The measurement of the
electron momentum before and after the interaction allows a probe
of the momentum distribution of the parton inside the nucleon. It
was found that with very large momentum transfer, the quarks
inside the hadron behave as if they were almost free~\cite{QCD}.
The strong coupling between quarks and gluons at large distances
and asymptotic freedom are the two remarkable features of QCD.
When the energy density is high enough either due to the high
temperature or high baryon density, the quark or gluon may be
deconfined from a hadron. The thermalized quark gluon system is
what we called quark-gluon plasma. Lattice QCD calculations,
considering two light quark flavors, predict a phase transition
from a confined phase, hadronic matter, to a deconfined phase, or
quark-gluon plasma (QGP), at a temperature of approximately
\begin{figure}[h]
\centering
\includegraphics[height=28pc,width=28pc]{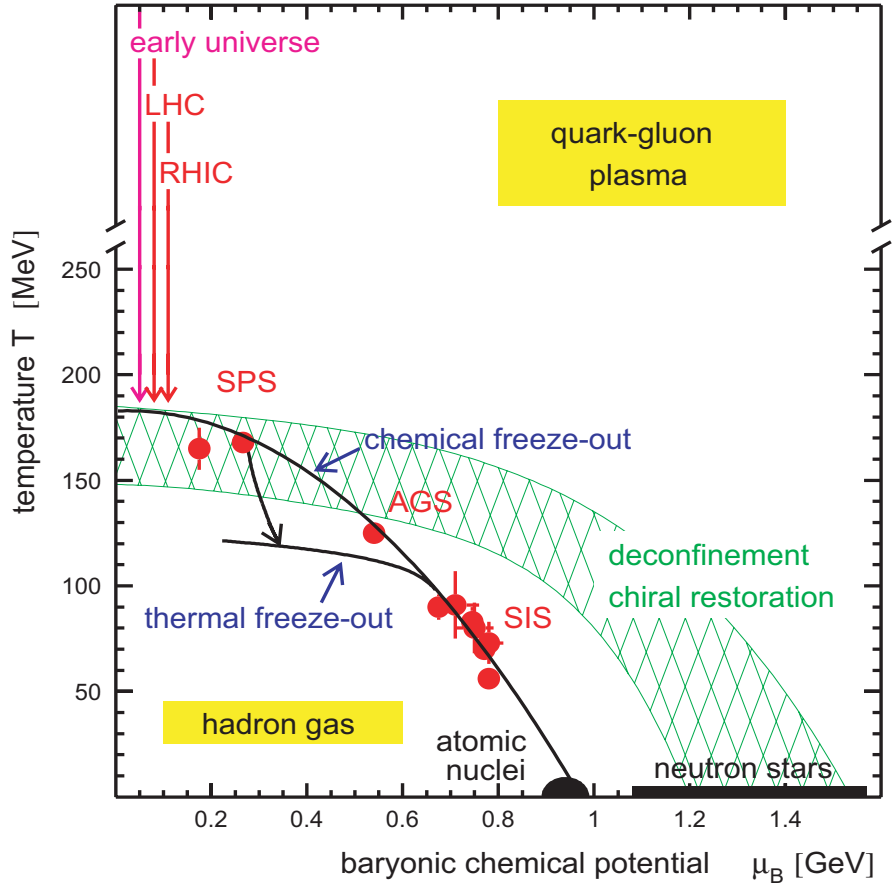}
\caption{Phase diagram of hadronic and partonic matter. Figure is
taken from~\cite{pbm:01}.}
\end{figure}
160 MeV~\cite{harris:98}. Figure 1.1 shows the phase diagram of
the hadronic and partonic matter. A phase transition from the
confined hadronic matter to the deconfined QGP matter is expected
to happen at either high temperature or large baryon chemical
potential $\mu_B$. Recent Lattice QCD calculations show that the
QGP is far from ideal below 3 $T_{c}$. The nonideal nature of this
strongly coupled QGP is also seen from the deviation of the
pressure, $P(T)$, and energy density $\epsilon(T)$ from the Stefan
Boltzmann limit as shown in Figure from~\cite{Fodor}.
\begin{figure}[h]
\centering
\includegraphics[height=16pc,width=28pc]{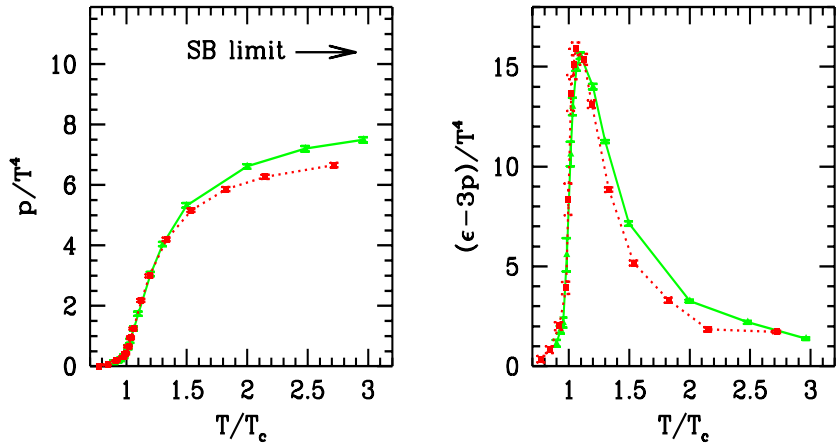}
\caption{A recent Lattice QCD calculation \protect{\cite{Fodor}}
of the pressure, $P(T)/T^4$, and a measure of the deviation
from the ideal Stefan-Boltzmann limit $(\epsilon(T)-3 P(T))/T^4$.}
\label{Fodorplot}
\end{figure}
Experiments on relativistic heavy ion collisions are designed to
search for and study the deconfined QGP matter.

\section{Relativistic Heavy Ion Collisions}
The experimental programs in relativistic heavy ions started in
1986 using the Alternating Gradient Synchrotron (AGS) at
Brookhaven National Lab (BNL) and the Super Proton Synchrotron
(SPS) at European laboratory for particle physics (CERN). At BNL,
ion beams of silicon and gold, accelerated to momenta of 14 and 11
GeV/c per nucleon, respectively, have been utilized in 10
fixed-target experiments. There have been 15 heavy ion experiments
at CERN utilizing beams of oxygen at 60 and 200 GeV/c per nucleon,
sulphur at 200 GeV/c per nucleon and Pb at 160 GeV/c
per nucleon~\cite{harris:98}.\\
The Relativistic Heavy Ion Collider (RHIC) at BNL is designed for
head-on Au+Au collisions at $\sqrt{s_{NN}}$ = 200 GeV. The first
RHIC run was performed in 2000 with Au+Au collisions at
$\sqrt{s_{NN}}$ = 130 GeV/c in four experiments, STAR, PHENIX,
PHOBOS and BRAHMS. The second RHIC run was in 2001 and 2002 with
Au+Au and p+p collisions at $\sqrt{s_{NN}}$ = 200 GeV. The third
RHIC run was in 2002 and 2003 with
d+Au and p+p collisions at $\sqrt{s_{NN}}$ = 200 GeV.\\
The above mentioned relativistic heavy ion collision experiments
are designed for the search and study of the possible deconfined
high energy density matter, quark-gluon plasma. In head-on
relativistic heavy ion collisions, two nuclei can be represented
as two thin disks approaching each other at high speed because of
the Lorentz contraction effect in the moving direction. During the
initial stage of the collisions, the energy density is higher than
the critical energy density from the Lattice QCD calculation, so
the quarks and gluons will be de-confined from nucleons and form
the quarks and gluons system. The large cross section of
interaction may lead to the thermalization of the quarks and
gluons system. That's what we called the formation of quark-gluon
plasma (QGP). In this stage, the high transverse momentum
($p_{T}$) jets and $c\bar{c}$ pair will be produced due to the
large momentum transfer. After that, the QGP will expand and cool
down and enter into the mixed-phase expansion. The chemical freeze
out point will be formed after the inelastic interactions stop.
That means that the particle yields and ratios will not change.
After the chemical freeze out, the elastic interactions between
hadrons will change the $p_{T}$ distribution of particles. The
particles will freeze out finally from the system after the
elastic interactions stop. That's what we called the kinetic
freeze out point. In the following the important results from RHIC
will be addressed.
\section{The experimental results at RHIC}
\subsection{Flow}
In non-central Au+Au collisions, the spatial space asymmetry will
be transferred into the momentum space asymmetry by the azimuthal
asymmetry of pressure gradients.
\begin{figure}[h]
\centering
\includegraphics[height=16pc,width=22pc]{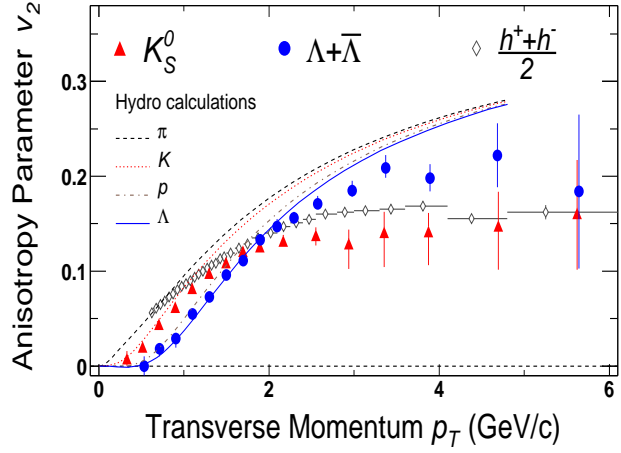}
\caption{The minimum-bias (0--80\% of the collision cross section)
$v_{2}(p_T)$ for $K_{S}^{0}$, $\Lambda+\overline{\Lambda}$ and
$h^{\pm}$. The error bars shown include statistical and
point-to-point systematic uncertainties from the background.  The
additional non-flow systematic uncertainties are approximately
-20\%. Hydrodynamical calculations of $v_2$ for pions, kaons,
protons and lambdas are also plotted~\cite{hydroPasi01}. Figure is
taken from~\cite{starv2raa}.} \label{KsLamv2}
\end{figure}
 The azimuthal particle distributions in
momentum space can be expanded in a form of Fourier series
\begin{equation}
E\frac{d^3N}{d^3p}=\frac{1}{2\pi}\frac{d^2N}{p_Tdp_Tdy}(1+
\sum^{\infty}_{n=1}2v_n\cos[n(\phi-\Psi_r)])
\end{equation}
where $\Psi_r$ denotes the reaction plane angle. The Fourier
expansion coefficient $v_n$ stands for the $n$th harmonic of the
event azimuthal anisotropies. $v_1$ is so called direct flow and
$v_2$ is the elliptic flow. The elliptic flow is generated mainly
during the highest density phase of the evolution before the
initial geometry asymmetry of the plasma disappears.
Hydrodynamical calculations~\cite{derekhydro} show most of $v_2$
is produced before 3 fm/c at RHIC.\\
Figure~\ref{KsLamv2} shows that The $v_2$ of $K_{S}^{0}$,
$\Lambda+\overline{\Lambda}$ and charged hadrons ($h^{\pm}$) as a
function of $p_T$ for 0--80\% of the collision cross
section~\cite{starv2raa}. Also shown are the $v_2$ of pions,
kaons, protons and lambdas from hydrodynamical
model~\cite{hydroPasi01}. The $v_2$ from hydrodynamical model
shows strong mass dependence, which fits the $K_{S}^{0}$ $v_2$ up
to $p_{T}\sim1$ GeV/c and fits the $\Lambda+\overline{\Lambda}$
$v_2$ up to $p_{T}\sim2.5$ GeV/c. Even though the $v_2$ from
hydrodynamical model shows consistency with data at low $p_T$,
however, the $v_2$ from experimental results show saturation at
intermediate $p_{T}$ while hydrodynamical predictions show rising
trend at the same $p_T$ range.

\subsection{High $p_{T}$ suppression and di-hadron azimuthal correlation}
The $v_2$ from hydrodynamical models show consistency with data at
lower $p_T$ and fail to reproduce data at higher $p_{T}$. At high
$p_{T}$, the suppression for charged hadron production was
observed in Au+Au collisions at RHIC energy. The comparison of the
spectra in Au+Au collisions through those in p+p collisions,
scaled by the number of binary nucleon nucleon collisions is the
nuclear modification factor $R_{AA}$.
\begin{equation}
R_{AA}(p_T)=\frac{d^2N^{AA}/dp_Td\eta}{T_{AA}d^2\sigma^{NN}/dp_Td\eta}
\end{equation}
where $T_{AA}=\langle N_{\text{bin}} \rangle
/\sigma^{NN}_{\text{inel}}$ accounts for the collision geometry,
averaged over the event centrality class. $\langle N_{\text{bin}}
\rangle$, the equivalent number of binary $NN$ collisions, is
calculated using a Glauber model. The $R_{AA}$ is an experimental
variable. The high $p_T$ hadron suppression in central Au+Au
collisions can also be investigated by comparing the hadron
spectra in central and peripheral Au+Au collisions. That's what we
called $R_{CP}$. $R_{CP}$ is defined as
\begin{equation}
R_{CP}=\frac{\langle N_{\text{bin}}^{\text{peripheral}} \rangle
d^2N^{\text{central}}/dp_Td\eta}{\langle
N_{\text{bin}}^{\text{central}} \rangle
d^2N^{\text{peripheral}}/dp_Td\eta}.
\end{equation}\\

\begin{figure}[h]
\centering
\includegraphics[height=20pc,width=26pc]{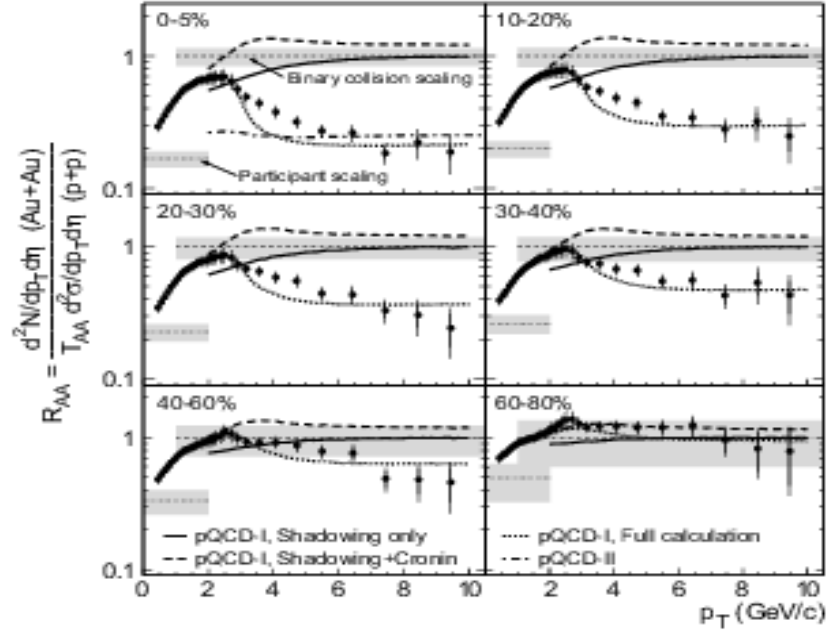}
\caption{$R_{AA}(p_T)$ of inclusive charged hadron for various
centrality bins. Figure is taken from ~\cite{starhighpt}.}
\label{raa200}
\end{figure}

Figure~\ref{raa200} shows $R_{AA}(p_T)$ of inclusive charged
hadron for various centrality bins in Au+Au collisions at
$\sqrt{s_{NN}}$=200 GeV. $R_{AA}(p_T)$ increases monotonically for
$p_T<$ 2 GeV/c at all centralities and saturates near unity for
$p_T>$ 2 GeV/c in the most peripheral bins. In contrast,
$R_{AA}(p_T)$ for the central bins reaches a maximum and then
decreases strongly above $p_T$ = 2 GeV/c, showing the suppression
of the charged hadron yield relative the $NN$
reference~\cite{starhighpt}.\\
Suppression of high $p_{T}$ hadron production in central Au+Au
collisions relative to p+p collisions
~\cite{starhighpt,phenixhighpt} has been interpreted as energy
loss of the energetic partons traversing the produced hot and
dense medium~\cite{jetquench}, that's so called jet quenching. If
a dense partonic matter is formed during the initial stage of a
heavy-ion collision with a large volume and a long life time
(relative to the confinement scale $1/\Lambda_{\rm QCD}$), the
produced large $E_T$ parton will interact with this dense medium
and will lose its energy via induced radiation. The energy loss
depends on the parton density of the medium. Therefore, the study
of parton energy loss can shed light on the properties of the
dense matter in the early stage of heavy-ion
collisions~\cite{jetquench}. At sufficiently high beam energy,
gluon saturation is also expected to result in a relative
suppression of hadron yield at high $p_{T}$ in A+A
collisions~\cite{cgc}. Also shown in the Figure~\ref{raa200} are
the results from perturbative QCD (pQCD) calculations. The
Full-pQCD calculations include the partonic energy loss, the
Cronin enhancement(due to initial multiple scattering) and the
nuclear shadowing effect. The suppression is predicted to be $p_T$
independent when $p_T$ is larger than 6 GeV/c, which is consistent
with our data. However, the discrepancy at 2-6 GeV/c was observed
between the prediction and the experimental data. This discrepancy
may be due to different mechanism for particle production at
intermediate $p_T$. The particle production at intermediate $p_T$
will be discussed later in this chapter.\\

\begin{figure}[tbh]
\begin{minipage}{0.49\textwidth}
\includegraphics[width=0.95\textwidth,angle=-90]{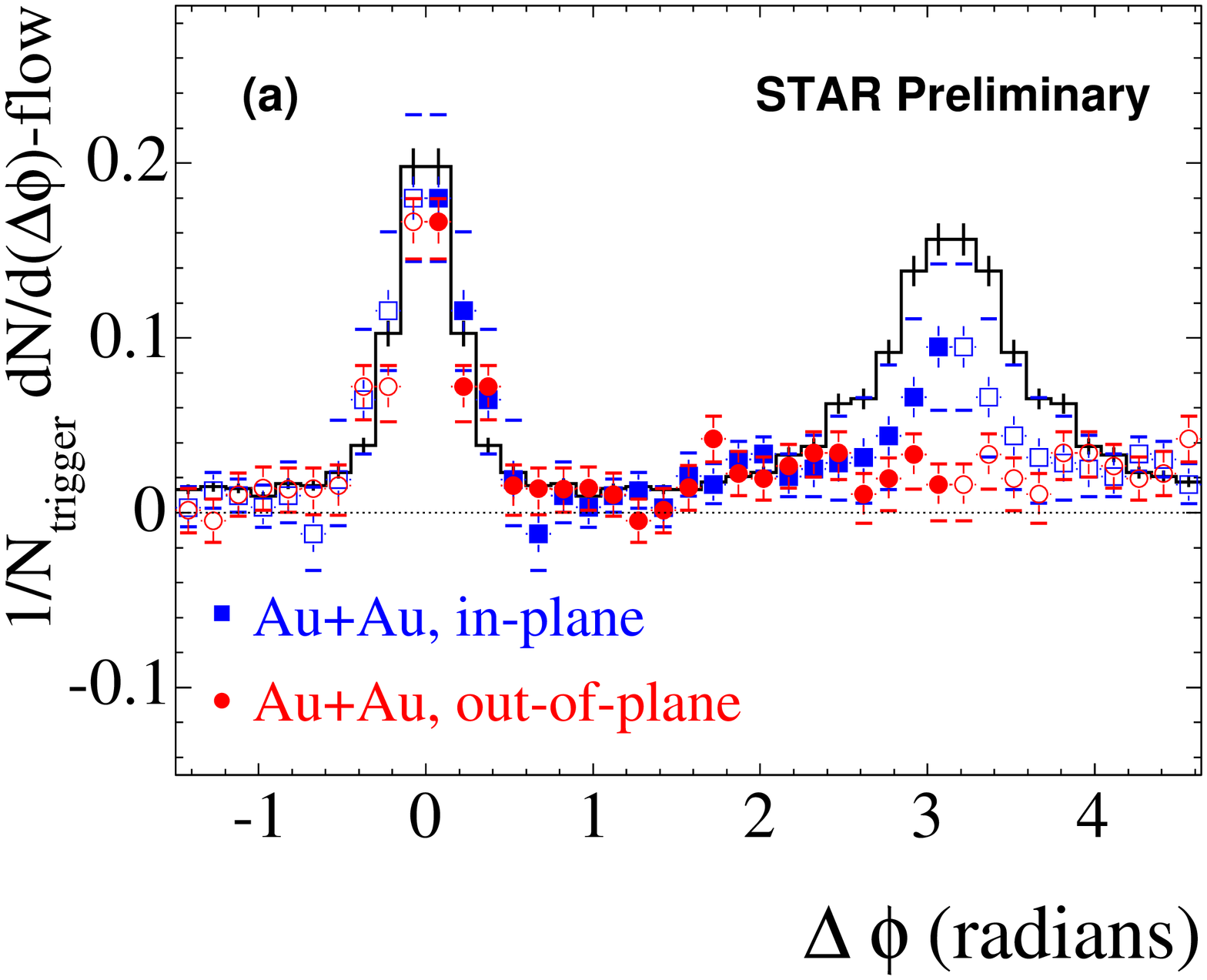}
\end{minipage}\hfill\hspace{2.5cm}
\begin{minipage}{0.49\textwidth}\vspace{-0.8cm}
\includegraphics[width=0.80\textwidth]{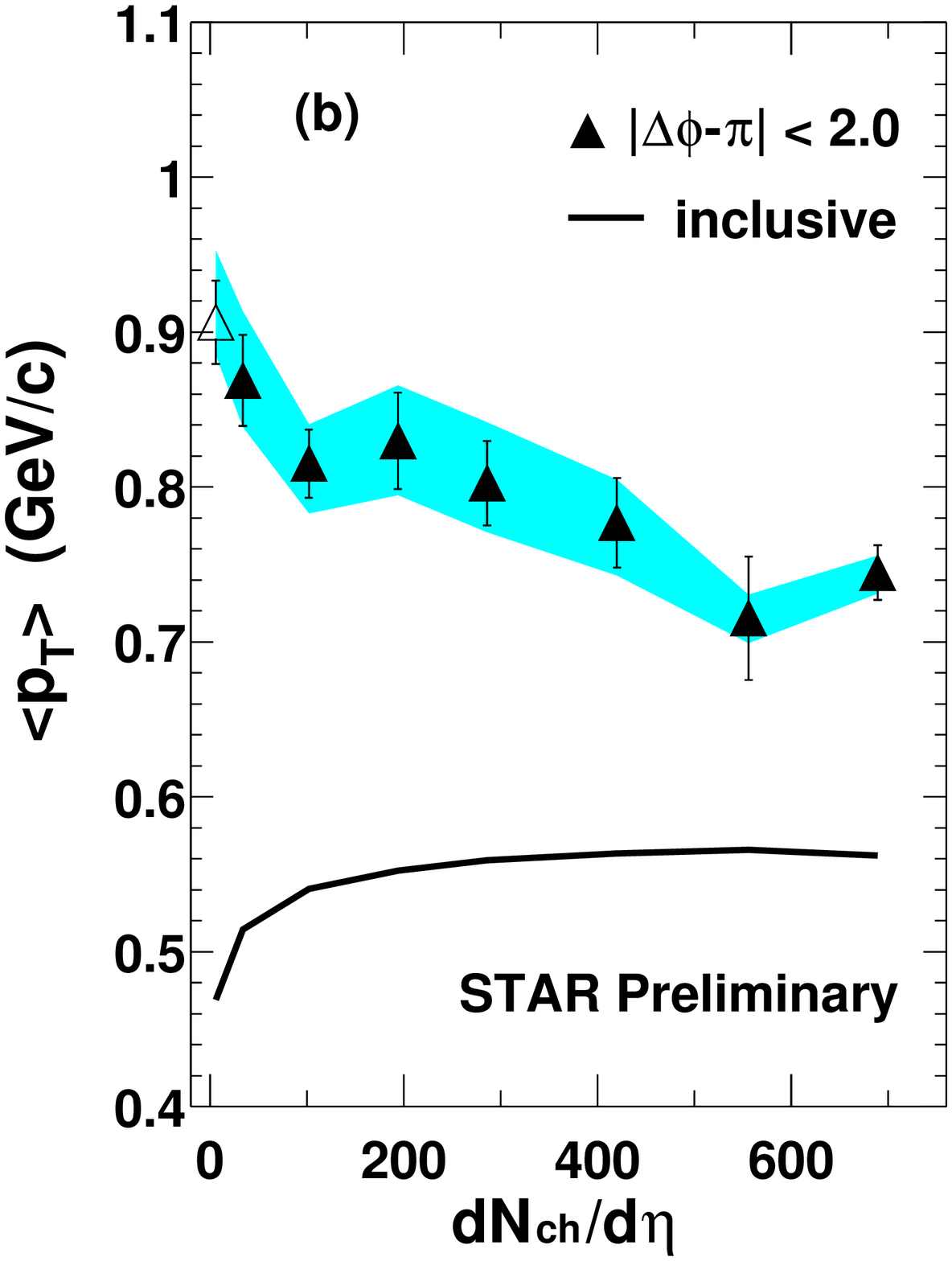}
\end{minipage}
\caption{(a) Azimuthal distribution of particles with respect to a
trigger particle for p+p collisions (solid line), and mid-central
Au+Au collisions within the reaction plane (squares) and
out-of-plane (circles) at 200 GeV~\cite{Aihong}. (b) Mean
transverse momentum for particles around the away-side region as a
function of number of charged particles~\cite{Fuqiang}. The solid
line shows the mean transverse momentum of inclusive hadrons.}
\label{jetplot}
\end{figure}

A more differential probe of parton energy loss is the measurement
of high $p_T$ di-hadron azimuthal correlation relative to the
reaction plane orientation. The trigger hadron is in the range
$4<p_{T}<6$ GeV/c and the associated particle is at $2<p_{T}<4$
GeV/c. Figure~\ref{jetplot} (left) shows the high $p_T$ di-hadron
correlation when the trigger particle is selected in the azimuthal
quadrants centered either in the reaction plane (in plane) or
orthogonal to it(out of plane). The near side di-hadron azimuthal
correlations in both cases were observed to be the same as that in
p+p collisions, while the suppression of back to back correlation
shows strong dependence on the relative angle between the
triggered high $p_T$ hadron and the reaction plane. This
systematic dependence is consistent with the picture of parton
energy loss: the path length for a dijet oriented out of plane is
longer than that for a dijet oriented in plane, leading to a
stronger suppression of parton energy loss in the out of plane.
The dependence of parton energy loss on the path length is
predicted to be substantially larger than
linear~\cite{jetquench}.\\
The energy lost by away side partons traversing the collision
matter must in the form of the excess of softer emerging particles
due to the transverse momentum conservation. An analysis of
azimuthal correlations between soft and hard particles has been
performed for both 200 GeV p+p and Au+Au collisions~\cite{Fuqiang}
at STAR as a first of attempt to trace the degree of the
degradation on the away side. With triggered hadron still in the
range $4<p_{T}^{trig}<$ 6 GeV/c, but the associated hadrons now
sought over $0.15<p_{T}<4$ GeV/c, combinatorial coincidences
dominate this correlation and they must be subtracted carefully by
mixed-event technique and also the elliptic flow effect was also
subtracted by hand~\cite{Fuqiang}. The results demonstrate that,
in comparison with the p+p and peripheral Au+Au collisions, the
momentum-balancing hadrons opposite to the high $p_T$ triggered
particle in central Au+Au are greater in number, much more widely
dispersed in azimuthal angle, and significantly softer in
momentum. Figure~\ref{jetplot} (right) shows the $\langle p_{T}
\rangle$ of the momentum-balancing hadrons opposite to the high
$p_T$ trigger as a function of centrality. The $\langle
p_{T}\rangle$ were observed to decrease from peripheral to central
Au+Au collisions. Also shown in the Figure~\ref{jetplot} (right)
is the $\langle p_{T}\rangle$ of the inclusive hadrons as a
function of centrality. This study will be extended to higher
$p_T$ trigger particle. The results may suggest that the
moderately hard parton traversing a significant path length
through the collision matter makes substantial progress toward
equilibrium with the bulk. The rapid attainment of thermalization
via multitude of softer parton-parton interactions in the earliest
collision stages would then not be so
surprising~\cite{starwhitepaper}.

\subsection{Particle composition in Au+Au at intermediate $p_{T}$}
\begin{figure}[h]
\begin{minipage}[t]{80mm}
\includegraphics[height=16pc,width=16pc]{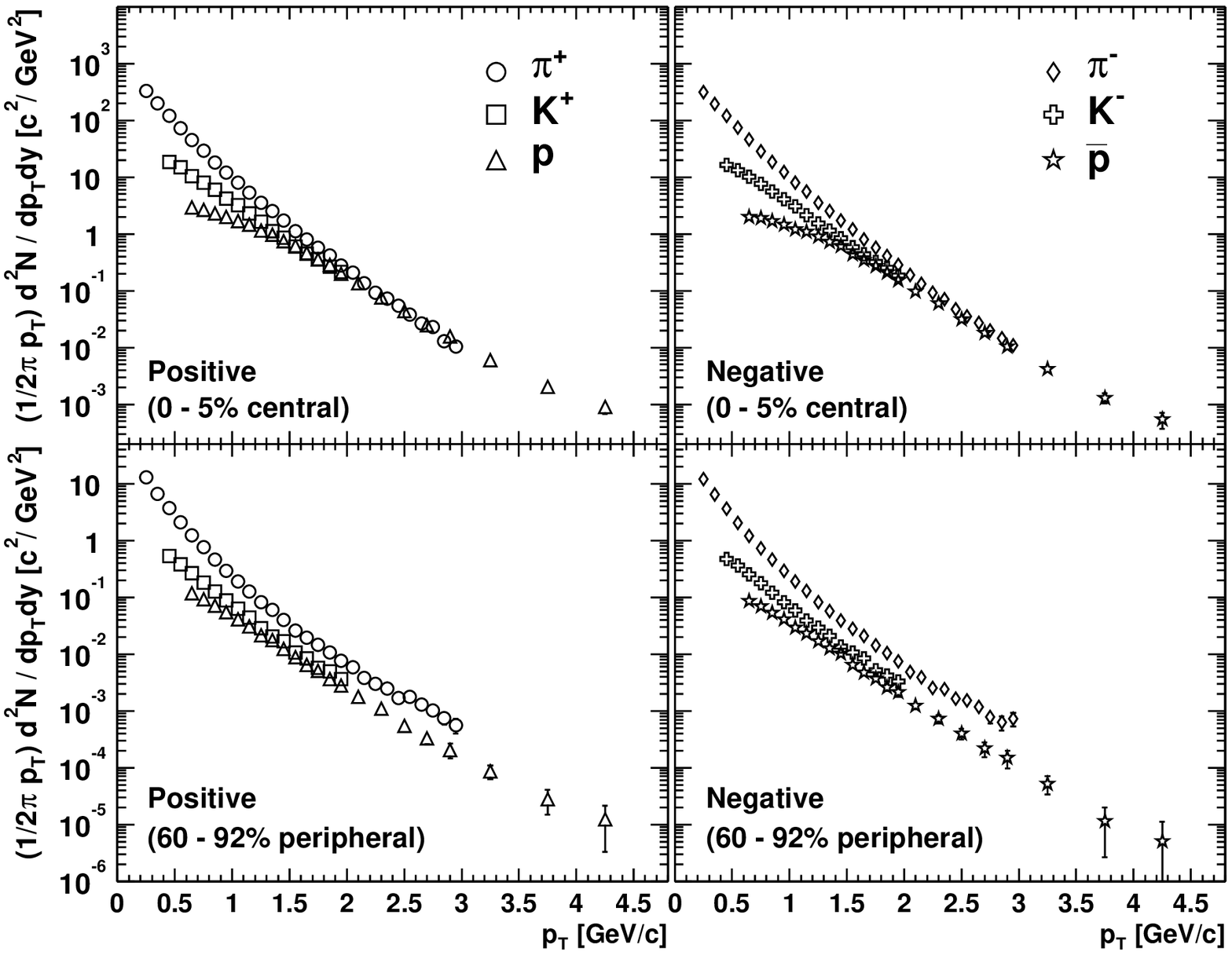}
\end{minipage}
\hspace{\fill}
\begin{minipage}[t]{80mm}
\includegraphics[height=16pc,width=16pc]{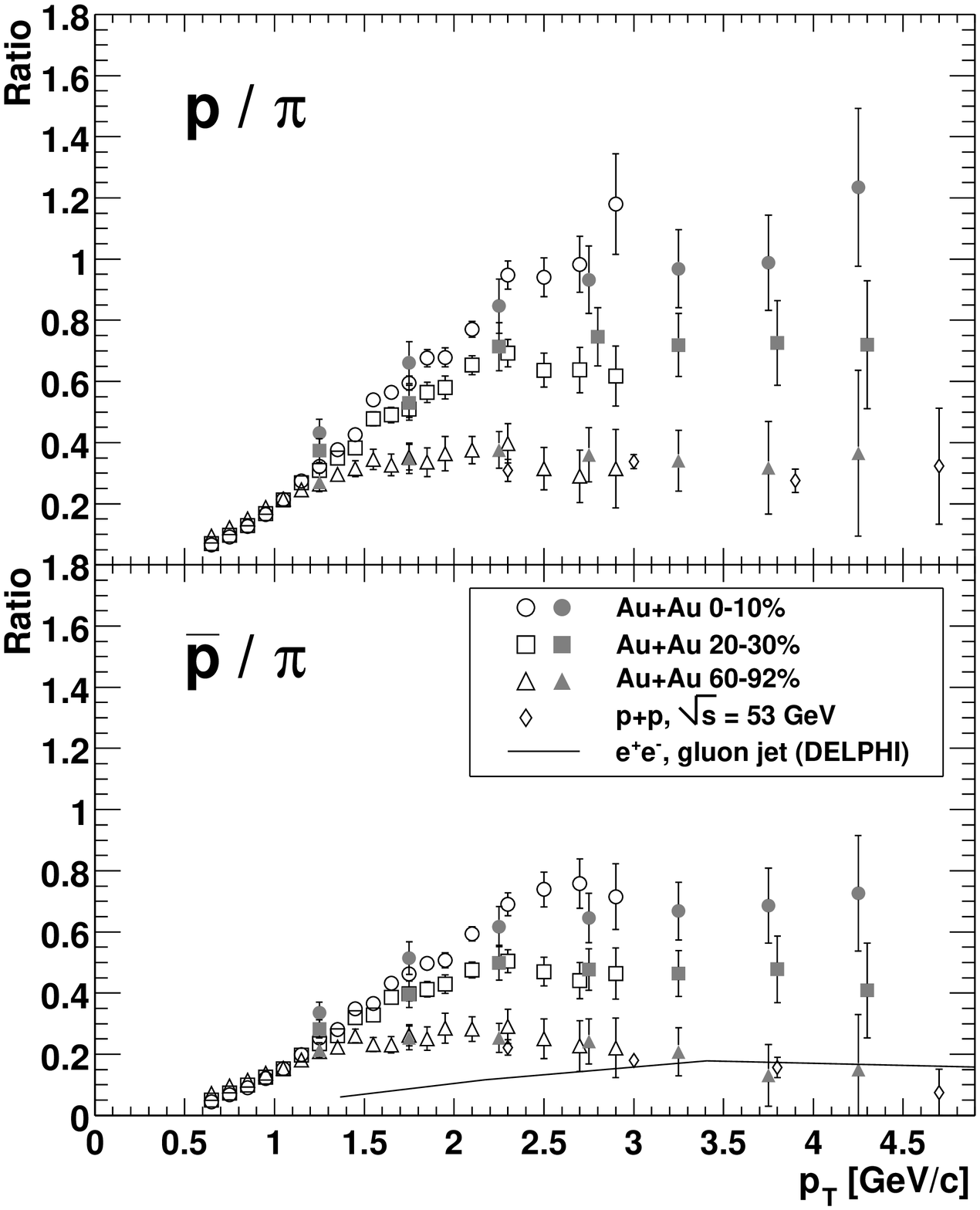}
\end{minipage}
\caption{(left) The $\pi$, K, p spectra in 0\%-5\% and 60\%-92\%
Au+Au 200 GeV collisions from~\cite{ex0307022}. (right)
Proton/pion (top) and anti-proton/pion (bottom) ratios for central
0--10\%, mid-central 20--30\% and peripheral 60--92\% in Au+Au
collisions at 200 GeV. Open (filled) points are for charged
(neutral) pions. The data at $\sqrt{s} = 53 $~GeV p+p
collisions~\cite{ISR} are also shown. The solid line is the
$(\bar{p} + p)/(\pi^{+} + \pi^{-})$ ratio measured in gluon
jets~\cite{DELPHI}. This figure is from~\cite{ex0307022}. }
\label{phenixAuAuspectra}
\end{figure}

\begin{figure}[tbph]
\centering\mbox{
\includegraphics[width=0.9\textwidth]{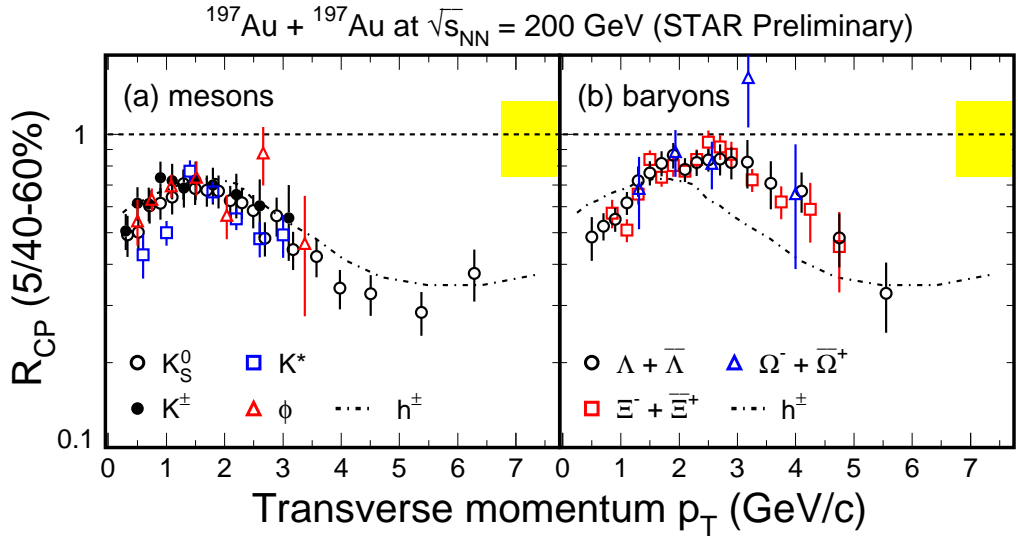}}
\caption{ The ratio $R_{CP}$ for identified mesons and baryons at
mid-rapidity calculated using centrality intervals, 0--5\% vs.
40--60\% of the collision cross section. The bands represent the
uncertainties in the model calculations of $\mathrm{N_{bin}}$. We
also show the charged hadron $R_{CP}$ measured by STAR for
$\sqrt{s_{_{NN}}}=200$~GeV~\cite{starhighpt}. This figure is
from~\cite{starhighlight}.} \label{KsLamRcp}
\end{figure}

As we have mentioned above, for $R_{AA}$, the pQCD model including
the parton energy loss, Cronin enhancement and nuclear shadowing
can qualitatively fit the trend of data at $2<p_{T}<6$ GeV/c,
however, the quantitative discrepancy between the model and the
data is also obvious. In the intermediate $p_{T}$, the mechanism
for particle production may be different from that at
high $p_{T}$.\\
Figure~\ref{phenixAuAuspectra} (left) shows the $\pi$, K, p
spectra in 0\%-5\% and 60\%-92\% Au+Au 200 GeV collisions
from~\cite{ex0307022}. It shows that the shapes of the spectra
show clear mass dependence. And in central collisions, the $\pi$,
K, p yields are close to each other at $p_{T}>2$ GeV/c while it's
not the case in peripheral collisions.
Figure~\ref{phenixAuAuspectra} (right) shows proton/pion (top) and
anti-proton/pion (bottom) ratios for central 0--10\%, mid-central
20--30\% and peripheral 60--92\% in Au+Au collisions at 200
GeV~\cite{ex0307022}. It shows that the $p(\bar{p})/\pi$ ratios
increase fast from peripheral to central collisions. In the 0-10\%
centrality bin, the proton yield is even larger than pion yield at
intermediate $p_T$. Figure~\ref{KsLamRcp} shows the ratio $R_{CP}$
for identified mesons and baryons at mid-rapidity calculated using
centrality intervals, 0--5\% vs. 40--60\% of the collision cross
section from STAR measurement~\cite{starhighlight}. It seems that
for meson, the $R_{CP}$ follows a common trend and for baryon, the
$R_{CP}$ also follows a common trend, which is different from that
for mesons. The $R_{CP}$ for baryons is observed to be larger than
that for mesons.\\

These experimental results suggest that the degree of suppression
depends on particle species(baryon/meson) at intermediate $p_T$.
The spectra of baryons (protons and lambdas) are less suppressed
than those of mesons (pions, kaons) ~\cite{starv2raa,phenixpid} in
the $p_{T}$ range $2<p_{T}<5$ GeV/c. The baryon content in the
hadrons at intermediate $p_{T}$ depends strongly on the impact
parameter (centrality) of the Au+Au collisions with about 40\% of
the hadrons being baryons in the minimum-bias collisions and 20\%
in very peripheral collisions~\cite{starv2raa,phenixpid}.
Hydrodynamics~\cite{derekhydro,pisahydro}, parton coalescence at
hadronization~\cite{hwa,fries,ko} and gluon
junctions~\cite{junction} have been suggested as explanations for
the observed particle-species dependence.

In these models, recombination/coalescence models successfully
reproduce $R_{AA}$ of baryons and mesons at intermediate $p_T$, as
well as showing consistency with the $v2$ measurement in the same
$p_T$ range.

\subsubsection{Recombination model}
The concept of quark recombination was introduced to describe
hadron production at forward rapidity in p+p
collisions~\cite{ref50:whitepaper}. At forward rapidity, this
mechanism allows a fast quark resulting from a hard parton
scattering to recombine with a slow anti-quark, which could be one
in the original sea in the incident hadron, or one incited by a
gluon~\cite{ref50:whitepaper}. If a QGP is formed in the
relativistic heavy ion collisions, then one might expect
coalescence of the abundant thermal partons to provide another
important hadron production mechanism, active over a wide range of
rapidity and transverse momentum~\cite{ref51:whitepaper}. In
particular, at moderate $p_T$ values(above the realm of
hydrodynamics applicability), the hadron production from
recombination of lower $p_T$ partons from thermal
bath~\cite{hwa,fries,ko} has been predicted to be competitive with
the production from fragmentation of higher $p_T$ scattered
partons. It has been suggested~\cite{ref53:whitepaper} that the
need for substantial recombination to explain the observed hadron
yield and flow may be taken as a signature of QGP
formation.\\
In order to explain the features of RHIC collisions, the
recombination models~\cite{ref51:whitepaper,hwa,fries,ko} make the
central assumption that coalescence proceeds via constituent
quarks, whose number in a hadron determines its production rate.
The constituent quarks are presumed to follow a thermal
(exponential) momentum spectrum and to carry a collective
transverse velocity distributions. This picture leads to clear
predicted effects on baryon and meson production rates, with the
former depending on the spectrum of thermal constituent quarks and
antiquarks at roughly one-third the baryon $p_T$, and the latter
determined by the spectrum at roughly one-half the meson $p_T$.
Indeed, the recombination model was recently was re-introduced at
RHIC context, precisely to explain the abnormal abundance of
baryon vs meson observed at intermediate
$p_T$~\cite{hwa,fries,ko}. If the observed saturated elliptic flow
values of hadrons in this momentum range result from coalescence
of collectively flowing constituent quarks, then one expect a
similarly simple baryon vs meson relationship~\cite{hwa,fries,ko}:
the baryon (meson) flow would be 3 (2) times the quark flow at
roughly one-third (one-half) the baryon (meson)
$p_T$~\cite{starwhitepaper}.

\subsection{Summary}
In summary, the several important results from RHIC have been
introduced. The elliptic flow $v2$ can be reproduced by
hydrodynamics at low $p_T$. At intermediate $p_T$, $v2$ from data
show saturation and deviate from hydrodynamical model predictions.
At the same time, $v2$ from data show baryon or meson species
dependence. High $p_T$ suppression can be reproduced by pQCD model
and gluon saturation model. The gluon saturation model is also
called color glass condensate model (CGC). The production rate
dependence on baryon or meson species has been observed at
intermediate $p_{T}$, which can be reproduced by the recombination
model.

\section{Cronin effect}
\subsection{Why we need d+Au run at RHIC}
In order to see the intermediate and high $p_T$ suppression is due
to the final-state effect or initial state effect, the
measurements from d+Au collisions will provide the essential
proof. Since the initial state in d+Au collisions is similar to
that in Au+Au collisions, and, it's believed that the quark-gluon
plasma doesn't exist in d+Au collisions, the results from d+Au
collisions will be very important for us to judge whether the
quark-gluon plasma exists in Au+Au collisions or not and to
understand the property of the dense matter created in Au+Au
collisions. Besides, if the identified particle spectra in d+Au
and p+p collisions are measured, they will not only provide the
reference for those in Au+Au collisions at 200 GeV, but also
provide a chance to see the mechanism of the Cronin effect itself
clearly at 200 GeV. Cronin effect was observed 30 years ago
experimentally and the study of this effect was only limited to
lower energy fixed target experiments. Before we go to the d+Au
collisions, let's look back on the p+A collisions at lower energy
fixed target experiment.

\subsection{Lower energy}
\begin{figure}[h]
\begin{minipage}[t]{80mm}
\includegraphics[height=18pc,width=18pc]{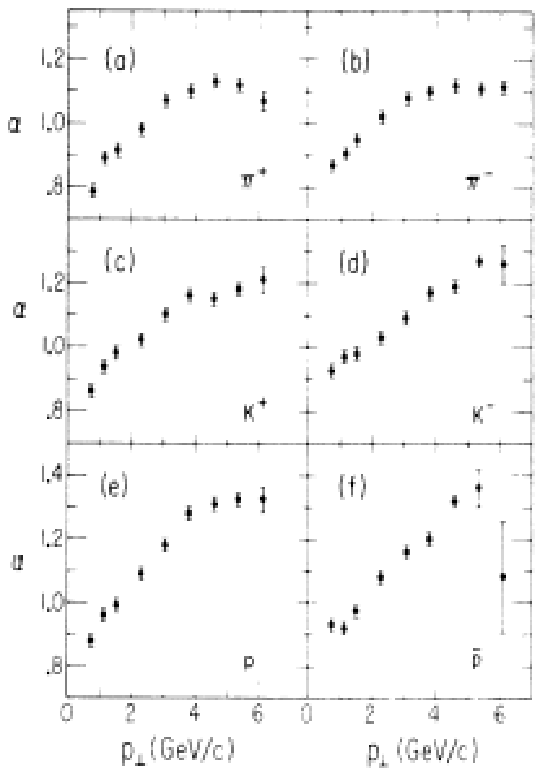}
\end{minipage}
\hspace{\fill}
\begin{minipage}[t]{80mm}
\includegraphics[height=18pc,width=18pc]{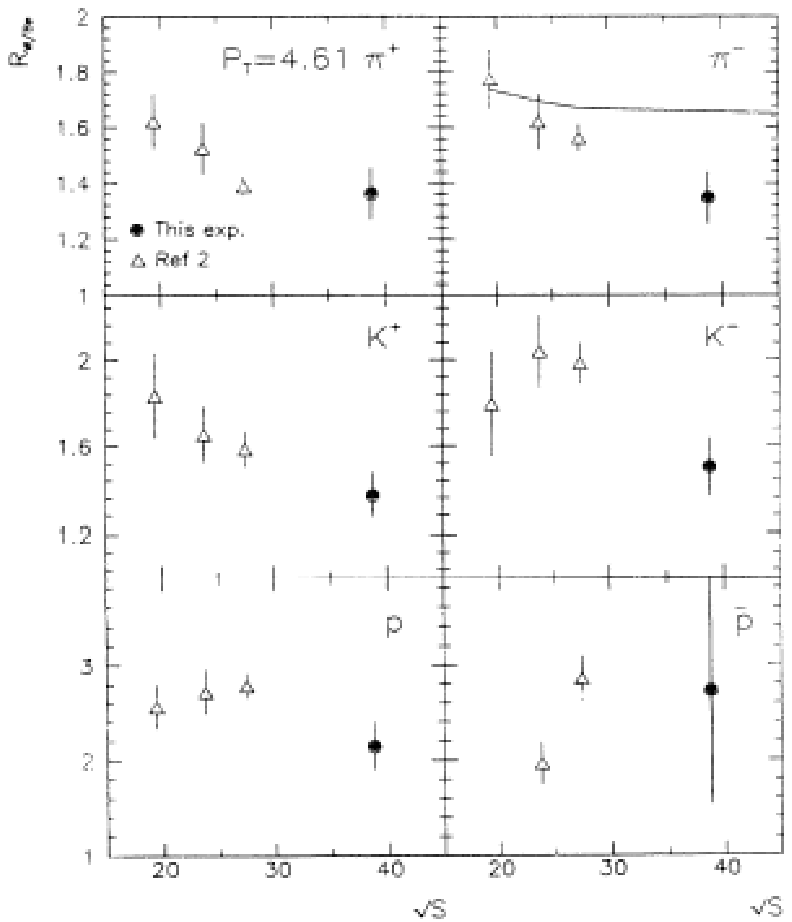}
\end{minipage}
\caption{(left) The power alpha of A dependence from 300 GeV
incident proton-fixed target experiment. This figure is
from~\cite{cronin}. (right) The Cronin ratio $R_{W/B_{e}}$ at
$p_{T}=4.61$ GeV/c versus energy. This plot is
from~\cite{cronin}.} \label{poweralphaplot}
\end{figure}

The hadron $p_{T}$ spectra have been observed to depend on the
target atomic weight ($A$) and the produced particle species in
lower energy p+A collisions~\cite{cronin}. This is known as the
``Cronin Effect'', a generic term for the experimentally observed
broadening of the transverse momentum distributions at
intermediate $p_{T}$ in p+A collisions as compared to those in p+p
collisions~\cite{cronin,petersson83,accardi}. The effect can be
characterized as a dependence of the yield on the target atomic
weight as $A^{\alpha}$.  At energies of $\sqrt{s} \simeq$ 30 GeV,
$\alpha$ depends on $p_{T}$ and is greater than unity at high
$p_{T}$~\cite{cronin}, indicating an enhancement of the production
cross section. As shown in Figure~\ref{poweralphaplot}, the
$\alpha$ is larger than 1 in the intermediate $p_{T}$ and shows
strong particle-species dependence. The $\alpha$ for proton and
antiproton are larger than those for kaon and pion. And $\alpha$
for kaon is larger than that for pion. This effect has been
interpreted as partonic scatterings at the initial
impact~\cite{petersson83,accardi}.
Besides, the lower energy data suggest the power $\alpha$
decreases with energy, as shown in Figure~\ref{poweralphaplot}.
However, the energy dependence study of Cronin effect is limited
to fixed target experiment at lower energy. What's the
extrapolation of Cronin effect at higher energy such as RHIC
energy 200 GeV. At higher energies, multiple parton collisions are
possible even in p+p collisions~\cite{e735kno}. This combined with
the hardening of the spectra with increasing beam energy would
reduce the Cronin effect~\cite{accardi}. There are several models
which give different predictions of Cronin effect at 200 GeV.
\subsection{Predictions: RHIC energy}
One of the models is the initial multiple parton scattering model.
In this model, the transverse momentum of the parton inside the
proton will be broadened when the proton traverses the Au nucleus
due to the multiple scattering between the proton and the nucleons
inside the Au nucleus. In these models, the Cronin ratio will
increases to a maximum value between 1 and 2 at 2.5$<p_T<$4.5
GeV/c and then decreases with $p_T$ increasing~\cite{accardi}. The
Cronin effect is predicted to be larger in central d+Au collisions
than in d+Au peripheral collisions~\cite{Vitev03}. Another model
is the gluon saturation model. At sufficiently high beam energy,
gluon saturation is expected to result in a relative suppression
of hadron yield at high $p_{T}$ in both p+A and A+A collisions and
in a substantial decrease and finally in the disappearance of the
Cronin effect~\cite{cgc}.
\begin{figure}
\centering
\includegraphics[width=0.5\textwidth]{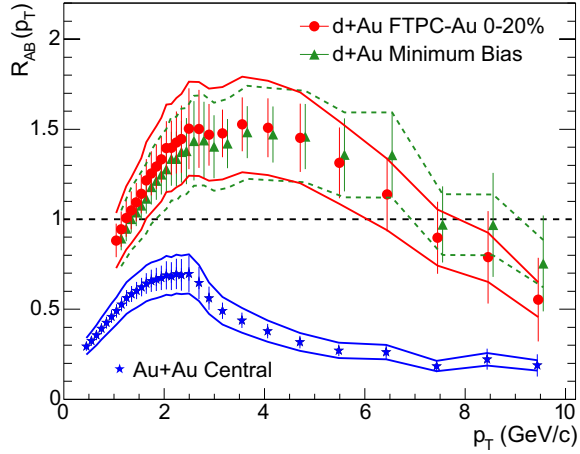}
\caption{ $R_{AB}$ for minimum bias and central d+Au collisions,
and central Au+Au collisions~\cite{starhighpt}. The minimum bias
d+Au data are displaced 100 MeV/c to the right for clarity. The
bands show the normalization uncertainties, which are highly
correlated point-to-point and between the two d+Au distributions.
This Figure is from~\cite{stardau}.} \label{FigThree}
\end{figure}
Figure~\ref{FigThree} shows the $R_{dAu}$ of charged hadron vs
$p_{T}$ from STAR. We can see that the Cronin ratio increases to a
maximum value around 1.5 at $3<p_{T}<4$ GeV/c and then decreases
again~\cite{stardau}. This is consistent with the initial multiple
parton scattering model~\cite{accardi}. These results on inclusive
hadron production from d+Au collisions indicate that hadron
suppression at intermediate and high $p_{T}$ in Au+Au collisions
is due to final state interactions in a dense and dissipative
medium produced during the collision and not due to the
initial state wave function of the Au nucleus~\cite{stardau,otherdau}.\\
Now we know that the hadron suppression at intermediate $p_{T}$ in
Au+Au collisions is due to final-state
effects~\cite{stardau,otherdau}. What's the effect on particle
composition at the same $p_{T}$ range in Au+Au collisions? Another
question is whether there is any Cronin effect dependence on
particle-species in d+Au collisions or not. In order to further
understand the mechanisms responsible for the particle dependence
of $p_{T}$ spectra in heavy ion collisions, and to separate the
effects of initial and final partonic rescatterings, we measured
the $p_{T}$ distributions of $\pi^{\pm}$, $K^{\pm}$, $p$ and
$\bar{p}$ from 200 GeV d+Au and p+p collisions. In this thesis, we
discuss the dependence of particle production on $p_{T}$,
collision energy, and target atomic weight. And we compare the
Cronin effect of $\pi^{\pm}$, $K^{\pm}$, $p$ and $\bar{p}$ with
models to address the mechanism for Cronin effect in d+Au
collisions at $\sqrt{s_{_{NN}}} = 200$ GeV.

\chapter{The STAR Experiment} \label{chp:star} \section{The RHIC
Accelerator} The Relativistic Heavy Ion Collider (RHIC) at
Brookhaven National Lab (BNL) is the first hadron accelerator and
collider consisting of two independent ring. It is designed to
operate at high collision luminosity over a wide range of beam
energies and particle species ranging from polarized proton to
heavy ion~\cite{rhic:01,rhic:02}, where the top energy of the
colliding center-of-mass energy per nucleon-nucleon pair is
$\sqrt{s_{NN}}$ = 200 GeV. The RHIC facility consists of two
super-conducting magnets, each with a circumference of 3.8 km,
which focus and guide the beams. \\
Figure 2.1 shows the BNL accelerator complex including the
accelerators used to bring the gold ions up to RHIC injection
energy. In the first, gold ions are accelerated to 15 MeV/nucleon
in the Tandem Van de Graaff facility. Then the beam is transferred
to the Booster Synchrotron and accelerated to 95 MeV/nucleon
through the Tandem-to-Booster line. Then the gold ions are
transferred to the Alternating Gradient Synchrotron (AGS) and
accelerated to 10.8 GeV/nucleon. Finally they are injected to RHIC
and accelerated to the collision energy 100 GeV/nucleon.\\
\begin{figure} \centering
\includegraphics[height=35pc,width=32pc]{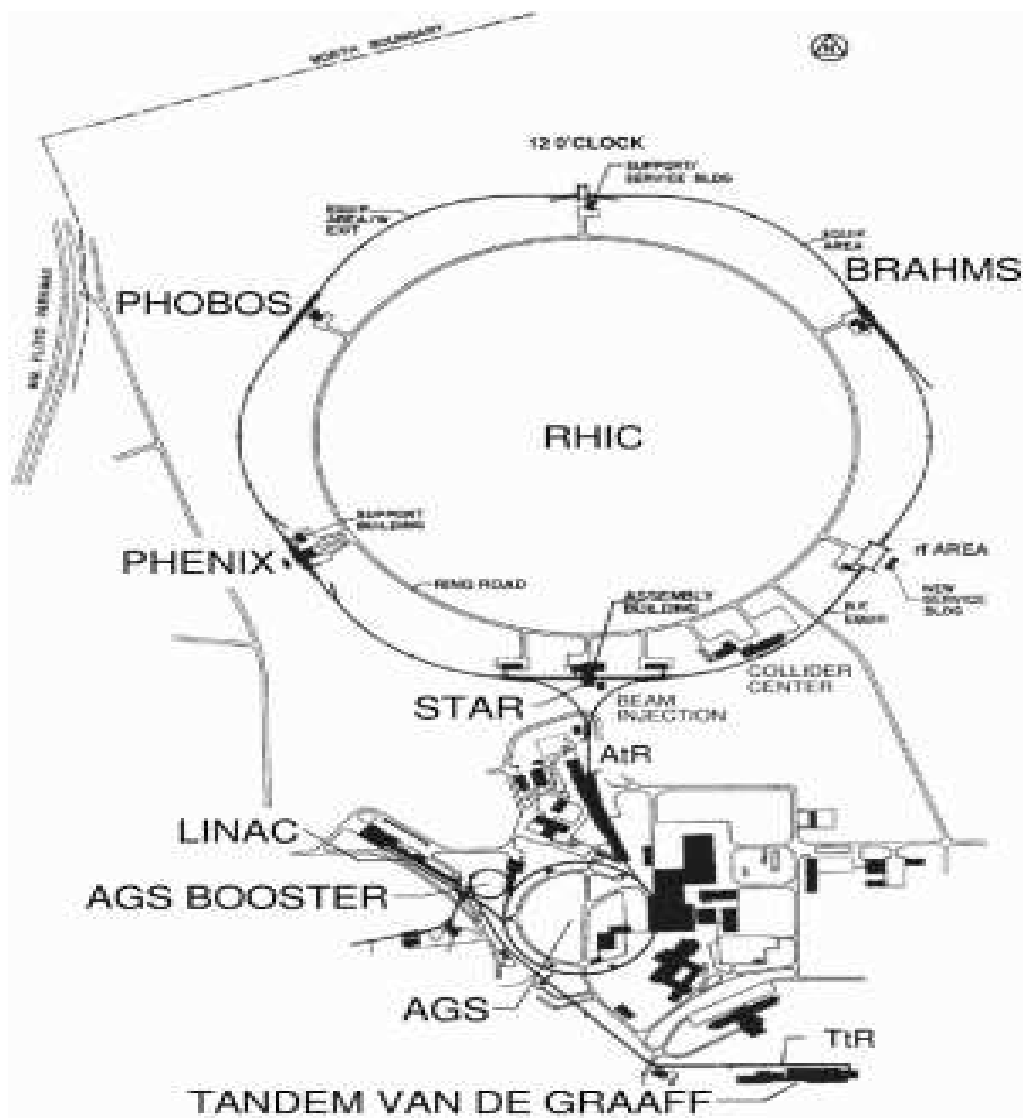} \caption{A
diagram of the Brookhaven National Laboratory collider complex
including the accelerators that bring the nuclear ions up to RHIC
injection energy (10.8 GeV/nucleon for $^{197}$Au). Figure is
taken from~\cite{sorenson:01,Haibin:03}.}
\end{figure}
RHIC's 3.8 km ring has six intersection points where its two rings
of accelerating magnets cross, allowing the particle beams to
collide. The collisions produce the fleeting signals that, when
captured by one of RHIC's experimental detectors, provide
physicists with information about the most fundamental workings of
nature. If RHIC's ring is thought of as a clock face, the four
current experiments are at 6 o'clock (STAR), 8 o'clock (PHENIX),
10 o'clock (PHOBOS) and 2 o'clock (BRAHMS). There are two
additional intersection points at 12 and 4 o'clock where future
experiments may be placed~\cite{rhic:01}.

\section{The STAR Detector}
\begin{figure}[h] \centering
\includegraphics[height=18pc,width=28pc]{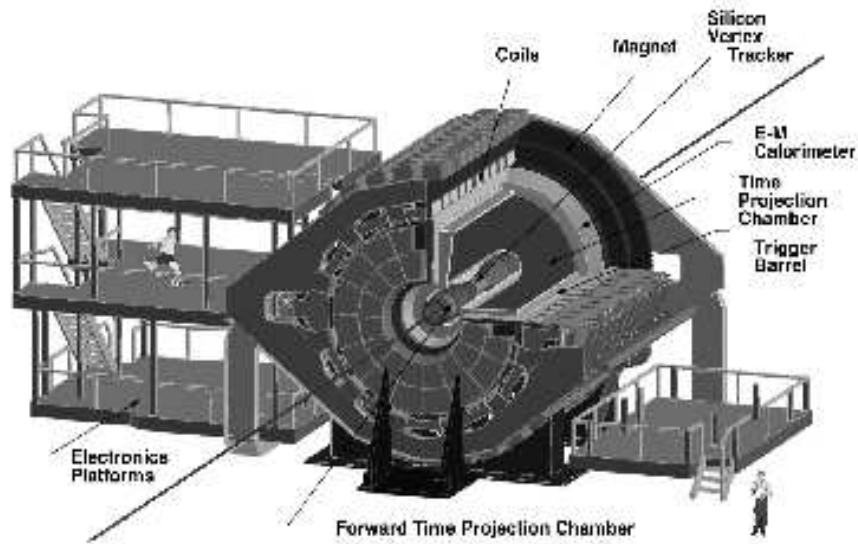}
\caption{Perspective view of the STAR detector, with a cutaway for
viewing inner detector
systems. Figure is taken
from ~\cite{detector:01}.}
\label{starfigure1}
\end{figure}

\begin{figure}[h]
\centering
\includegraphics[height=22pc,width=28pc]{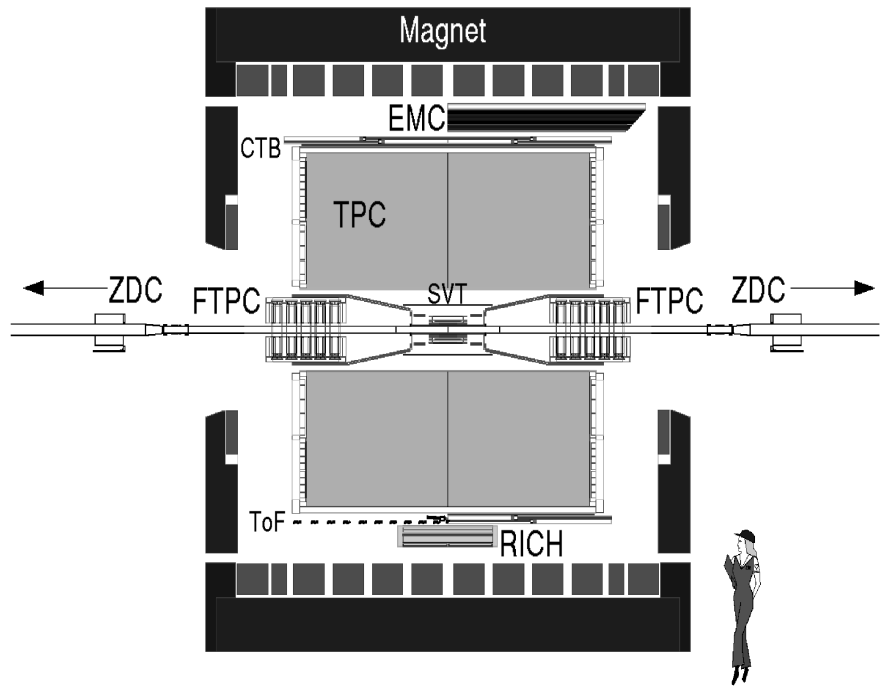}
\caption{Cutaway side view
of the STAR detector as
configured in 2001. Figure
is taken
from~\cite{detector:01}.}
\label{starfigure2}
\end{figure} The Solenoidal
Tracker at RHIC (STAR) is one of the two large detector systems
constructed at the Relativistic Heavy Ion Collider (RHIC) at
Brookhaven National Laboratory. STAR was constructed to
investigate the behavior of strongly interacting matter at high
energy density and to search  for signatures of quark-gluon plasma
(QGP) formation.  Key features  of the nuclear  environment at
RHIC are  a large  number of produced  particles  (up  to
approximately  one  thousand  per  unit pseudo-rapidity) and  high
momentum particles  from hard parton-parton scattering.   STAR can
measure  many  observables simultaneously  to study signatures of
a possible  QGP phase transition and to understand the space-time
evolution    of   the    collision    process in
ultra-relativistic  heavy ion  collisions.  The  goal is  to
obtain a fundamental  understanding  of  the  microscopic
structure  of  these hadronic interactions at high energy
densities. In order to accomplish this, STAR was designed
primarily for measurements of hadron production over a large solid
angle, featuring detector systems for high precision tracking,
momentum analysis, and particle identification at the center of
mass (c.m.) rapidity.  The large acceptance of STAR makes it
particularly well suited for event-by-event characterizations of
heavy ion collisions and for the detection of hadron jets~\cite{detector:01}.\\
The layout of the STAR experiment~\cite{STAR CDR} is shown in
Figure~\ref{starfigure1}. A cutaway side view of the STAR detector
as configured for the RHIC 2001 run is displayed in
Figure~\ref{starfigure2}.  A room temperature solenoidal
magnet~\cite{brown} with a maximum magnetic field of 0.5 T
provides a uniform magnetic field for charged particle momentum
analysis. Charged particle tracking close to the interaction
region is accomplished by a Silicon Vertex Tracker~\cite{bellwied}
(SVT). The Silicon Drift Detectors~\cite{baudot} (SDD) installed
after 2001 is also for the inner tracking. The silicon detectors
cover a pseudo-rapidity range $\mid {\eta }\mid \leq 1$ with
complete azimuthal symmetry ($\Delta \phi = 2\pi$). Silicon
tracking close to the interaction allows precision localization of
the primary interaction vertex and identification of secondary
vertices from weak decays of, for example, $\Lambda$, $\Xi$, and
$\Omega$.  A large volume Time Projection
Chamber~\cite{wieman,tpc} (TPC) for charged particle tracking and
particle identification is located at a radial distance from 50 to
200 cm from the beam axis. The TPC is 4 meters long and it covers
a pseudo-rapidity range $\mid {\eta}\mid \leq 1.8$ for tracking
with complete azimuthal symmetry ($\Delta \phi = 2\pi$). Both the
SVT and TPC contribute to particle identification using ionization
energy loss, with an anticipated combined energy loss resolution
(dE/dx) of 7 \% ($\sigma$).  The momentum resolution of the SVT
and TPC reach a value of $\delta $p/p = 0.02 for a majority of the
tracks in the TPC.  The $\delta $p/p resolution improves as the
number of hit points along the track increases and
as the particle's momentum decreases, as expected~\cite{detector:01}. \\
To extend the tracking to the forward region, a radial-drift TPC
(FTPC)~\cite{eckardt} is installed covering $2.5<\mid{\eta }\mid <
4$, also with complete azimuthal coverage and symmetry.  To extend
the particle identification in STAR to larger momenta over a small
solid angle for identified single-particle spectra at
mid-rapidity, a ring imaging Cherenkov detector
~\cite{ALICE_HMPID} covering $\mid\eta\mid < 0.3$ and $\Delta \phi
= 0.11\pi$, and a time-of-flight patch (TOFp)~\cite{pVPD} covering
$-1<\eta <0$ and $\Delta\phi = 0.04\pi $ (as shown in
Figure~\ref{starfigure2}) was installed at STAR in
2001~\cite{detector:01}. In 2003, a time-of-flight tray (TOFr)
based on multi-gap resistive plate chamber (MRPC)
technology~\cite{startof} was installed in STAR detector, covering
$-1<\eta <0$ and $\Delta\phi = \pi/30 $. For the time-of-flight
system, the Pseudo-Vertex Position Detectors (pVPD) was installed
as the start-timing detector, which was 5.4 m away from TPC center
and covers $4.4<|\eta|<4.9$ with the azimuthal coverage
19\%~\cite{pVPD} in 2003.\\
The fast detectors that provide input to the trigger system are a
central trigger barrel (CTB) at $|\eta|<1$ and two zero-degree
calorimeters (ZDC) located in the forward directions at $\theta<2$
mrad. The CTB surrounds the outer cylinder of the TPC, and
triggers on the flux of charged particles in the mid-rapidity
region. The ZDCs are used for determining the energy in neutral
particles remaining in the forward directions~\cite{detector:01}.
A minimum bias  trigger was obtained by selecting  events with a
pulse height larger than that of one neutron in each of the
forward ZDCs, which corresponds to 95 percent of the geometrical
cross section~\cite{detector:01}.

\subsection{The Time Projection
Chamber} 
\begin{figure}[htb]
\includegraphics[width=14cm]{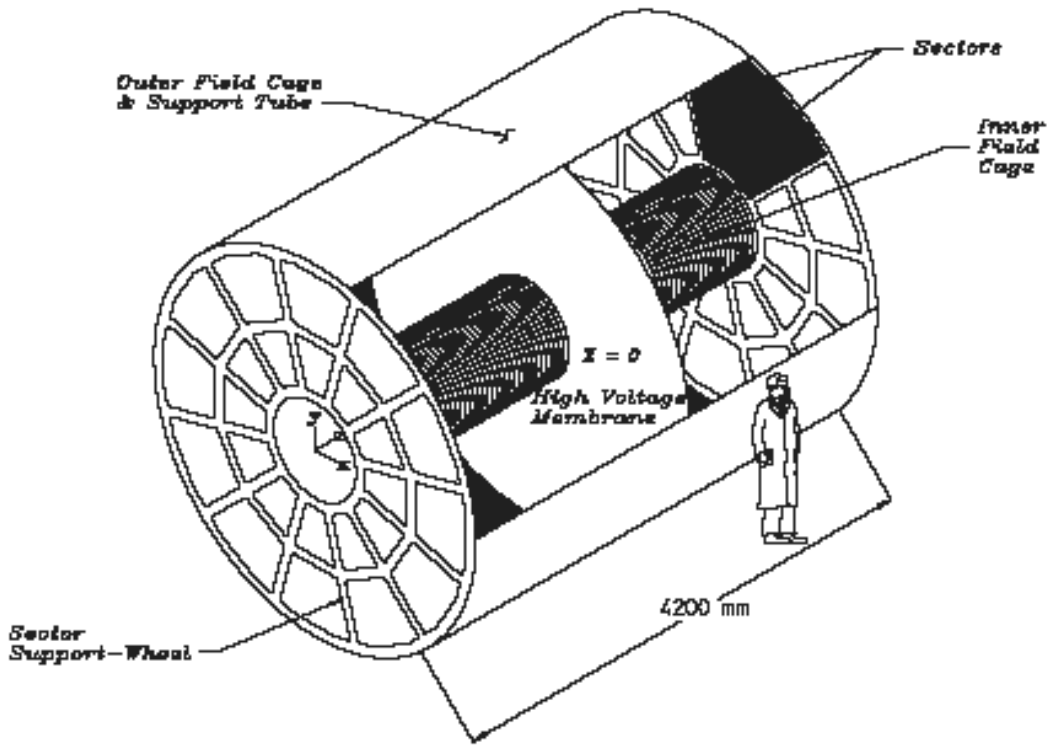}
\caption{The STAR TPC surrounds a beam-beam interaction region at
RHIC. The collisions take place near the center of the TPC.}
\label{tpcman}
\end{figure}

\begin{figure}[h]
\centering
\includegraphics[height=14pc,width=18pc]{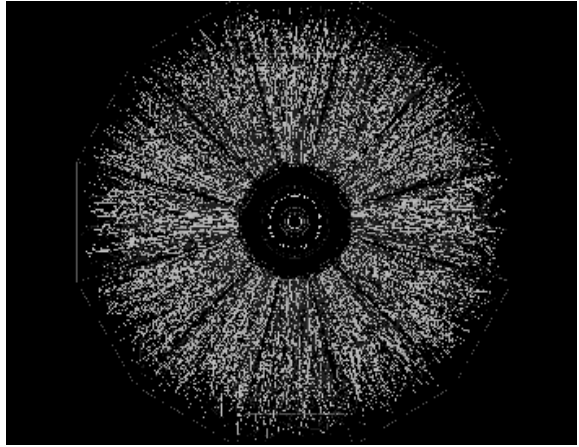} \caption{Beam's
eye view of a central Au+Au collision event in the STAR Time
Projection Chamber. This event was drawn by the STAR online
display. Figure is taken from~\cite{detector:01}.}
\label{eventshow} \end{figure}

The  STAR  detector~\cite{STAR CDR} uses  the  TPC  as
its primary tracking  device. The  TPC  records the tracks  of
particles,  measures their  momenta, and identifies the particles
by measuring their  ionization energy loss  ($dE/dx$). Particles
are identified over a momentum  range from 100 MeV/c  to greater
than 1  GeV/c and momenta are measured over a range of 100
MeV/c to 30 GeV/c~\cite{tpc}.\\
The STAR TPC is shown schematically in Figure~\ref{tpcman}. It is
a volume of gas in a well defined uniform electric field of
$\approx$ 135 V/cm. The working gas of TPC is P10 gas (10\%
methane, 90\% argon) regulated at 2 mbar  above atmospheric
pressure\cite{gas}. This gas has long been used in TPCs. Its
primary attribute is a fast drift velocity which peaks at a low
electric field. Operating on the peak of the velocity curve  makes
the  drift velocity stable and  insensitive  to  small variations
in temperature and pressure~\cite{tpc}. The paths of primary
ionizing particles passing through the  gas volume are
reconstructed  with high precision from the  released secondary
electrons  which drift to the readout end  caps at the ends  of
the chamber. The drift velocity of electrons is 5.45 cm/$\mu$s.
The uniform electric field  which is required to drift the
electrons is defined by a thin conductive Central Membrane (CM) at
the center of  the TPC, concentric field cage cylinders and  the
read out end caps~\cite{tpc}. The readout system is based on Multi
Wire Proportional Chambers (MWPC) with  readout pads. The drifting
electrons  avalanche in the  high fields at the 20 $\mu$m anode
wires providing an amplification of 1000 to  3000. The induced
charge from  an avalanche is  shared  over  several adjacent pads,
so  the original track position can be reconstructed to a small
fraction of a pad width.  There are a total of 136,608 pads in the
readout system~\cite{tpc}, which give $x$-$y$ coordinate
information. The $z$ position
information is provided by 512 time buckets.\\
At the Data Acquisition (DAQ) stage, raw events containing
millions of ADC values and TDC values were recorded. Raw data were
then reconstructed into hits, tracks, vertices, and the collision
vertex through the reconstruction chain of TPC~\cite{starsoftware}
by Kalman method. The collision vertex are called the primary
vertex. The tracks are called the global tracks. If the
3-dimensional distance of closest approach (DCA/dca) of the global
track to the primary vertex is less than 3 cm, this track will be
chosen for a re-fit by forcing a new track helix ending at the
primary vertex. These newly reconstructed helices are called
primary tracks~\cite{Haibin:03}. As expected, the vertex
resolution decreases  as the  square root of the number of tracks
used  in the calculation.  The vertex resolution is 350 $\mu$m
when there are more than 1,000 tracks~\cite{tpc}.
Figure~\ref{eventshow} shows the beam's eye view of a central
Au+Au collision event in the STAR TPC.

\subsubsection{Particle Identification (PID) of TPC by
dE/dx}
\begin{figure}[h]
\centering
\includegraphics[width=22pc]{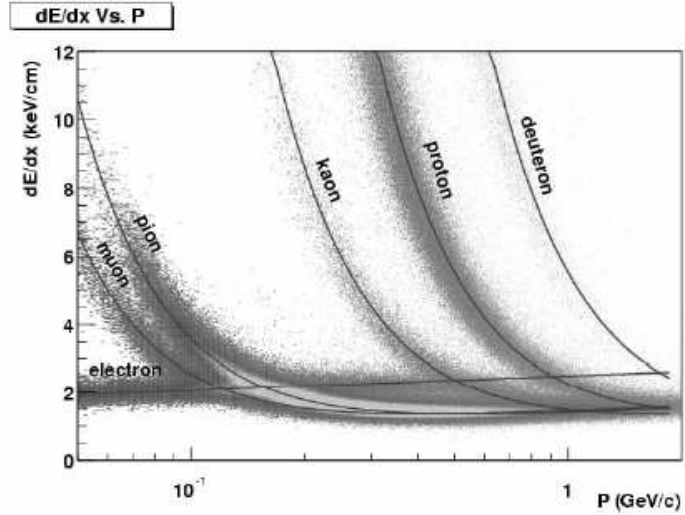}
\caption{The  energy  loss   distribution  for  primary  and
secondary particles in  the STAR TPC as a  function of the $p_T$
of the primary particle. This figure is taken from~\cite{tpc}.}
\label{fdedx}
\end{figure}

Energy lost in the TPC gas is a valuable tool for
identifying particle species. It  works especially well for low
momentum particles but as the particle energy rises, the energy
loss becomes less mass-dependent and it is  hard to separate
particles with velocities $v>0.7$c~\cite{tpc}. For a particle with
charge $z$ (in units of $e$) and speed $\beta=v/c$ passing through
a medium with density $\rho$, the mean energy loss it suffers can
be described by the Bethe-Bloch formula
\begin{equation} \langle \frac{dE}{dx} \rangle = 2\pi
N_0r_e^2m_ec^2\rho\frac{Zz^2}{A\beta^2}
[\text{ln}\frac{2m_e\gamma^2v^2E_M}{I^2}-2\beta^2] \end{equation}
where $N_0$ is Avogadro's number, $m_e$ is the electron mass,
$r_e$ ($=e^2/m_e$) is the classical electron radius, $c$ is the
speed of light, $Z$ is the atomic number of the absorber, $A$ is
the atomic weight of the absorber, $\gamma=1/\sqrt{1-\beta^2}$,
$I$ is the mean excitation energy, and $E_M$
($=2m_ec^2\beta^2/(1-\beta^2)$) is the maximum transferable energy
in a single collision~\cite{tang:01,Haibin:03}. From the above
equation, we can see that different charged particles (electron,
muon, pion, kaon, proton or deuteron) with the same momentum $p$
passing through the TPC gas can result in different energy loss.
Figure~\ref{fdedx} shows the energy loss for particles in the TPC
as a function of the  particle momentum, which includes both
primary  and  secondary  particles. We can see that charged pions
and kaons can be identified up to about transverse momentum 0.75
GeV/c and protons and anti-protons can be identified to 1.1 GeV/c.

In order to quantitatively describe the particle identification,
we define the variable $N_{\sigma\pi}$ (in the case of charged
pion identification) as
\begin{equation}
N_{\sigma\pi}=[\frac{dE}{dx}_{meas.}-\langle\frac{dE}{dx}\rangle_\pi]/
[\frac{0.55}{\sqrt{N}}\frac{dE}{dx}_{meas.}] \end{equation} in
which $N$ is the number of hits for a track in the TPC,
$\frac{dE}{dx}_{meas.}$ is the measured energy loss of a track and
$\langle\frac{dE}{dx}\rangle_\pi$ is the mean energy loss for
charged pions. In order to identify charged kaons, protons and
anti-protons, we can have similar definition of $N_{\sigma K}$ and
$N_{\sigma p}$. Thus we can cut on the variables $N_{\sigma\pi}$,
$N_{\sigma K}$ and $N_{\sigma p}$ to select different particle
species~\cite{Haibin:03}.\\
A specific part of the particle
identification is the topological identification of neutral
particles, such as the $K_S^0$ and $\Lambda$. These neutral
particles can be reconstructed by identifying the secondary
vertex, commonly called V0 vertex, of their charged daughter decay
modes, $K_S^0\rightarrow\pi^+\pi^-$ and $\Lambda\rightarrow p
\pi^-$~\cite{Haibin:03}.

\subsection{The time-of-flight tray based on MRPC technology}
\begin{figure}[h] \centering
\includegraphics[height=24pc,width=32pc]{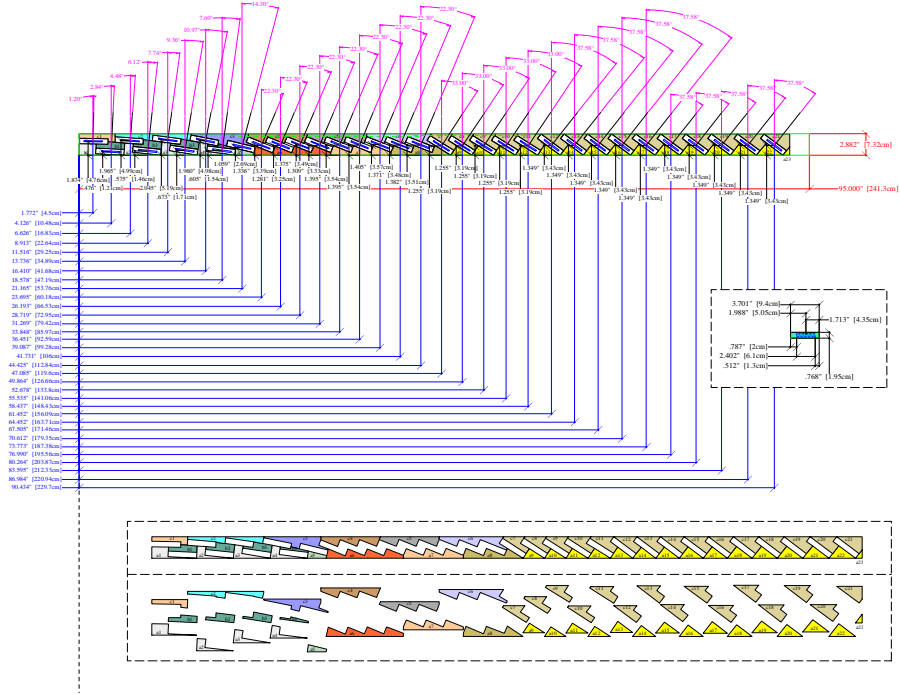}
\caption{Tray structure. Figure is taken from
~\cite{tofproposal}.} \label{tofrtray}
\end{figure}
In 2003, the time-of-flight tray (TOFr) based on multi-gap
resistive plate chamber (MRPC) technology~\cite{startof} was
installed in STAR detector. It extends particle identification up
to $p_{T}\sim3$ GeV/c for $p$ and $\bar{p}$. This tray was
installed on the Au beam outgoing direction. MRPC technology was
first developed by the CERN ALICE group~\cite{williams} to provide
a cost-effective solution for large-area time-of-flight coverage.
For full time-of-flight coverage at STAR, there will be 120 trays,
with 60 on east side and 60 on west side. For each tray, there
will be 33 MRPCs. For each MRPC, there are 6 read-out channels.
Figure~\ref{tofrtray} shows the tray which indicates the position
of each MRPC module. The MRPCs are tilted differently so that each
MRPC is most projective to the average primary vertex location at
Z=0. In 2003 d+Au and p+p run, only 28 MRPCs were installed in the
tray and 12 out of 28 were instrumented with the electronics,
representing 0.3\% of TPC coverage. If we number the 33 MRPCs in
the tray from 1 to 33, with 1 close to TPC center and 33 far from
TPC center, the numbers of 12 modules instrumented with the
electronics in 2003 are 3,4,5,7,9,10,11,12,13,14,26 and 32.

\subsubsection{The introduction of MRPC}
\begin{figure}[h]
\begin{minipage}[t]{1.0\linewidth}
\includegraphics[height=16pc,width=32pc]{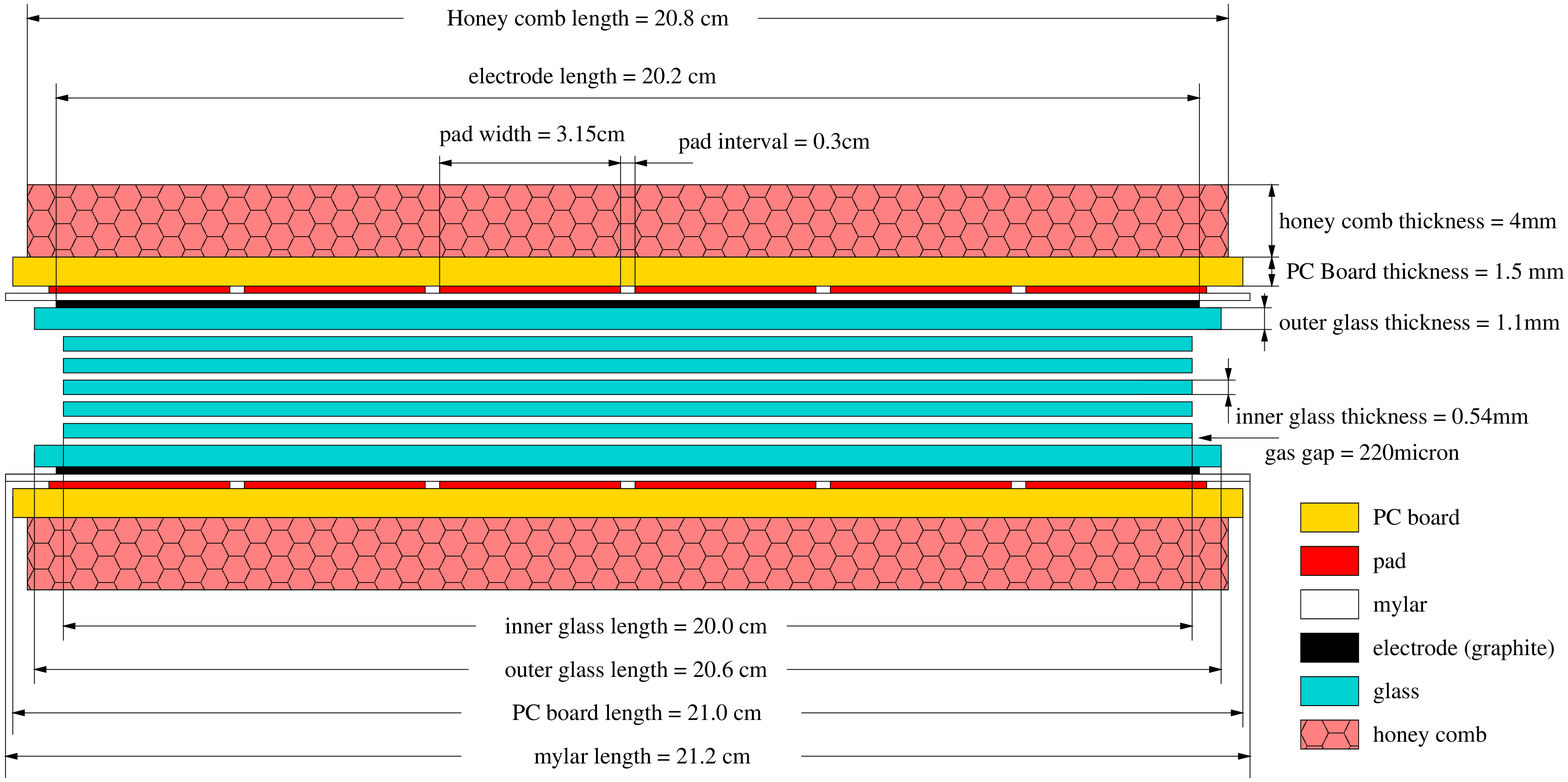}
\end{minipage}
\hspace{\fill}
\begin{minipage}[t]{1.0\linewidth}
\includegraphics[height=11.5pc,width=32pc]{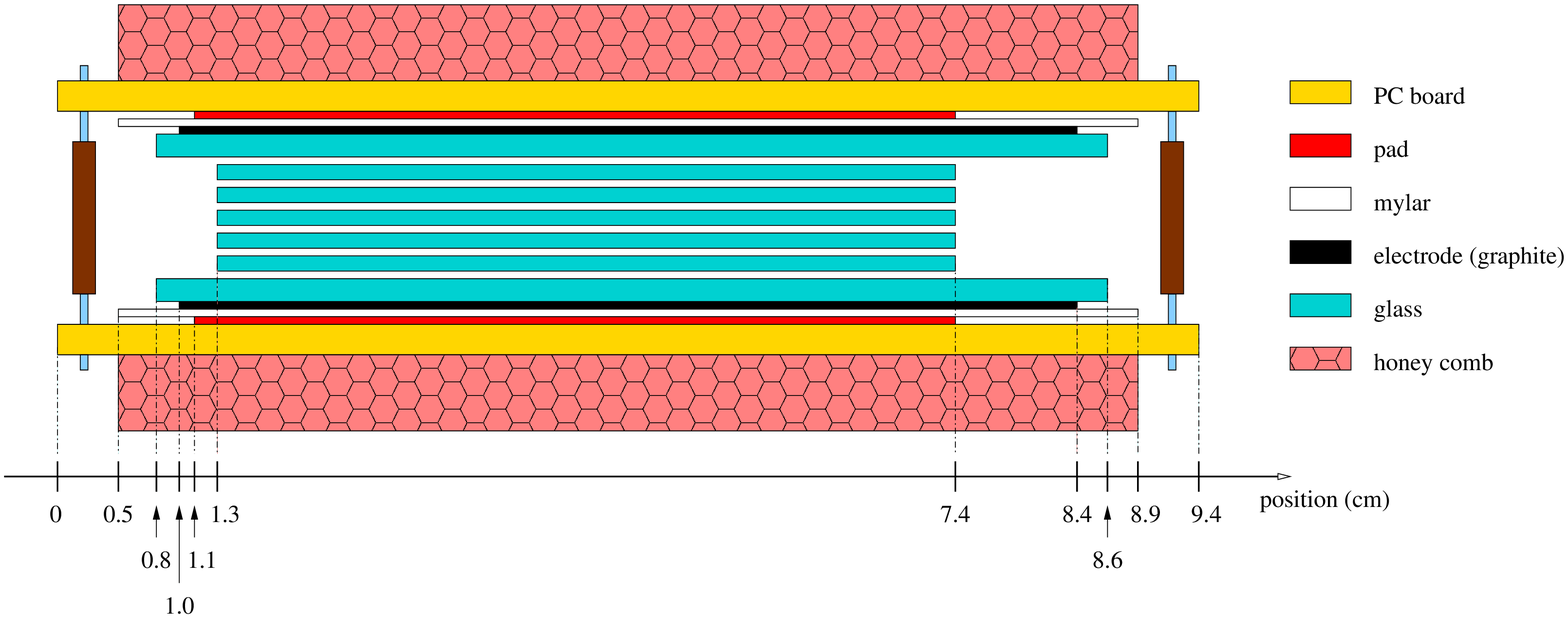}
\end{minipage}
\caption{Two side views of MRPC. The upper (lower) is for long
(short) side view. The two plots are not at the same scale. Figure
is taken from ~\cite{tofproposal}.} \label{mrpcstru}
\end{figure}

\begin{figure}[h]
\centering
\includegraphics[height=12pc,width=24pc]{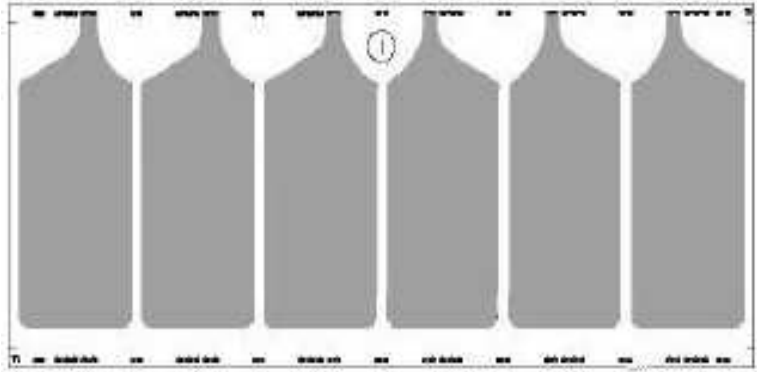}
\caption{The shape of the 6 read-out strips for each MRPC.}
\label{readout}
\end{figure}

Resistive Plate Chambers (RPCs) were developed in
1980s~\cite{mysimu:01}, and were originally operated in streamer
mode. This operation mode allows us to get high detection
efficiency ($>$95\%) and time resolution (~1 ns), with low fluxes
of incident particles. At higher fluxes ($>$200 $Hz/cm^2$), RPCs
begin to lose their efficiency. A way to overcome this problem is
to operate RPCs in avalanche mode. The Multi-gap Resistive Plate
Chamber (MRPC) was developed less than 10 years
ago~\cite{mysimu:02}. It consists of a stack of resistive plates,
spaced one from the other with equal sized spacers creating a
series of gas gaps. Electrodes are connected to the outer surfaces
of the stack of resistive plates while all the internal plates are
left electrically floating. Initially the voltage on these
internal plates is given by electrostatics, but they are kept at
the correct voltage due to the flow of electrons and ions created
in the avalanches. Figure~\ref{mrpcstru} shows the structure of
MRPC detector. For each MRPC, there are 6 read-out strips.
Figure~\ref{readout} shows the shape of the read-out strip.
The detailed production process can be found at Appendix B.\\
MRPC, as a new kind of detector for time of fight system, operated
in avalanche mode with a non flammable gas mixture of 90\% F134A,
5\% isobutane, 5\% SF6, can fulfill all these requirements: high
efficiency ($>$95\%), excellent intrinsic time resolution ($<$100
ps)~\cite{mysimu:13,startof,mysimu:15,mysimu:16,mysimu:17}, high
rate capability (~500 $Hz/cm^2$), high modularity and simplicity
for construction, good uniformity of response, high
granularity/low occupancy, and large acceptance.

\subsubsection{Simulation: the work principle of this chamber}
A detailed description of the model used in the simulation was
reported in these
papers~\cite{mysimu:03,mysimu:04,mysimu:05,mysimu:06}, here just
the main items will be repeated. The program starts from
considering an ionizing particle which crosses the gas gaps and
generates a certain number of clusters of ion-electron pairs. The
electrons contained in the clusters drift towards the anode and,
if the electric field is sufficiently high, give rise to the
avalanche processes. \\
The primary cluster numbers and the avalanche growth are assumed
to follow, respectively, simple Poisson statistics and the usual
exponential law. Avalanche gain fluctuations have been taken into
account using a Polya distribution~\cite{mysimu:07}. After the
simulation of the drifting avalanches, the program computes, by
means of Ramo~\cite{mysimu:08} theorem, the charge $q_{ind}$
induced on the external pick-up electrodes (strips or pads) by the
avalanche motion. Under certain approximations, this is given by
the formula
\begin{equation} q_{ind} =
\frac{q_{e}}{\eta{d}}\triangle{V}_{w}{\sum_{j=1}^{n_{cl}}{n_{j}M(e^{\eta{(d-x_{j})}}-1)}}
\end{equation}  where $q_e$ is the
electron charge, $\eta$ 1st effective Townsend coefficient
$\eta=\alpha-\beta$, $\alpha$ is the Townsend coefficient, $\beta$
is the attachment coefficient, $x_j$ the $j_{th}$ cluster initial
distances from the anode, $d$ the gap width, $n_j$ the number of
initial electrons in the considered $j_{th}$ cluster, $M$ the
avalanche gain fluctuations factor, and $\triangle{V}_{w}/d=E_{w}$
is the normalized weighting field. In addition to $q_{ind}$, the
current $i_{ind}(t)$ induced on the same electrodes by the
drifting charge $q_d(t)$ may be computed as
\begin{equation} i_{ind}(t) =
\triangle{V}_{w}\frac{v_{d}}{d}q_d(t)Me^{\eta{v_{d}t}}
\end{equation}, where $v_d$ is the electron drift velocity. The computation of
$i_{ind}$ allows us to reproduce the whole information coming out
from MRPC, such as time
distribution.\\
\textbf{Charge Spectrum Simulation}: The almost Gaussian charge
distribution obtained with the MRPC is a key ingredient to its
performance. If the avalanches grew following Townsend's formula
the charge distribution would be exponential in shape. Thus the
space charge effects must be considered in the simulation. \\
The input parameters for the simulation program are: the Townsend
coefficient $\alpha$, the attachment coefficient $\beta$, the
average distance between clusters $\lambda$ and the probability
distribution of the number of electrons per cluster. These pieces
of information can be obtained, for a given gas mixture and given
conditions (pressure and temperature) and electric field, by the
programs HEED~\cite{mysimu:09} and
MAGBOLTZ~\cite{mysimu:10,mysimu:11}. In addition, a maximum number
of electrons in an avalanche (cutoff value) is specified. \\
In a given gap, we generate a number of clusters with distances
exponentially distributed with average distance $\lambda$. For
each cluster, we then generate a certain number of electrons,
according to the distribution obtained by the program HEED. Each
electron from the primary cluster will give rise to a number of
electrons, generated according to an exponential probability
law.\\
\begin{figure}[h]
\centering
\includegraphics[height=18pc,width=24pc]{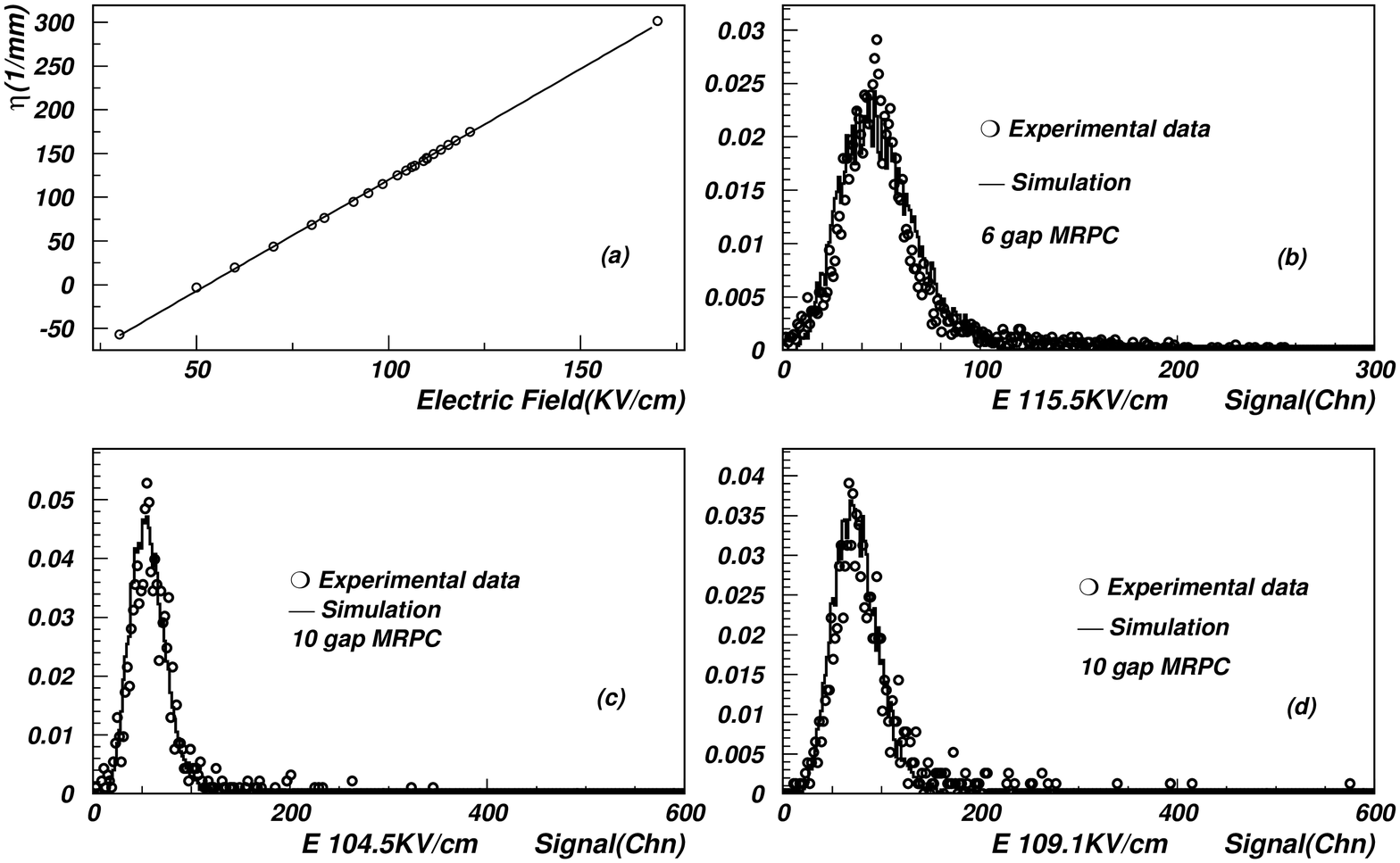}
\caption{Simulated 1st effective Townsend coefficient curve and
normalized charge distribution for a 6 and 10 gap MRPC.}
\label{chargedistribution}
\end{figure}
For each cluster, the avalanche growth is stopped when the total
charge reaches a certain cutoff value, as originally suggested in
ref.~\cite{mysimu:12} to take into account space charge effects in
the avalanche development. This cutoff value has been set to be
$1.6\times10^7$ electrons.\\
In Figure~\ref{chargedistribution} we show the results of
simulations, Figure~\ref{chargedistribution}(a) is the simulated
curve of the 1st effective Townsend coefficient $\eta$ versus the
electric field, which is generated by Magboltz. The curve shows
that the correlation between $\eta$ and the electric field is
almost linear when MRPC is operated at high electric field for the
gas mixture. Figure~\ref{chargedistribution}(b) is the charge
spectrum for a 6 gap chamber and (c) (d) for a 10 gap chamber
compared to experimental data~\cite{mysimu:13,startof}, and the
number under each plot shows the electric field $E$ in the gas gap
for MRPC. In both cases the gap size is 220 $\mu{m}$. The gas
mixture was 90\% F134A, 5\% isobutane and 5\% SF6 in normal
conditions of pressure and temperature. The value of $\lambda$
used was 0.1 $mm$, derived from HEED program. \\
The charge distribution has an almost Gaussian form, especially
for the 10 gap MRPC. The left side of the distribution (very few
events at values near zero) is due to the fact that the MRPC
operates at high gain $\eta \times{d}\sim30$. This means that
avalanches starting in the middle of the gap width, which only
avalanche over half the distance, give a detectable signal. The
charge distribution is the superposition of several probability
distributions which, according to the central limit theorem, will
tend to a Gaussian form. The right side of the charge distribution
(the fact that the tails are not very long) indicates that indeed
the space charge effects stop the development of the avalanche.\\
\begin{figure}[h]
\centering
\includegraphics[height=18pc,width=24pc]{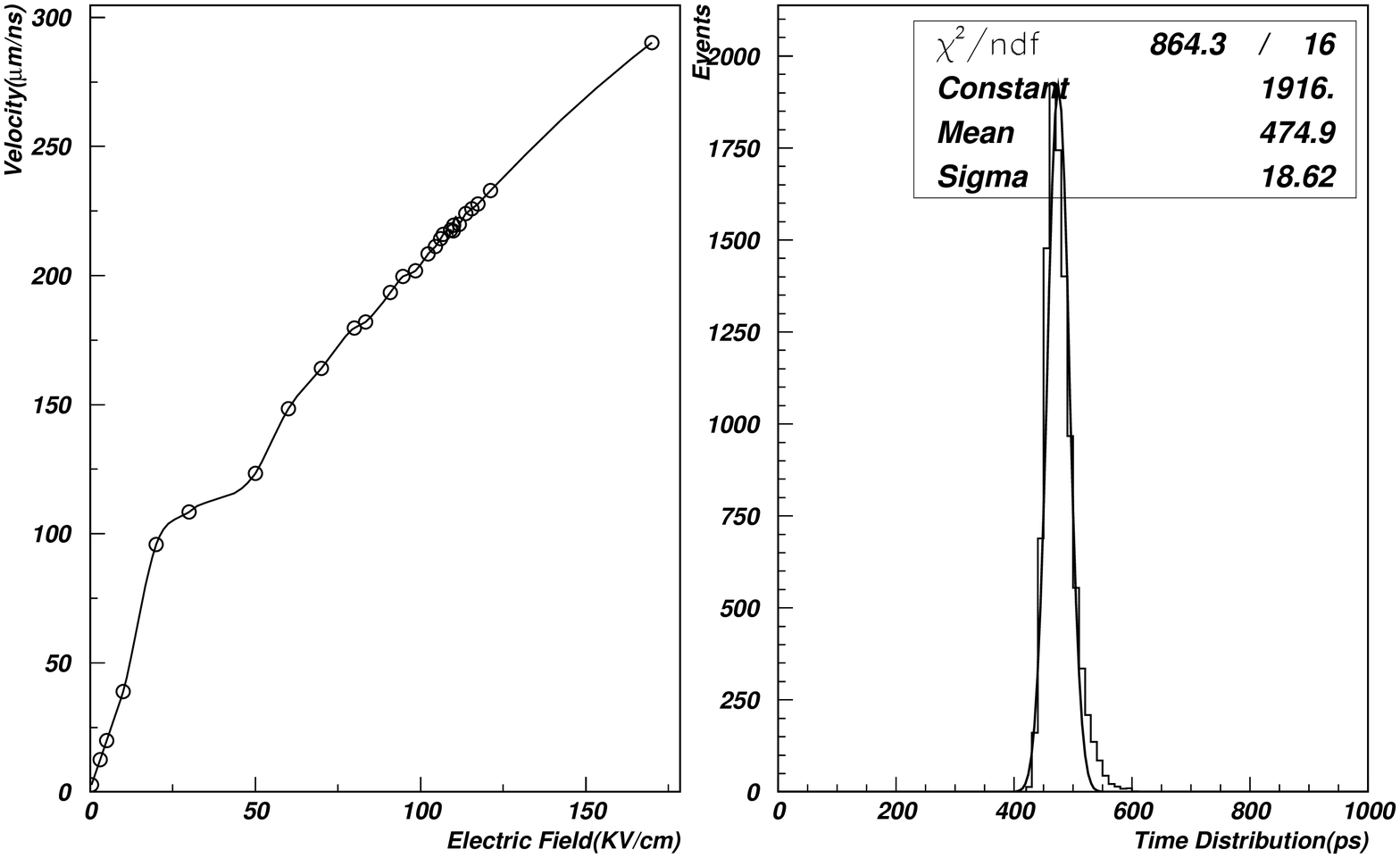}
\caption{Simulated results of a 6 gap MRPC.}
\label{timedistribution}
\end{figure}
\textbf{Time Distribution Simulation:} We then proceed to simulate
the time distribution of these same chambers. The electron drift
velocity can be obtained from HEED. When the total induced charge
signal is over threshold, the time is recorded. In this paper, the
threshold is 13 fc for the 6 gap MRPC and 26 fc for the 10 gap
MRPC. Fig.2 is the simulated results for a 6 gap chamber.
Figure~\ref{timedistribution} (a) is the simulated curve of the
electron drift velocity versus the electric field, which is
generated by Magboltz.  Figure~\ref{timedistribution} (b) is the
time distribution of a 6 gap MRPC. The intrinsic time resolution
is only 19 ps or so. If we consider other contributions, such as
front-end electronics 30 ps, TDC resolution 25 ps, fanout start
signal 10 ps, beam size (1cm) 15 ps, we can get the MRPC
resolution is $\sqrt{20^2+30^2+25^2+10^2+15^2}=47$ ps. This value
is similar to the experimental result~\cite{mysimu:13,startof}.
For a 10 gap MRPC, the intrinsic time resolution is about 15
ps.\\
From the simulation, we can get the bottom line of MRPC time
resolution $\sim20$ ps. And we need to keep control of all these
contributions to ensure best time resolution.
\subsubsection{MRPC for this tray installed in 2003}
In 2003, for the MRPCs in the TOFr, the inner glass thickness is
0.54 mm, the outer glass is 1.1 mm. The gas gap is 0.22 mm. Both
the volume resistivity ($10^{12-13} ohm.cm$) of the glass plates
and the surface resistivity(2M  ohm per square) of carbon layer at
room temperature are presented in~\cite{resistivity}. It is found
the volume resistivity of the plate decreases with the temperature
increasing. And the radiation will decrease the volume resistivity
of the plate~\cite{resistivity}. In order not to pollute the
working gas of TPC, SF6 is not used as part of the working gas of
TOFr. The working gas of MRPC-TOFr at STAR is 95\% freon and 5\%
iso-butane at normal atmospheric pressure. The high voltage
applied to the electrodes is 14.0 kV.

\chapter{Analysis Methods}
\label{chp:analysis}

\section{Trigger}

  The detector used for these studies was the Solenoidal Tracker at
  RHIC (STAR).  The main tracking device is the Time Projection
  Chamber (TPC) which provides momentum information and particle
  identification for charged particles up to $p_{T}\sim1.1$ GeV/c by
  measuring their ionization energy loss ({\it dE/dx})~\cite{tpc}.
  Detailed descriptions of the TPC and d+Au run conditions have been
  presented in Ref.~\cite{stardau,tpc}.  A prototype time-of-flight detector
  (TOFr) based on multi-gap resistive plate chambers
  (MRPC)~\cite{startof} was installed in STAR for the d+Au and p+p
  runs. It extends particle identification up to $p_{T}\sim3$ GeV/c
  for $p$ and $\bar{p}$.

 TOFr covers $\pi/30$ in azimuth and $-1\!<\!\eta\!<\!0$ in
  pseudorapidity at a radius of $\sim220$ cm. It contains 28 MRPC
  modules which were partially instrumented during the 2003 run.
Since the acceptance of TOFr is small, a special trigger selected
  events with a valid pVPD coincidence and at least one TOFr hit. A
  total of 1.89 million and 1.08 million events were used for the
  analysis from TOFr triggered d+Au and non-singly diffractive (NSD)
  p+p collisions, representing an integrated luminosity of about 40
  $\mathrm{{\mu}b}^{-1}$ and 30 $\mathrm{nb}^{-1}$, respectively.
  Minimum-bias d+Au and p+p collisions that did not require pVPD and
  TOFr hits were also used to study the trigger bias and enhancement,
  and the TOFr efficiency and acceptance. The d+Au minimum-bias
  trigger required an equivalent energy deposition of about 15 GeV in
  the Zero Degree Calorimeter in the Au beam
  direction~\cite{stardau}. The trigger efficiency was determined to
  be $95\pm3\%$. Minimum-bias p+p events were triggered by the
  coincidence of two beam-beam counters (BBC) covering $3.3<
  |\eta|<5.0$~\cite{starhighpt}. The NSD cross section was measured to
  be $30.0\pm3.5$ mb by a van der Meer scan and PYTHIA~\cite{pythia}
  simulation of the BBC acceptance~\cite{starhighpt}.

\subsection{Centrality tagging}

Centrality tagging of d+Au collisions was based on the charged
particle multiplicity in $-3.8<\eta<-2.8$, measured by the Forward
Time Projection Chamber in the Au beam
direction~\cite{stardau,ftpc}. The TOFr triggered d+Au events were
divided into three centralities: most central $20\%$, $20-40\%$
and $40-\sim100\%$  of the hadronic cross section.  The average
number of binary collisions $\langle N_{bin}\rangle$ for each
centrality class and for the combined minimum-bias event sample is
derived from Glauber model calculations and listed in
Table~\ref{centrality}.


Table~\ref{centrality} also lists the uncorrected FTPC east
reference multiplicity ranges for centrality definitions.
\begin{table}[h]
  \centering
  \begin{tabular}{|c|c|c|c|}
    \hline
        Centrality Bin & Uncorr. FTPCRefMult Range & Uncorr. $N_{charge}$ & $N_{bin}$ \\ \hline
        M.B. &     & 10.2 & $7.5\pm0.4$ \\ \hline
        0\%-20\% & FTPCRefMult $\geq$ 17 & 17.58 & $15.0\pm1.1$ \\ \hline
        20\%-40\% & 10 $\leq$ FTPCRefMult $<$ 17 & 12.55 & $10.2\pm1.0$ \\ \hline
        40\%-100\% & 0 $\leq$ FTPCRefMult $<$ 10 & 6.17 & $4.0\pm0.3$ \\ \hline
  \end{tabular}
\caption{Centrality definitions for different uncorrected FTPC
east reference multiplicity ranges. Uncorrected $N_{charge}$
stands for the average value of uncorrected reference multiplicity
in certain centrality bin. The fourth column represents the number
of binary collisons $\langle N_{bin}\rangle$ calculated from
Glauber model.} \label{centrality}
\end{table}

\subsection{Trigger bias study}
\begin{figure}[h]
\centering
\includegraphics[height=18pc,width=24pc]{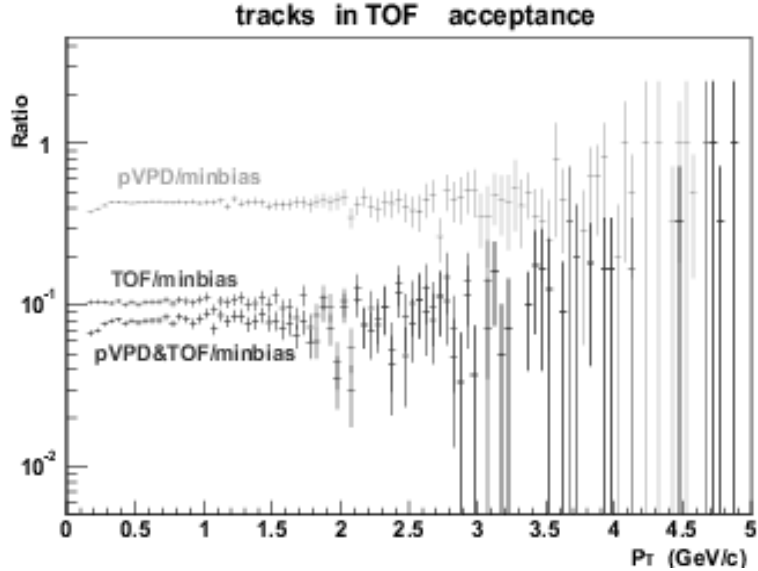}
\caption{The $p_{T}$ dependence plot of the trigger bias.}
\label{PtRatioRealTOFAcceptance}
\end{figure}

Since we set up a special trigger which selected events with a
valid pVPD coincidence and at least one TOFr hit, the study of
$p_{T}$ dependence of trigger bias is necessary.
Figure~\ref{PtRatioRealTOFAcceptance} shows there is negligible
trigger bias on $p_{T}$ dependence at $p_{T}>$ 0.3 GeV/c from
simulation. In this figure, pVPD means that pVPD is required to
fire in minimum-bias collisions. TOF means that TOFr is required
to fire in minimum-bias collisions, and pVPD $\&$ TOF means that
pVPD and TOFr are required to fire in minimum-bias collisions.
From this figure, if we required pVPD and TOFr to fire, we can see
the ratio is flat with $p_{T}$ when $p_{T}$ is larger than 0.3
GeV/c by comparison through the $p_{T}$ distribution in
minimum-bias collisions. That means the trigger bias for $p_T$
distribution is negligible at $p_{T}>$ 0.3 GeV/c.

\begin{figure}[h]
\centering
\includegraphics[height=18pc,width=24pc]{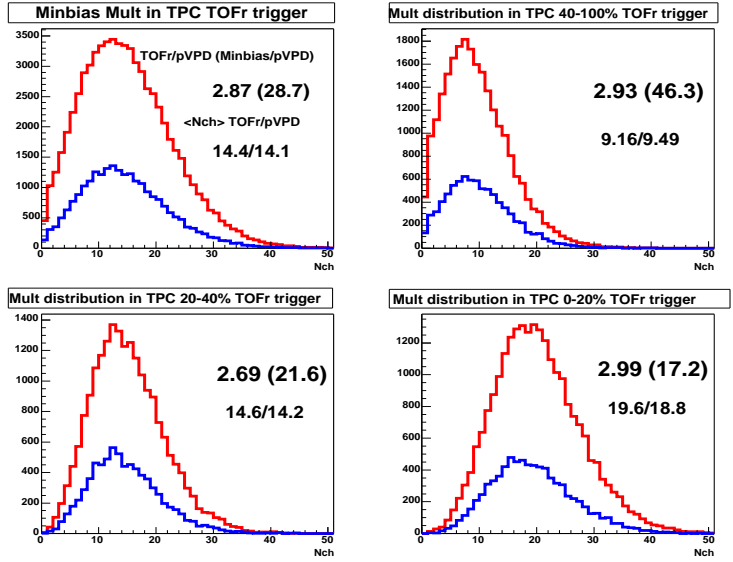}
\caption{The enhancement factor and $\langle N_{ch}\rangle$ bias
in minimum-bias and centrality selected d+Au collisions. }
\label{NchbiasNew}
\end{figure}

Minimum-bias d+Au and p+p collisions are used to study the trigger
bias and enhancement. Figure~\ref{NchbiasNew} shows the trigger
bias and enhancement in d+Au minimum-bias collisions and three
centrality bins. In this figure, TOFr means that TOFr is required
to fire in minimum-bias events. pVPD means that TOFr and pVPD are
required to fire in minimum-bias events. Minbias means the
minimum-bias triggered events. For enhancement study, TOFr/pVPD is
the ratio of the number of events in which TOFr is required to
fire over the number of events in which TOFr and pVPD are required
to fire, and Minbias/pVPD is the ratio of the number of
minimum-bias triggered events over the number of events in which
TOFr and pVPD are required to fire. The enhancement factor for
TOFr is (Minbias/pVPD)/(TOFr/pVPD). For example, in minimum-bias
collision, Minbias/pVPD is equal to 28.7, while TOFr/pVPD is 2.87,
so in minimum-bias collisions, the enhancement of TOFr trigger is
10. For $\langle N_{ch}\rangle$ bias study, TOFr/pVPD is the ratio
of $\langle N_{ch}\rangle$ in the events where TOFr is required to
fire over the $\langle N_{ch}\rangle$ in the events where TOFr and
pVPD are required to fire. Since in our triggered events, TOFr and
pVPD are required to fire, TOFr/pVPD is our $\langle
N_{ch}\rangle$ bias factor. The curves in this figure show the
charged particle multiplicity at mid-rapidity in TOFr events and
in TOFr and pVPD events individually. Table~\ref{triggerbiastable}
lists the enhancement factor and trigger bias in minimum-bias,
centrality selected d+Au collisions and minimum-bias p+p
collisions.
\begin{table}[h]
  \begin{tabular}{|c|c|c|c|}
    \hline
        Centrality Bin & TOFr triggered events & enhancement factor & $\langle N_{ch}\rangle$ bias  \\ \hline
        0\%-100\%       & 1.80 M     & 10.0 & 1.02 \\ \hline
        0\%-20\%   & 0.523 M & 5.75 & 1.04 \\ \hline
        20\%-40\%  & 0.500 M  & 8.03 & 1.03 \\ \hline
        40\%-100\% & 0.479 M & 15.8 &  0.965 \\ \hline
        p+p        & 0.995 M& 37.4  &  1.19 \\ \hline
  \end{tabular}
\caption{Trigger bias study. The $\langle N_{ch}\rangle$ bias and
enhancement factor in minimum-bias, centrality selected d+Au
collisions and minimum-bias p+p collisions.}
\label{triggerbiastable}
\end{table}

\section{Track selection and calibration}
The TPC and TOFr are two independent systems. In the analysis,
hits from particles traversing the TPC were reconstructed as
tracks with well defined geometry, momentum, and {\it dE/dx}
~\cite{tpc}.
 The particle trajectory was then extended outward to the TOFr
detector plane. The pad with the largest signal within one pad
distance to the projected point was associated with the track for
further time-of-flight and velocity ($\beta$) calculations.

\subsection{Calibration}
\subsubsection{pVPD calibration}
\begin{figure}[h]
\begin{minipage}[t]{80mm}
\includegraphics[height=13pc,width=18pc]{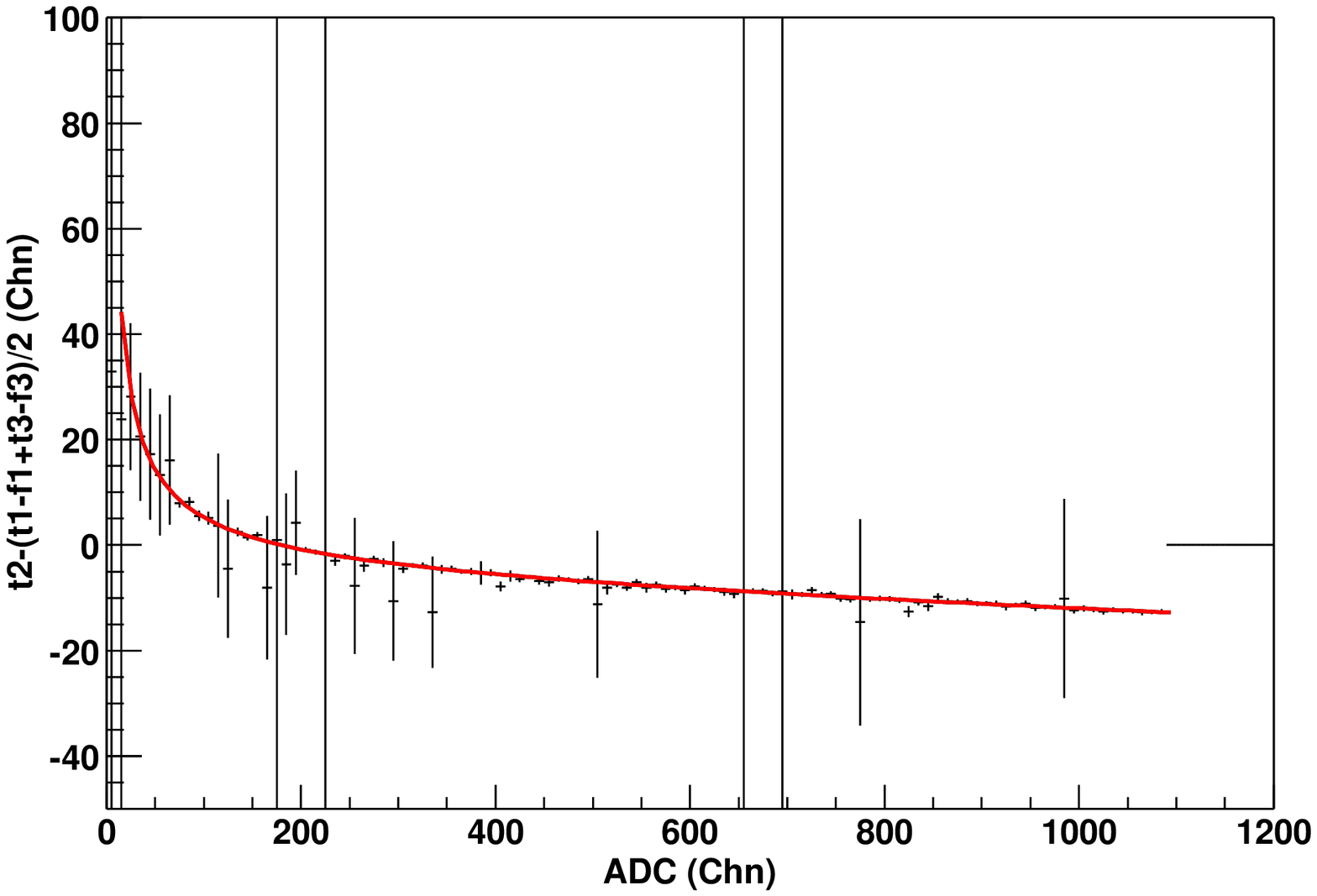}
\end{minipage}
\hspace{\fill}
\begin{minipage}[t]{80mm}
\includegraphics[height=13pc,width=18pc]{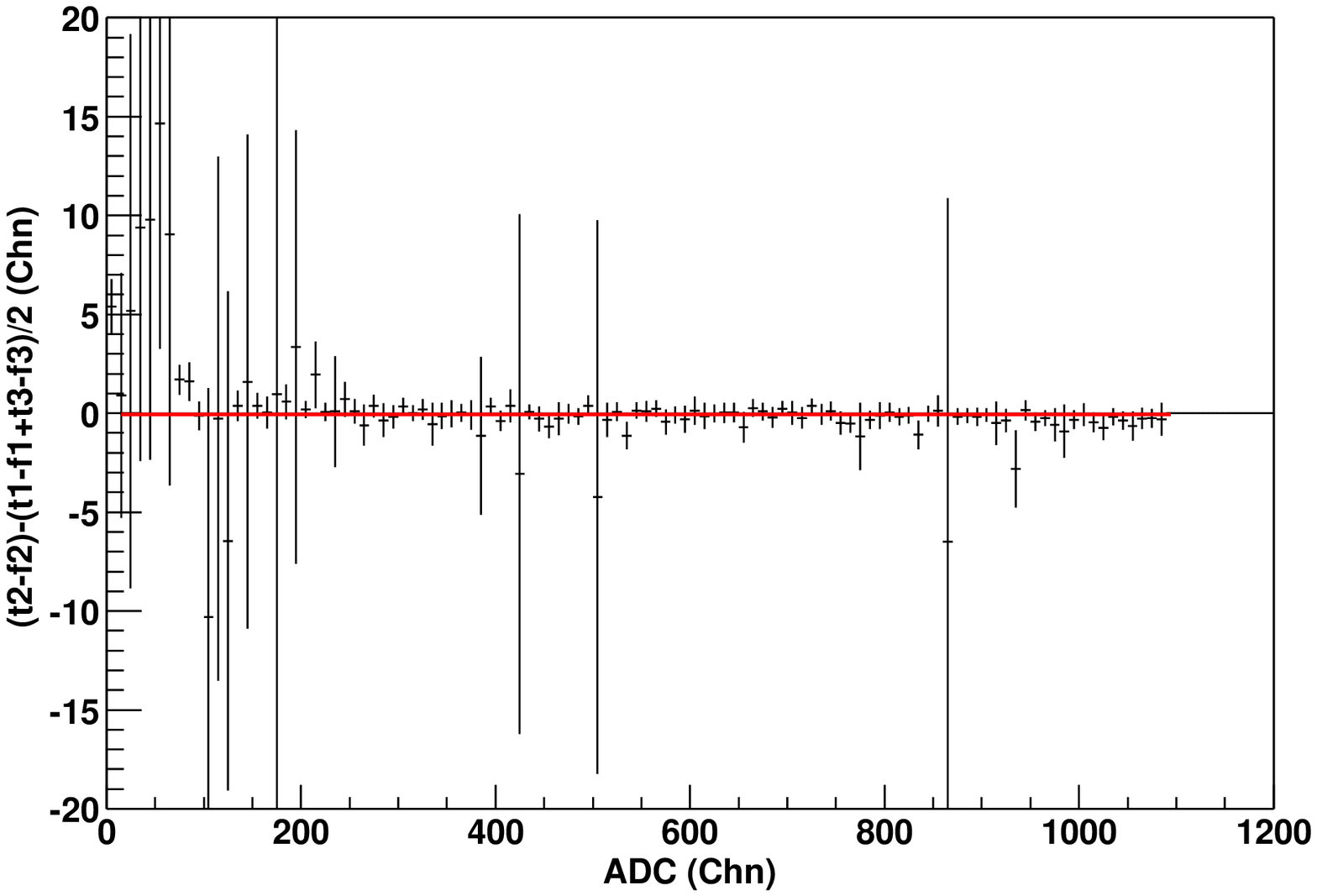}
\end{minipage}
\caption{pVPD slewing correction.}
\label{pvpdslewingplotforthesis}
\end{figure}

For TOFr, we use pVPD as our start-timing detector. In d+Au and
p+p collisions, at least one east pVPD and one west pVPD were
required to fire. In d+Au collisions, to calibrate east pVPD, we
required 3 east pVPD to fire; to calibrate west pVPD, we required
3 west pVPD to fire. In p+p collisions, to calibrate east pVPD, we
required 2 east pVPD to fire; to calibrate west pVPD, we required
2 west pVPD to fire. Let's take the east pVPD calibration in d+Au
collisions as an example. The label for 3 pVPD are pVPD1, pVPD2,
pVPD3, the adc and tdc value for pVPD1 are $a1$, $t1$, and the
slewing correction function is $f1$; the adc and tdc value for
pVPD2 are $a2$, $t2$, and the slewing correction function is $f2$;
the adc and tdc value for pVPD3 are $a3$, $t3$, and the slewing
correction function is $f3$. We use $t1-((t2-f2)+(t3-f3))/2$ vs
$a1$ to get the slewing correction for pVPD1; use
$t2-((t3-f3)+(t1-f1))/2$ vs $a2$ to get the slewing correction for
pVPD2; use $t3-((t1-f1)+(t2-f2))/2$ vs $a3$ to get the slewing
correction for pVPD3. At the beginning, $f1=f2=f3=0$, we got 3
curves of $t1-((t2-f2)+(t3-f3))/2$ vs $a1$,
$t2-((t3-f3)+(t1-f1))/2$ vs $a2$ and $t3-((t1-f1)+(t2-f2))/2$ vs
$a3$. The 3 curves corresponded to the 3 slewing functions $f(a1),
f(a2), f(a3)$; For the second step, $f1=f(a1), f2=f(a2),
f3=f(a3)$, also plot  $t1-((t2-f2)+(t3-f3))/2$ vs $a1$,
$t2-((t3-f3)+(t1-f1))/2$ vs $a2$ and $t3-((t1-f1)+(t2-f2))/2$ vs
$a3$. And we got the new three slewing curves $f'(a1), f'(a2),
f'(a3)$. For the third step, $f1=f'(a1), f2=f'(a2), f3=f'(a3)$,
also plot $t1-((t2-f2)+(t3-f3))/2$ vs $a1$,
$t2-((t3-f3)+(t1-f1))/2$ vs $a2$ and $t3-((t1-f1)+(t2-f2))/2$ vs
$a3$. And we got another new three slewing curves $f''(a1),
f''(a2), f''(a3)$. And so on and so forth till the resolution of
$t1-f1-((t2-f2)+(t3-f3))/2, t2-f2-((t3-f3)+(t1-f1))/2$ and
$t3-f3-((t1-f1)+(t2-f2))/2$ converged. The looping method is to
subtract the correlation of different pVPD tubes in the same
direction. The function for the slewing correction we use is
$y=par[0]+par[1]/\sqrt{x}+par[2]/x+par[3]\times{x}$. In
Figure~\ref{pvpdslewingplotforthesis}, the left plot shows the
pVPD2 slewing plot and the right plot shows that the timing is
independent on the ADC value after the slewing correction.

After the slewing correction, we got the corrected timing of east
pVPD and west pVPD. For each side, the timing difference should be
shifted to zero. That's to say the mean value in the distribution
of $t1-f1-(t2-f2)$ and $t1-f1-(t3-f3)$ were shifted to zero. Also
we need to correct for the effect caused by the different numbers
of fired pVPD in different events. What we did was shifting the
mean value of the distribution of
($\sum{te})/Ne-(\sum{tw})/Nw-2.\times{Vz/c}$ to zero, where the
$\sum{te}$, $\sum{tw}$ means the sum of the corrected timing of
east fired pVPD and west fired pVPD respectively, $Ne, Nw$ means
the number of east fired pVPD and west fired pVPD, $Vz$ is the $z$
value of primary vertex of the event, and $c$ is the light
velocity.

\subsubsection{TOFr calibration}

After the slewing correction for pVPD, we use this variable as our
start timing:
\begin{equation}
T_{start}=\frac{{\sum_{i=1}^{Ne}{te}}+{\sum_{i=1}^{Nw}{tw}}-(Ne-Nw)\times{Vz}/c}{Ne+Nw}
\end{equation}

\begin{figure}[h]
\centering
\includegraphics[height=18pc,width=24pc]{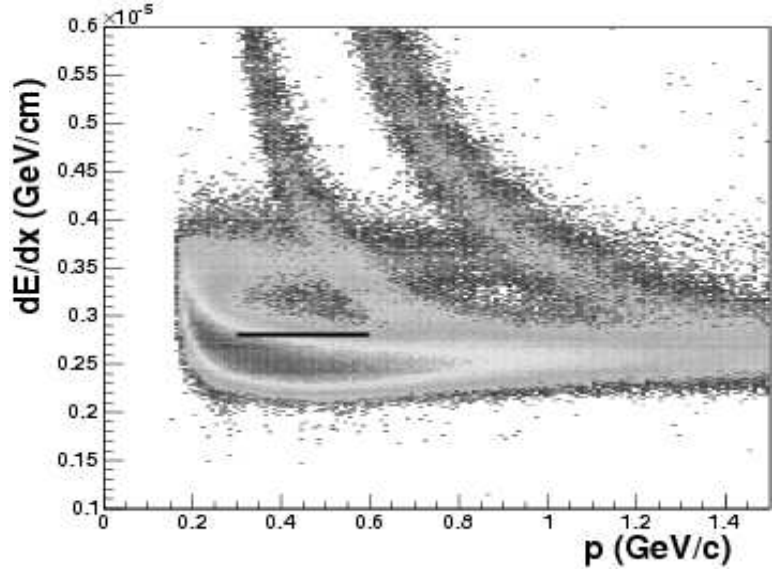}
\caption{dE/dx vs $p$ plot from d+Au collisions. The line
represents that $dE/dx=0.028\times{10^{-4}}$ GeV/cm in this
momentum range $0.3<p<0.6$ GeV/c.} \label{dAudedxplot}
\end{figure}

\begin{figure}[h]
\centering
\includegraphics[height=18pc,width=24pc]{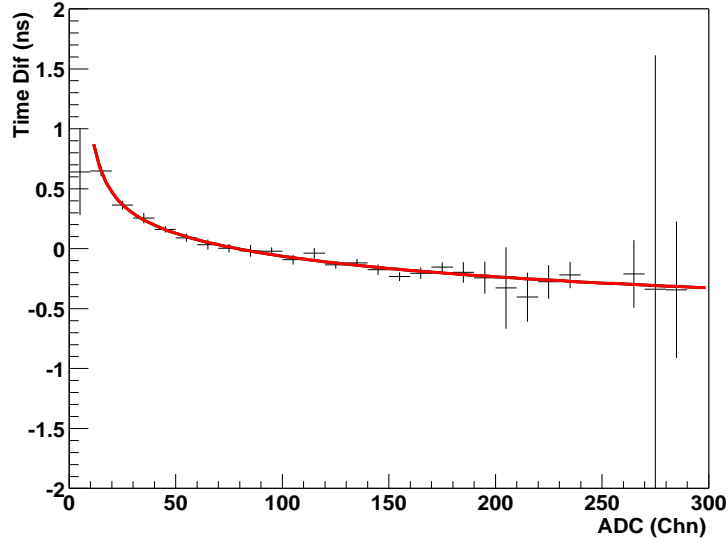}
\caption{The slewing correction.} \label{slewingplot}
\end{figure}

\begin{figure}[h]
\centering
\includegraphics[height=18pc,width=24pc]{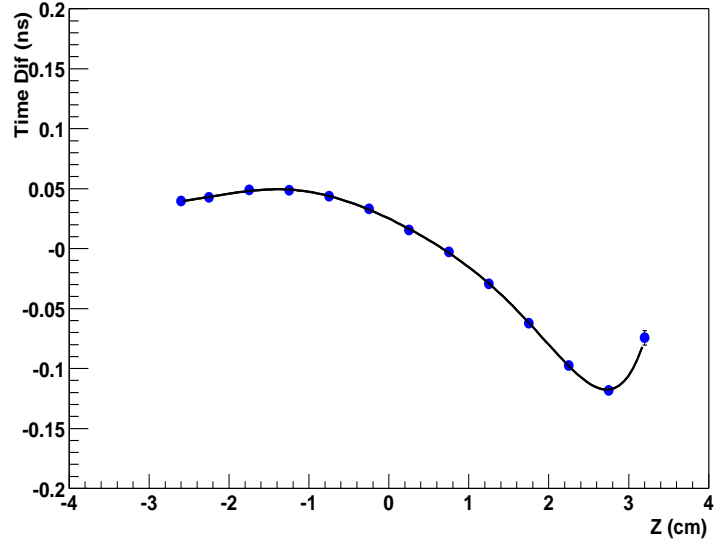}
\caption{The z position correction.} \label{ZFit_forthesis}
\end{figure}

\begin{figure}[h]
\begin{minipage}[t]{80mm}
\includegraphics[height=13pc,width=18pc]{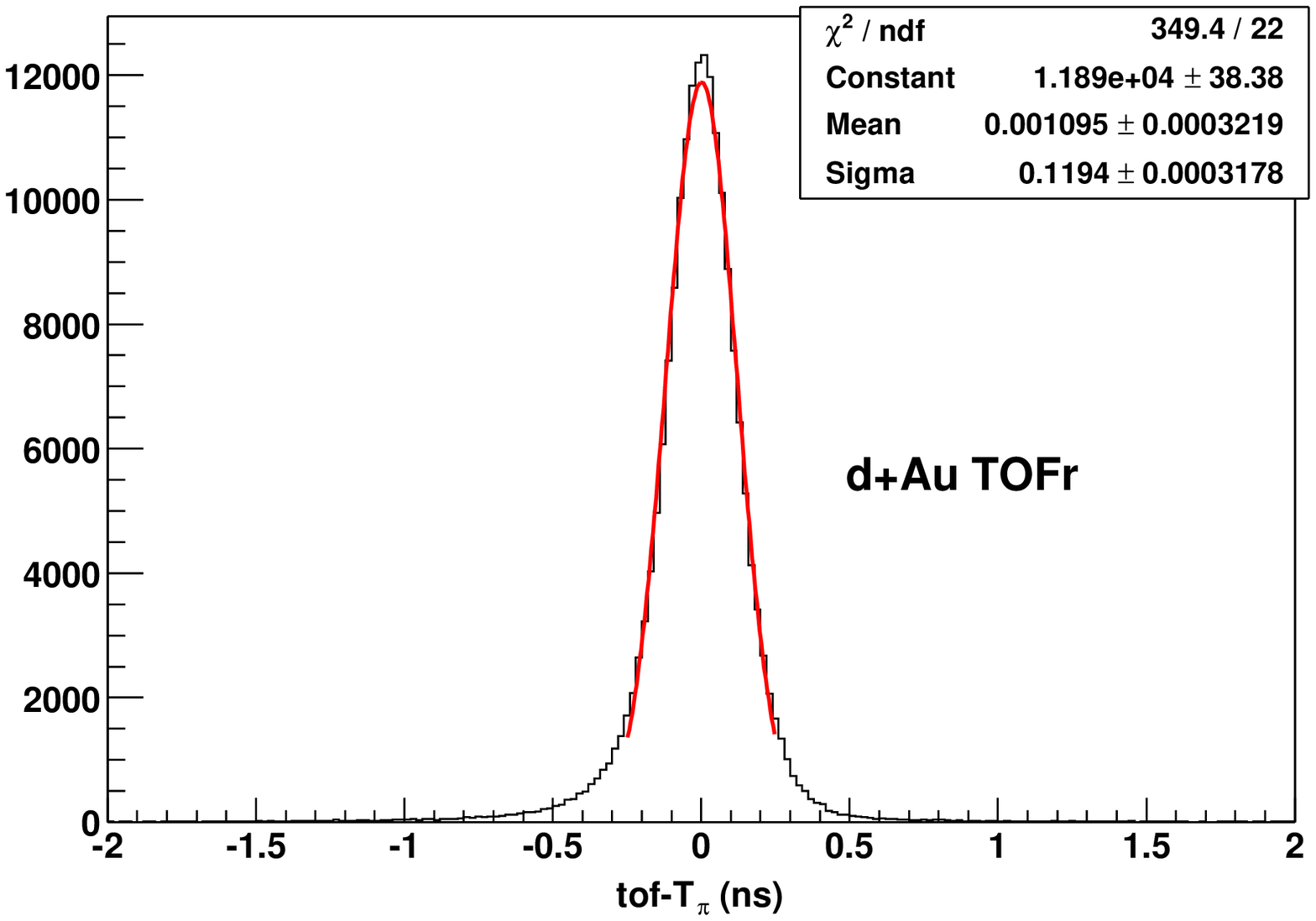}
\end{minipage}
\hspace{\fill}
\begin{minipage}[t]{80mm}
\includegraphics[height=13pc,width=18pc]{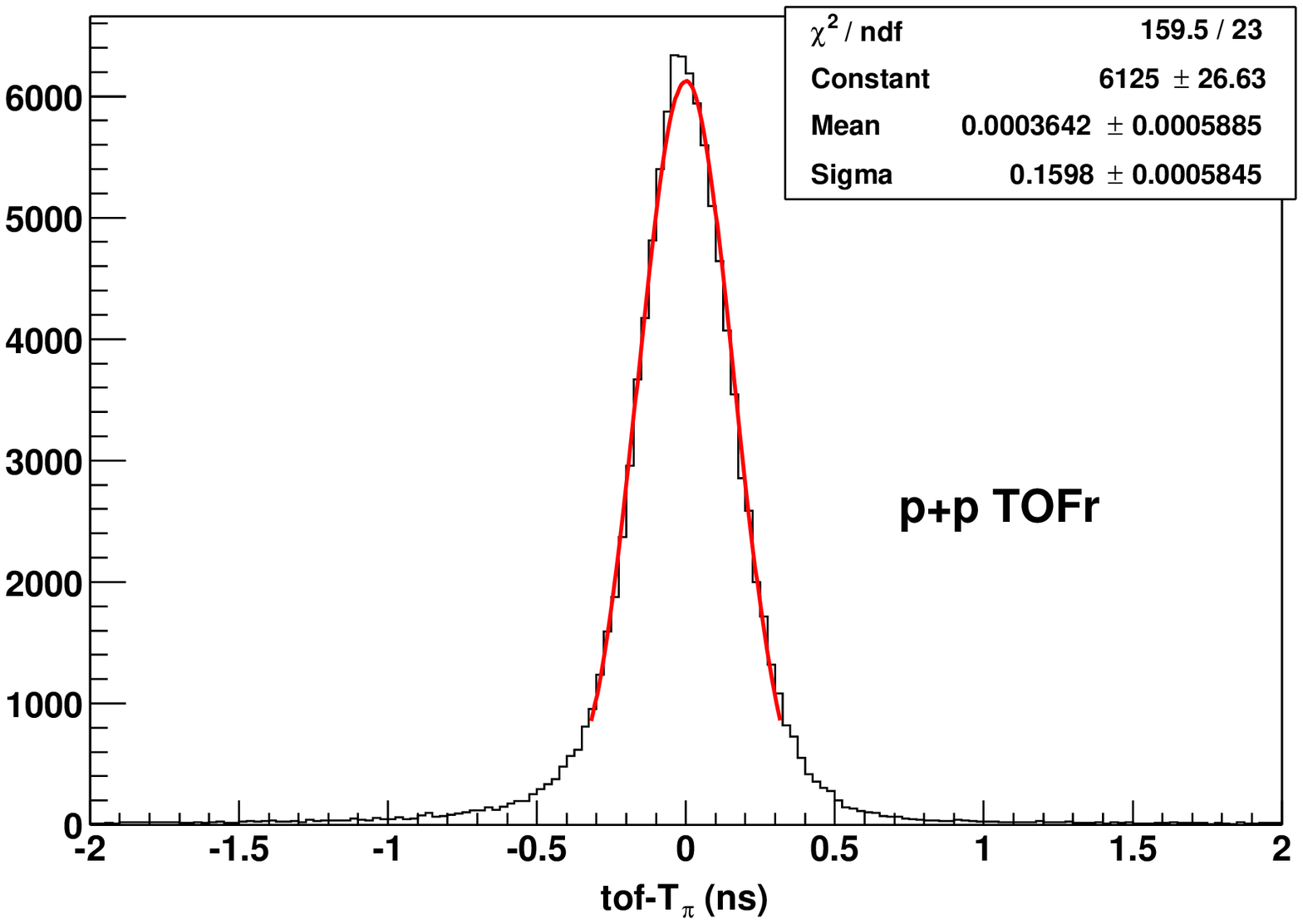}
\end{minipage}
\caption{The overall timing resolution after the calibration.}
\label{timeresolution}
\end{figure}

The difference between TOFr timing $T_{tofr}$ and start timing
$T_{start}$ is our time of flight $tof=T_{tofr}-T_{start}$. To
calibrate the $tof$, the pure pion sample was chosen by selecting
the particle energy loss $dE/dx$ in TPC at
$dE/dx<0.028\times{10^{-4}}$ GeV/cm in the momentum range
$0.3<p<0.6$ GeV/c. Figure~\ref{dAudedxplot} shows dE/dx vs $p$
plot from d+Au collisions. Firstly the so called $T_{0}$
correction was done due to the different cable lengths for
different read-out channels, which was done by shifting the mean
value of the distribution of $tof-T_{\pi}$ to zero channel by
channel, where $T_{\pi}$ is the calculation timing assuming the
particle was pion particle. Secondly, the slewing correction due
to correlation between timing and signal amplitude of the
electronics was done by getting the curve of $tof'-T_{\pi}$ vs
$adc$ for each channel, where the $tof'$ was the time of flight
after the $T_{0}$ correction and $adc$ was the ADC value of TOFr.
The slewing curve is like the plot shown in
Figure~\ref{slewingplot}.  The function of the slewing correction
is
$y=par[0]+par[1]/\sqrt{x}+par[2]/x+par[3]/\sqrt{x}/x+par[4]/x/x$.

The z position correction was also done since the different hit
positions on the read-out strip will generate different
transmission timing. This was done by getting the function of
$tof''-T_{\pi}$ versus $Z_{local}$, where the $tof''$ is the time
of flight after the $T_{0}$ and slewing correction, and
$Z_{local}$ is the the hit local z position of the TOFr. The
function for the z position correction is
$y=\sum_{i=0}^{7}{(par[i]\times{x^{i}})}$. The z position
correction for all the channels is shown in
Figure~\ref{ZFit_forthesis}. After the z position was done, the
calibration for TOFr was finished. The overall resolution of TOFr
was 120 ps and 160 ps in d+Au and p+p collisions respectively,
where the effective timing resolution of the pVPDs was 85 ps and
140 ps, respectively. Figure~\ref{timeresolution} shows the
overall resolution of TOFr in d+Au and p+p collisions.

\section{Raw yield}

\begin{figure}[h]
\centering
\includegraphics[height=18pc,width=24pc]{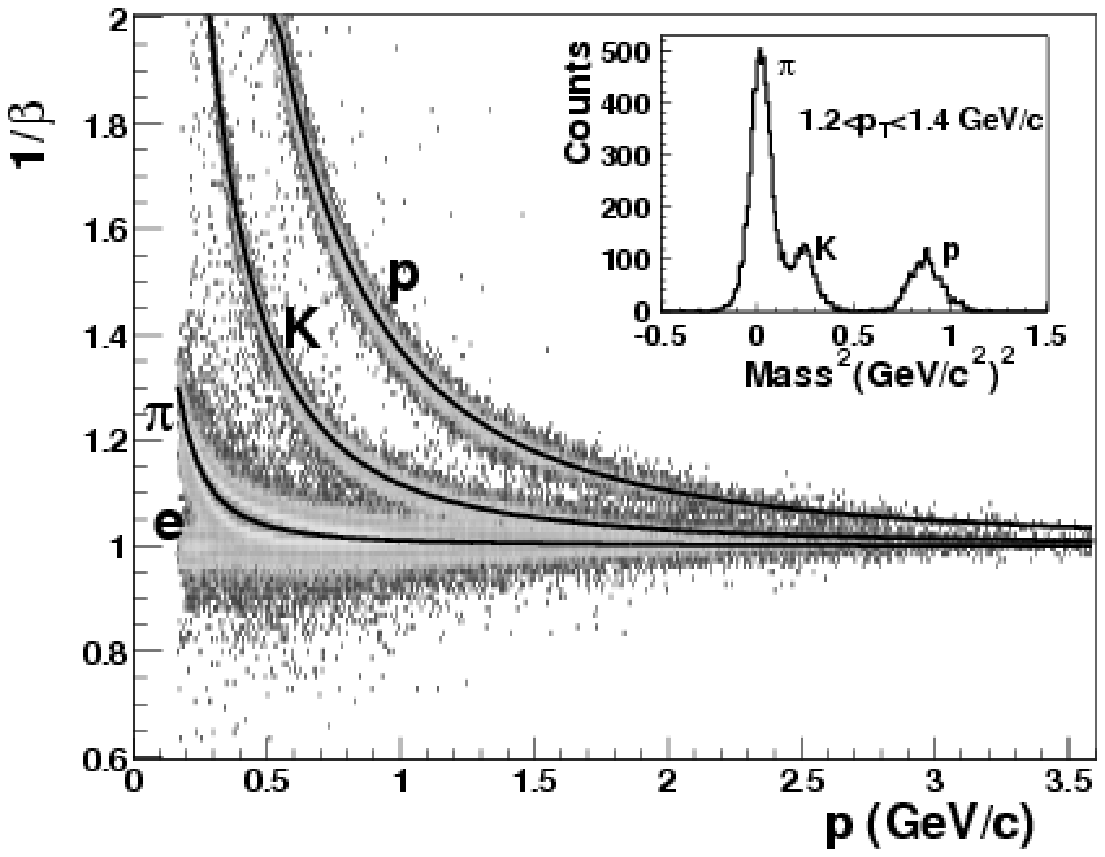}
\caption{$1/\beta$ vs. momentum for $\pi^{\pm}$, $K^{\pm}$, and
$p(\bar{p})$ from 200 GeV d+Au collisions. Separations between
pions and kaons, kaons and protons are achieved up to
$p_{T}\simeq1.6$ and $3.0$ GeV/c, respectively. The insert shows
$m^{2}=p^{2}(1/\beta^{2}-1)$ for $1.2<p_{T}<1.4$ GeV/c. Clear
separation of $\pi$, $K$ and $p$ is seen.} \label{beta}
\end{figure}
From the timing information $t$ from TOFr after the calibration
and the pathlength $L$ from TPC, the velocity $\beta$ of the
particle can be easily got by $\beta=L/t/c$. Figure~\ref{beta}
shows $1/\beta$ from TOFr measurement as a function of momentum
($p$) calculated from TPC tracking in TOFr triggered d+Au
collisions. The raw yields of $\pi^{\pm}$, $K^{\pm}$, $p$ and
$\bar{p}$ are obtained from Gaussian fits to the distributions in
$m^{2}=p^{2}(1/\beta^{2}-1)$ in each $p_{T}$ bin.

\subsection{$\pi$ raw yield extraction}
For $\pi^{\pm}$, the rapidity range is $-0.5<y_{\pi}<0.$. After
$|N_{\sigma\pi}|<2$ was required, the mass squared
$m^{2}=p^{2}(1/\beta^{2}-1)$ distributions in different $p_{T}$
bin in d+Au minimum-bias collisions are shown is
Figure~\ref{pionplusrawyieldplot} and
Figure~\ref{pionminusrawyieldplot}. At $p_{T}<0.8$ GeV/c, the
single Gaussian function was used to fit the distribution of
$m^{2}$ to get the raw yield. At the same time, the counting
result by counting the track number at the range $-0.1<m^{2}<0.1$
$(GeV/c^{2})^2$ was also used to compare with the raw yield from
the fitting method. The difference between them was found in one
sigma range. The raw yield we quote is from the fitting method. At
$p_{T}>0.8$ GeV/c, the double Gaussian function was used to
extract the raw yield. The raw signals in each $P_{T}$ bin are
shown in Table~\ref{pionplustable} and Table~\ref{pionminustable}.
Also shown in the tables are those in centrality selected d+Au
collisions and minimum-bias p+p collisions.

\subsection{$K$ raw yield extraction}

For $K^{\pm}$, the rapidity range is $-0.5<y_{K}<0$. After
$|N_{\sigma K}|<2$ was required, the mass squared
$m^{2}=p^{2}(1/\beta^{2}-1)$ distributions in different $p_{T}$
bin in d+Au minimum-bias collisions are shown is
Figure~\ref{kaonplusrawyieldplot} and
Figure~\ref{kaonminusrawyieldplot}. At $p_{T}<0.8$  GeV/c, the
single Gaussian function was used to fit the distribution of
$m^{2}$ to get the raw yield. At the same time, the counting
result by counting the track number at the range $0.16<m^{2}<0.36$
$(GeV/c^{2})^2$ was also used to compare with the raw yield from
the fitting method. The difference between them was found in one
sigma range. The raw yield we quote is from the fitting method. At
$p_{T}>0.8$ GeV/c, the double Gaussian function was used to
extract the raw yield. The raw signals in each $P_{T}$ bin are
shown in Table~\ref{kaonplustable} and Table~\ref{kaonminustable}.
Also shown in the tables are those in centrality selected d+Au
collisions and minimum-bias p+p collisions.

\subsection{$p$ and $\bar{p}$ raw yield extraction}
\begin{figure}[h]
\centering
\includegraphics[height=18pc,width=24pc]{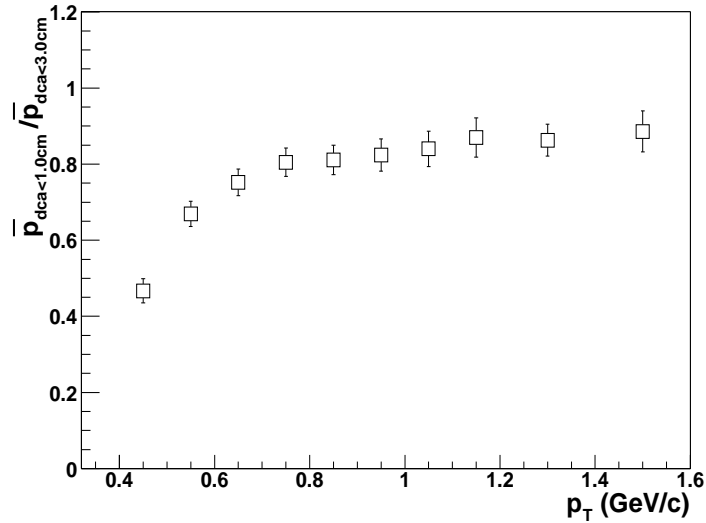}
\caption{the ratio of $\bar{p}$ at $dca<1.0$ cm over $\bar{p}$ at
$dca<3.0$ cm.} \label{pbardcaratio}
\end{figure}
For $\bar{p}$, the rapidity range is $-0.5<y_{\bar{p}}<0$. After
$|N_{\sigma p}|<2$ was required, the mass squared
$m^{2}=p^{2}(1/\beta^{2}-1)$ distributions in different $p_{T}$
bin in d+Au minimum-bias collisions are shown is
Figure~\ref{pbarrawyieldplot}. At $p_{T}<1.6$ GeV/c, the single
Gaussian function was used to fit the distribution of $m^{2}$ to
get the raw yield. At the same time, the counting result by
counting the track number at the range $0.64<m^{2}<1.44$
$(GeV/c^{2})^2$ was also used to compare with the raw yield from
the fitting method. The difference between them was found in one
sigma range. The raw yield we quote is from the fitting method. At
$p_{T}>1.6$ GeV/c, the double Gaussian function was used to
extract the raw yield. The raw signals in each $P_{T}$ bin are
shown in Table~\ref{pbartable}. For the $p$, the raw yield
extraction method is the same as $\bar{p}$ except that at
$p_{T}<1.6$
 GeV/c, we use the method
$Np=Np_{dca<1.cm}\times{(N\bar{p}_{dca<3.cm}/N\bar{p}_{dca<1.cm})
}$ to reject the background, where $Np$ and $N\bar{p}$ are the
number of the $p$ and $\bar{p}$ tracks individually, and
$N\bar{p}_{dca<1.cm}/N\bar{p}_{dca<3.cm}$ is the ratio of
$\bar{p}$ tracks at $dca<1.0$ cm over those at $dca<3.0$ cm. In
Figure~\ref{protonrawyieldplot}, the first 10 $p_{T}$ bins are for
$dca<1.0$ cm, the last 4 $p_{T}$ bins are for $dca<3.0$ cm.
Figure~\ref{pbardcaratio} shows the ratio of $\bar{p}$ tracks at
$dca<1.0$ cm over those at $dca<3.0$ cm. After this correction of
$Np=Np_{dca<1.cm}\times{(N\bar{p}_{dca<3.cm}/N\bar{p}_{dca<1.cm})
}$, the $p$ raw signals in each $P_{T}$ bin are shown in
Table~\ref{protontable}.

\section{Efficiency and acceptance correction}

\begin{figure}[h]
\begin{minipage}[t]{80mm}
\includegraphics[height=13pc,width=18pc]{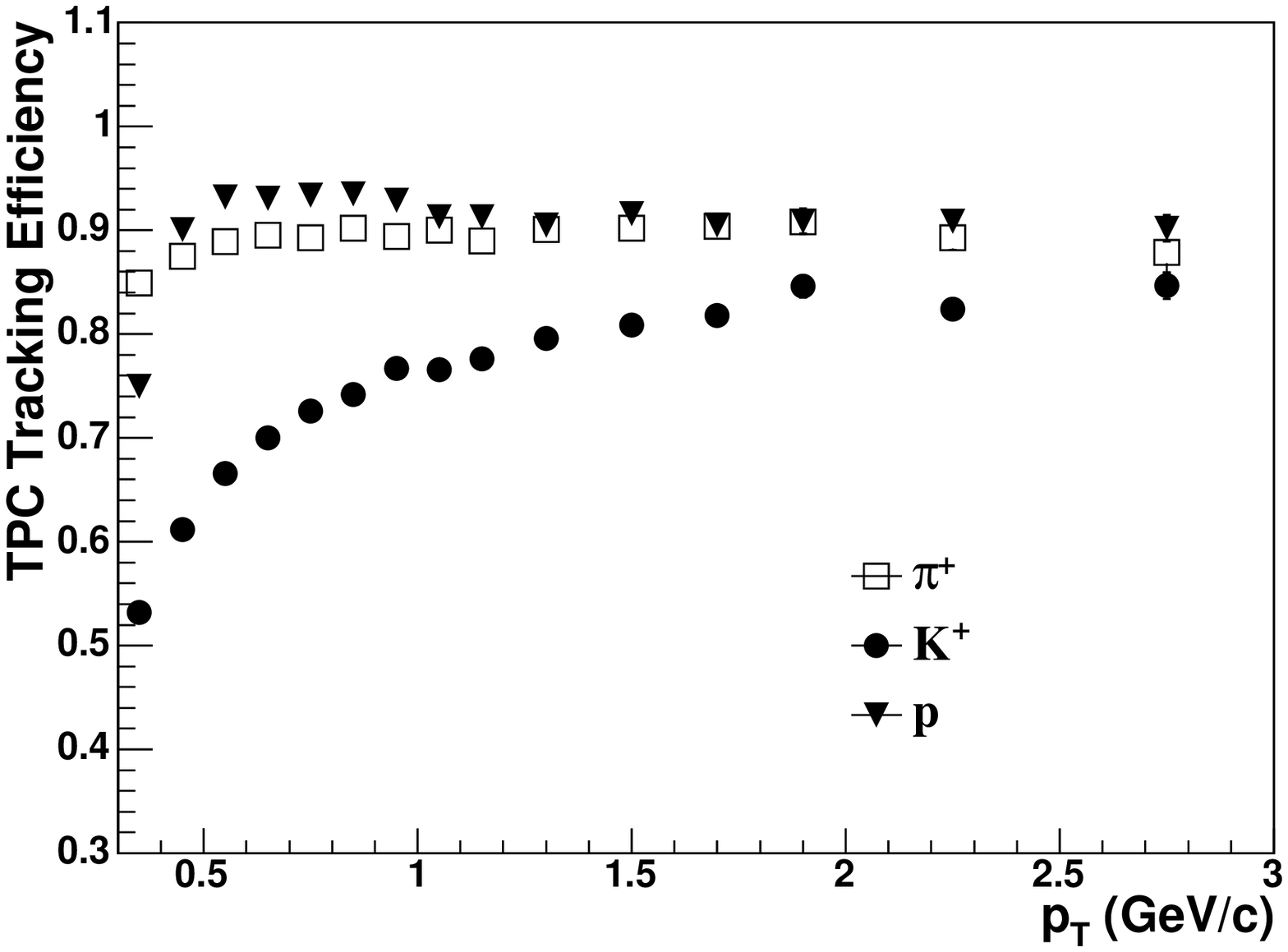}
\end{minipage}
\hspace{\fill}
\begin{minipage}[t]{80mm}
\includegraphics[height=13pc,width=18pc]{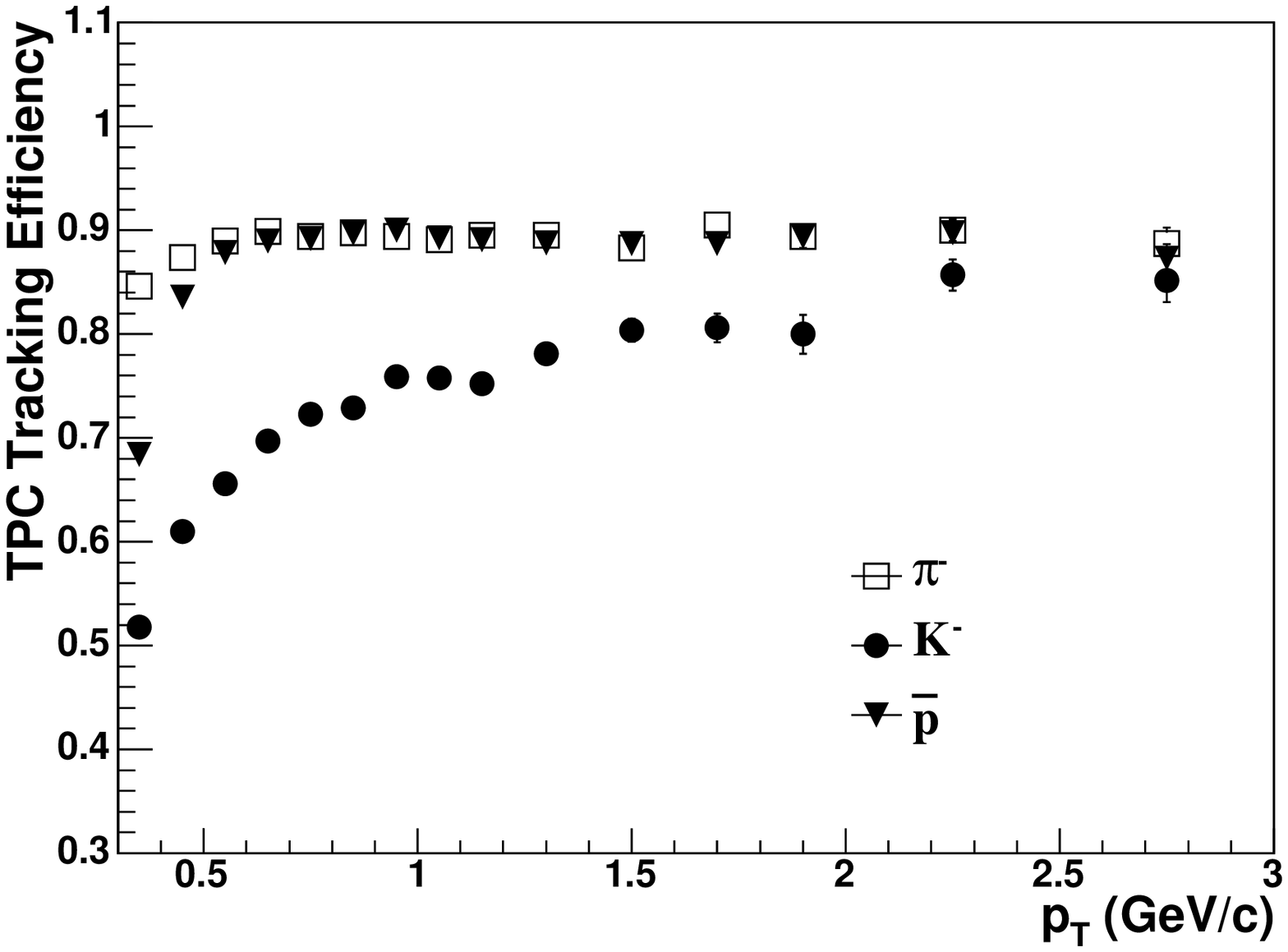}
\end{minipage}
\caption{TPC reconstruction efficiency of $\pi^{\pm}$, $K^{\pm}$,
$p$ and $\bar{p}$ as a function of $p_{T}$. The left plot for
charged plus particle and the right for charged minus particle. }
\label{tpceff}
\end{figure}

\begin{figure}[h]
\begin{minipage}[t]{80mm}
\includegraphics[height=13pc,width=18pc]{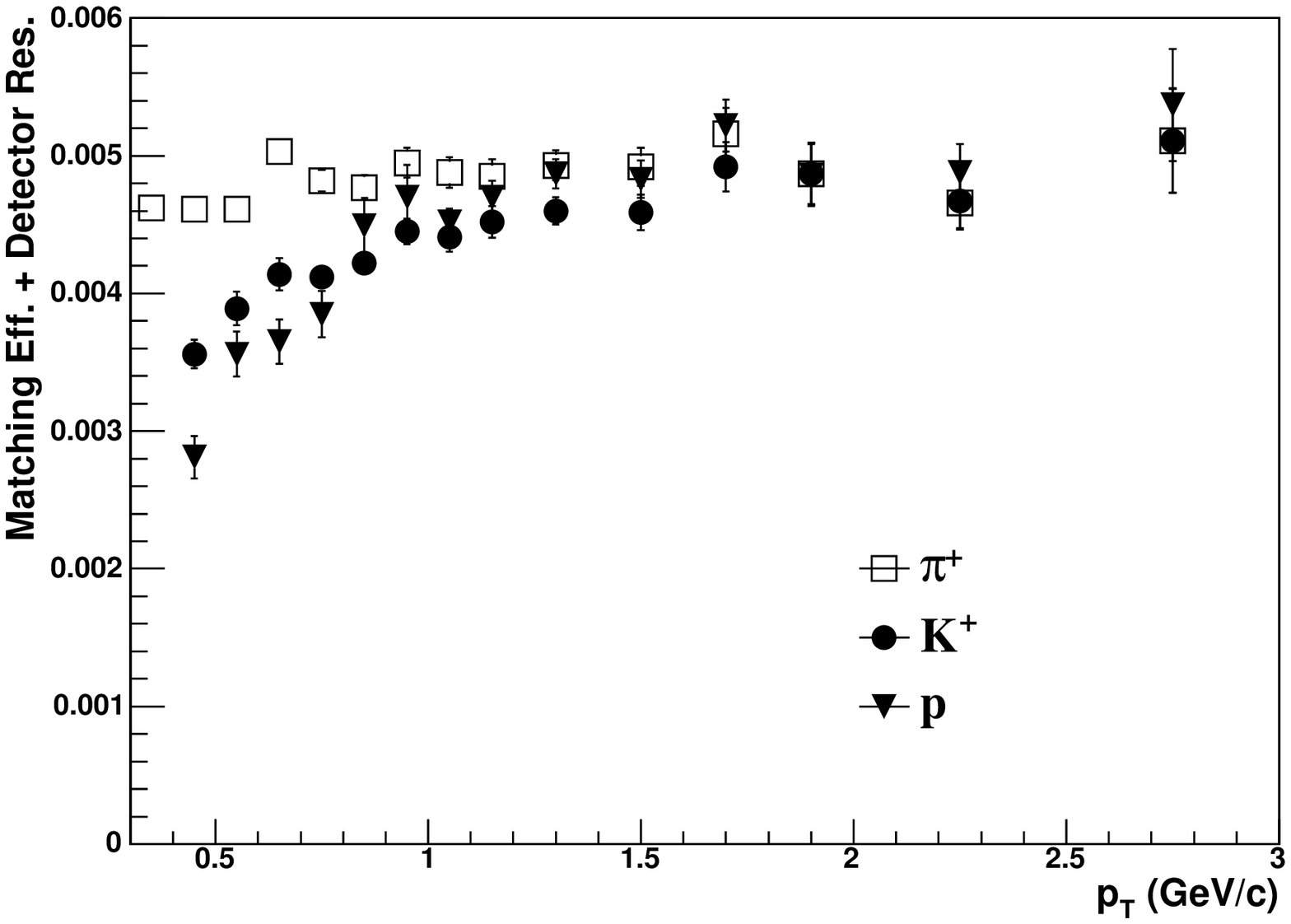}
\end{minipage}
\hspace{\fill}
\begin{minipage}[t]{80mm}
\includegraphics[height=13pc,width=18pc]{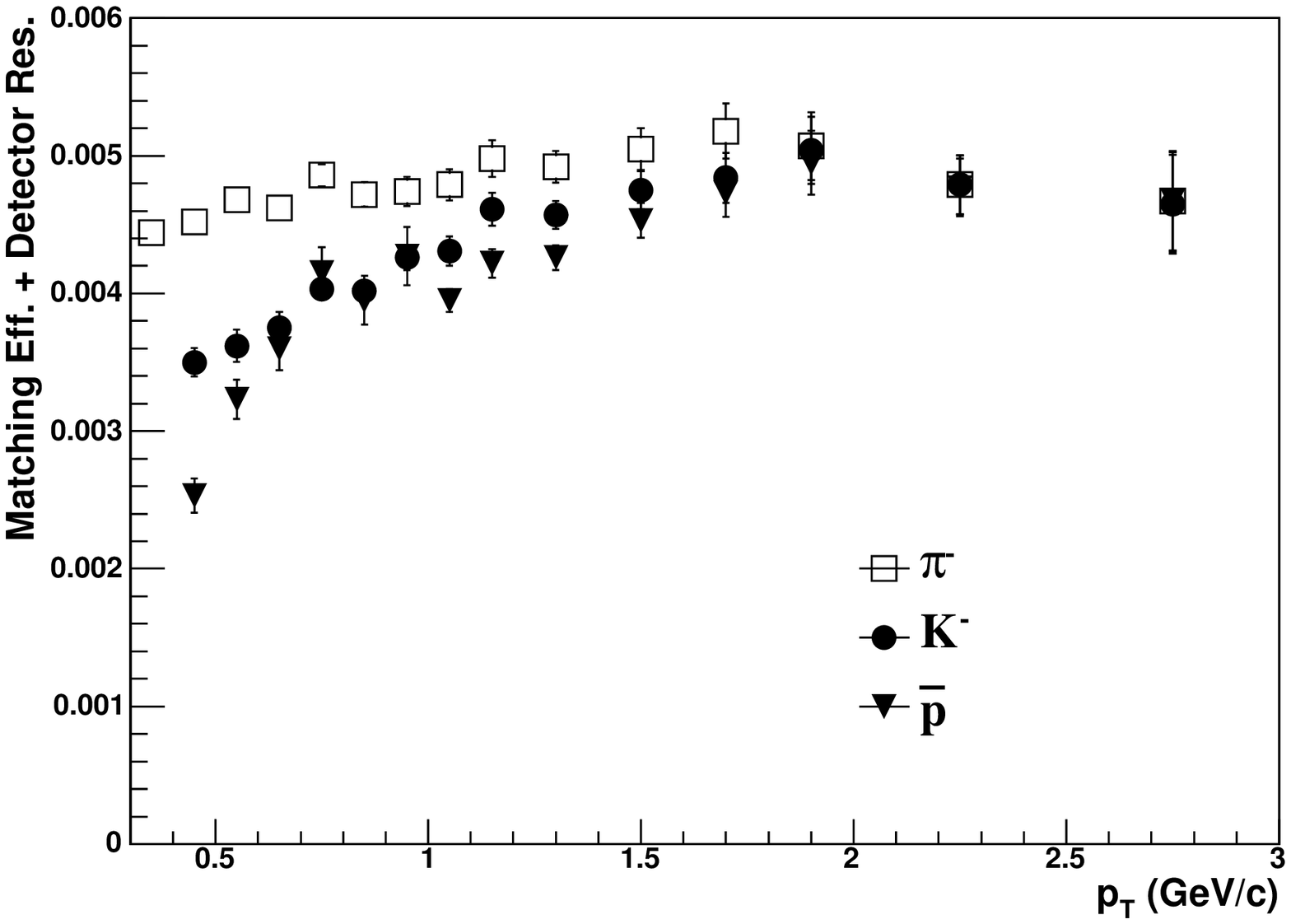}
\end{minipage}
\caption{Matching efficiency from TOFr to TPC of $\pi^{\pm}$,
$K^{\pm}$, $p$ and $\bar{p}$ as a function of $p_{T}$, including
detector response. The left plot for charged plus particle and the
right for charged minus particle.} \label{matcheff}
\end{figure}

Acceptance and efficiency were studied by Monte Carlo simulations
and by matching TPC track and TOFr hits in real data.
TPC tracking efficiency was studied by Monte Carlo simulations.
The simulated $\pi^{\pm}$, $K^{\pm}$, $p$ and $\bar{p}$ are
generated using a flat $p_T$ and a flat $y$ distribution and pass
through GSTAR~\cite{long:01} (the framework software package to
run the STAR detector simulation using
GEANT~\cite{geant:01,geant:02}) and TRS (the TPC Response
Simulator~\cite{long:01}). The simulated $\pi^{\pm}$, $K^{\pm}$,
$p$ and $\bar{p}$ are then combined with a real raw event and we
call this combined event a simulated event. This simulated event
is then passed through the standard STAR reconstruction chain and
we call this event after reconstruction a reconstructed event. The
reconstructed information of those particles in the reconstructed
event is then associated with the Monte-Carlo information in the
simulated event. And then we get the total number of simulated
$\pi^{\pm}$, $K^{\pm}$, $p$ and $\bar{p}$ from simulated events in
a certain transverse momentum bin. Also we can get the total
number of associated tracks in the reconstructed events in this
transverse momentum bin~\cite{Haibin:03}. In the end, take the
ratio of the number of associated $\pi^{\pm}$, $K^{\pm}$, $p$ and
$\bar{p}$ over the number of simulated $\pi^{\pm}$, $K^{\pm}$, $p$
and $\bar{p}$ and this ratio is the TPC reconstruction efficiency
for a certain transverse momentum bin in the mid-rapidity range.
Figure~\ref{tpceff} shows the TPC reconstruction efficiency of
$\pi^{\pm}$, $K^{\pm}$, $p$ and $\bar{p}$ as a function of
$p_{T}$.

\begin{figure}[h]
\centering
\includegraphics[height=18pc,width=24pc]{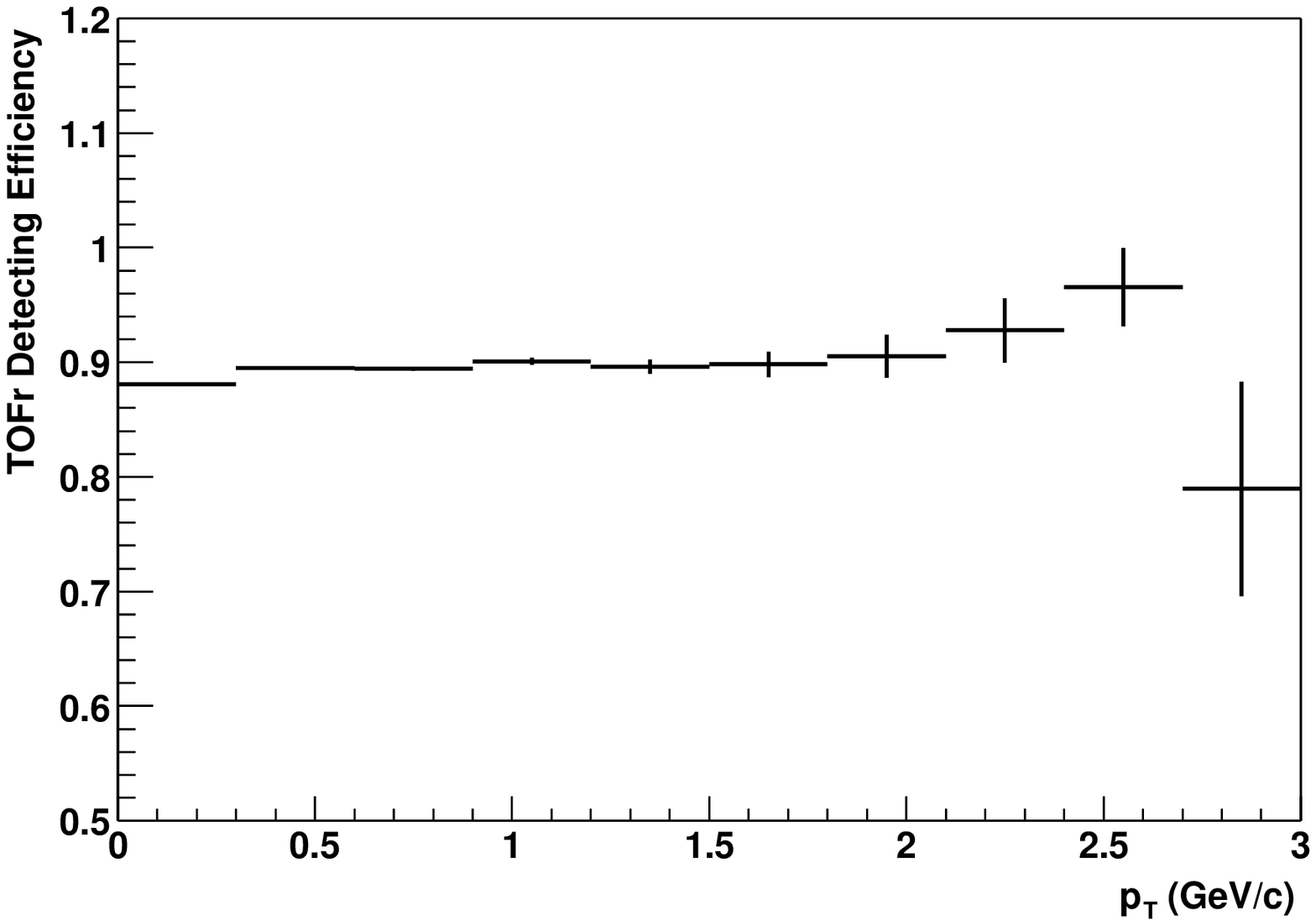}
\caption{The TOFr response efficiency as a function of $p_{T}$.}
\label{detectorresponse}
\end{figure}

 The Matching Efficiency from TPC to TOFr were studied in
real data, and the formula are
\begin{equation}
Eff_{Match}=\frac{TofrMatchedTracks/dAuTOFrEvents}
{(MinBiasTracks/MinBiasEvents)_{pVPD}\times{factor1}\times{factor2}}
\end{equation}
where the $TofrMatchedTracks/dAuTOFrEvents$ is the number of TOFr
matched tracks per dAuTOFr trigger event,
$(MinBiasTracks/MinBiasEvents)_{pVPD}$ is the number of
minimum-bias tracks per minimum-bias event by requiring the pVPD
to fire, $factor1$ is the enhancement factor of dAuTOFr trigger,
and $factor2$ is the other factors such as the TOFr trip factor.
The $Eff_{Match}$ includes the detector response efficiency.
Figure~\ref{matcheff} shows the matching efficiency of different
particle species including the detector response versus $p_{T}$.
The detector response efficiency, including the material
absorption and scattering effect between TPC and TOFr, as a
function of $p_{T}$ is shown in Figure~\ref{detectorresponse},
which is around 90\% at $p_{T}>$ 0.3 GeV/c. After the material
absorption and scattering effect correction, the detector response
efficiency is around 95\%.

\section{Background correction}

\begin{figure}[h]
\centering
\includegraphics[height=18pc,width=18pc]{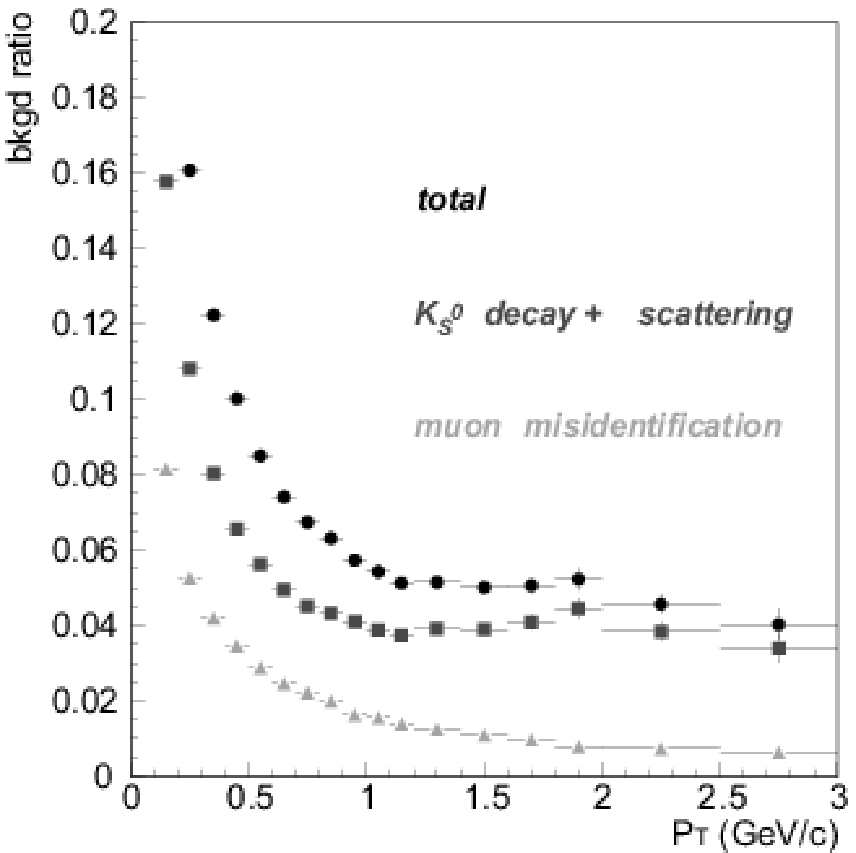}
\caption{$\pi$ background contribution as a function of $p_{T}$.
The circled symbols represent the total $\pi$ background
contribution including feed-down and $\mu$ misidentification. The
squared and triangled symbols represent the week-decay and $\mu$
misidentification contributions individually.}
\label{pionbackground}
\end{figure}

Weak-decay feeddown (e.g. $K_{s}^{0}\rightarrow\pi^{+}\pi^{-}$) to
pions is $\sim12\%$ at low $p_{T}$ and $\sim5\%$ at high $p_{T}$,
and was corrected for using PYTHIA~\cite{pythia} and
HIJING~\cite{hijing} simulations, as shown in
Figure~\ref{pionbackground}. For $\pi$ spectra, the $\mu$
misidentification was also corrected for, which is also shown in
Figure~\ref{pionbackground}.
\begin{figure}[h]
\centering
\includegraphics[height=18pc,width=24pc]{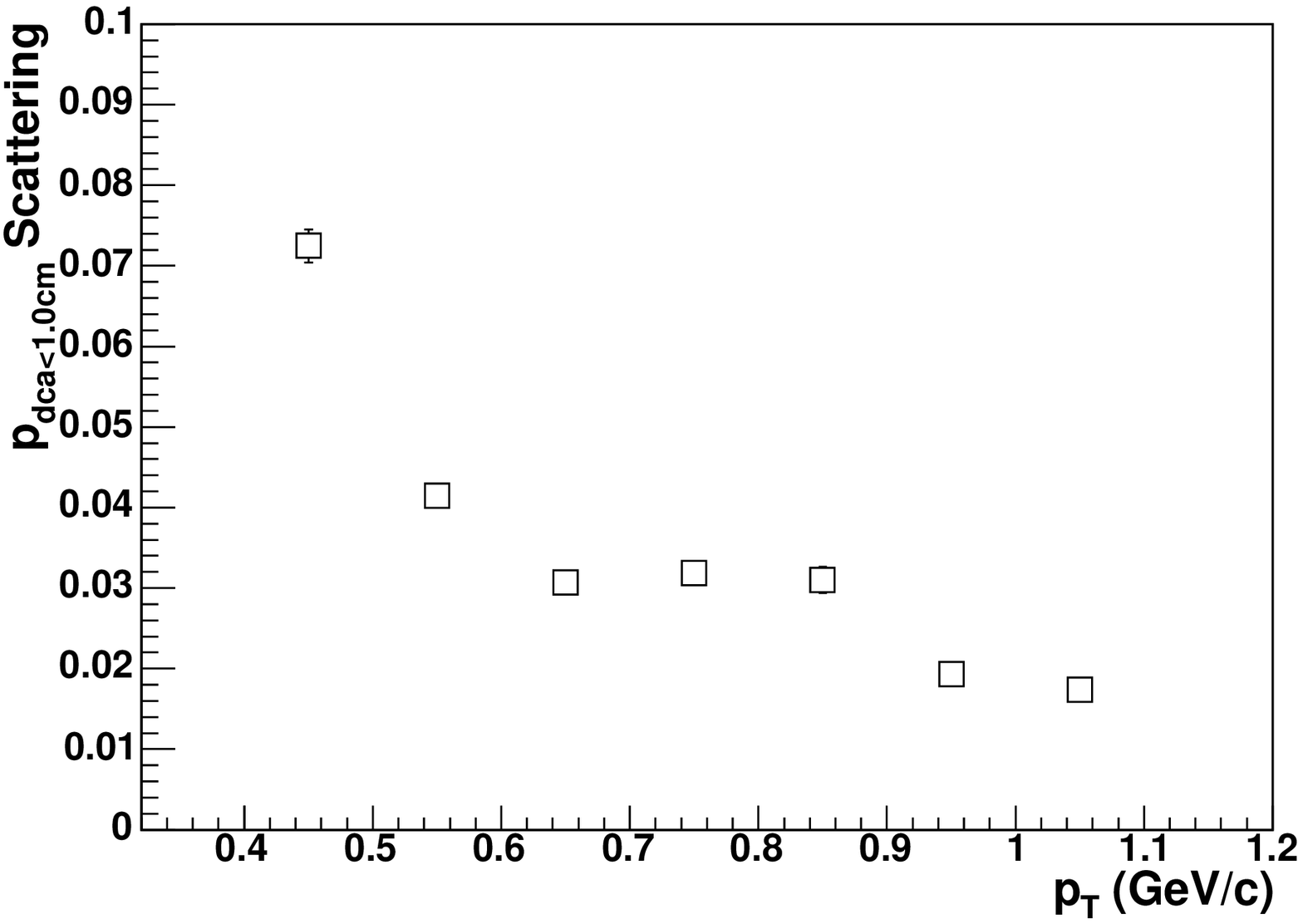}
\caption{The $p$ scattering effect contribution when we cut
$dca<1.0$ cm.} \label{protonScatter}
\end{figure}

Inclusive $p$ and $\bar{p}$ production is presented without
hyperon feeddown correction. $p$ and $\bar{p}$ from hyperon decays
have the same detection efficiency as primary $p$ and
$\bar{p}$~\cite{antiproton} and contribute about 20\% to the
inclusive $p$ and $\bar{p}$ yield, as estimated from the
simulation. However, for $p$, there is still some scattering
contribution which comes from the beam pipe interaction after the
cut of $dca<1.0$ cm. Figure~\ref{protonScatter} shows the
contribution of scattering effect for proton when we cut $dca<1.0$
cm. The correction is done at $p_{T}<$ 1.1 GeV/c and negligible at
higher $p_{T}$.

\section{Energy loss correction}
\begin{figure}[h]
\begin{minipage}[t]{80mm}
\includegraphics[height=13pc,width=18pc]{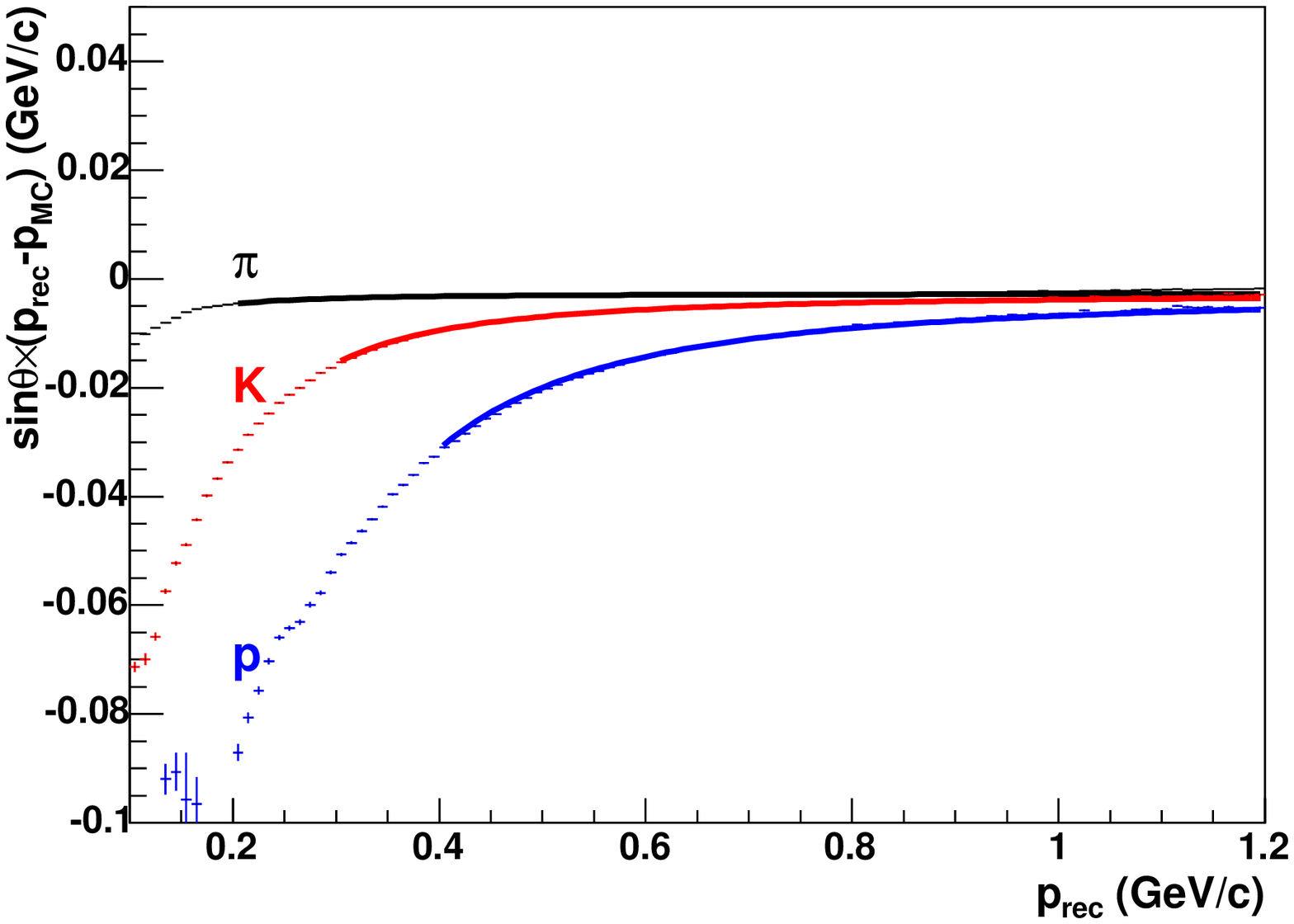}
\end{minipage}
\hspace{\fill}
\begin{minipage}[t]{80mm}
\includegraphics[height=13pc,width=18pc]{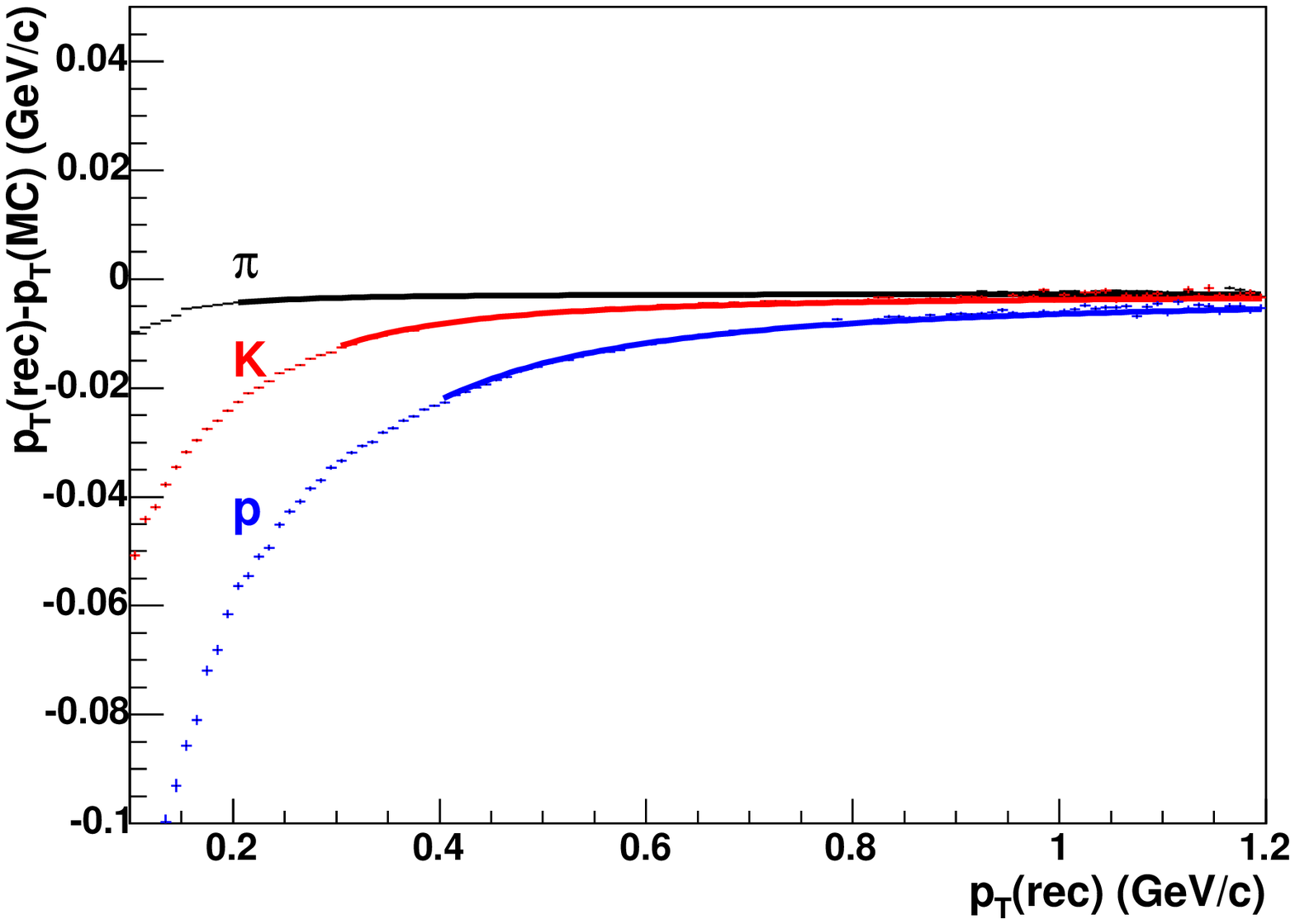}
\end{minipage}
\caption{(left) p energy loss correction of different particle
species as a function of p. $p_{rec}$ is the reconstructed
momentum before the energy loss correction, $p_{MC}$ is the
momentum after energy loss correction from simulation, $\theta$ is
the angle between the reconstructed momentum and beam line.
(right) $p_{T}$ energy loss correction of different particle
species as a function of $p_{T}$. $p_{T}(rec)$ is the
reconstructed transverse momentum before the energy loss
correction, $p_{T}(MC)$ is the transverse momentum after energy
loss correction from simulation. } \label{eloss}
\end{figure}

\begin{figure}[h]
\hspace{-3pc}
\includegraphics[height=40pc,width=40pc]{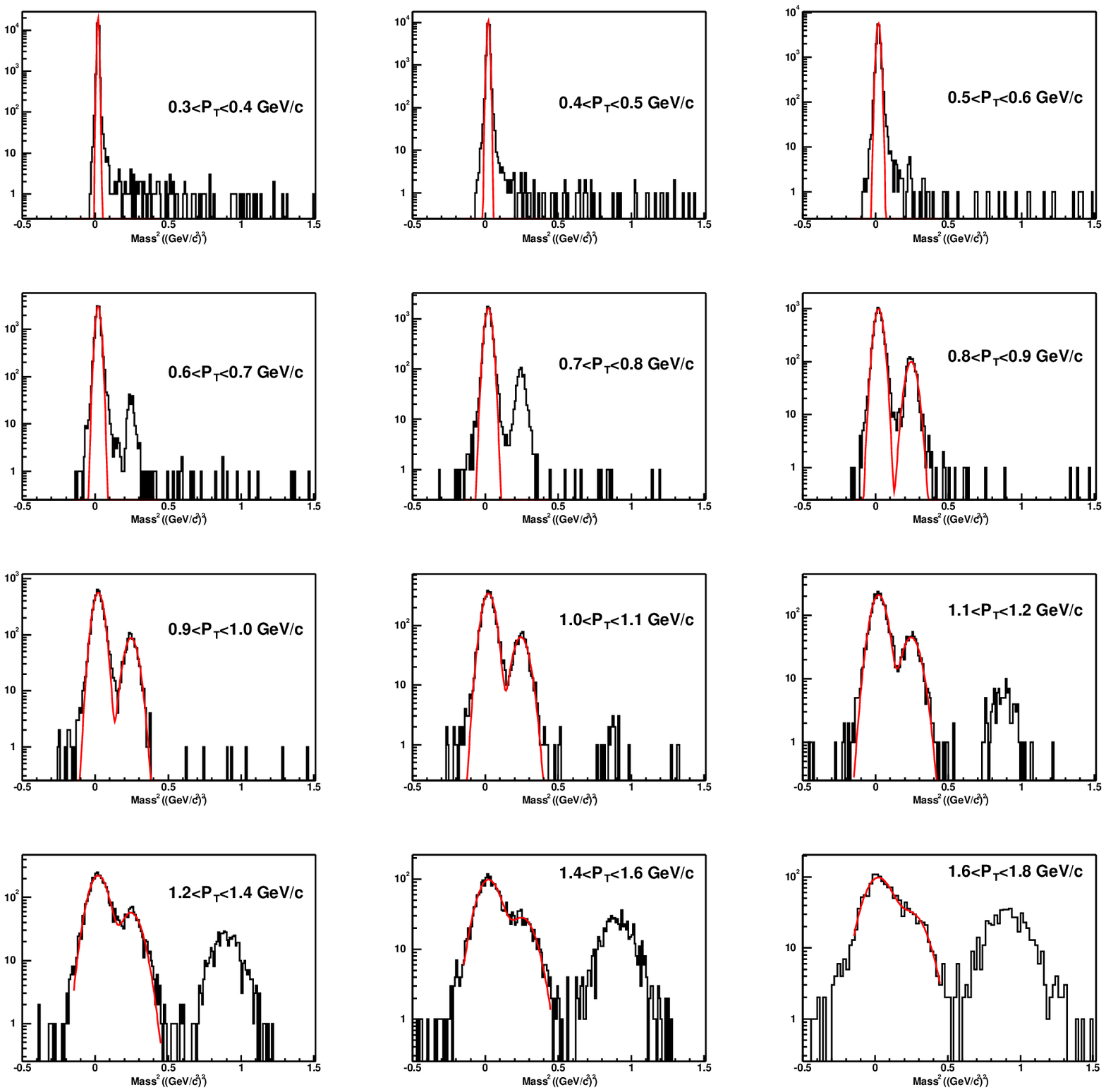}
\caption{$\pi^{+}$ raw yields versus mass squared distribution.
The histograms are our data. The curves are Gaussian fits.}
\label{pionplusrawyieldplot}
\end{figure}

\begin{figure}[h]
\hspace{-3pc}
\includegraphics[height=40pc,width=40pc]{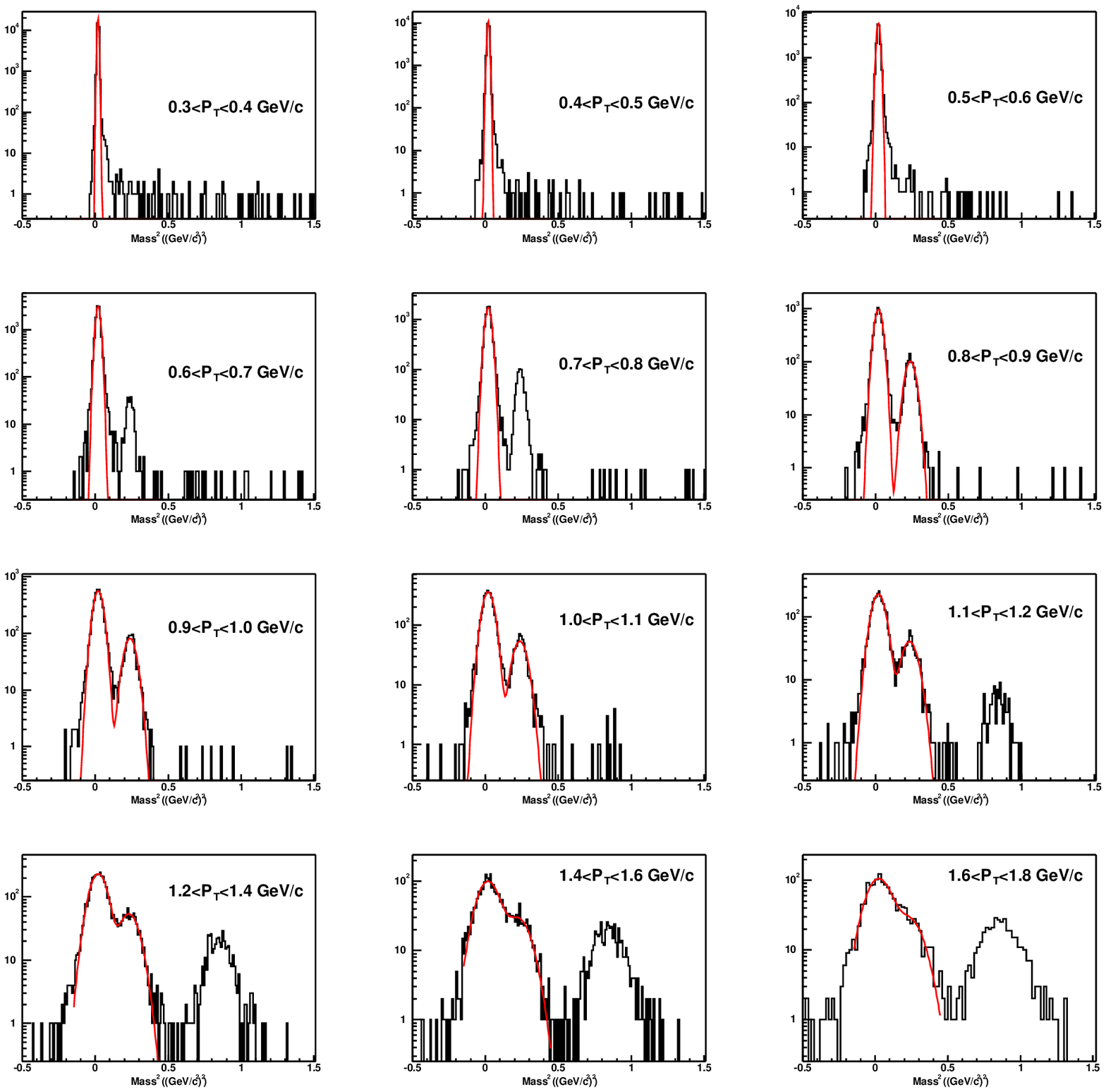}
\caption{$\pi^{-}$ raw yields versus mass squared distribution.
The histograms are our data. The curves are Gaussian fits.}
\label{pionminusrawyieldplot}
\end{figure}

\begin{figure}[h]
\hspace{-3pc}
\includegraphics[height=40pc,width=40pc]{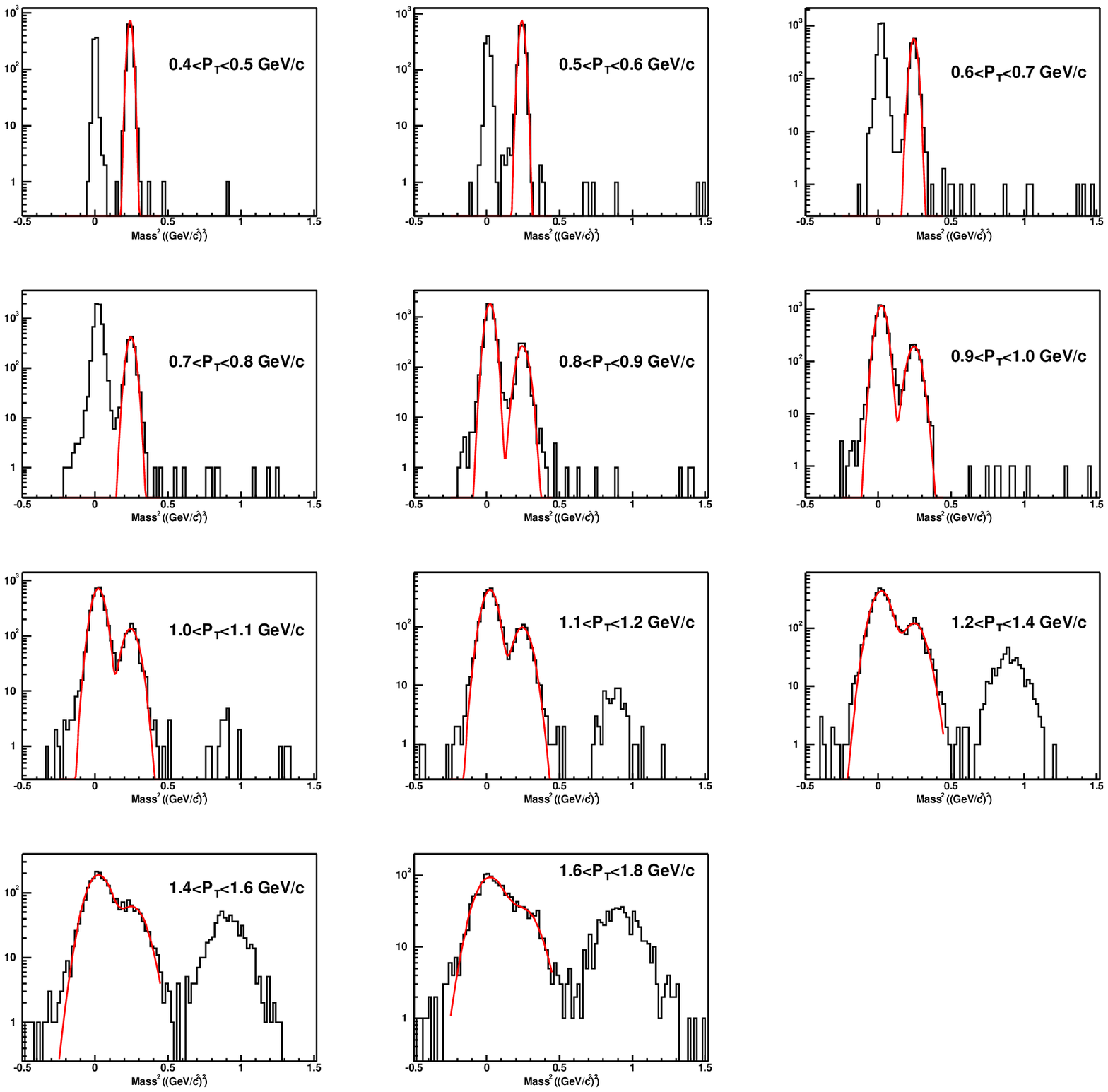}
\caption{$K^{+}$ raw yields versus mass squared distribution. The
histograms are our data. The curves are Gaussian fits.}
\label{kaonplusrawyieldplot}
\end{figure}

\begin{figure}[h]
\hspace{-3pc}
\includegraphics[height=40pc,width=40pc]{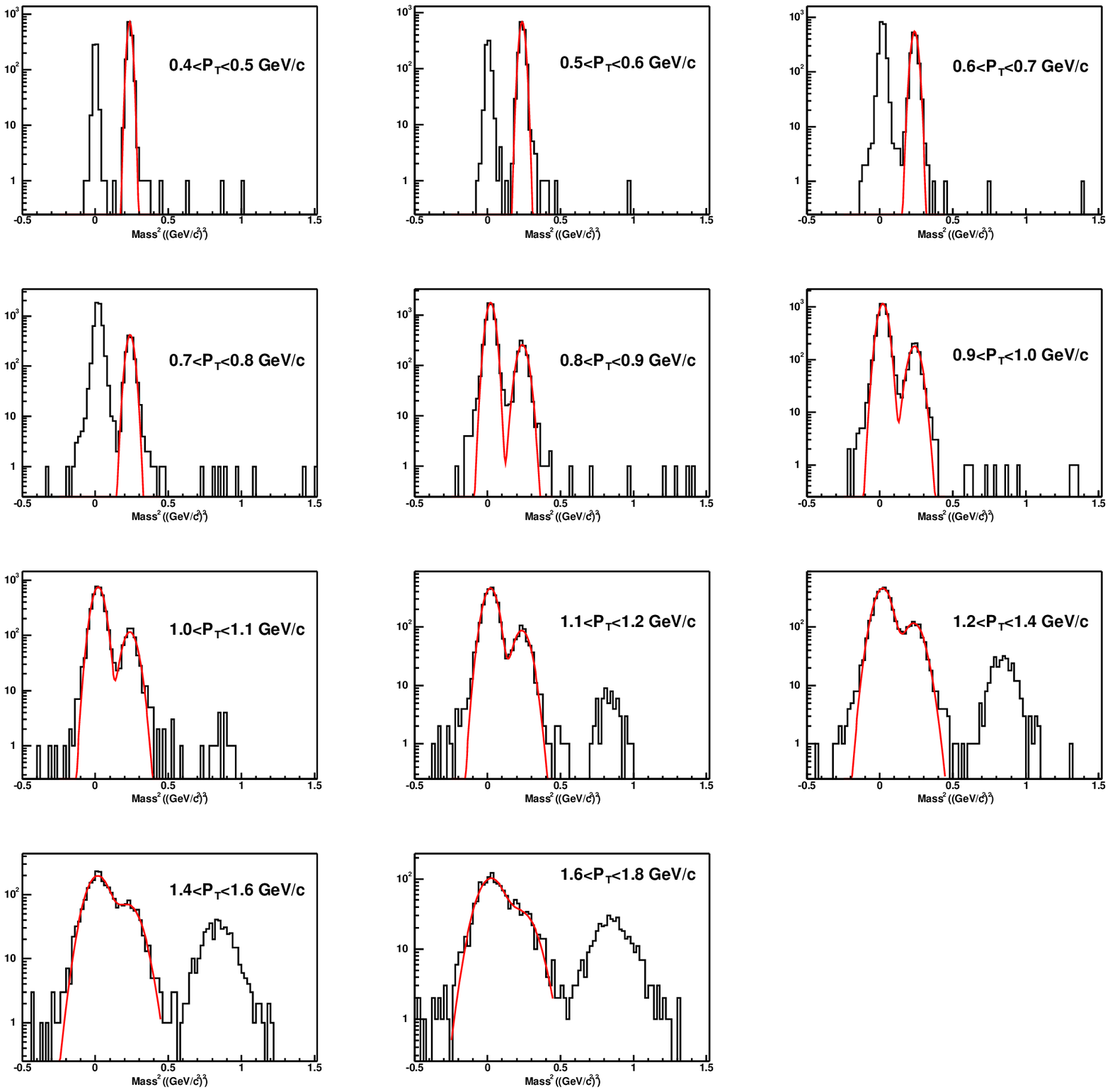}
\caption{$K^{-}$ raw yields versus mass squared distribution. The
histograms are our data. The curves are Gaussian fits.}
\label{kaonminusrawyieldplot}
\end{figure}

\begin{figure}[h]
\hspace{-3pc}
\includegraphics[height=40pc,width=40pc]{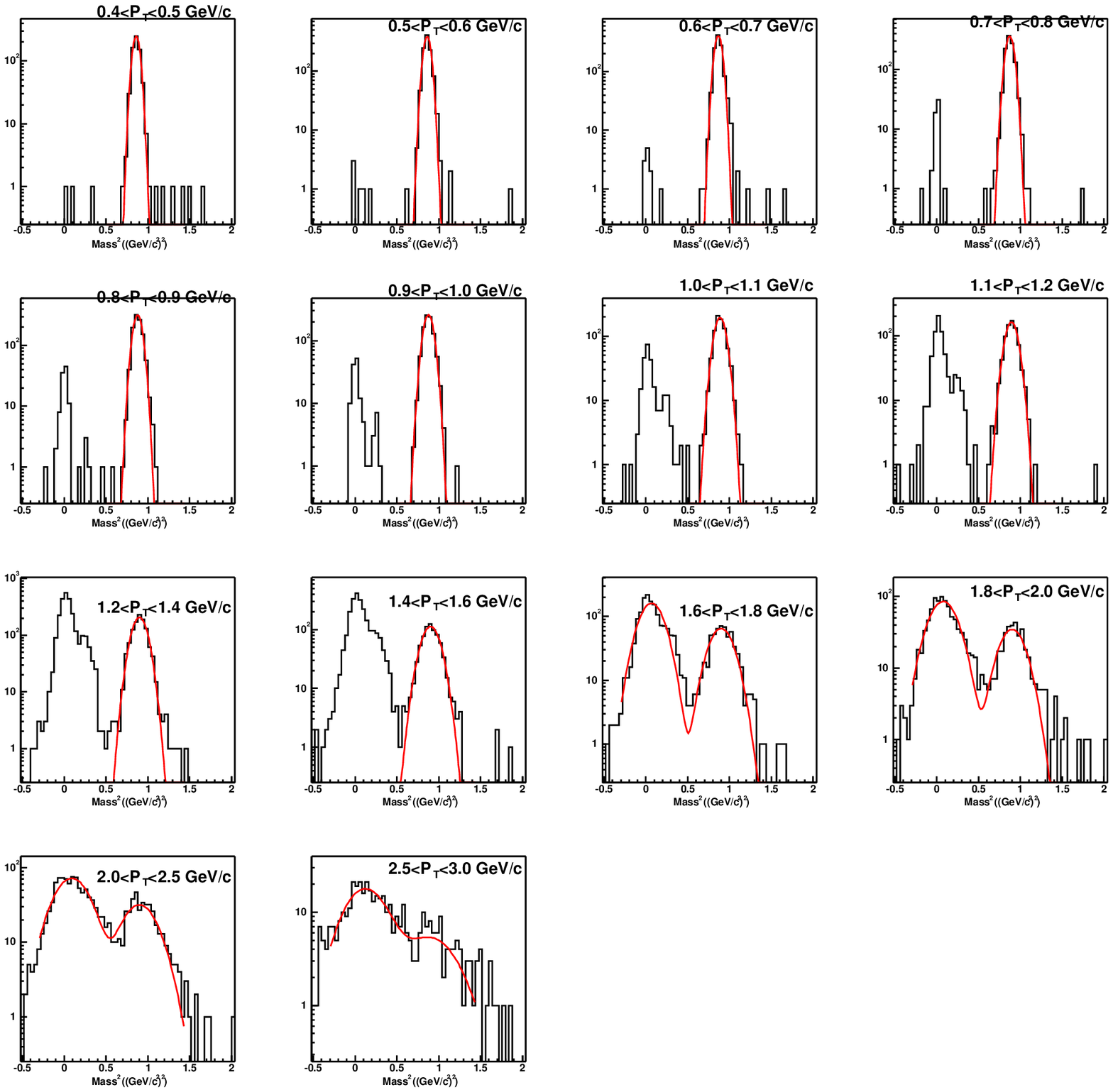}
\caption{$p$ raw yields versus mass squared distribution. The
histograms are our data. The curves are Gaussian fits.}
\label{protonrawyieldplot}
\end{figure}

\begin{figure}[h]
\hspace{-3pc}
\includegraphics[height=40pc,width=40pc]{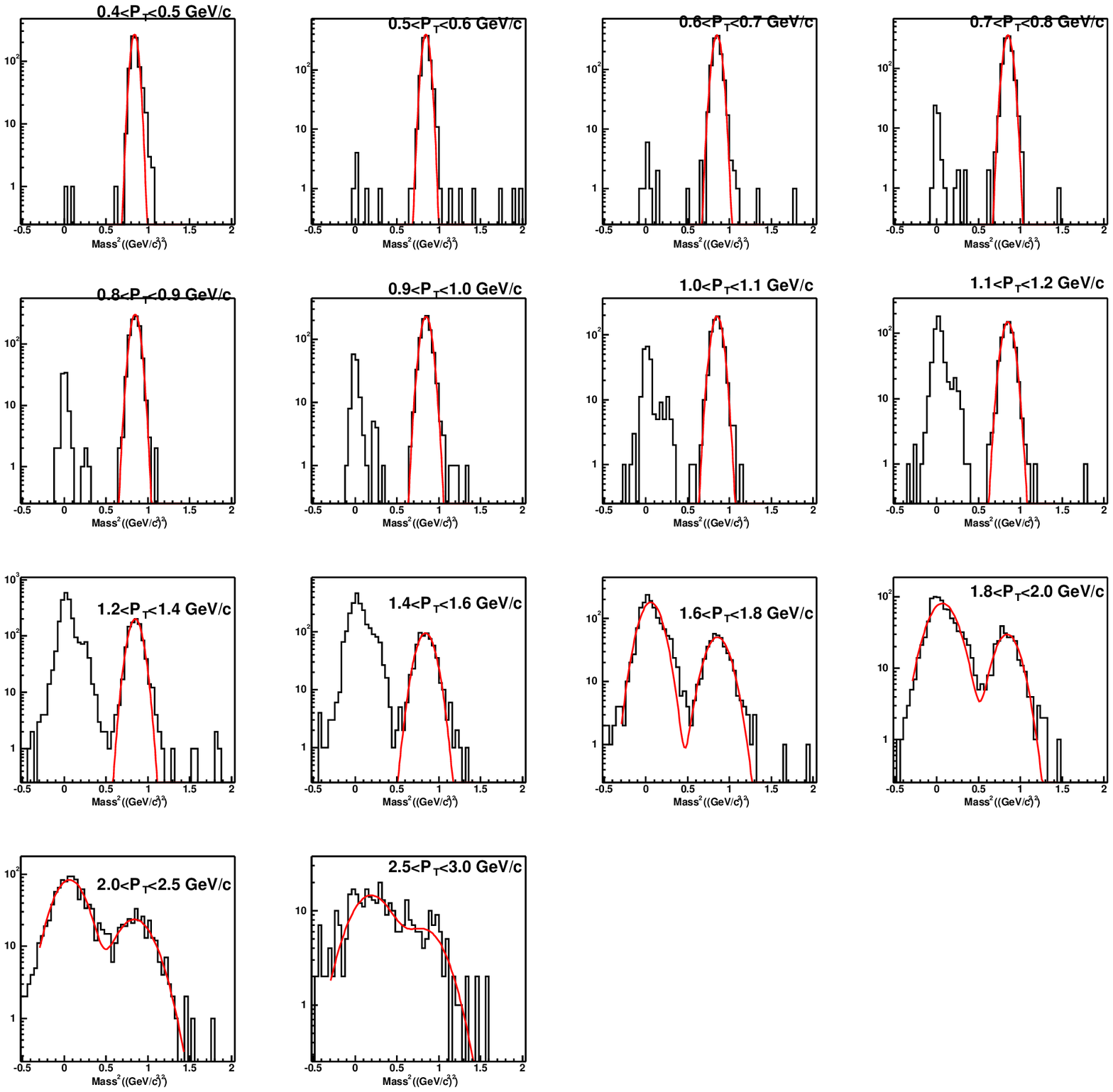}
\caption{$\bar{p}$ raw yields versus mass squared distribution.
The histograms are our data. The curves are Gaussian fits.}
\label{pbarrawyieldplot}
\end{figure}

The energy loss effect due to the interaction with the detector
material was also corrected for. This was studied by simulation.
Figure~\ref{eloss} shows the momentum and transverse momentum
correction for energy loss effect. At $p_{T}>$0.35 GeV/c, for
$\pi$, the energy loss effect is negligible while for kaon and
proton, the energy loss correction is non-negligible at lower
$p_{T}$ and negligible at higher $p_{T}$. The correction was done
by shifting the position of $p_{T}$ in the $p_{T}$ spectra.

\section{Normalization}
The efficiency including vertex efficiency and trigger efficiency
is 91\% in d+Au minimum-bias collisions and 85\% in p+p and
40-100\% d+Au collisions. In 0\%-20\% and 20\%-40\% d+Au
collisions, the efficiency is 100\%. Since the statistic of p+p
minimum-bias events in run 3 is not good enough for us to get very
precise enhancement factor and $N_{ch}$ bias factor. We compare
the $\pi$ spectra in the first 5 $p_{T}$ bin with those from the
paper~\cite{olga} and get the additional normalization factor for
p+p collisions.

\begin{table}[h]
\begin{scriptsize}
  \centering
  \begin{tabular}{|c|c|c|c|c|c|}
    \hline
        $p_{T}$ (GeV/c) & d+Au Trigger & 0\%-20\%  & 20\%-40\% & 40\%-100\% & p+p \\ \hline
  0.3-0.4 & $2.929e+04\pm171.4$ & $9219\pm96.21$ & $8735\pm93.6$ & $8604\pm92.84$ & $1.806e+04\pm134.4$ \\ \hline
  0.4-0.5 & $2.185e+04\pm147.8$ & $6894\pm83.03$ & $6657\pm81.59$ & $6325\pm79.53$ & $1.274e+04\pm114.3$ \\ \hline
  0.5-0.6 & $1.592e+04\pm126.2$ & $5162\pm71.85$ & $4901\pm70.3$ & $4534\pm67.34$ & $9180\pm95.81$ \\ \hline
  0.6-0.7 & $1.166e+04\pm108$ & $3832\pm62.19$ & $3556\pm59.64$ & $3311\pm57.54$ & $6531\pm80.82$ \\ \hline
  0.7-0.8 & $8556\pm92.5$ & $2909\pm53.93$ & $2628\pm51.26$ & $2368\pm48.67$ & $4447\pm66.74$ \\ \hline
  0.8-0.9 & $6198\pm78.86$ & $2099\pm45.85$ & $1936\pm44.33$ & $1693\pm41.17$ & $2973\pm54.57$ \\ \hline
  0.9-1 & $4520\pm67.25$ & $1487\pm38.57$ & $1361\pm36.9$ & $1276\pm35.74$ & $2132\pm46.31$ \\ \hline
    1-1.1 & $3312\pm57.61$ & $1147\pm33.9$ & $1033\pm32.17$ & $845.9\pm29.15$ & $1386\pm37.71$ \\ \hline
  1.1-1.2 & $2406\pm49.35$ & $788.6\pm28.19$ & $752.5\pm27.58$ & $652.7\pm25.7$ & $959.2\pm31.93$ \\ \hline
  1.2-1.4 & $3227\pm58.17$ & $1132\pm34.28$ & $934.4\pm30.98$ & $831.5\pm29.82$ & $1183\pm40.11$ \\ \hline
  1.4-1.6 & $1756\pm45.2$ & $573.7\pm26.2$ & $543.7\pm24$ & $412.8\pm21.83$ & $625.5\pm30.16$ \\ \hline
  1.6-1.8 & $1046\pm39$ & $337.9\pm20.42$ & $309.8\pm18.62$ & $234\pm16.58$ & $364.5\pm32.64$ \\ \hline
  \end{tabular}
\caption{$\pi^{+}$ raw signal table in minimum-bias, centrality
selected d+Au collisions and minimum-bias p+p collisions.}
\label{pionplustable}
\end{scriptsize}
\end{table}

\begin{table}[h]
\begin{scriptsize}
  \centering
  \begin{tabular}{|c|c|c|c|c|c|}
    \hline
        $p_{T}$ (GeV/c) & d+Au Trigger & 0\%-20\%  & 20\%-40\% & 40\%-100\% & p+p \\ \hline
  0.3-0.4 & $2.861e+04\pm169.4$ & $8922\pm94.66$ & $8507\pm92.38$ & $8519\pm92.4$ & $1.715e+04\pm131$ \\ \hline
  0.4-0.5 & $2.139e+04\pm146.3$ & $6805\pm82.49$ & $6458\pm80.35$ & $6306\pm79.41$ & $1.28e+04\pm113.1$ \\ \hline
  0.5-0.6 & $1.611e+04\pm126.9$ & $5327\pm72.98$ & $4873\pm69.8$ & $4605\pm67.86$ & $9189\pm95.86$ \\ \hline
  0.6-0.7 & $1.166e+04\pm108$ & $3831\pm61.9$ & $3550\pm59.5$ & $3355\pm57.92$ & $6362\pm79.69$ \\ \hline
  0.7-0.8 & $8447\pm91.91$ & $2837\pm53.64$ & $2540\pm50.4$ & $2387\pm48.86$ & $4154\pm64.5$ \\ \hline
  0.8-0.9 & $5950\pm77.17$ & $2076\pm45.61$ & $1780\pm42.2$ & $1646\pm40.68$ & $2899\pm53.87$ \\ \hline
  0.9-1 & $4284\pm65.46$ & $1446\pm38.03$ & $1317\pm36.31$ & $1171\pm34.29$ & $1924\pm44.01$ \\ \hline
    1-1.1 & $3296\pm57.47$ & $1123\pm33.55$ & $1014\pm31.9$ & $897.5\pm30.02$ & $1372\pm37.77$ \\ \hline
  1.1-1.2 & $2464\pm49.88$ & $812.4\pm28.58$ & $762.4\pm27.69$ & $650.8\pm25.72$ & $1005\pm33.16$ \\ \hline
  1.2-1.4 & $3136\pm57.28$ & $1027\pm32.54$ & $972.9\pm31.71$ & $828.8\pm29.65$ & $1243\pm39.78$ \\ \hline
  1.4-1.6 & $1716\pm45.79$ & $612.5\pm25.87$ & $539.3\pm25.67$ & $422.7\pm24.22$ & $603.8\pm30.49$ \\ \hline
  1.6-1.8 & $1033\pm39.74$ & $375.3\pm21.21$ & $306.1\pm19.59$ & $239.9\pm48.55$ & $337.1\pm28.82$ \\ \hline
  \end{tabular}
\caption{$\pi^{-}$ raw signal table in minimum-bias, centrality
selected d+Au collisions and minimum-bias p+p collisions.}
\label{pionminustable}
\end{scriptsize}
\end{table}

\begin{table}[h]
\begin{scriptsize}
  \centering
  \begin{tabular}{|c|c|c|c|c|c|}
    \hline
        $p_{T}$ (GeV/c) & d+Au Trigger & 0\%-20\%  & 20\%-40\% & 40\%-100\% & p+p \\ \hline
  0.4-0.5 & $1410\pm37.54$ & $417.2\pm20.44$ & $420.9\pm20.52$ & $354.9\pm18.84$ & $753.2\pm27.44$ \\ \hline
  0.5-0.6 & $1588\pm39.85$ & $486.3\pm22.06$ & $461\pm21.48$ & $435\pm20.87$ & $729.3\pm27$ \\ \hline
  0.6-0.7 & $1499\pm38.71$ & $465.9\pm21.59$ & $445.5\pm21.17$ & $395\pm19.87$ & $710.1\pm26.65$ \\ \hline
  0.7-0.8 & $1346\pm36.69$ & $423.9\pm20.59$ & $419.8\pm20.62$ & $335.7\pm18.32$ & $579\pm24.06$ \\ \hline
  0.8-0.9 & $1105\pm33.59$ & $369.7\pm19.3$ & $317.9\pm18.43$ & $282.5\pm17.03$ & $496.2\pm22.44$ \\ \hline
  0.9-1 & $969.1\pm31.19$ & $283.9\pm16.86$ & $305.1\pm17.52$ & $258.7\pm16.23$ & $381.2\pm19.87$ \\ \hline
    1-1.1 & $799.3\pm28.41$ & $278.6\pm16.79$ & $224\pm15.04$ & $192.3\pm14.04$ & $301.5\pm18.39$ \\ \hline
  1.1-1.2 & $656.7\pm26.24$ & $199.1\pm14.33$ & $186.2\pm14$ & $155.7\pm12.94$ & $267.8\pm18.71$ \\ \hline
  1.2-1.4 & $1013\pm34.43$ & $335\pm19.51$ & $283.1\pm17.71$ & $234.1\pm17.68$ & $421.1\pm29.14$ \\ \hline
  1.4-1.6 & $605.9\pm30.14$ & $191.1\pm17.77$ & $174.1\pm14.77$ & $148.8\pm15.48$ & $238.8\pm13.97$ \\ \hline
  1.6-1.8 & $382.9\pm28.9$ & $---$ & $---$ & $---$ & $---$ \\ \hline
  \end{tabular}
\caption{$K^{+}$ raw signal table in minimum-bias, centrality
selected d+Au collisions and minimum-bias p+p collisions.}
\label{kaonplustable}
\end{scriptsize}
\end{table}

\begin{table}[h]
\begin{scriptsize}
  \centering
  \begin{tabular}{|c|c|c|c|c|c|}
    \hline
        $p_{T}$ (GeV/c) & d+Au Trigger & 0\%-20\%  & 20\%-40\% & 40\%-100\% & p+p \\ \hline
  0.4-0.5 & $1341\pm36.62$ & $411.6\pm20.29$ & $367.6\pm19.19$ & $378\pm19.44$ & $682.6\pm26.13$ \\ \hline
  0.5-0.6 & $1498\pm38.7$ & $460\pm21.45$ & $411.5\pm20.32$ & $420.5\pm20.51$ & $740.7\pm27.21$ \\ \hline
  0.6-0.7 & $1410\pm37.55$ & $436.2\pm20.89$ & $398.4\pm19.97$ & $361.9\pm19.02$ & $616.8\pm24.83$ \\ \hline
  0.7-0.8 & $1207\pm34.74$ & $366\pm19.14$ & $349.7\pm18.7$ & $350\pm18.75$ & $557.4\pm23.61$ \\ \hline
  0.8-0.9 & $1057\pm32.66$ & $317.4\pm18.12$ & $332.4\pm18.37$ & $268.2\pm16.66$ & $432.9\pm20.93$ \\ \hline
  0.9-1 & $863.7\pm29.42$ & $256.9\pm16.09$ & $267.2\pm16.43$ & $223.8\pm15.03$ & $368.9\pm19.59$ \\ \hline
    1-1.1 & $635.2\pm25.35$ & $198.2\pm14.35$ & $183.1\pm13.61$ & $187.9\pm13.88$ & $320.4\pm19.42$ \\ \hline
  1.1-1.2 & $543\pm23.92$ & $166.4\pm13.14$ & $143\pm12.22$ & $154.1\pm12.89$ & $248.5\pm18.84$ \\ \hline
  1.2-1.4 & $895\pm32.45$ & $302.1\pm18.3$ & $258\pm17.06$ & $206.7\pm16.17$ & $377.1\pm26.67$ \\ \hline
  1.4-1.6 & $645.5\pm31.61$ & $202.4\pm16.78$ & $141\pm15.58$ & $166.6\pm17.87$ & $237.6\pm20.52$ \\ \hline
  1.6-1.8 & $351.3\pm29.79$ & $---$ & $---$ & $---$ & $---$ \\ \hline

  \end{tabular}
\caption{$K^{-}$ raw signal table in minimum-bias, centrality
selected d+Au collisions and minimum-bias p+p collisions.}
\label{kaonminustable}
\end{scriptsize}
\end{table}

\begin{table}[h]
\begin{scriptsize}
  \centering
  \begin{tabular}{|c|c|c|c|c|c|}
    \hline
        $p_{T}$ (GeV/c) & d+Au Trigger & 0\%-20\%  & 20\%-40\% & 40\%-100\% & p+p \\ \hline
  0.4-0.5 & $1377\pm107.5$ & $412.6\pm40.73$ & $403.5\pm40.11$ & $422.7\pm41.42$ & $657.5\pm64.94$ \\ \hline
  0.5-0.6 & $1527\pm89.71$ & $492.2\pm36.58$ & $428.6\pm33.09$ & $424.8\pm32.89$ & $752.6\pm59.91$ \\ \hline
  0.6-0.7 & $1456\pm80.28$ & $437.9\pm31.47$ & $420\pm30.56$ & $401.3\pm29.6$ & $704.9\pm54.1$ \\ \hline
  0.7-0.8 & $1336\pm73.68$ & $410.6\pm29.43$ & $387.9\pm28.28$ & $385\pm28.14$ & $670\pm55.36$ \\ \hline
  0.8-0.9 & $1278\pm72.45$ & $387.6\pm28.57$ & $371.4\pm27.72$ & $312\pm24.57$ & $498.8\pm45.17$ \\ \hline
  0.9-1 & $1124\pm68.87$ & $349.1\pm27.39$ & $322.1\pm25.87$ & $288.1\pm23.94$ & $448.3\pm44.89$ \\ \hline
    1-1.1 & $954.5\pm62.39$ & $288.2\pm24.38$ & $285.3\pm24.23$ & $265.9\pm23.11$ & $365\pm40.02$ \\ \hline
  1.1-1.2 & $832.1\pm58.34$ & $257.2\pm23.06$ & $245.4\pm22.25$ & $213.3\pm20.25$ & $273.8\pm34.08$ \\ \hline
  1.2-1.4 & $1268\pm72.57$ & $441.9\pm31.19$ & $362.6\pm27.1$ & $320.2\pm24.77$ & $393.8\pm41.64$ \\ \hline
  1.4-1.6 & $806.8\pm57.58$ & $306.9\pm26.38$ & $221.5\pm20.77$ & $190.8\pm18.71$ & $210.3\pm31.59$ \\ \hline
  1.6-1.8 & $540.8\pm23.27$ & $170.7\pm13.06$ & $146.2\pm12.14$ & $116.3\pm10.81$ & $126\pm11.31$ \\ \hline
  1.8-2 & $314.2\pm17.8$ & $119.2\pm10.97$ & $81.7\pm9.764$ & $68.35\pm9.093$ & $93.98\pm10.15$ \\ \hline
    2-2.5 & $388.1\pm21.21$ & $148.4\pm12.48$ & $135.7\pm12.01$ & $89.74\pm10.02$ & $109\pm12.33$ \\ \hline
  2.5-3 & $109.1\pm12.92$ & $36.33\pm6.809$ & $34.3\pm8.488$ & $30.64\pm7.487$ & $24.22\pm5.422$ \\ \hline
  3-4 & $82.18\pm12.30$ & $---$ & $---$ & $---$ & $---$ \\ \hline

  \end{tabular}
\caption{$p$ raw signal table in minimum-bias, centrality selected
d+Au collisions and minimum-bias p+p collisions.}
\label{protontable}
\end{scriptsize}
\end{table}

\begin{table}[h]
\begin{scriptsize}
  \centering
  \begin{tabular}{|c|c|c|c|c|c|}
    \hline
        $p_{T}$ (GeV/c) & d+Au Trigger & 0\%-20\%  & 20\%-40\% & 40\%-100\% & p+p \\ \hline
  0.4-0.5 & $692.6\pm26.33$ & $215.1\pm14.67$ & $183\pm13.53$ & $202.8\pm14.26$ & $421.7\pm20.56$ \\ \hline
  0.5-0.6 & $1009\pm31.76$ & $310.8\pm17.63$ & $304\pm17.43$ & $268.5\pm16.39$ & $526.6\pm22.95$ \\ \hline
  0.6-0.7 & $1098\pm33.17$ & $317.9\pm17.84$ & $305.6\pm17.51$ & $327\pm18.11$ & $561.5\pm23.71$ \\ \hline
  0.7-0.8 & $1062\pm32.59$ & $340.2\pm18.44$ & $307\pm17.53$ & $285.5\pm16.9$ & $435.1\pm20.86$ \\ \hline
  0.8-0.9 & $992.2\pm31.5$ & $315.1\pm17.81$ & $284.4\pm16.96$ & $244.4\pm15.63$ & $376.4\pm19.4$ \\ \hline
  0.9-1 & $827.5\pm28.76$ & $288.9\pm17$ & $225\pm15.01$ & $202.6\pm14.24$ & $310.4\pm17.62$ \\ \hline
    1-1.1 & $724\pm26.91$ & $240.2\pm15.5$ & $181.5\pm13.48$ & $192.2\pm13.87$ & $246.4\pm15.7$ \\ \hline
  1.1-1.2 & $608.5\pm24.67$ & $161.3\pm12.7$ & $184.3\pm13.61$ & $149.8\pm12.25$ & $192\pm13.87$ \\ \hline
  1.2-1.4 & $914.9\pm30.24$ & $301.5\pm17.36$ & $269.7\pm16.42$ & $214.1\pm14.63$ & $269.6\pm16.42$ \\ \hline
  1.4-1.6 & $575.8\pm24$ & $204.9\pm14.32$ & $160.9\pm12.71$ & $120.5\pm10.98$ & $138.6\pm12.01$ \\ \hline
  1.6-1.8 & $407.2\pm20.18$ & $127.3\pm11.29$ & $108.1\pm10.43$ & $89.9\pm9.497$ & $100.6\pm10.63$ \\ \hline
  1.8-2 & $257.3\pm16.26$ & $73.85\pm8.992$ & $92.22\pm9.69$ & $46.81\pm7.802$ & $71.23\pm8.92$ \\ \hline
    2-2.5 & $305.6\pm18.45$ & $114\pm11.01$ & $83.43\pm9.464$ & $77.84\pm9.16$ & $64.02\pm10.44$ \\ \hline
  2.5-3 & $111\pm12.79$ & $28.91\pm6.26$ & $29.29\pm8.869$ & $20.55\pm7.198$ & $25.71\pm6.856$ \\ \hline
  3-4 & $67.05\pm11.87$ & $---$ & $---$ & $---$ & $---$ \\ \hline

  \end{tabular}
\caption{$\bar{p}$ raw signal table in minimum-bias, centrality
selected d+Au collisions and minimum-bias p+p collisions.}
\label{pbartable}
\end{scriptsize}
\end{table}

\chapter{Results}
\label{chp:results}

\section{$\pi, K, p$ and $\bar{p}$ spectra in d+Au and p+p collisions at mid-rapidity}
\begin{figure}[h]
\centering
\includegraphics[height=18pc,width=24pc]{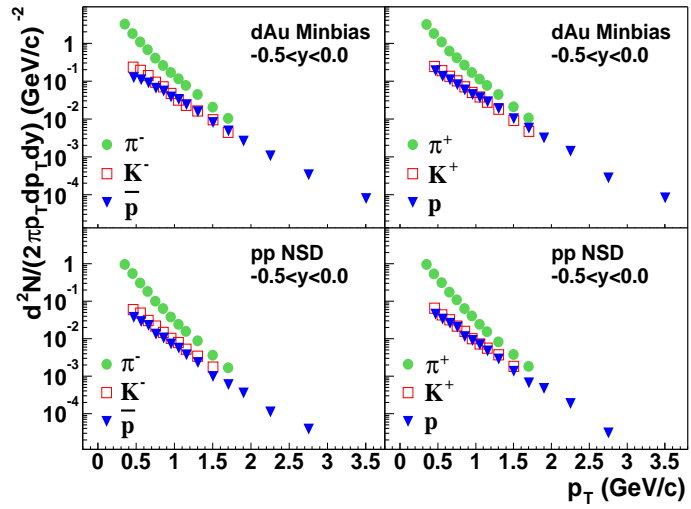}
\caption{The invariant yields of pions (filled circles), kaons
(open squares), protons (filled triangles) and their
anti-particles as a function of $p_{T}$ from d+Au and NSD p+p
events at 200 GeV. The rapidity range was $-0.5<y<0.0$ with the
direction of the outgoing Au ions as negative rapidity.  Errors
are statistical.} \label{spectra}
\end{figure}

The invariant yields $\frac{1}{2\pi p_T}\frac{d^2N}{dydp_T}$ of
$\pi^{\pm}$, $K^{\pm}$, $p$ and $\bar{p}$ from both NSD p+p and
minimum-bias d+Au events at mid-rapidity $-0.5<y<0$ are shown in
Figure\ref{spectra}, where $N$ is the corrected signal number per
minimum-bias event in each $p_{T}$ bin.
$N=\frac{N_{raw}\times{factor3}\times{factor4}\times{factor5}}{N_{total}\times{factor1}\times{factor2}}$,
where $N_{raw}$ is the raw signal number in each $p_{T}$ bin,
$N_{total}$ is the total TOFr triggered events, $factor1$ is the
enhancement factor of TOFr trigger, $factor2$ is the TPC
efficiency times TOFr matching efficiency, $factor3$ is the
background correction factor, $factor4$ is the $\langle N_{ch}
\rangle$ bias factor, and $factor5$ is the vertex efficiency times
trigger efficiency and normalization factor.

\subsection{Systematic uncertainty}
For the invariant yield of $\pi^{\pm}$, $K^{\pm}$, $p$ and
$\bar{p}$, the average bin-to-bin systematic uncertainty was
estimated to be of the order of 8\%. The systematic uncertainty is
dominated by the uncertainty in the detector response in Monte
Carlo simulations ($\pm7\%$). Additional factors contributing to
the total systematic uncertainty include the background correction
($\pm3\%$), the small $\eta$ acceptance of the TOFr ($\pm2\%$),
TOFr response ($\pm2\%$), the correction for energy loss in the
detector (${}^{<}_{\sim}10\pm10\%$ at $p_{T}<0.6$ GeV/c for the
$p$ and $\bar{p}$, much smaller for other species and negligible
at higher $p_{T}$), absorption of $\bar{p}$ in the material
($\pm3\%$), and the momentum resolution correction
($\simeq5\pm2\%$). The normalization uncertainties in d+Au
minimum-bias and p+p NSD collisions are $10\%$ and $14\%$,
respectively~\cite{starhighpt,stardau}. The charged pion yields
are consistent with $\pi^0$ yields measured by the PHENIX
collaboration in the overlapping $p_{T}$
range~\cite{phenixhighpt,otherdau}. The invariant yields of
$\pi^{\pm}$, $K^{\pm}$, $p$ and $\bar{p}$ in minimum-bias,
centrality selected d+Au and minimum-bias p+p collisions, are
listed in the tables in Appendix A with statistical errors and
systematic uncertainties.

\section{Cronin effect}
\begin{figure}[h]
\centering
\includegraphics[height=18pc,width=24pc]{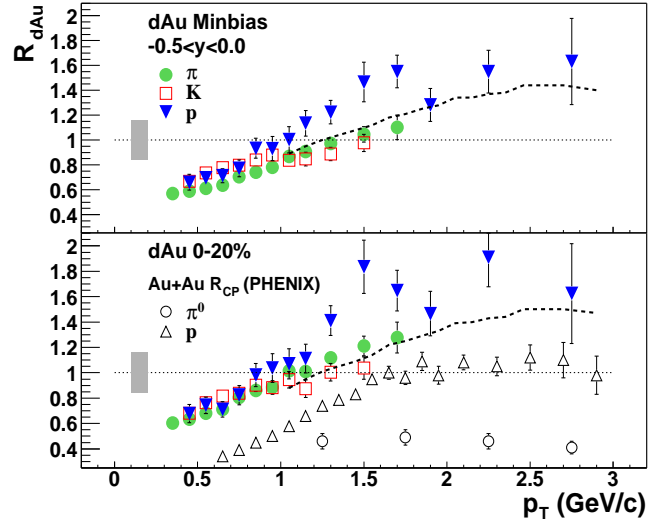}
\caption{The identified particle $R_{dAu}$ for minimum-bias and
top 20\% d+Au collisions. The filled triangles are for
$p+\bar{p}$, the filled circles are for $\pi^{+}+\pi^{-}$ and the
open squares are for $K^{+}+K^{-}$. Dashed lines are $R_{dAu}$ of
inclusive charged hadrons from~\cite{stardau}. The open triangles
and open circles are $R_{CP}$ of $p+\bar{p}$ and $\pi^{0}$ in
Au+Au collisions measured by PHENIX~\cite{phenixpid}.  Errors are
statistical. The gray band represents the normalization
uncertainty of 16\%.} \label{Rdau}
\end{figure}
Nuclear effects on hadron production in d+Au collisions are
measured through comparison to the p+p spectrum, scaled by the
number of underlying nucleon-nucleon inelastic collisions using
the ratio
\[R_{dAu}=\frac{d^{2}N/(2{\pi}p_{T}dp_{T}dy)}{T_{dAu}d^{2}\sigma^{pp}_{inel}/(2{\pi}p_{T}dp_{T}dy)}  ,\]
where $T_{dAu}={\langle N_{bin}\rangle}/\sigma^{pp}_{inel}$
describes the nuclear geometry, and
$d^{2}\sigma^{pp}_{inel}/(2{\pi}p_{T}dp_{T}dy)$ for p+p inelastic
collisions is derived from the measured p+p NSD cross section. The
difference between NSD and inelastic differential cross sections
at mid-rapidity, as estimated from PYTHIA~\cite{pythia}, is $5\%$
at low $p_{T}$ and negligible at $p_{T}>1.0$ GeV/c.
Figure.~\ref{Rdau} shows $R_{dAu}$ of $\pi^{+}+\pi^{-}$,
$K^{+}+K^{-}$ and $p+\bar{p}$ for minimum-bias and central d+Au
collisions. The systematic uncertainties on $R_{dAu}$ are of the
order of 16\%, dominated by the uncertainty in normalization.  The
$R_{dAu}$ of the same particle species are similar between
minimum-bias and top 20\% d+Au collisions. In both cases, the
$R_{dAu}$ of protons rise faster than $R_{dAu}$ of pions and
kaons. We observe that the spectra of $\pi^{\pm}$, $K^{\pm}$, $p$
and $\bar{p}$ are considerably harder in d+Au than those in p+p
collisions. The $R_{dAu}$ of the identified particles has
characteristics of the Cronin effect~\cite{cronin,accardi} in
particle production with $R_{dAu}$ less than unity at low $p_{T}$
and above unity at $p_{T}{}^{>}_{\sim} 1.0$ GeV/c.

\section{$p+\bar{p}/h$ ratio in d+Au and p+p collisions at
middle pseudo-rapidity}

\begin{figure}[h]
\centering
\includegraphics[height=18pc,width=24pc]{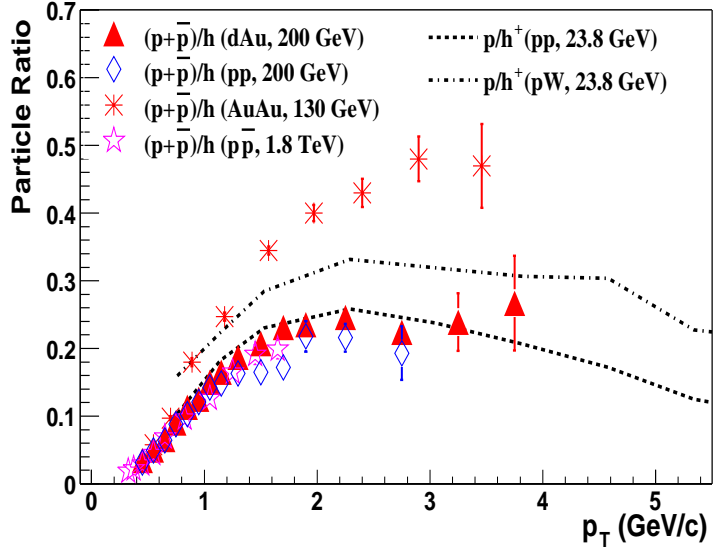}
\caption{Minimum-bias ratios of ($p+\bar{p}$) over charged hadrons
at $-0.5\!<\!\eta\!<\!0.0$ from $\sqrt{s_{_{NN}}} =200$ GeV p+p
(open diamonds), d+Au (filled triangles) and $\sqrt{s_{_{NN}}}
=130$ GeV Au+Au~\cite{phenixpid} (asterisks) collisions. Results
of $\mathrm{p+\bar{p}}$ collisions at $\sqrt{s_{_{NN}}} = 1.8$
TeV~\cite{e735} are shown as open stars. Dashed lines are results
of $p/h^{+}$ ratios from $\sqrt{s_{_{NN}}} = 23.8$ GeV p+p
(short-dashed lines) and p+W (dot-dashed)
collisions~\cite{cronin}. Errors are statistical. }
\label{bnchratio}
\end{figure}

Figure~\ref{bnchratio} depicts $(p+\bar{p})/h$, the ratio of
$p+\bar{p}$ over inclusive charged hadrons as a function of
$p_{T}$ in d+Au and p+p minimum-bias collisions at
$\sqrt{s_{_{NN}}} = 200$ GeV, and $p/h^{+}$ ratios in p+p and p+W
minimum-bias collisions at $\sqrt{s_{_{NN}}} = 23.8$
GeV~\cite{cronin}.  Although the relative yields of particles and
anti-particles are very different at $\sqrt{s}<40$ GeV due to the
valence quark effects from target and projectile, the Cronin
effects are similar. The systematic uncertainties on these ratios
were estimated to be of the order of 10\% for
$p_{T}{}^{<}_{\sim}1.0$ GeV/c, decreasing to 3\% at higher
$p_{T}$. At RHIC energies, the anti-particle to particle ratios
approach unity ($\bar{p}/p=0.81\pm0.02\pm0.04$ in d+Au
minimum-bias collisions) and their nuclear modification factors
are similar. The difference between $R_{dAu}$ at $\sqrt{s_{_{NN}}}
= 200$ GeV for $p+\bar{p}$ and $h$ can be obtained from the
$(p+\bar{p})/h$ ratios in d+Au and p+p collisions.
Table~\ref{pbarpnchratio} shows $R_{dAu}^{p+\bar{p}}/R_{dAu}^{h}$
determined by averaging over the bins within $1.2<p_{T}<3.0$
GeV/c. At lower energy, the $\alpha$ parameter in the power law
dependence on target atomic weight $A^{\alpha}$ of identified
particle production falls with $\sqrt{s}$~\cite{cronin}. From the
ratios of $R_{dAu}$ between $p+\bar{p}$ and $h$, we may further
derive the $\alpha_{p}-\alpha_{\pi}$ for $1.2< p_{T}< 3.0$ GeV/c
to be $0.041\pm0.010$(stat)$\pm0.006$(syst) under the assumptions
that $\alpha_{K}\simeq\alpha_{\pi}$ and that $(p+\bar{p})/{\pi}$
and $K/{\pi}$ are between 0.1 and 0.4 in p+p collisions. This
result is significantly smaller than the value $0.095\pm0.004$ in
the same $p_{T}$ range found at lower energies~\cite{cronin}.\\
\begin{table}[h]
\caption{\label{pbarpnchratio}$\langle N_{bin}\rangle$ from a
Glauber model calculation, $(p+\bar{p})/h$ averaged over the bins
within $1.2<p_{T}<2.0$ GeV/c (left column) and within
$2.0<p_{T}<3.0$ GeV/c (right column) and the $R_{dAu}$ ratios
between $p+\bar{p}$ and $h$ averaged over $1.2<p_{T}<3.0$ GeV/c
for minimum-bias, centrality selected d+Au collisions and
minimum-bias p+p collisions.  A p+p inelastic cross section of
$\sigma_{inel}=42$ mb was used in the calculation. For $R_{dAu}$
ratios, only statistical errors are shown and the systematic
uncertainties are 0.03 for all centrality bins. } {\centering
{\begin{tabular}{c|c|c|c|c} \hline \hline centrality &
 $\langle N_{bin}\rangle$ & \multicolumn{2}{c|} {$(p+\bar{p})/h$} &
 ${R_{dAu}^{p+\bar{p}}}/{R_{dAu}^h}$\\ \hline min. bias & $7.5\pm0.4$
 &$0.21\pm0.01$ &$0.24\pm0.01$ & $1.19\pm0.05$\\ 0--20\% &
 $15.0\pm1.1$ &$0.21\pm0.01$ &$0.24\pm0.02$ & $1.18\pm0.06$\\ 20--40\%
 & $10.2\pm1.0$ &$0.20\pm0.01$ &$0.24\pm0.02$ & $1.16\pm0.06$\\
 40--$\sim$100\% & $4.0^{+0.8}_{-0.3}$ &$0.20\pm0.01$ &$0.23\pm0.02$ &
 $1.13\pm0.06$\\ \hline p+p & $1.0$ &$0.17\pm0.01$ &$0.21\pm0.02$ &
 --- \\ \hline \hline
   \end{tabular}
 }
\par}
\label{Tab:D}
\end{table}

\begin{figure}[h]
\centering
\includegraphics[height=18pc,width=24pc]{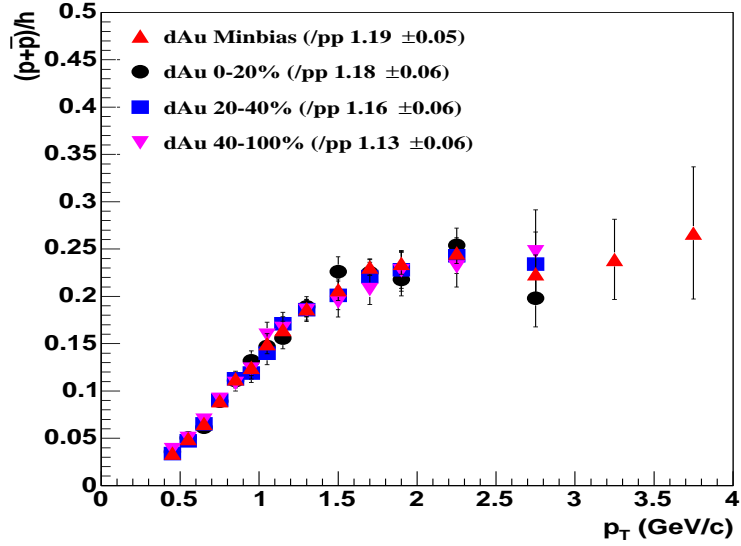}
\caption{Minimum-bias ratios of ($p+\bar{p}$) over charged hadrons
at $-0.5\!<\!\eta\!<\!0.0$ from $\sqrt{s_{_{NN}}} =200$ GeV
minimum-bias and centrality selected d+Au collisions. Errors are
statistical. } \label{bnchratiocentrality}
\end{figure}
Also shown is $(p+\bar{p})/h$ ratio from the Au+Au minimum-bias
collisions at $\sqrt{s_{_{NN}}} = 130$ GeV~\cite{phenixpid}. The
$(p+\bar{p})/h$ ratio from minimum-bias Au+Au
collisions~\cite{phenixpid} at a similar energy is about a factor
of 2 higher than that in d+Au and p+p collisions for
$p_{T}{}^{>}_{\sim}2.0$ GeV/c.  This enhancement is most likely
due to final-state effects in Au+Au
collisions~\cite{jetquench,junction,derekhydro,pisahydro,fries,ko}.
The ratios show little centrality dependence in d+Au collisions,
as shown in Table~\ref{Tab:D} and
Figure~\ref{bnchratiocentrality}. For $p_{T}<2.0$ GeV/c, the ratio
in $\mathrm{p+\bar{p}}$ collisions at $\sqrt{s_{_{NN}}} = 1.8$
TeV~\cite{e735} is very similar to those in d+Au and p+p
collisions at $\sqrt{s_{_{NN}}} = 200$ GeV.

\section{$K/\pi$, $p/\pi$ and anti-particle to particle ratios}
\begin{figure}[h]
\begin{minipage}[t]{50mm}
\includegraphics[height=11pc,width=13pc]{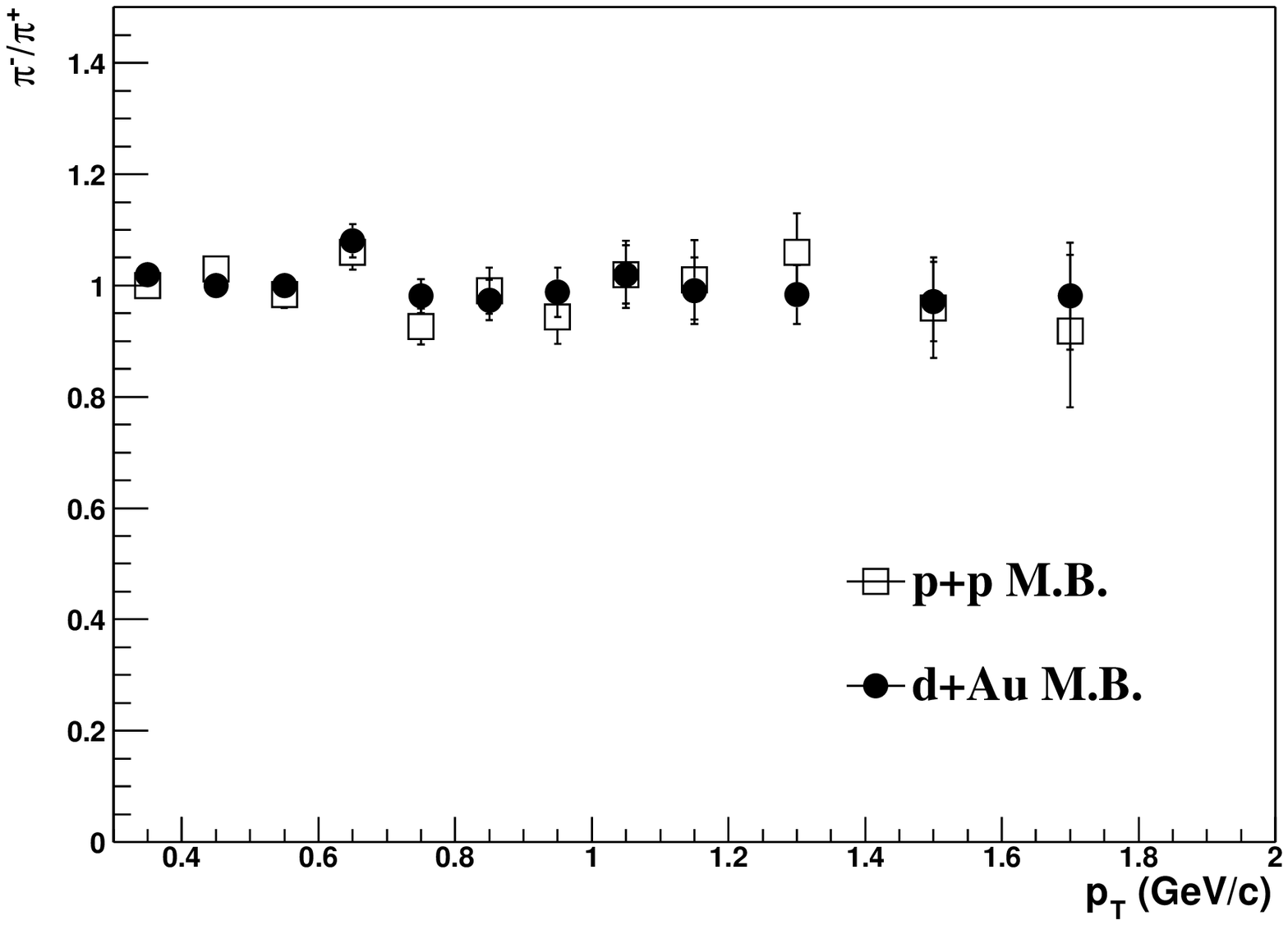}
\end{minipage}
\begin{minipage}[t]{50mm}
\includegraphics[height=11pc,width=13pc]{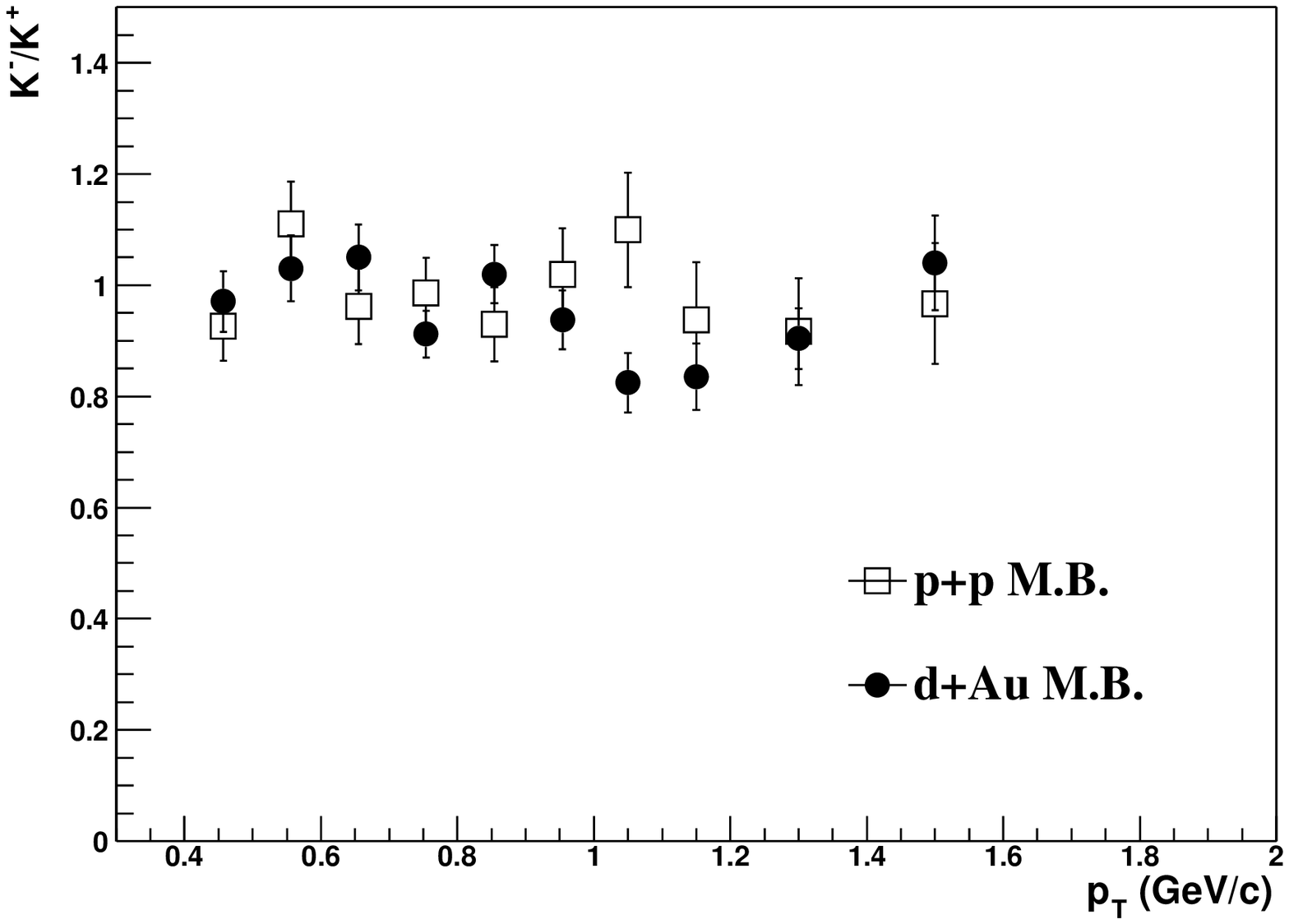}
\end{minipage}
\begin{minipage}[t]{50mm}
\includegraphics[height=11pc,width=13pc]{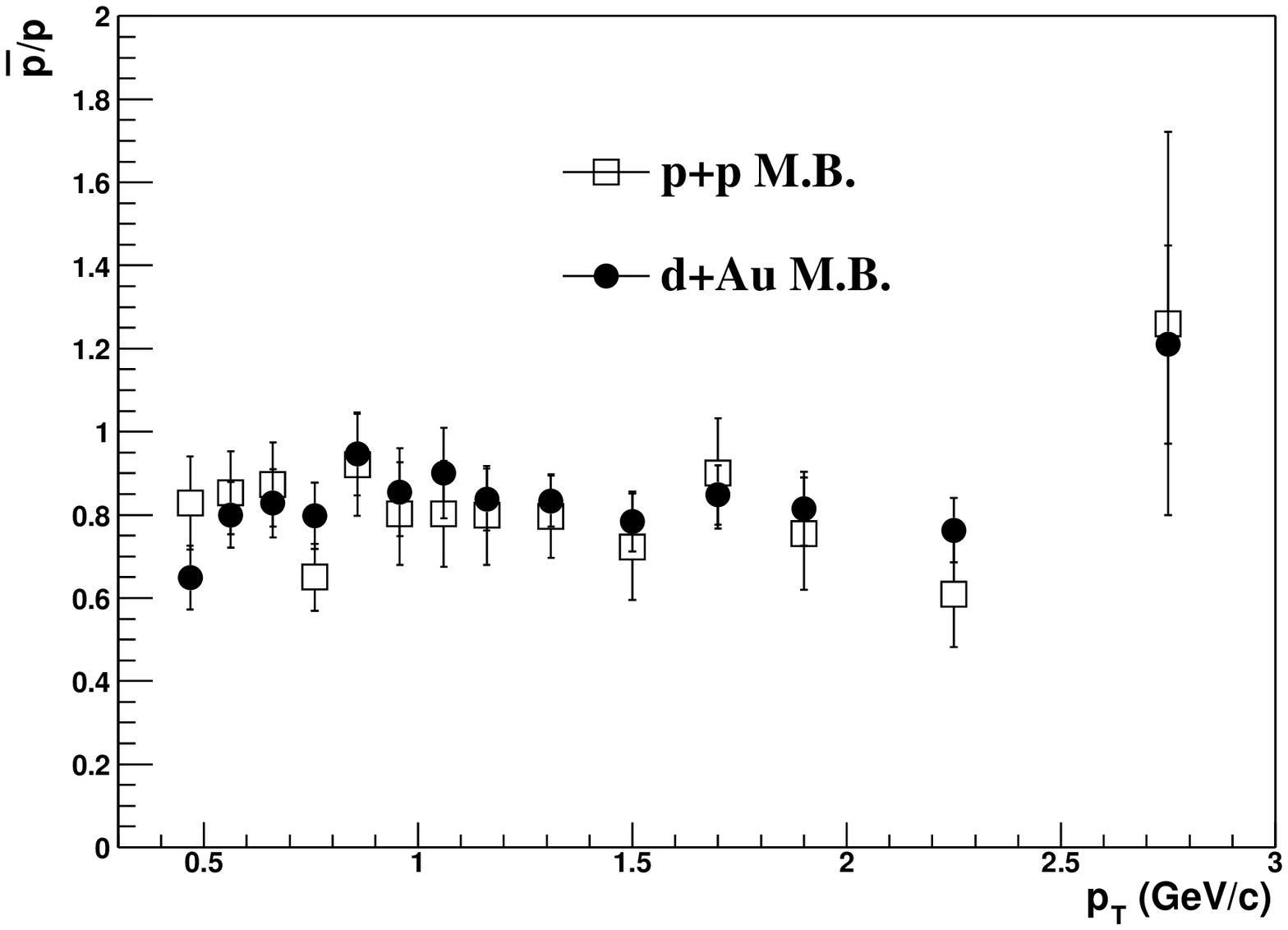}
\end{minipage}
\caption{$\pi^{-}$/$\pi^{+}$, $K^{-}$/$K^{+}$ and $\bar{p}/p$
ratios as a function of $p_{T}$ in d+Au and p+p minimum-bias
collisions. The open symbols are for p+p collisions and the solid
symbols for d+Au collisions. Errors are statistical.}
\label{antiparticleratio}
\end{figure}
\begin{table}[h]
\begin{scriptsize} \centering
  \begin{tabular}{|c|c|c|c|c|c|c|}
    \hline
        Centrality Bin & $\pi^{-}$/$\pi^{+}$  & $X^{2}/ndf$ & $K^{-}$/$K^{+}$ & $X^{2}/ndf$ & $\bar{p}/p$ & $X^{2}/ndf$ \\ \hline
        d+Au M.B.      & $1.01\pm0.01$ & $0.88$ & $0.94\pm0.02$ & $1.78$ & $0.81\pm0.02$ & $0.85$\\ \hline
        0\%-20\% & $1.01\pm0.01$ & $0.80$ & $0.93\pm0.03$ & $1.43$ & $0.80\pm0.03$ & $0.70$  \\ \hline
        20\%-40\% & $1.00\pm0.01$ & $0.98$ & $0.91\pm0.03$ & $1.19$ & $0.79\pm0.03$ & $1.14$  \\ \hline
        40\%-100\% & $1.02\pm0.01$ & $0.81$ & $1.02\pm0.03$ & $0.45$ & $0.78\pm0.03$ & $0.70$  \\ \hline
        p+p & $1.00\pm0.01$ & $1.24$ & $0.98\pm0.02$ & $0.71$ & $0.79\pm0.03$ & $0.73$  \\ \hline
  \end{tabular}
\caption{$\pi^{-}$/$\pi^{+}$, $K^{-}$/$K^{+}$ and $\bar{p}/p$
ratios in p+p and d+Au minimum-bias collisions. Also shows in the
table are the ratios in centrality selected d+Au collisions.
Errors are statistical. } \label{antiparticletoparticleratio}
\end{scriptsize}
\end{table}

Figure~\ref{antiparticleratio} shows the $\pi^{-}/\pi^{+}$,
$K^{-}/K^{+}$ and $\bar{p}/p$ ratios as a function of $p_{T}$ in
d+Au and p+p minimum-bias collisions. It shows the anti-particle
to particle ratios are flat with $p_{T}$. The zero order
polynominal function was used to fit the data and get the
anti-particle to particle ratios. The results are list in
Table~\ref{antiparticletoparticleratio}. In centrality selected
d+Au collisions, the anti-particle to particle ratios are also
flat with $p_{T}$ and show little centrality dependence. The
results are also shown in the Table
~\ref{antiparticletoparticleratio}.

\begin{figure}[h]
\begin{minipage}[t]{80mm}
\includegraphics[height=13pc,width=14pc]{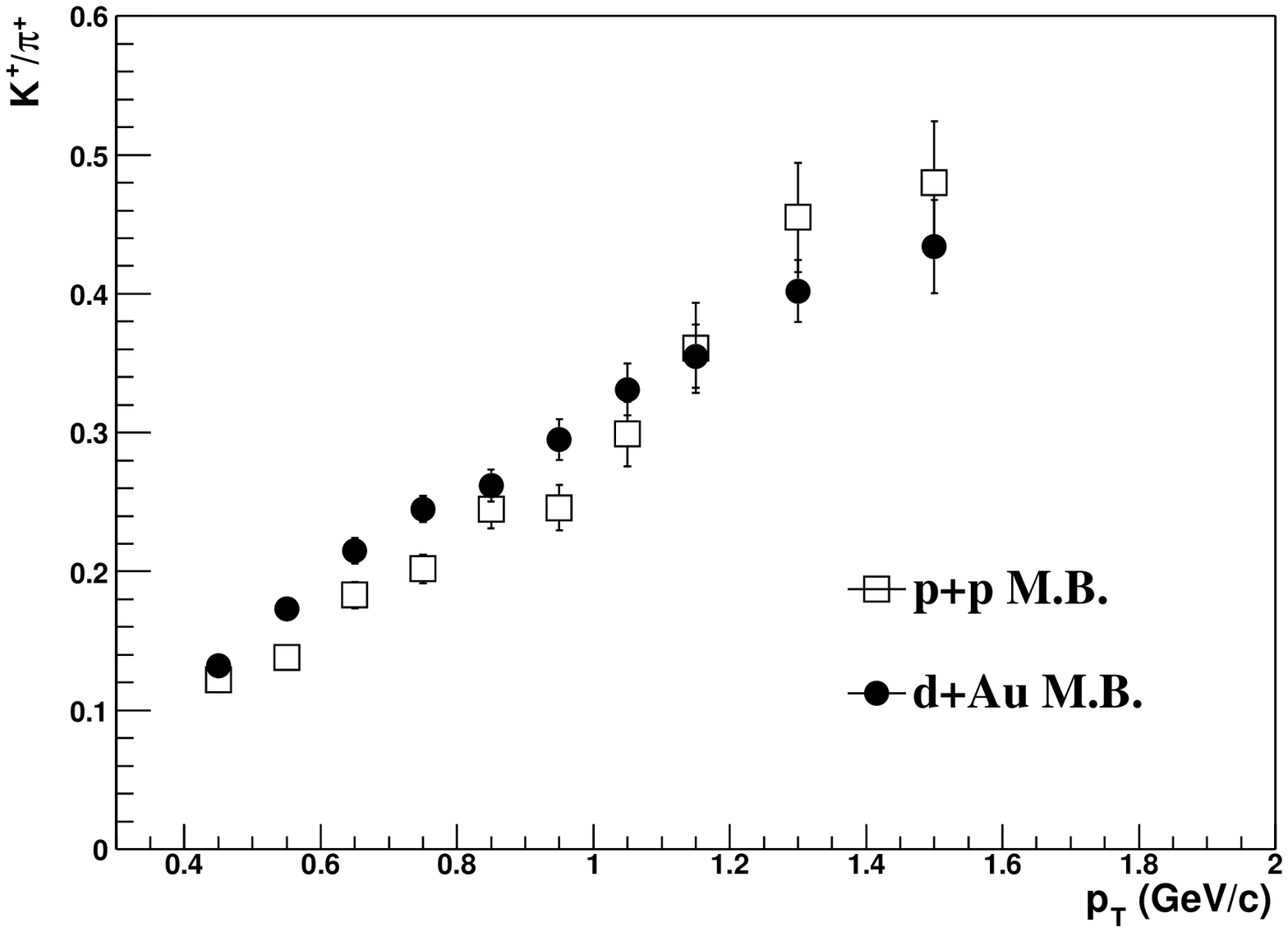}
\end{minipage}
\begin{minipage}[t]{80mm}
\includegraphics[height=13pc,width=14pc]{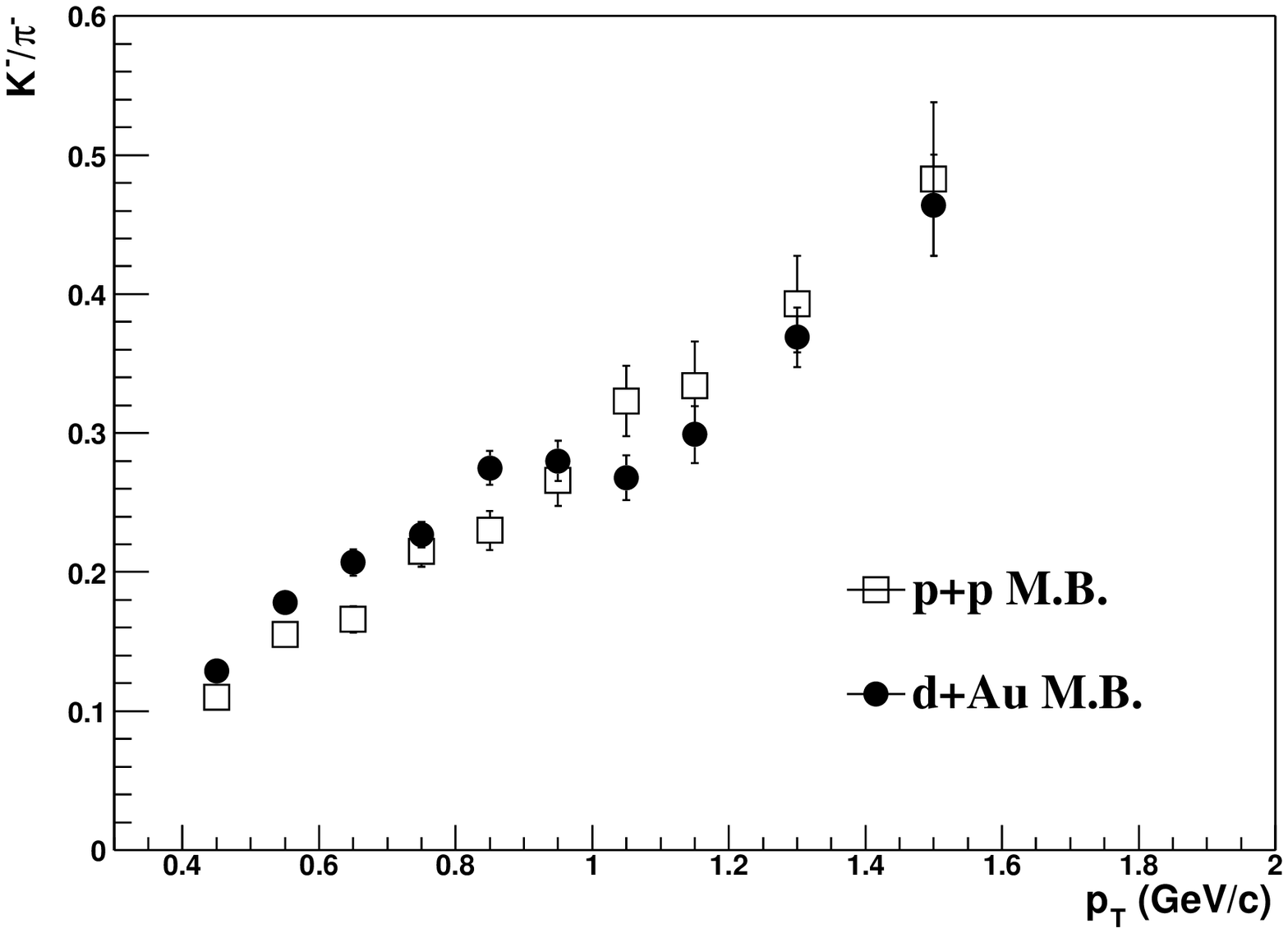}
\end{minipage}
\caption{$K/\pi$ ratios as a function of $p_{T}$ in d+Au and p+p
minimum-bias collisions. The open symbols are for p+p collisions
and the solid symbols for d+Au collisions. Errors are
statistical.} \label{kpiratio}
\end{figure}

\begin{figure}[h]
\begin{minipage}[t]{80mm}
\includegraphics[height=13pc,width=14pc]{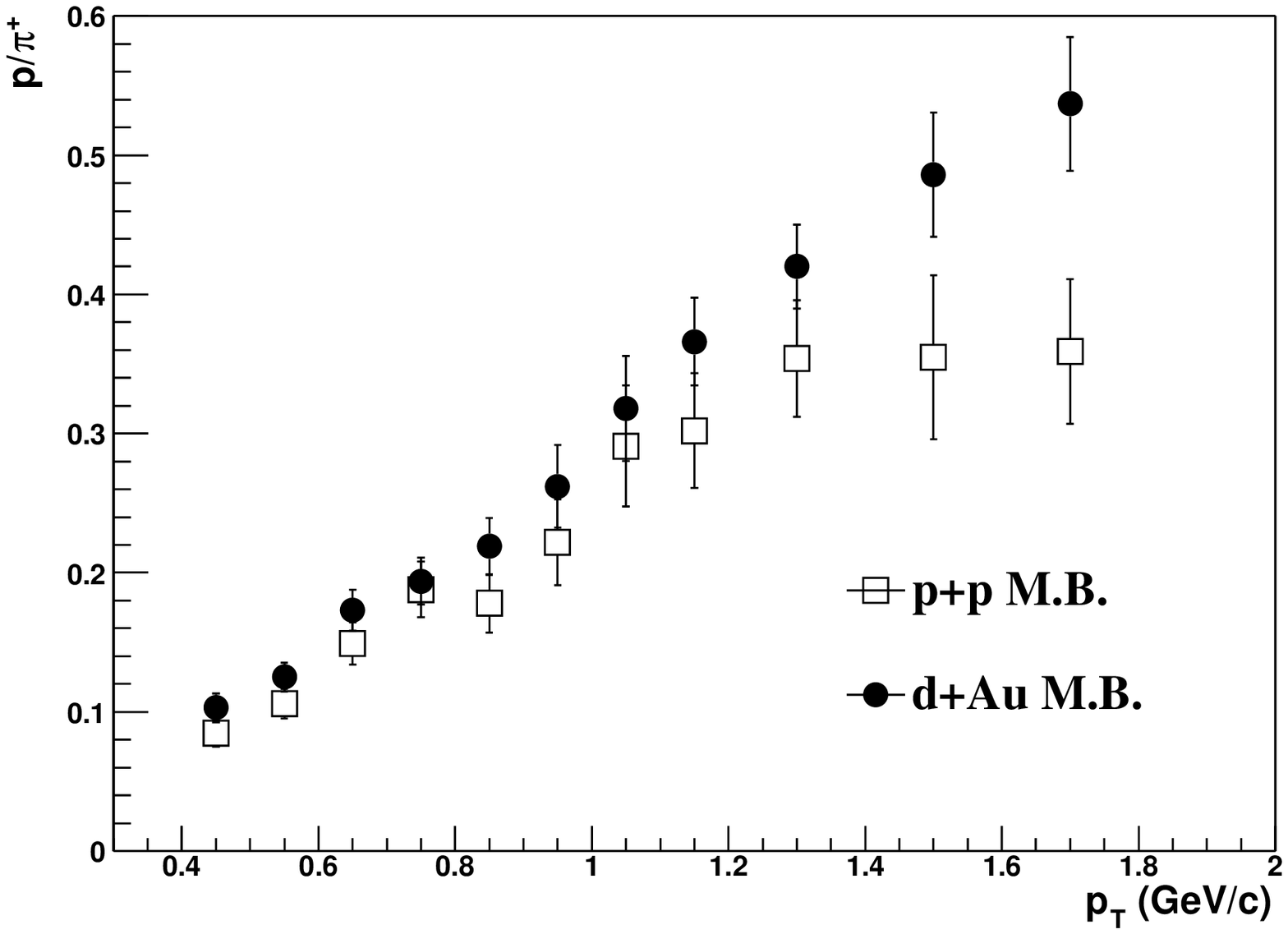}
\end{minipage}
\begin{minipage}[t]{80mm}
\includegraphics[height=13pc,width=14pc]{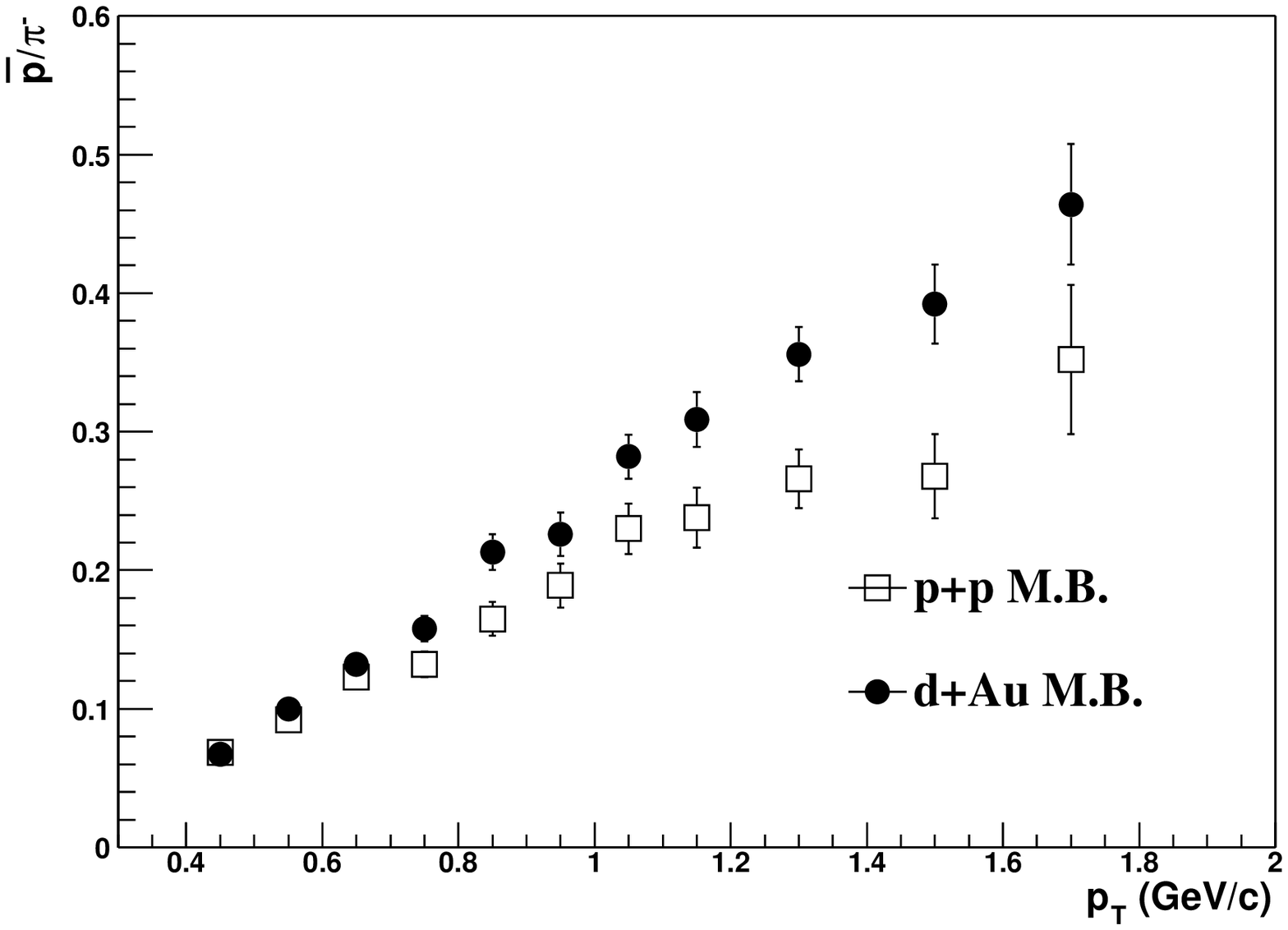}
\end{minipage}
\caption{$p(\bar{p})/\pi$ ratios as a function of $p_{T}$ in d+Au
and p+p minimum-bias collisions. The open symbols are for p+p
collisions and the solid symbols for d+Au collisions. Errors are
statistical.} \label{ppiratio}
\end{figure}

The $K/\pi$ and $p/\pi$ ratios are shown in Figure~\ref{kpiratio}
and Figure~\ref{ppiratio} individually. From the plots,  the
$K/\pi$ ratios increase with $p_{T}$ in both d+Au and p+p
collisions and the increasing trend is the same within our errors.
The $p/\pi$ ratios increase with $p_{T}$ in both d+Au and p+p
collisions and the increasing in d+Au collisions is faster than
that in p+p collisions. The trends of the $K/\pi$ and $p/\pi$ as a
function of $p_{T}$ show little centrality dependence in d+Au
collisions.

\section{$dN/dy$, $\langle p_T \rangle$, and model fits}
\begin{figure}[h]
\begin{minipage}[t]{80mm}
\includegraphics[height=17pc,width=18pc]{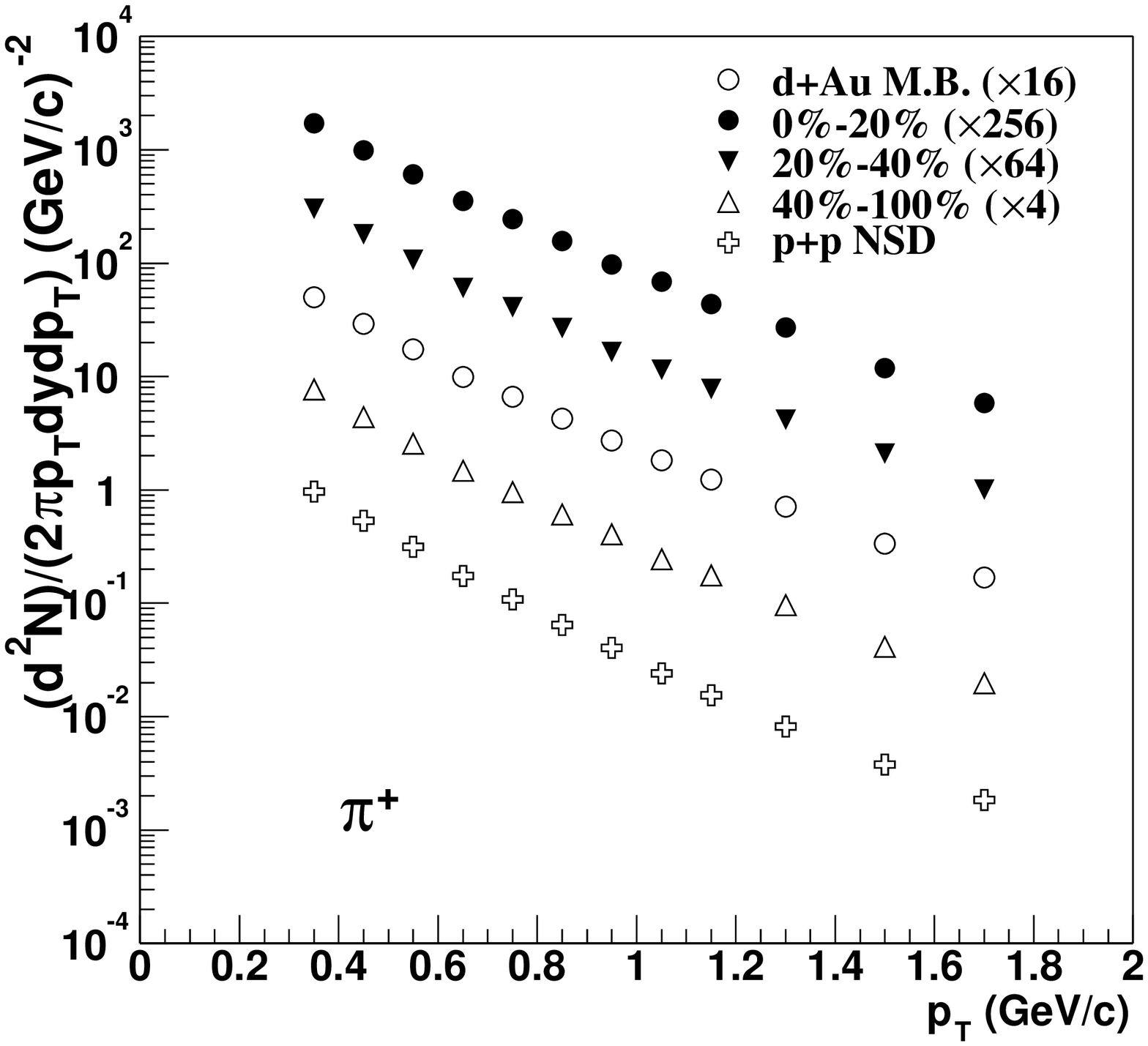}
\end{minipage}
\hspace{\fill}
\begin{minipage}[t]{80mm}
\includegraphics[height=17pc,width=18pc]{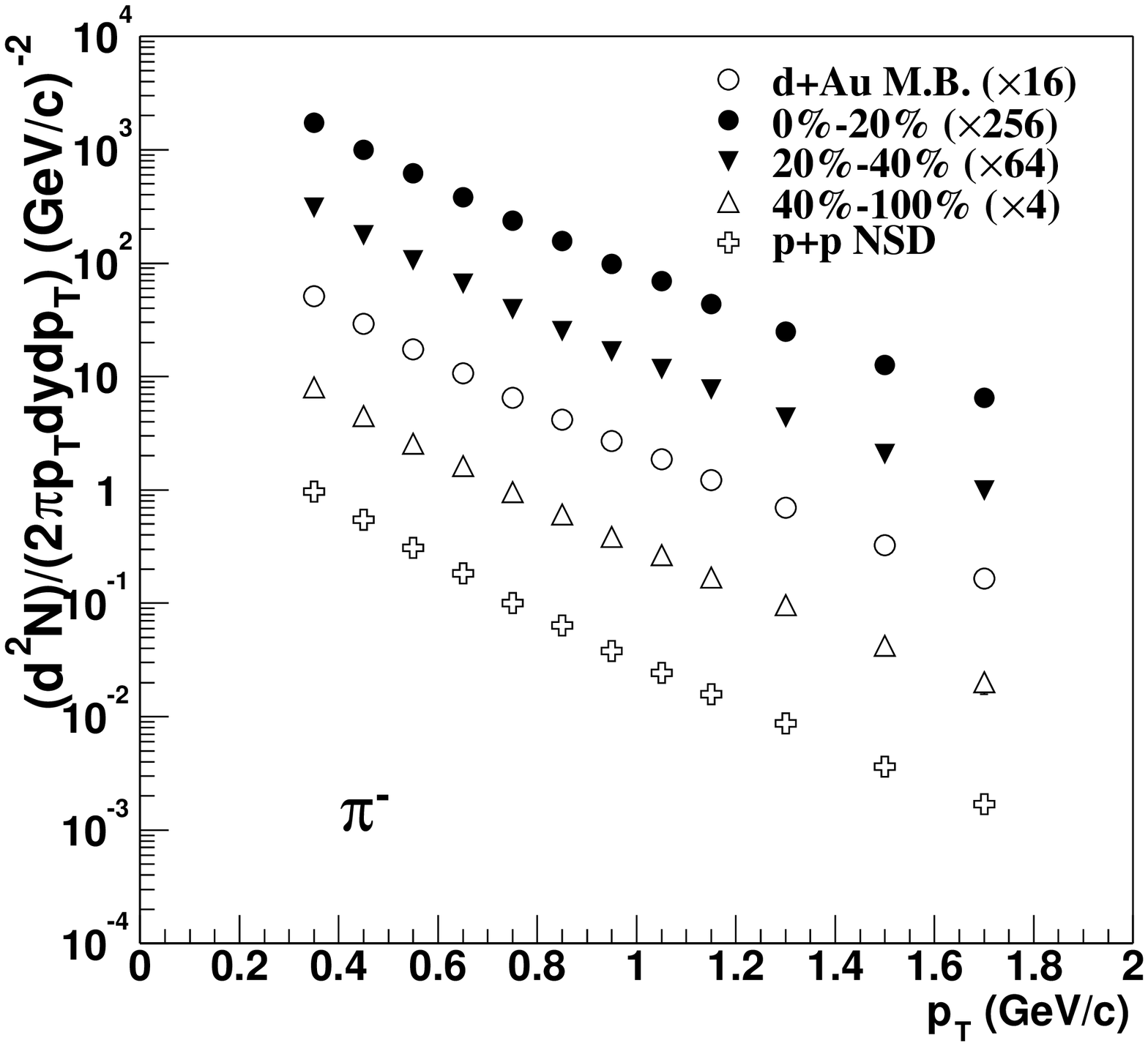}
\end{minipage}
\caption{The re-scaled $\pi^{+}$ and $\pi^{-}$ spectra in
minimum-bias, centrality selected d+Au collisions and also in p+p
collisions. The errors are statistical.} \label{pionspectra}
\end{figure}

\begin{figure}[h]
\begin{minipage}[t]{80mm}
\includegraphics[height=17pc,width=18pc]{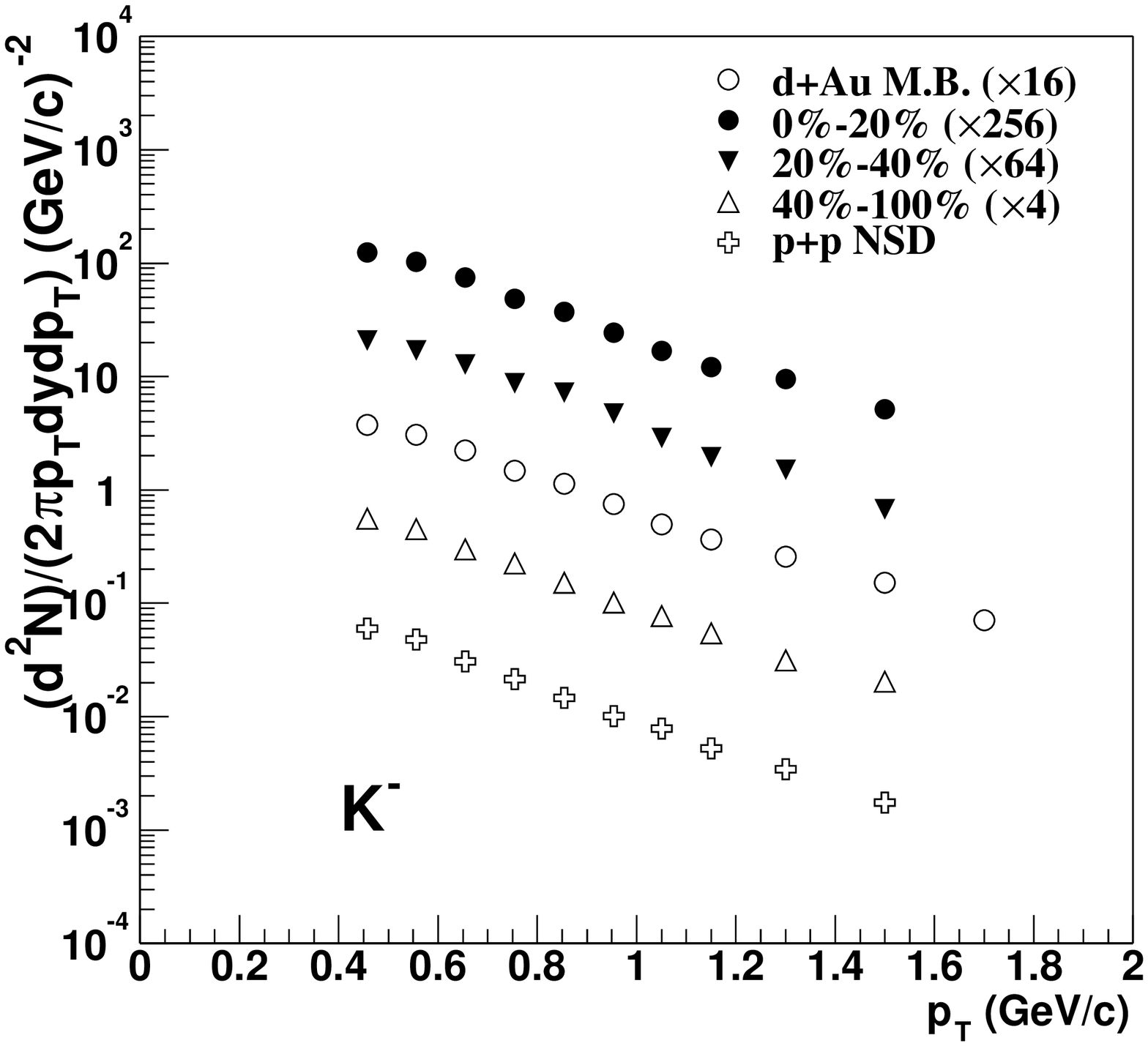}
\end{minipage}
\hspace{\fill}
\begin{minipage}[t]{80mm}
\includegraphics[height=17pc,width=18pc]{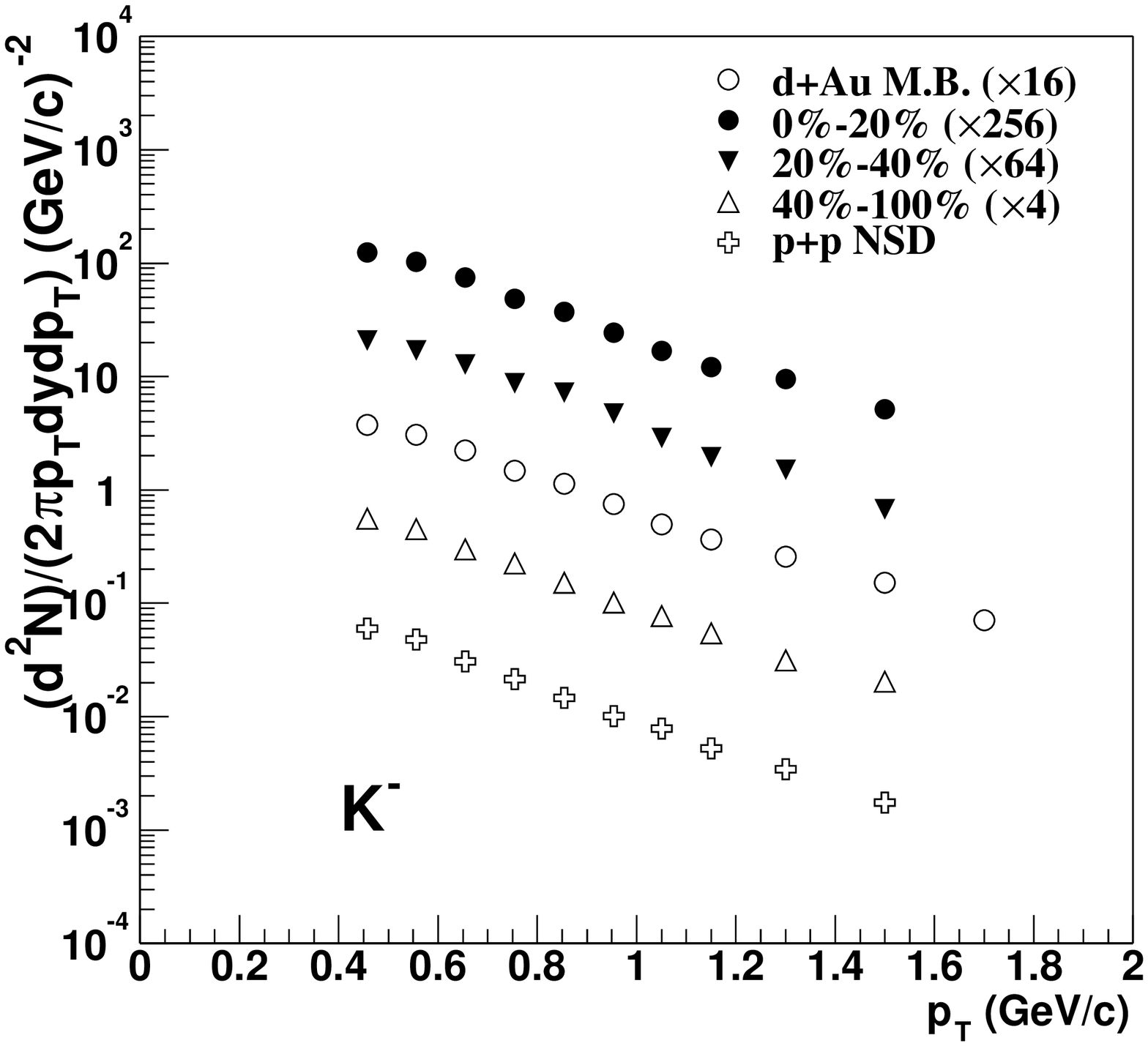}
\end{minipage}
\caption{The re-scaled $K^{+}$ and $K^{-}$ spectra in
minimum-bias, centrality selected d+Au collisions and also in p+p
collisions. The errors are statistical.} \label{kaonspectra}
\end{figure}
\begin{figure}[h]
\begin{minipage}[t]{80mm}
\includegraphics[height=17pc,width=18pc]{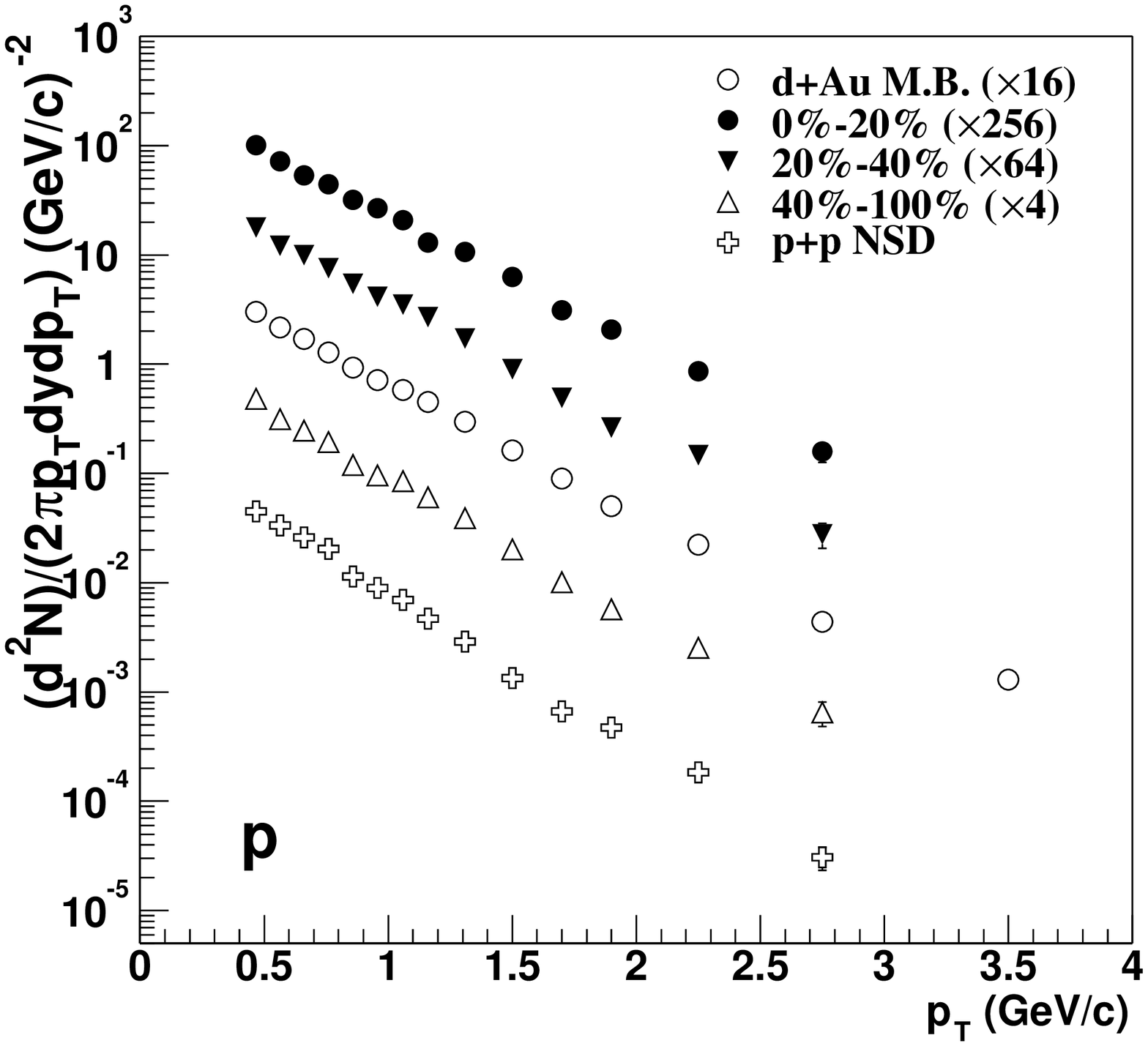}
\end{minipage}
\hspace{\fill}
\begin{minipage}[t]{80mm}
\includegraphics[height=17pc,width=18pc]{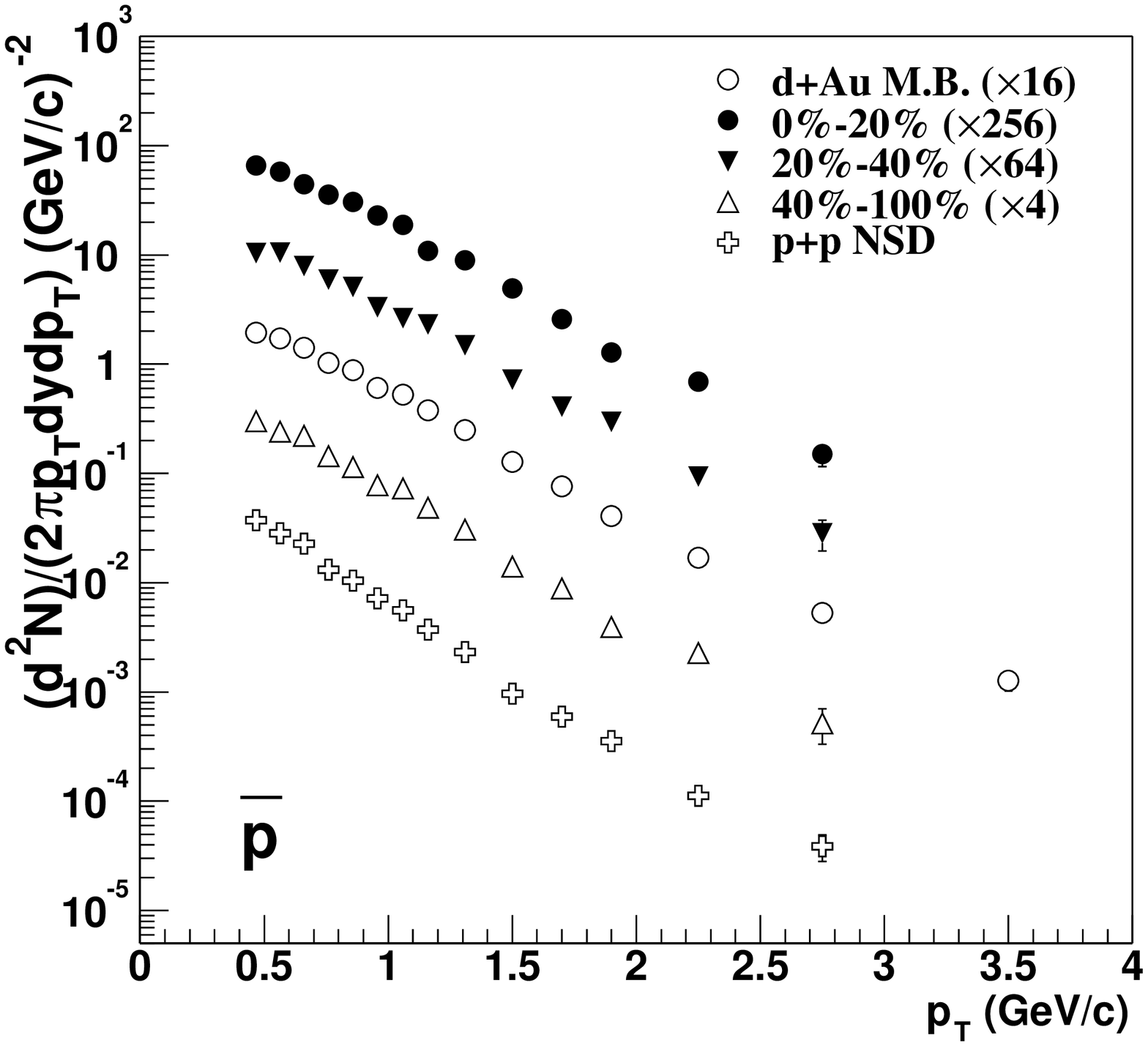}
\end{minipage}
\caption{The re-scaled $p$ and $\bar{p}$ spectra in minimum-bias,
centrality selected d+Au collisions and also in p+p collisions.
The errors are statistical.} \label{protonspectra}
\end{figure}
\begin{figure}[h]
\centering
\includegraphics[height=18pc,width=24pc]{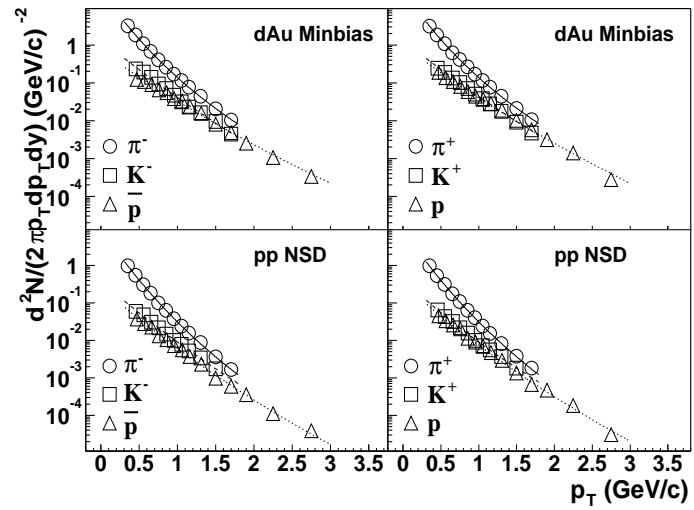}
\caption{The spectra of $\pi^{-}$, $\pi^{+}$, $K^{-}$, $K^{+}$,
$\bar{p}$ and $p$ in d+Au and p+p minimum-bias collisions. The
curves are from power law fit.} \label{powerlawfit}
\end{figure}
\begin{figure}[h]
\centering
\includegraphics[height=24pc,width=24pc]{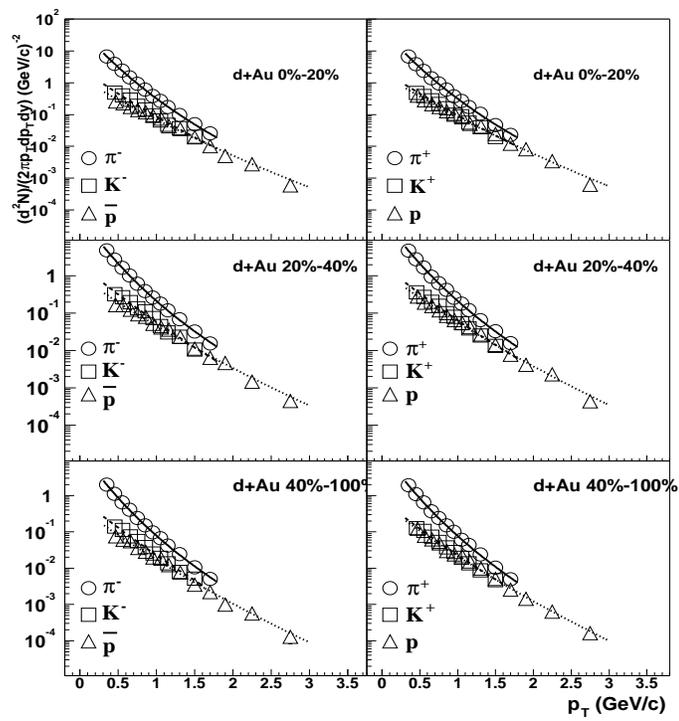}
\caption{The spectra of $\pi^{-}$, $\pi^{+}$, $K^{-}$, $K^{+}$,
$\bar{p}$ and $p$ in three centrality selected d+Au collisions.
The curves are from power law fit.} \label{3centralitypowerlawfit}
\end{figure}

The spectra in minimum-bias and centrality selected d+Au
collisions and also in p+p collisions are shown in
Figure~\ref{pionspectra},
 Figure~\ref{kaonspectra} and Figure~\ref{protonspectra}. The spectra
show little centrality dependence for each particle in d+Au
collisions but harder than those in p+p collisions. The power law
function was used to fit the spectra and get the $dN/dy$ and
$\langle p_T \rangle$. The power law fit function is:
\begin{equation}
\frac{1}{2\pi p_T}\frac{d^2N}{dydp_T}=a(1+\frac{p_T}{\langle p_T
\rangle \frac{n-3}{2}})^{-n}
\end{equation}
Where the parameter $a$ is a constant value proportional to the
mid-rapidity yield $dN/dy$, the parameter $n$ is the order of the
power law and $\langle p_T \rangle$ is the mean value of the
transverse momentum which is extracted from the fit.
Figure~\ref{powerlawfit} shows power law fit to the spectra of
minimum-bias d+Au and p+p collisions.
Figure~\ref{3centralitypowerlawfit} shows power law fit to the
spectra of 3 centrality selected d+Au collisions. The power law
fit results are listed in Table~\ref{dndypowerlawfit} and
Table~\ref{meanptpowerlawfit} individually. The thermal
model~\cite{thermal} was also used to fit the spectra. The final
$dN/dy$ and $\langle p_T \rangle$ are shown in
Table~\ref{finaldndy} and Table~\ref{finalmeanpt} respectively,
which were obtained by averaging the results from the power law
fit and thermal fit. Half of the differences in them are taken as
the systematic errors due to the extrapolation to low $p_{T}$
region. The errors in this table include the systematic
uncertainties and statistical errors.
\begin{table}[h]
\begin{scriptsize}
\centering
  \begin{tabular}{|c|c|c|c|c|c|c|}
    \hline
        Centrality Bin & $\pi^{-}$ & $\pi^{+}$ & $K^{-}$ & $K^{+}$ & $\bar{p}$ & $p$ \\ \hline
        d+Au M.B.      & $0.403\pm0.004$ & $0.405\pm0.004$ & $0.609\pm0.009$ & $0.629\pm0.009$ & $0.714\pm0.008$ & $0.677\pm0.010$\\ \hline
        0\%-20\% & $0.421\pm0.004$ & $0.421\pm0.004$ & $0.626\pm0.018$ & $0.658\pm0.016$ & $0.727\pm0.013$ & $0.705\pm0.014$  \\ \hline
        20\%-40\% & $0.408\pm0.004$ & $0.411\pm0.004$ & $0.604\pm0.015$ & $0.625\pm0.015$ & $0.725\pm0.013$ & $0.691\pm0.015$  \\ \hline
        40\%-100\% & $0.387\pm0.005$ & $0.391\pm0.004$ & $0.589\pm0.016$ & $0.616\pm0.016$ & $0.667\pm0.013$ & $0.646\pm0.014$  \\ \hline
        p+p & $0.357\pm0.004$ & $0.361\pm0.004$ & $0.571\pm0.013$ & $0.571\pm0.013$ & $0.567\pm0.010$ & $0.569\pm0.012$  \\ \hline
  \end{tabular}
\caption{$\langle p_T \rangle$ of $\pi^{-}$, $\pi^{+}$, $K^{-}$,
$K^{+}$, $\bar{p}$ and $p$ from power law fit in minimum-bias,
centrality selected d+Au collisions and also in p+p collisions.
The errors are from the power law fit. The unit of $p_{T}$ is
GeV/c.} \label{meanptpowerlawfit}
\end{scriptsize}
\end{table}

\begin{table}[h]
\begin{scriptsize}
\centering
  \begin{tabular}{|c|c|c|c|c|c|c|}
    \hline
        Centrality Bin & $\pi^{-}$ & $\pi^{+}$ & $K^{-}$ & $K^{+}$ & $\bar{p}$ & $p$ \\ \hline
        d+Au M.B.      & $5.078\pm0.080$ & $5.032\pm0.080$ & $0.685\pm0.013$ & $0.703\pm0.012$ & $0.466\pm0.009$ & $0.594\pm0.019$\\ \hline
        0\%-20\% & $10.657\pm0.190$ & $10.521\pm0.187$ & $1.448\pm0.085$ & $1.453\pm0.035$ & $0.972\pm0.026$ & $1.222\pm0.045$  \\ \hline
        20\%-40\% & $7.631\pm0.148$ & $7.515\pm0.139$ & $0.988\pm0.028$ & $1.051\pm0.027$ & $0.651\pm0.018$ & $0.842\pm0.033$  \\ \hline
        40\%-100\% & $3.153\pm0.069$ & $3.024\pm0.060$ & $0.399\pm0.012$ & $0.379\pm0.011$ & $0.261\pm0.008$ & $0.338\pm0.014$  \\ \hline
        p+p & $1.524\pm0.027$ & $1.504\pm0.027$ & $0.166\pm0.004$ & $0.173\pm0.009$ & $0.113\pm0.003$ & $0.137\pm0.007$  \\ \hline
  \end{tabular}
\caption{$dN/dy$ of $\pi^{-}$, $\pi^{+}$, $K^{-}$, $K^{+}$,
$\bar{p}$ and $p$ from power law fit in minimum-bias, centrality
selected d+Au collisions and also in p+p collisions. The errors
are from the power law fit.} \label{dndypowerlawfit}
\end{scriptsize}
\end{table}

\begin{table}[h]
\begin{scriptsize}
\centering
  \begin{tabular}{|c|c|c|c|c|c|c|}
    \hline
        Centrality Bin & $\pi^{-}$ & $\pi^{+}$ & $K^{-}$ & $K^{+}$ & $\bar{p}$ & $p$ \\ \hline
        d+Au M.B.      & $0.420\pm0.019$ & $0.422\pm0.019$ & $0.613\pm0.025$ & $0.625\pm0.025$ & $0.761\pm0.056$ & $0.739\pm0.069$\\ \hline
        0\%-20\% & $0.435\pm0.017$ & $0.436\pm0.017$ & $0.627\pm0.025$ & $0.646\pm0.028$ & $0.774\pm0.056$ & $0.761\pm0.063$  \\ \hline
        20\%-40\% & $0.425\pm0.019$ & $0.427\pm0.018$ & $0.610\pm0.025$ & $0.622\pm0.025$ & $0.766\pm0.052$ & $0.744\pm0.061$  \\ \hline
        40\%-100\% & $0.405\pm0.020$ & $0.408\pm0.019$ & $0.591\pm0.024$ & $0.608\pm0.026$ & $0.715\pm0.056$ & $0.703\pm0.063$  \\ \hline
        p+p & $0.377\pm0.021$ & $0.379\pm0.020$ & $0.565\pm0.023$ & $0.565\pm0.023$ & $0.627\pm0.065$ & $0.634\pm0.070$  \\ \hline
  \end{tabular}
\caption{The final $\langle p_T \rangle$ of $\pi^{-}$, $\pi^{+}$,
$K^{-}$, $K^{+}$, $\bar{p}$ and $p$ in minimum-bias, centrality
selected d+Au collisions and also in p+p collisions. The unit of
$p_{T}$ is GeV/c.} \label{finalmeanpt}
\end{scriptsize}
\end{table}

\begin{table}[h]
\begin{scriptsize}
\centering
  \begin{tabular}{|c|c|c|c|c|c|c|}
    \hline
        Centrality Bin & $\pi^{-}$ & $\pi^{+}$ & $K^{-}$ & $K^{+}$ & $\bar{p}$ & $p$ \\ \hline
        d+Au M.B.      & $4.731\pm0.359$ & $4.668\pm0.356$ & $0.662\pm0.040$ & $0.684\pm0.039$ & $0.425\pm0.044$ & $0.531\pm0.067$\\ \hline
        0\%-20\% & $10.063\pm0.628$ & $9.932\pm0.621$ & $1.383\pm0.095$ & $1.418\pm0.079$ & $0.896\pm0.084$ & $1.114\pm0.117$  \\ \hline
        20\%-40\% & $7.137\pm0.514$ & $7.074\pm0.464$ & $0.952\pm0.059$ & $1.020\pm0.059$ & $0.603\pm0.054$ & $0.765\pm0.083$  \\ \hline
        40\%-100\% & $2.925\pm0.236$ & $2.829\pm0.203$ & $0.387\pm0.023$ & $0.371\pm0.020$ & $0.238\pm0.025$ & $0.304\pm0.036$  \\ \hline
        p+p & $1.411\pm0.116$ & $1.400\pm0.108$ & $0.163\pm0.009$ & $0.168\pm0.010$ & $0.099\pm0.015$ & $0.120\pm0.018$  \\ \hline
  \end{tabular}
\caption{The final $dN/dy$ of $\pi^{-}$, $\pi^{+}$, $K^{-}$,
$K^{+}$, $\bar{p}$ and $p$ in minimum-bias, centrality selected
d+Au collisions and also in p+p collisions.} \label{finaldndy}
\end{scriptsize}
\end{table}

\section{System comparison}
\begin{figure}[h]
\centering
\includegraphics[height=24pc,width=24pc]{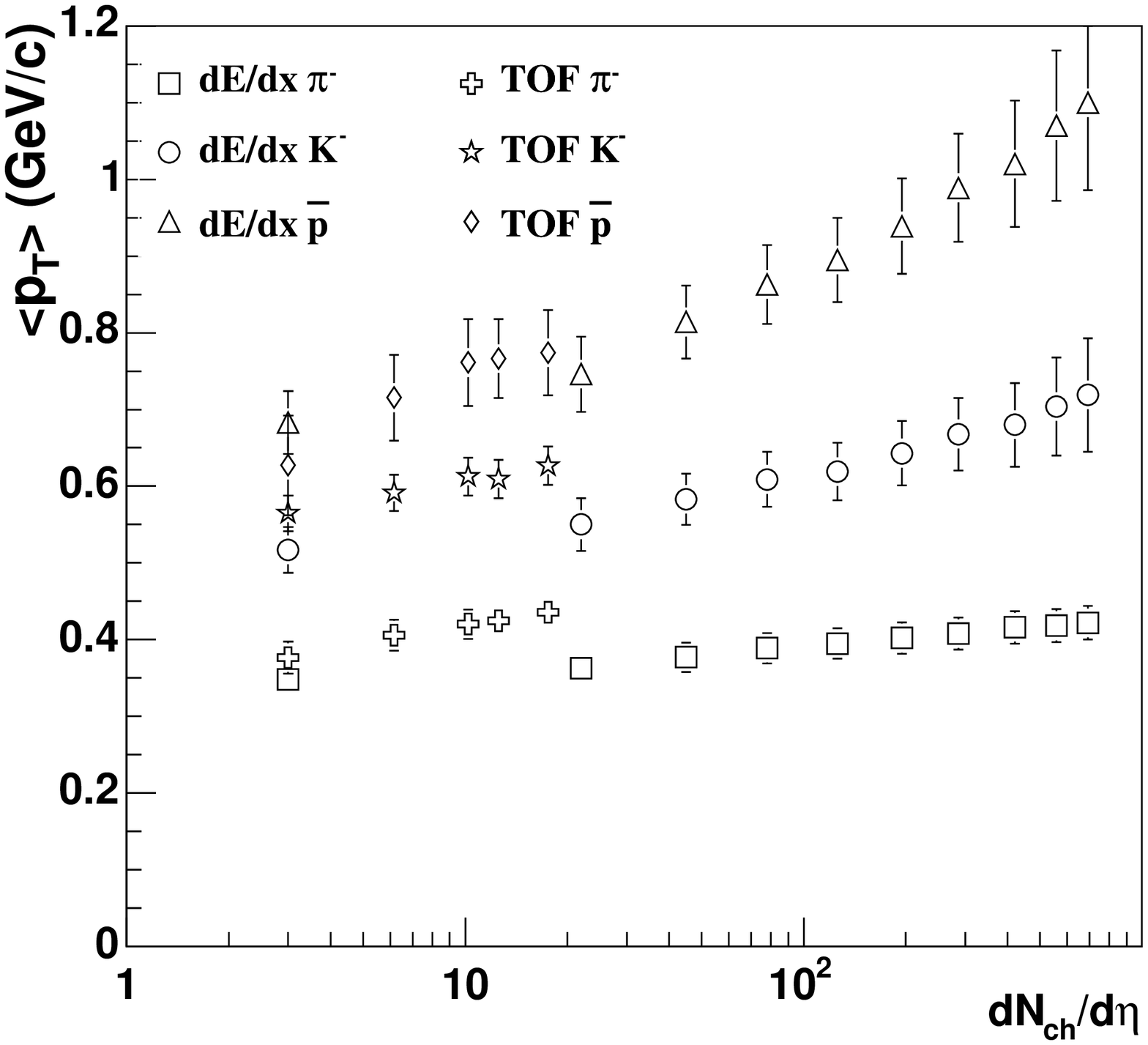}
\caption{$\langle p_T \rangle$ as a function of $dN/d\eta$. The
squared, circled and triangled symbols are from~\cite{olga} in p+p
and Au+Au collisions. The cross, star and diamond are our data
points in p+p and d+Au collisions. Statistic errors and systematic
uncertainties have been added in quadrature.} \label{meanptNch}
\end{figure}
Figure~\ref{meanptNch} shows the $\langle p_T \rangle$ of
$\pi^{-}$, $K^{-}$ and $\bar{p}$ as a function of charged particle
multiplicity at mid-rapidity. From p+p to d+Au collisions, the
$\langle p_T \rangle$ increase with charged particle multiplicity
smoothly. We observed the $\langle p_T \rangle$ in 0\%-20\% d+Au
collisions are larger than those in peripheral Au+Au collisions.
\begin{figure}[h]
\centering
\includegraphics[height=24pc,width=24pc]{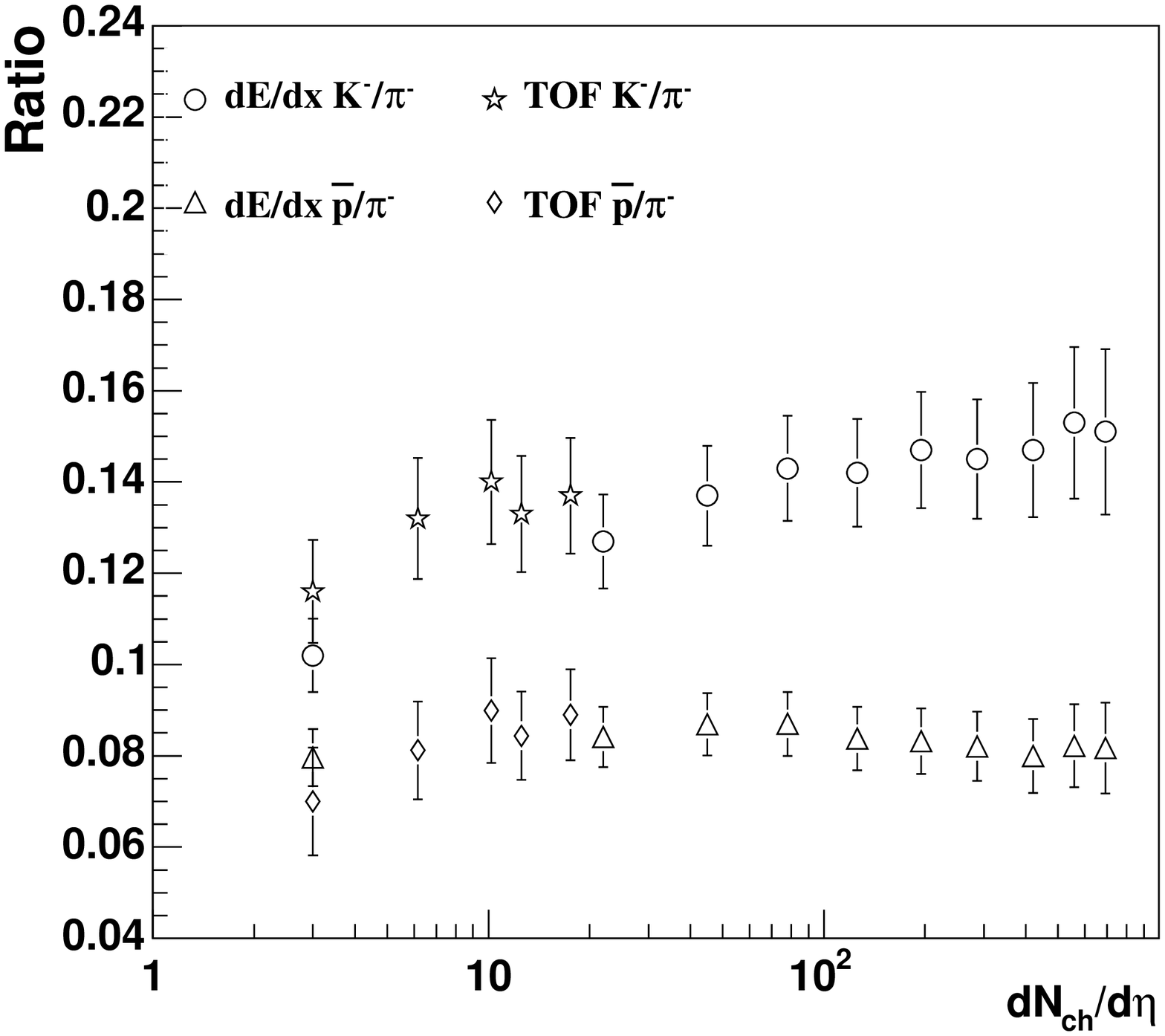}
\caption{$K^{-}/\pi^{-}$ and $\bar{p}/\pi^{-}$ as a function of
$dN/d\eta$.  The circled and triangled symbols are
from~\cite{olga} in p+p and Au+Au collisions. The star and diamond
are our data points in p+p and d+Au collisions. Statistic errors
and systematic uncertainties have been added in quadrature.}
\label{kaonpbarpionratio}
\end{figure}
The $K^{-}/\pi^{-}$ and $\bar{p}/\pi^{-}$ as a function of charged
particle multiplicity at mid-rapidity are shown in
Figure~\ref{kaonpbarpionratio}. The $K^{-}/\pi^{-}$ and
$\bar{p}/\pi^{-}$ ratios were derived by taking the ratios of the
dN/dy of $K^{-}$ or $\bar{p}$ over the dN/dy of $\pi^{-}$ in
table~\ref{finaldndy}. These ratios increase with charged particle
multiplicity from p+p, d+Au to Au+Au collisions smoothly. The
kinetic freeze out temperature $T_{kin}$ and flow velocity
$\langle \beta \rangle$ from thermal fit as a function of charged
particle multiplicity are shown in Figure~\ref{freezeoutT}. We can
see the $T_{kin}$ is flat from p+p to d+Au and then decreases from
d+Au to Au+Au collisions and the $\langle \beta \rangle$ increases
from p+p, d+Au to Au+Au collisions.
\begin{figure}[h]
\begin{minipage}[t]{80mm}
\includegraphics[height=17pc,width=18pc]{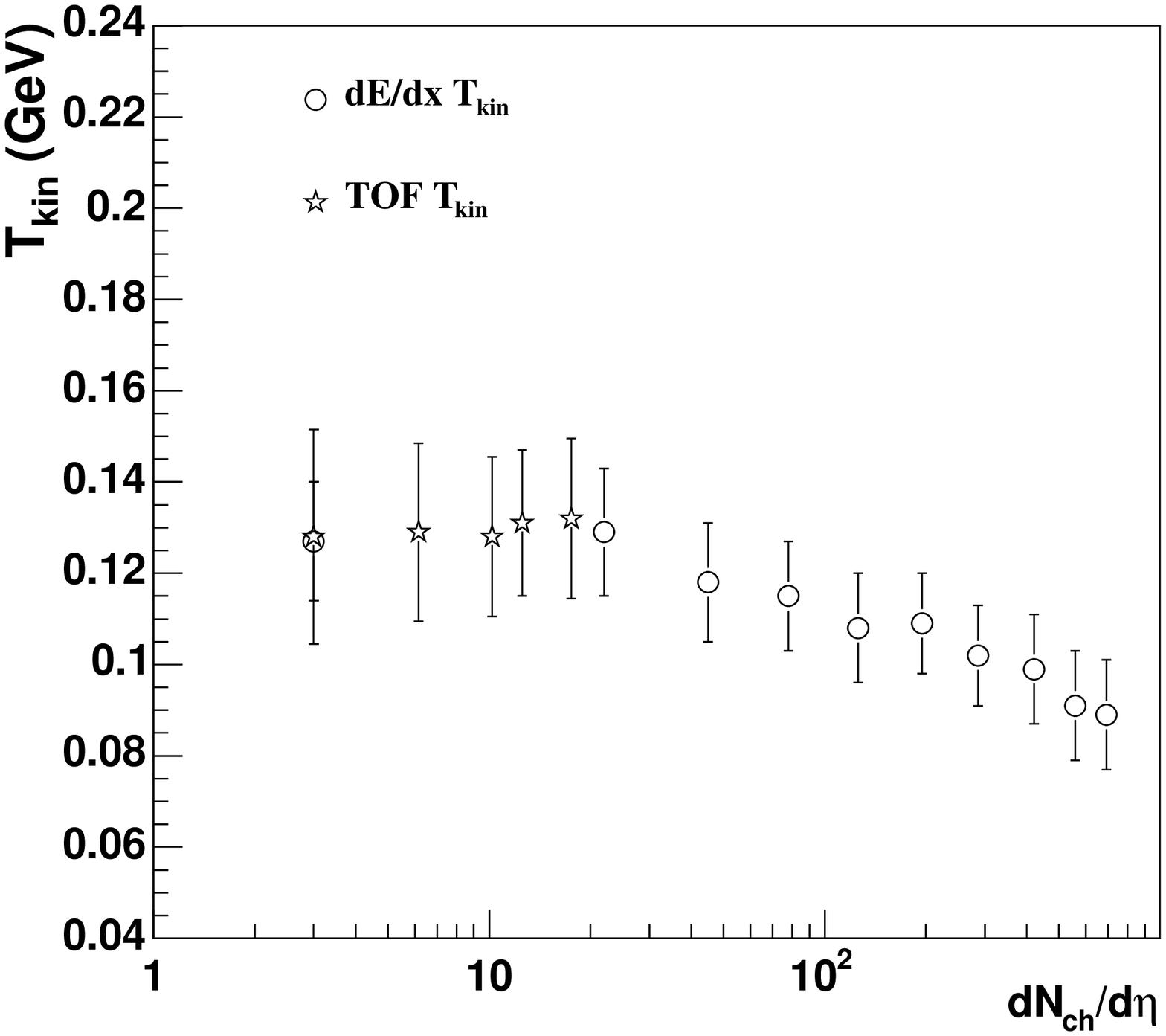}
\end{minipage}
\hspace{\fill}
\begin{minipage}[t]{80mm}
\includegraphics[height=17pc,width=18pc]{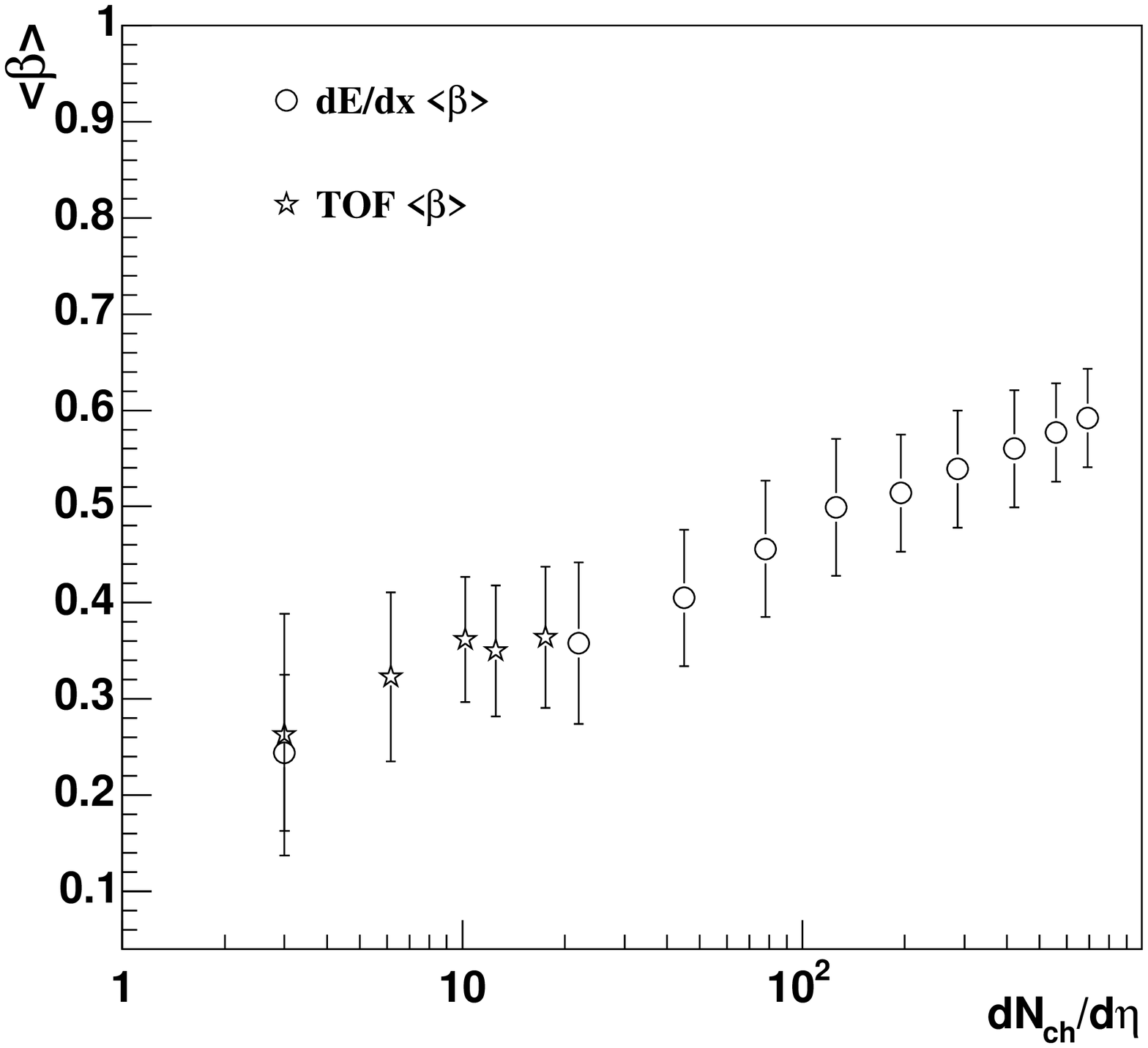}
\end{minipage}
\caption{The kinetic freeze out temperature $T_{kin}$ (left) and
flow velocity $\langle \beta \rangle$ (right) from thermal fit as
a function of charged particle multiplicity. The circled symbols
are from ~\cite{olga} in p+p and Au+Au collisions. The star are
our data points in p+p and d+Au collisions. Errors are
systematic.} \label{freezeoutT}
\end{figure}

\chapter{Discussion}
\label{chp:discussion}
\section{Cronin effect}
The identified particle spectra in d+Au and p+p collisions not
only provide the reference for those in Au+Au collisions at 200
GeV, but also provide a chance to see the mechanism of the Cronin
effect itself clearly. Cronin effect was observed 30 years
ago~\cite{cronin}. It is the enhancement of particle production at
high $p_{T}$. The enhancement was explained by initial multiple
parton scattering. Also the recent experimental results of Cronin
effect on inclusive charged hadron are consistent with the
predictions based on initial multiple parton
scattering~\cite{accardi}. It suggests the suppression at
intermediate $p_{T}$ in Au+Au collisions is due to final state
effects. However, the initial multiple parton scattering with the
independent fragmentation function will result in the same Cronin
effect for $p(\bar{p})$ and for pions, while experimentally the
Cronin effect for $p(\bar{p})$ is larger than that for $\pi$.
That's to say the initial multiple scattering with the independent
fragmentation function can't account for the Cronin effect
observed. Maybe in the initial multiple parton scattering, the
broadening for gluon and for quark/antiquark are not the
same~\cite{xinniancommu}. Or maybe the fragmentation processes in
p+A collisions are not the same as those in p+p
collisions~\cite{qiucommu}. Whether the Cronin effect is initial
state effect or final state effect will be discussed below.

\subsection{Model comparison: initial state effect?}
\begin{figure}[h]
\begin{minipage}[t]{80mm}
\includegraphics[height=18pc,width=14pc]{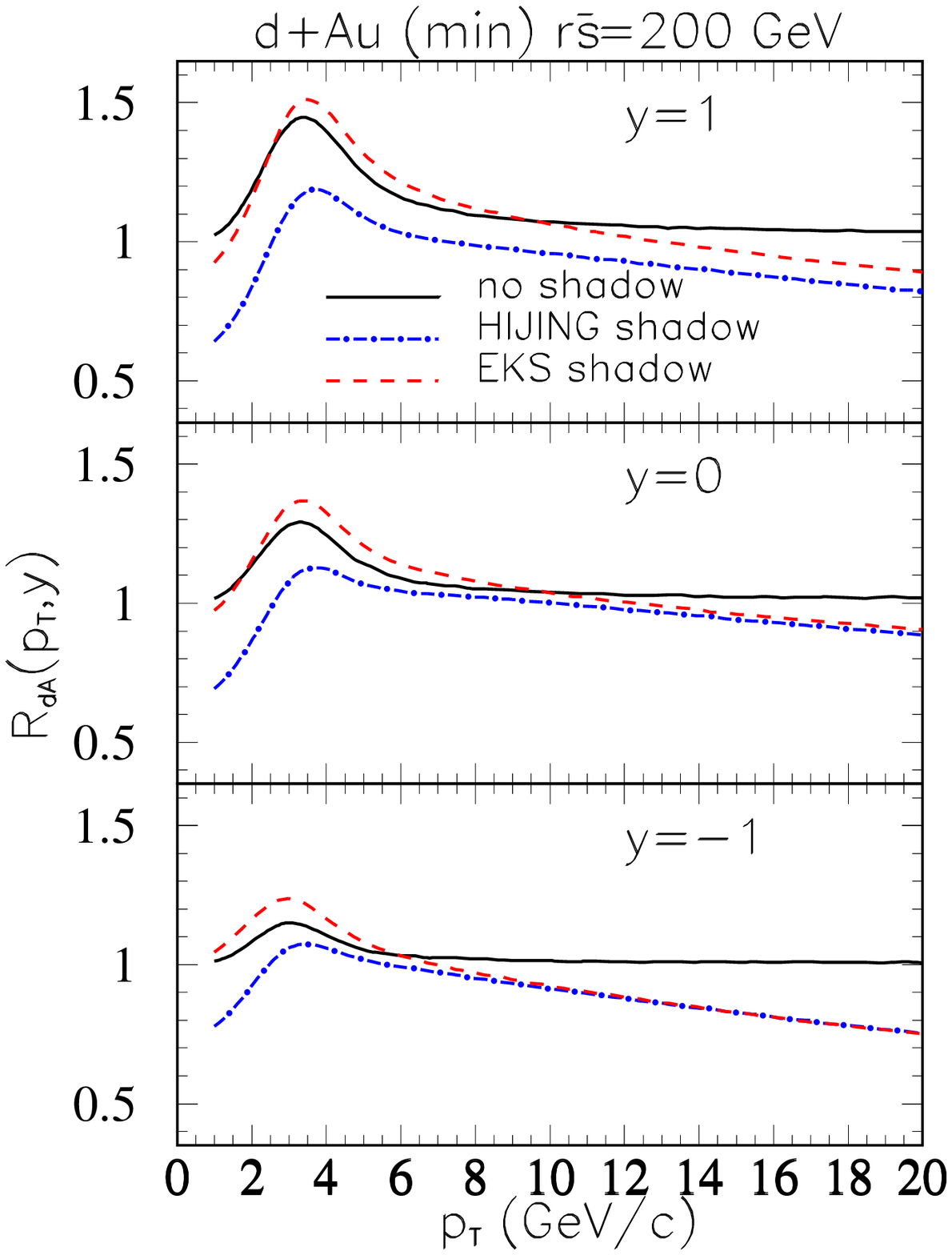}
\end{minipage}
\hspace{-2cm}
\begin{minipage}[t]{80mm}
\includegraphics[height=17pc,width=24pc]{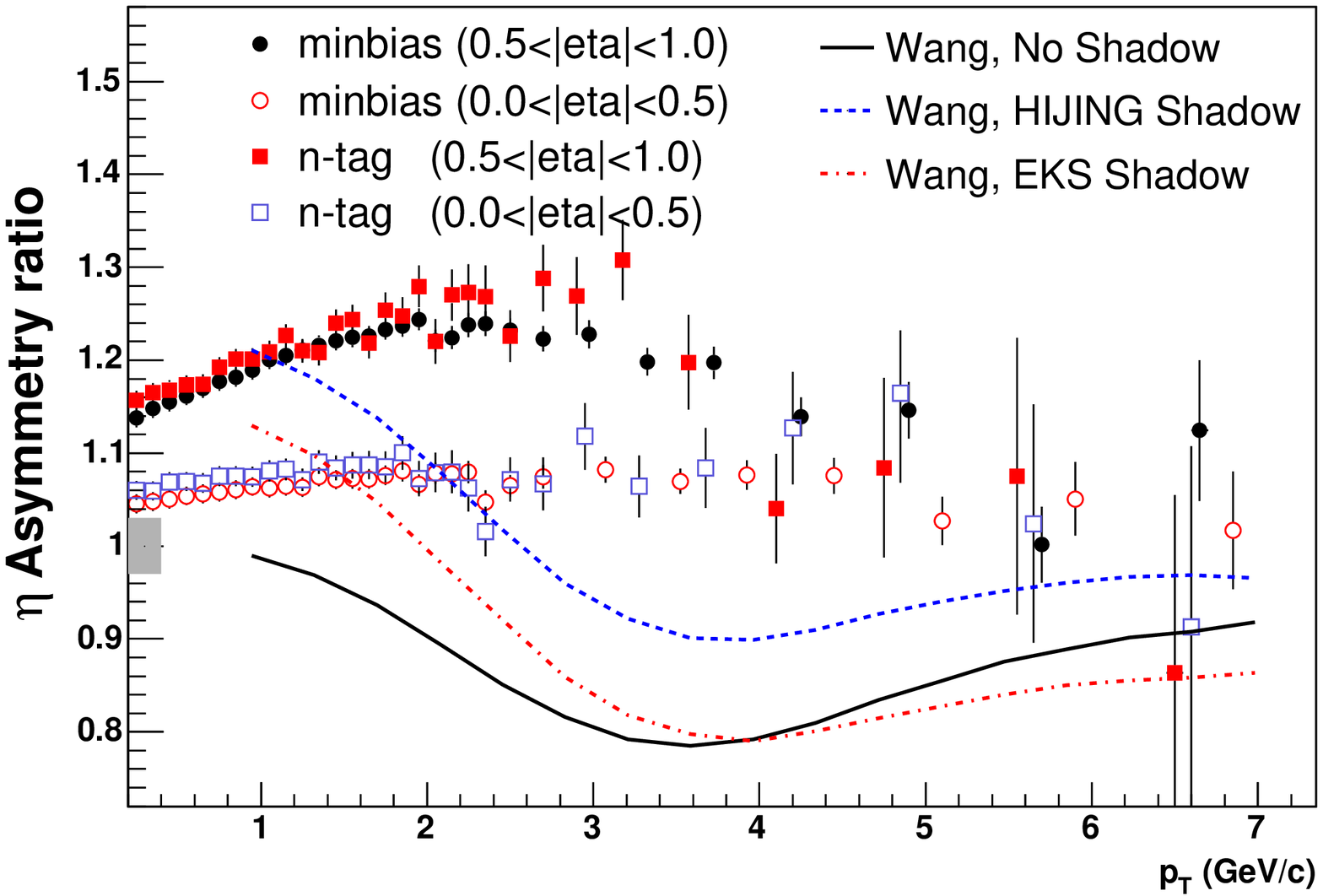}
\end{minipage}
\caption{(left) The Cronin effect at different rapidity as a
function of $p_{T}$. The different curve in each panel shows the
different shadowing. This figure is from ~\cite{xinnian:dAu}.
(right) The $\eta$ asymmetry of the Cronin effect: the ratio of
Cronin effect in Au beam outgoing direction over the Cronin effect
in deuteron beam outgoing direction. This figure is from
~\cite{johan}.} \label{etaasymmetry}
\end{figure}

The initial multiple parton scattering model predicts that the
Cronin effect on deuteron beam outgoing side is larger than that
on Au beam outgoing side since the deuteron traverses a much
larger nucleus~\cite{xinnian:dAu}. Figure~\ref{etaasymmetry}
(left) shows the predictions for the Cronin effect at different
rapidity range. The different curves correspond to the prediction
results from different shadowings. The $y=1$ is on the deuteron
beam outgoing side. The $y=-1$ is on the Au beam outgoing side.
The $y=0$ is at mid-rapidity. We can see that the $R_{dAu}$ on
deuteron beam side ($y=1$) increases faster than that on Au beam
side ($y=-1$). If we take the ratio of $R_{dAu}$ on Au beam side
over $R_{dAu}$ on deuteron beam side, it will result in a minimum
value at $p_{T}\sim3.5$ GeV/c, as shown in the curves on the right
plot of Figure~\ref{etaasymmetry}. The solid symbol on the right
plot of Figure~\ref{etaasymmetry} represents the data
points~\cite{johan}, which is the ratio of $R_{dAu}$ on Au beam
side at $-1<\eta<-0.5$ over $R_{dAu}$ on deuteron beam side at
$0.5<\eta<1$. We observe the $\eta$ asymmetry from experiment
reaches a maximum value firstly and then decreases. This is
different from the predictions. That means, the model based on
initial multiple parton scattering only, can't reproduce the
experimental results. Recently, Qiu and Vitev have come up with
the idea of coherent multiple scattering and applied it to the
RHIC experiments~\cite{coherent}. In this picture, the hard probe
may interact coherently with many low x parton inside different
nucleons inside the nucleus. As a result, this process will lead
to the suppression of the total cross section. This coherent
effect will play an important role in p+A collisions at forward
rapidity. In the deuteron outgoing beam direction, the coherent
effect is non-negligible since the Au nucleus is big while on the
Au side, the coherent effect is not big since the deuteron is of a
small size. This will result in bigger suppression on the deuteron
side than on the Au side. It may qualitatively reproduce the data.
This coherent multiple scattering is a final-state effect.

As we all known, in Au+Au collisions, the suppression at
intermediate $p_{T}$ can be reproduced by the initial multiple
scattering and jet quenching
qualitatively~\cite{starhighpt,jetquench}. However, the model
based on the initial multiple scattering, jet quenching and
independent fragmentation will result in the same suppression for
baryons and mesons at intermediate $p_{T}$ in Au+Au collisions.
Experimentally $R_{cp}$ for baryons are larger than $R_{cp}$ for
mesons at intermediate $p_{T}$. This difference can be reproduced
by coalescence or recombination models~\cite{hwa,fries,ko}.
Recently the recombination model~\cite{hwayang} has been applied
to d+Au system to see whether it can reproduce the Cronin effect
or not.

With the help of Prof. C.B. Yang~\cite{yang}, I also compare our
pion and proton spectra in d+Au collisions with the recombination
model~\cite{hwayang}. In the following the recombination
model~\cite{hwayang} will be discussed and the comparison between
the data and the model will be presented in detail.
\subsection{Model comparison: recombination}
The inclusive distribution for the production of pions can be
written in the recombination model~\cite{hwayang}, when mass
effects are negligible, in the invariant form
\begin{eqnarray} p{dN_{\pi}  \over  dp} = \int {dp_1 \over
p_1}{dp_2 \over p_2}F_{q\bar{q}} (p_1, p_2) R_{\pi}(p_1, p_2, p) ,
\label{1}
\end{eqnarray} where $F_{q\bar{q}} (p_1, p_2)$ is the joint
distribution of a $q$ and $\bar q$ at $p_1$ and $p_2$, and
$R_{\pi}(p_1, p_2, p)$ is the recombination function for forming a
pion at $p$:  $R_{\pi}(p_1, p_2, p) = (p_1p_2/p)\delta (p_1+p_2-
p)$. $F_{q\bar{q}}$ depends on the colliding hadron/nuclei.  In
general, $F_{q\bar{q}}$ has four contributing components
represented schematically by
\begin{eqnarray} F_{q\bar{q}} = {\cal TT} + {\cal TS} + ({\cal SS})
_1 + ({\cal SS})_2
\end{eqnarray} where $\cal{ T}$ denotes thermal distribution and
$\cal{S}$ shower distribution.  $({\cal SS})_1$ signifies two
shower partons in the same hard-parton jet, while $({\cal SS})_2$
stands for two shower partons from two nearby jets~\cite{hwayang}.

\begin{figure}[h]
\begin{minipage}[t]{80mm}
\includegraphics[height=18pc,width=18pc]{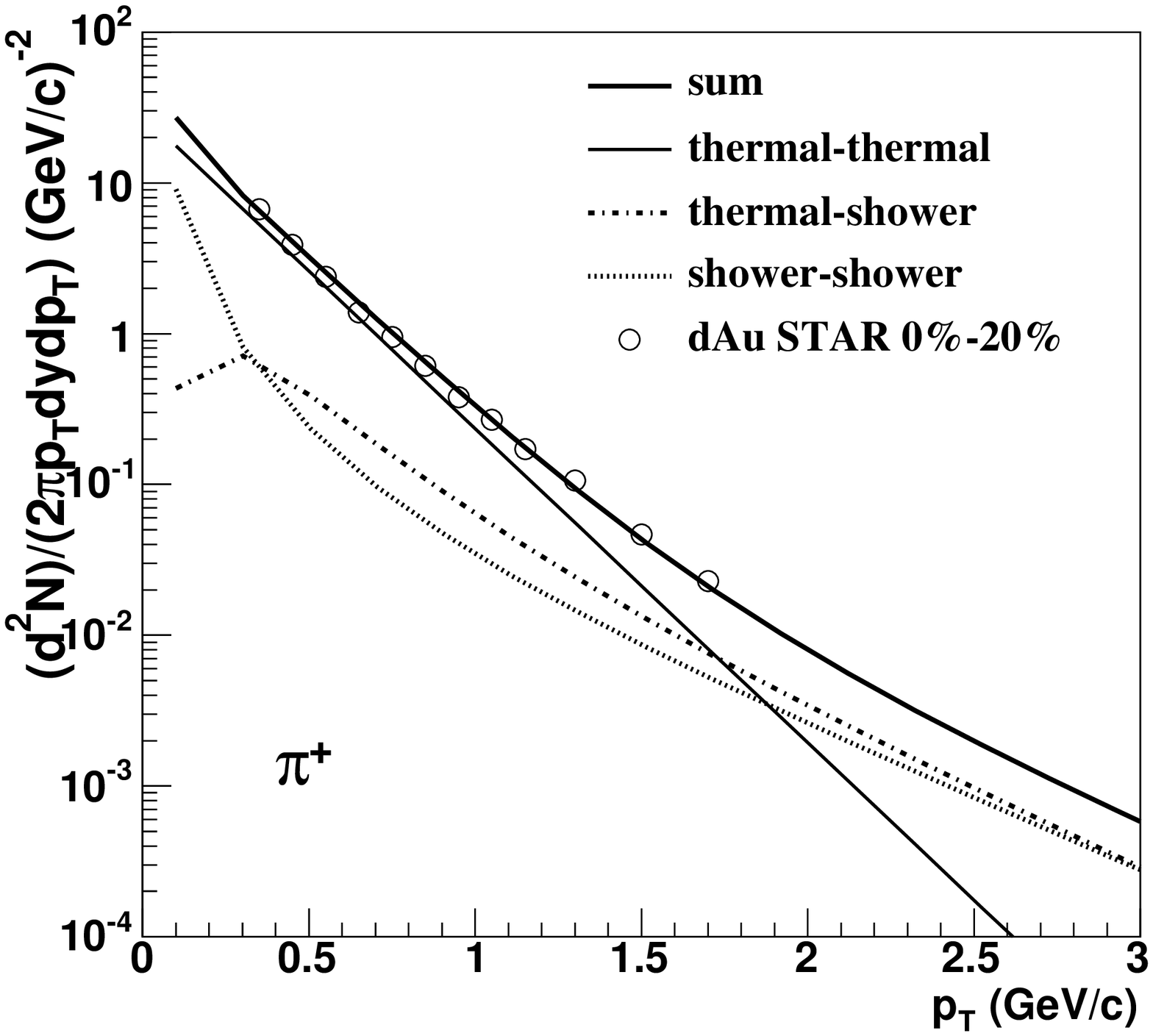}
\end{minipage}
\hspace{0mm}
\begin{minipage}[t]{80mm}
\includegraphics[height=18pc,width=18pc]{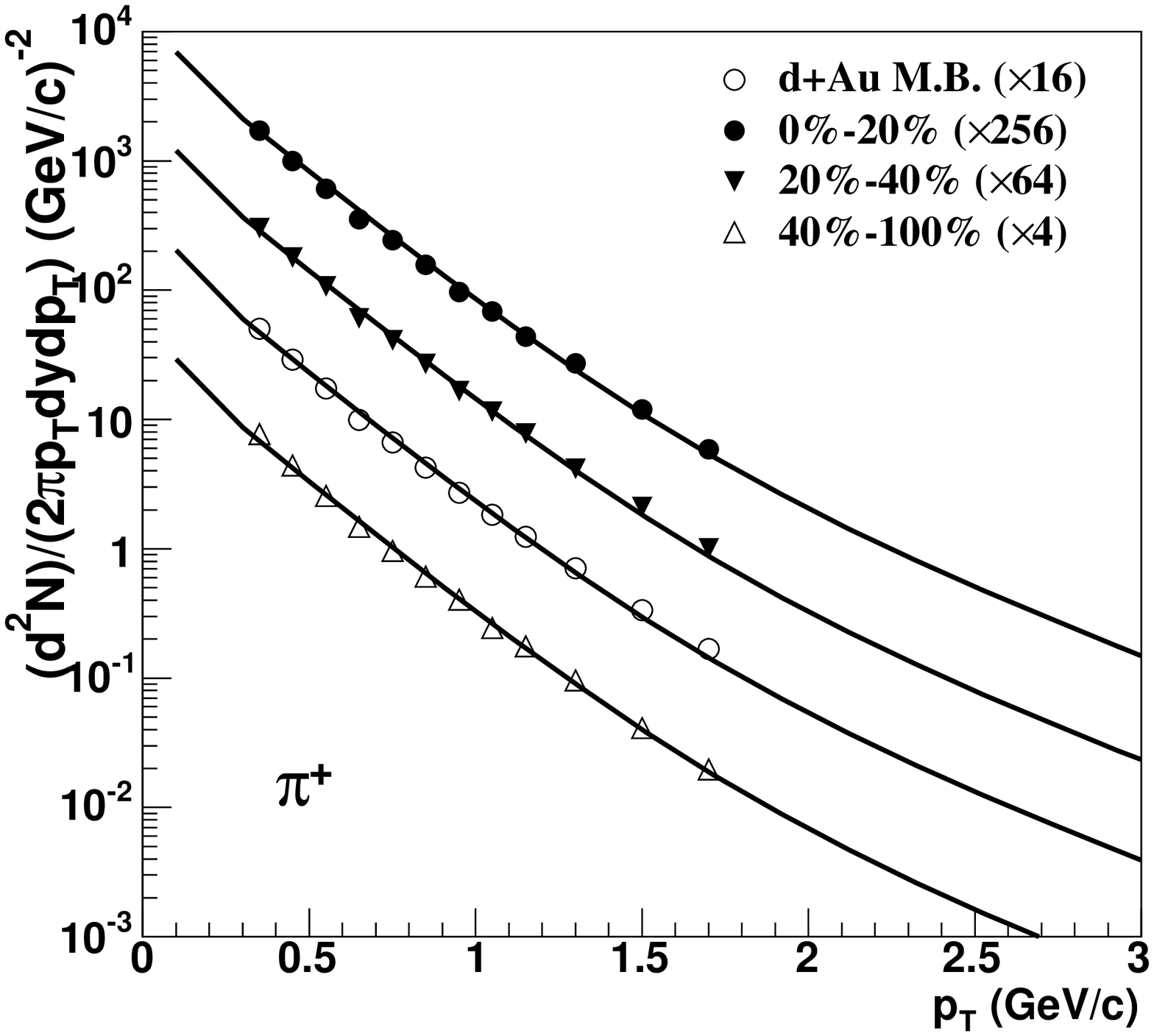}
\end{minipage}
\caption{(left) The invariant yield for $\pi^{+}$ at 0\%-20\% d+Au
collisions as a function of $p_{T}$. The open circles are our data
points. The curves are the calculation results from recombination
model. Sum represents the total contribution from recombination
model. Thermal-thermal represents the soft contribution. The
thermal-shower represents the contribution from the interplay
between soft and hard components. The shower-shower represents the
hard contribution. (right) The invariant yields for $\pi^{+}$ in
minimum-bias and centrality selected d+Au collisions as a function
of $p_{T}$. The symbols represent our data points. The curves on
the top of the symbols are the corresponding calculation results
from recombination model. } \label{pirecombination}
\end{figure}

For $p+A$ collisions it may not be appropriate to refer to any
partons as thermal in the sense of a hot plasma as in heavy-ion
collisions. Here in d+Au collisions, the symbol $\cal{ T}$
represents the soft parton distribution at low $k_T$. At low $p_T$
the observed pion distribution is exponential; we identify it with
the contribution of the ${\cal TT}$ term~\cite{hwayang}.
\begin{eqnarray}
{dN^{{\cal TT}}_{\pi}  \over  pdp} =  {C^2 \over 6} exp (-p/T)
\end{eqnarray}
where $T$ is the inverse slope. We shall determine $C$ and $T$ by
fitting the d+Au data at low $p_T$. The pion spectra for different
centralities can be calculated from thermal-thermal ($pion_{tt}$),
thermal-shower ($pion_{ts}$) and shower-shower ($pion_{ss}$)
contributions by using parameters $C$ and $N_{bin}$:
$dN/p_Tdp_T=C\times C\times pion_{tt}+2.5 \times C \times N_{bin}
\times pion_{ts}+2.5 \times N_{bin} \times pion_{ss}$, where C is
determined by fitting the d+Au data at $0.4<p_T<1.0$ GeV/c, and
$N_{bin}$ is the number of binary collisions. The data points of
$pion_{tt}, pion_{ts}$ and $pion_{ss}$ are from Prof. C.B.
Yang~\cite{yang}. The $C$ values for minimum-bias, 0-20\%, 20-40\%
and 40-$\sim$100\% d+Au collisions are 8.85, 13.08, 10.96 and 6.84
individually. The $T$ value of 0.21 GeV is used in the low $p_{T}$
fit. Figure~\ref{pirecombination} shows the $\pi^{+}$ spectra in
d+Au collisions as well as those from recombination model. This
figure shows that the recombination model can reproduce the
spectra of pion in minimum-bias and centrality selected d+Au
collisions.

\begin{figure}[h]
\begin{minipage}[t]{80mm}
\includegraphics[height=18pc,width=18pc]{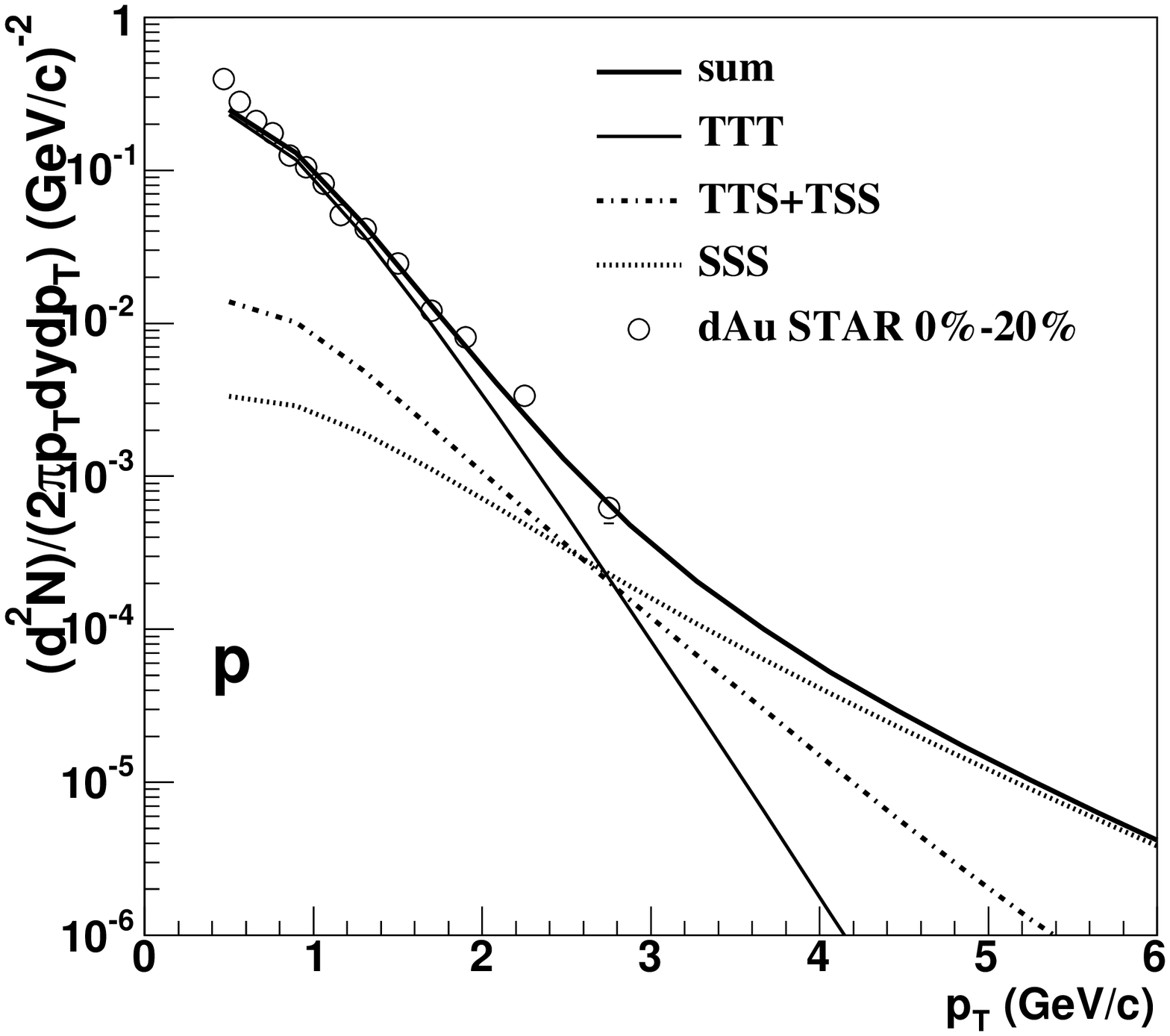}
\end{minipage}
\hspace{0mm}
\begin{minipage}[t]{80mm}
\includegraphics[height=18pc,width=18pc]{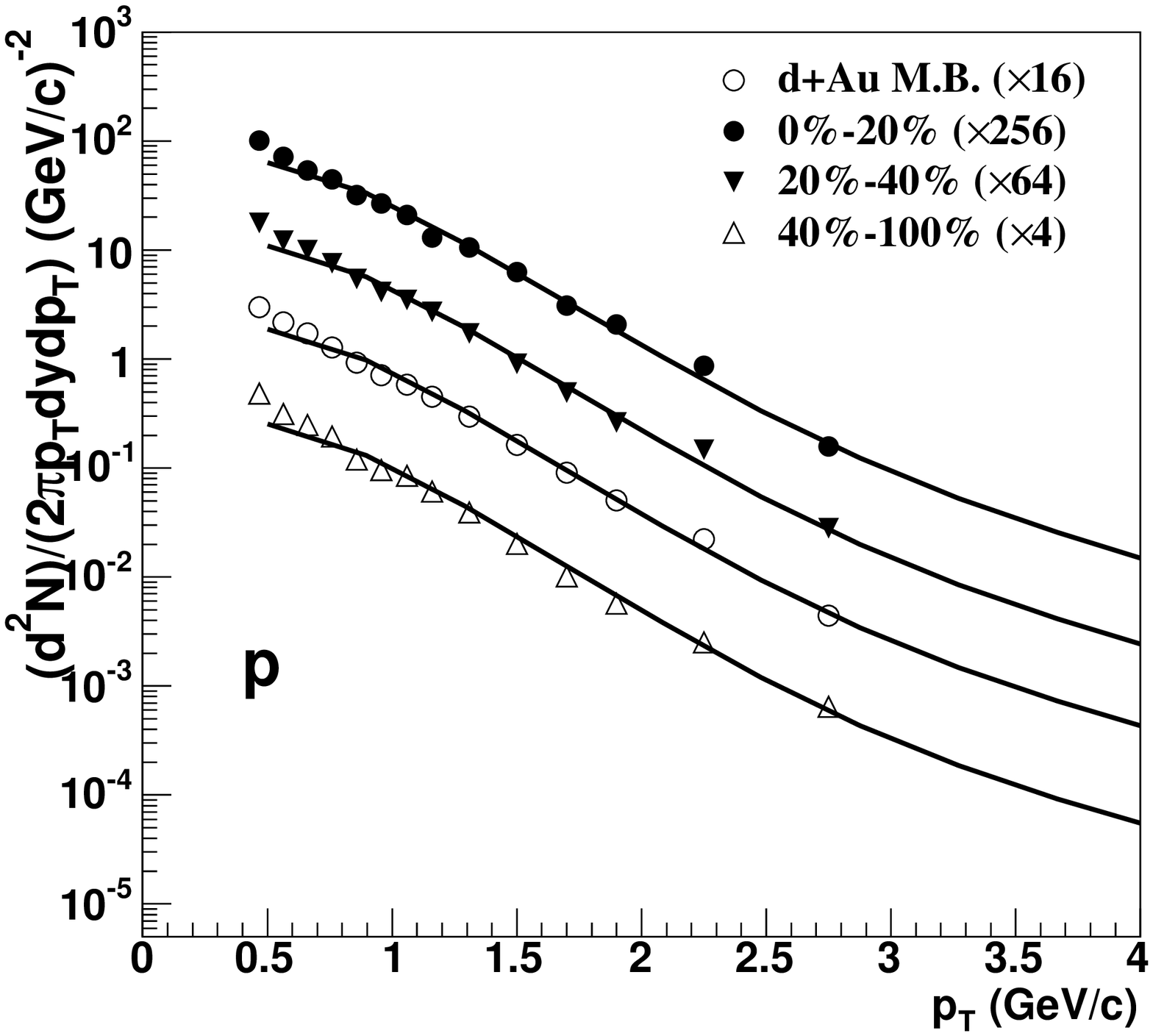}
\end{minipage}
\caption{(left) The invariant yield for $p$ at 0\%-20\% d+Au
collisions as a function of $p_{T}$. The open circles are our data
points. The curves are the calculation results from recombination
model. Sum represents the total contribution from recombination
model. TTT represents the soft contribution. The TTS+TSS
represents the contribution from the interplay between soft and
hard components. The SSS represents the hard contribution. (right)
The invariant yields for $p$ in minimum-bias and centrality
selected d+Au collisions as a function of $p_{T}$. The symbols
represent our data points. The curves on the top of the symbols
are the corresponding calculation results from recombination
model. } \label{precombination}
\end{figure}

The invariant inclusive distribution for proton formation at
midrapidity in the recombination model ~\cite{hwayang}
\begin{eqnarray}
p^0{dN_p  \over  dp} = \int {dp_1 \over p_1}{dp_2 \over p_2} F
(p_1, p_2, p_3) R_p(p_1, p_2, p_3, p)
\end{eqnarray}
where all momentum variables $p_i$ and $p$ are transverse momenta,
and $p^0$ denotes the energy of the proton. $F (p_1, p_2, p_3)$ is
the joint distribution of $u, u,$ and $d$ quarks at $p_1, p_2$ and
$p_3$, respectively.  $R_p(p_1, p_2, p_3, p)$ is the recombination
function  for a proton with momentum $p$. We write schematically
\begin{eqnarray}
F = {\cal TTT} + {\cal TTS} + {\cal TSS}  + {\cal SSS}
\end{eqnarray}
where all the shower partons $\cal{S}$ are from one hard parton
jet. Shower partons from different jets are ignored here for RHIC
energies. In d+Au collisions, $\cal{ T}$ denotes the soft partons
that are not associated with the shower components of a hard
parton. The ${\cal SSS}$ term is regarded as the fragmentation of
a hard parton into a proton.  The ${\cal TTT}$ term comes entirely
from the soft partons, while ${\cal TTS}$ and ${\cal TSS}$
accounts for the interplay between the soft and shower partons.
The soft contribution to the proton spectrum arising from ${\cal
TTT}$ recombination is
\begin{eqnarray}
{dN^{\rm th}_{proton}  \over  pdp} =  {C^3 \over 6} {p^2 \over
p^0} e^{-p/T} { B (\alpha + 2,  \gamma +2) B (\alpha + 2,\alpha +
\gamma +4) \over B (\alpha + 1, \gamma +1) B (\alpha + 1,\alpha +
\gamma +2) }
\end{eqnarray}.
Where $C$ and $T$ are determined by fitting the proton spectra at
low $p_{T}$, the $\alpha$ is equal to 1.75, $\gamma$ is equal to
1.05, $B(x,y)$ is the beta function~\cite{hwayang}. For the
invariant yield of proton, there are 4 different contributions:
soft-soft-soft ($proton_{ttt}$), soft-soft-shower
($proton_{tts}$), soft-shower-shower ($proton_{tss}$), and
shower-shower-shower ($proton_{sss}$). The total contributions are
$dN/p_Tdp_T=C \times C \times C \times proton_{ttt}+C \times C
\times N_{bin} \times proton_{tts}+C \times N_{bin} \times
proton_{tss}+N_{bin} \times proton_{sss}$, where $C$ is determined
by fitting the d+Au data at $0.5<p_T<1.5$ GeV/c, and $N_{bin}$ is
the number of binary collisions. The data points of $proton_{ttt},
proton _{tts}, proton_{tss}$ and $proton_{sss}$ are from Prof.
C.B. Yang~\cite{yang}. The $C$ values for minimum-bias, 0-20\%,
20-40\% and 40-$\sim$100\% d+Au collisions are 9.67, 12.34, 10.92
and 7.91 individually. The $T$ value of 0.21 GeV is used in the
low $p_{T}$ fit. Figure~\ref{precombination} shows the proton
spectra in d+Au collisions as well as those from recombination
model. This figure shows that the recombination model can
reproduce the spectra of proton in minimum-bias and centrality
selected d+Au collisions. From the comparison between our data and
the calculation results from the recombination model, we know that
the recombination model actually can reproduce both the proton and
pion spectra in d+Au collisions, while as we have mentioned above,
the initial multiple parton scattering model~\cite{accardi} with
independent fragmentation can't reproduce the difference of Cronin
effect between proton and pion. Besides, the initial multiple
parton scattering model with independent fragmentation can't
reproduce the $\eta$ asymmetry of the Cronin effect. In the
recombination model~\cite{hwayang}, the number of such soft
partons on the Au outgoing side is larger than that on the
deuteron outgoing side. This will result in the Cronin effect on
the Au side larger than that on the deuteron side~\cite{hwayang}.
Qualitatively the recombination model can reproduce the $\eta$
asymmetry of the Cronin effect. As we know, the recombination
model is a final-state effect model. These all seem to indicate
that the Cronin effect is not initial-state effect only. The
final-state effect plays an important role too. To directly
confirm the Cronin effect is initial or final state effect, it's
necessary for us to compare the Cronin effect of Drell-Yan process
with those of pion, kaon and proton. I will come to this later.


\subsection{Integral yield $R_{dAu}$: shadowing effect?}
\begin{figure}[h]
\centering
\includegraphics[height=24pc,width=24pc]{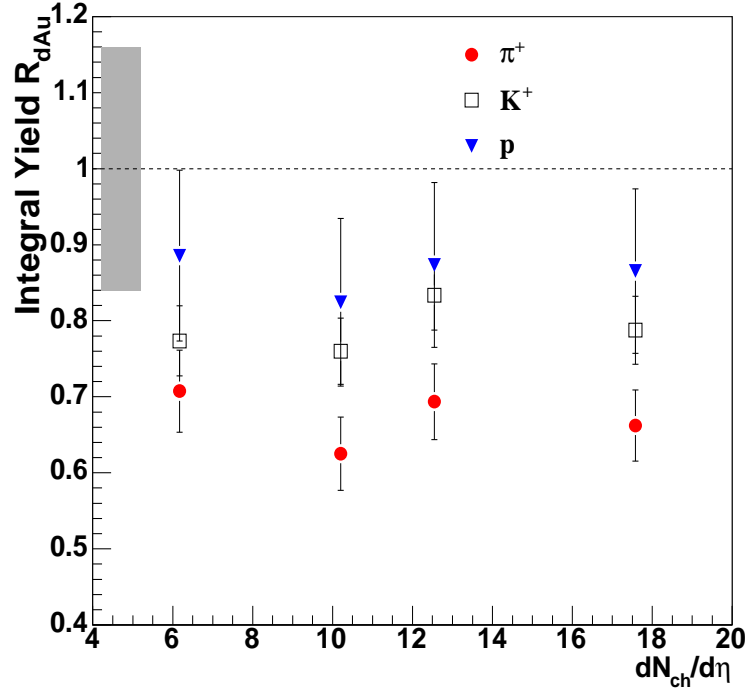}
\caption{Integral yield $R_{dAu}$ as a function of $dN/d\eta$ in
minimum-bias and centrality selected d+Au collisions at
mid-rapidity. Statistic errors and systematic uncertainties have
been added in quadrature. The shadowing represents the
normalization uncertainty.} \label{integralrdau}
\end{figure}

The initial multiple elastic scattering only changes the $p_{T}$
distribution while the total cross section should not change. Thus
we can look at the integral yield $dN/dy$ $R_{dAu}$, which are
measured through comparison to the integral yield $dN/dy$ in p+p
collisions, scaled by the number of binary collisions $N_{bin}$.
Figure~\ref{integralrdau} shows that integral yield $R_{dAu}$ of
pion, kaon and proton as a function of $dN/d\eta$ in minimum-bias
and centrality selected d+Au collisions at mid-rapidity. The
integral yield $R_{dAu}$ of pion and kaon are less than 1 while
that of proton is close to 1. This may be the indication of
shadowing effect at 200 GeV. The integral yield $R_{dAu}$ for
proton is larger than that for kaon and a little bit more larger
than that for pion. This may be the indication that the shadowing
effect is mass dependent at 200 GeV.

\subsection{Initial or final state effect: Drell-Yan process}
\begin{figure}[h]
\centering
\includegraphics[height=24pc,width=24pc]{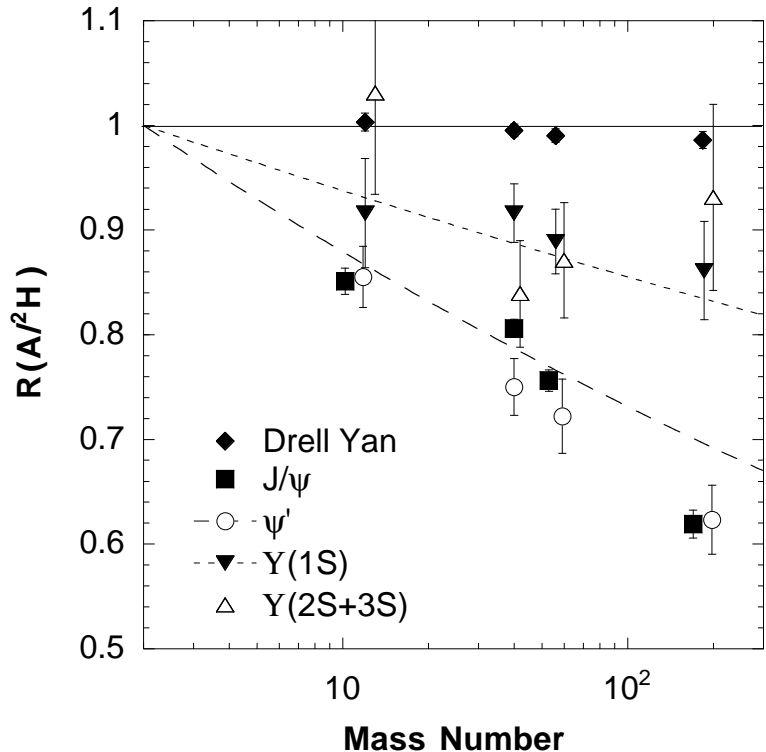}
\caption{Integral yield Cronin ratio as a function of atomic
weight in p+A fixed target experiment for Drell-Yan process, etc.
This figure is from~\cite{drell}.} \label{drellyan}
\end{figure}

In order to see the Cronin effect is initial or final state
effect, we may look into the Drell-Yan process since there is
little final state effect in Drell-Yan process. If there is no
enhancement at high $p_{T}$ for Drell-Yan process, the enhancement
for $\pi, K, p$ is due to final-state effect.
Figure~\ref{drellyan} shows the integral yield Cronin ratio as a
function of atomic weight at p-A fixed target
experiment~\cite{drell}. The proton incident energy is 800 GeV. We
can see there is no enhancement for Drell-Yan process. However,
this is the total cross section while what we want to compare is
Cronin ratio as a function of $p_{T}$. It will be better if we
have the $p_{T}$ dependence of Cronin ratio of Drell-Yan process.
However, at the same $p_T$ range with the same proton incident
energy, the Cronin ratio of Drell-Yan is not available in p+A
collisions. It's hard to compare the Cronin ratio of Drell-Yan
process with those of $\pi, K, p$.

\section{Baryon excess in Au+Au collisions}
Now let's come to another important physics from d+Au collisions.
We know that the $(p+\bar{p})/h$ ratio from minimum-bias Au+Au
collisions~\cite{phenixpid} at a similar energy is about a factor
of 2 higher than that in d+Au and p+p collisions for
$p_{T}{}^{>}_{\sim}2.0$ GeV/c.  This enhancement is most likely
due to final-state effects in Au+Au collisions. There are many
models trying to explain this baryon excess in Au+Au
collisions~\cite{jetquench,junction,derekhydro,pisahydro,fries,ko}.
In the following baryon production mechanism will be discussed.

\subsection{$\bar{p}/p$ ratio vs $p_T$}

\begin{figure}[h]
\centering
\includegraphics[height=18pc,width=24pc]{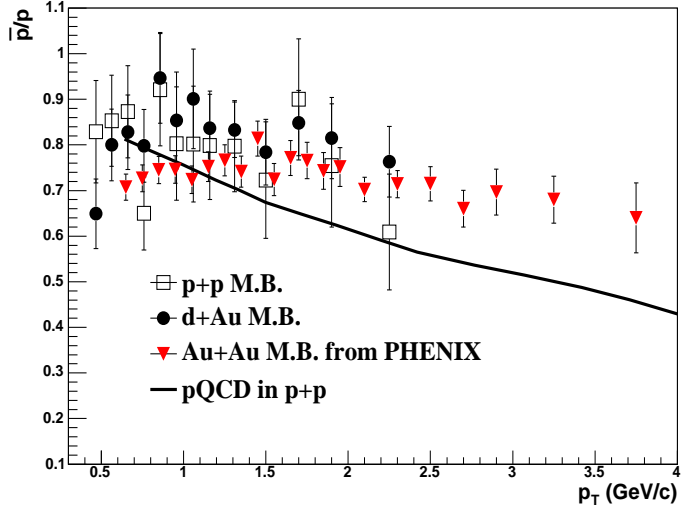}
\caption{$\bar{p}/p$ ratio as a function of $p_{T}$ in d+Au and
p+p minimum-bias collisions. The open squared symbols are for p+p
collisions and the solid circled symbols for d+Au collisions. The
triangled symbols represent the result from Au+Au minimum-bias
collision~\cite{ex0307022}. The curve is the pQCD calculation
results from~\cite{junction} in p+p collisions.  Errors are
statistical.} \label{pbarpdiscussion}
\end{figure}

In 200 GeV Au+Au collisions, $\bar{p}/p$ ratio was observed to be
flat with $p_{T}$ till intermediate $p_{T}$
range~\cite{ex0307022}, as shown in Figure~\ref{pbarpdiscussion}.
The baryon junction model~\cite{junction} tried to explain it by
using junction anti-junction production with jet quenching, on the
basis of pQCD calculation~\cite{junction} where the $\bar{p}/p$
ratio decreases with $p_{T}$ in p+p collisions. The curve from
pQCD calculation~\cite{junction} is also shown in
Figure~\ref{pbarpdiscussion}. However, $\bar{p}/p$ ratios in d+Au
and p+p collisions in our data show to be flat with $p_{T}$ within
errors. Anyway, the precise measurement with more statistics in
p+p and d+Au collisions is needed to address this issue.

\subsection{Baryon production at RHIC: multi-gluon dynamics?}

\begin{figure}[h]
\centering
\includegraphics[height=18pc,width=18pc]{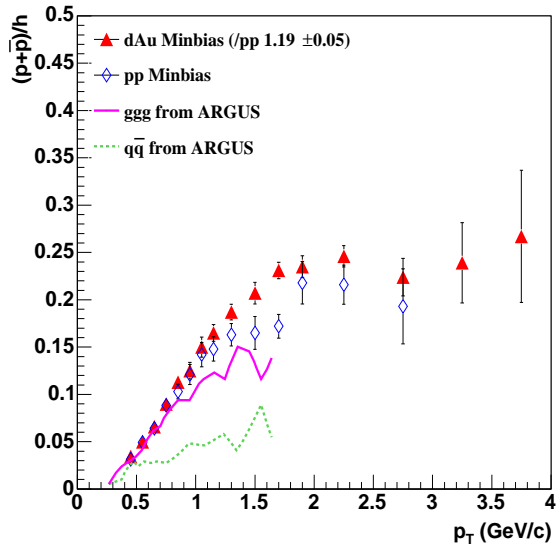}
\caption{Minimum-bias ratios of ($p+\bar{p}$) over charged hadrons
at $-0.5\!<\!\eta\!<\!0.0$ from $\sqrt{s_{_{NN}}} =200$ GeV p+p
(open diamonds) and d+Au (filled triangles) collisions. Also shown
are the $(p+\bar{p})/h$ ratios in $e^{+}e^{-}$ collisions at
ARGUS~\cite{argus}. The solid line represents the $(p+\bar{p})/h$
ratio from three gluon hadronization while the dashed line for the
ratio from quark and antiquark fragmentation~\cite{argus}. Errors
are statistical.} \label{pbarpnchcomparison}
\end{figure}

Let's compare the $(p+\bar{p})/h$ ratio in p+p, d+Au and Au+Au
collisions at RHIC energy 200 GeV with the ratio in $e^{+}e^{-}$
collisions at ARGUS~\cite{argus}. Using the ARGUS detector at the
$e^{+}e^{-}$ storage ring DORIS II, the inclusive production of
pion, kaon and proton in multihadron events at 9.98 GeV and in
direct decays of the $\Upsilon(1S)$ meson were
investigated~\cite{argus}. Multihadron final states in
$e^{+}e^{-}$ annihilation are produced via quark and antiquark
fragmentation, and those from direct $\Upsilon(1S)$ decays
originate from the hadronization of three gluons~\cite{argus}.
Figure~\ref{pbarpnchcomparison} shows the $(p+\bar{p})/h$ ratio in
200 GeV p+p collisions together with the ratio in $e^{+}e^{-}$
collisions at ARGUS~\cite{argus}. The plot shows that the
$(p+\bar{p})/h$ ratio from three gluon hadronization is a factor
of 3 higher than that from quark and antiquark fragmentation at
ARGUS. Our data from 200 GeV p+p collisions is close to
$(p+\bar{p})/h$ ratio from 3 gluon hadronization. This may be the
indication that in the heavy ion collisions at RHIC energy,
multi-gluon hadronization plays an important role for the particle
production.

\chapter{Conclusion and Outlook}
\label{chp:conclusion}
\section{Conclusion}
In summary, we have reported the identified particle spectra of
pions, kaons, protons and anti-protons at mid-rapidity from 200
GeV minimum-bias, centrality selected d+Au collisions and NSD p+p
collisions. The time-of-flight detector, based on novel multi-gap
resistive plate chamber technology, was used for particle
identification. This is the first time that MRPC detector was
installed to take data as a time-of-flight detector in the
collider experiment. The calibration method was set up in the STAR
experiment for the first time and has been applied to the data
taken later successfully. The intrinsic timing resolution of the
MRPC was 85 ps after the calibration. In 2003 run, the pion/kaon
can be separated up to transverse momentum 1.6 GeV/c while proton
can be identified up to 3.0 GeV/c.

The spectra of $\pi^{\pm}$, $K^{\pm}$, $p$ and $\bar{p}$ in d+Au
and p+p collisions provide an important reference for those in
Au+Au collisions. The initial state in d+Au collisions is similar
to that in Au+Au collisions, and, it's believed that the
quark-gluon plasma doesn't exist in d+Au collisions. These results
from d+Au collisions are very important for us to judge whether
the quark-gluon plasma exists in Au+Au collisions or not and to
understand the property of the dense matter created in Au+Au
collisions. We observe that the spectra of $\pi^{\pm}$, $K^{\pm}$,
$p$ and $\bar{p}$ are considerably harder in d+Au than those in
p+p collisions. In $\sqrt{s_{_{NN}}} = 200$ GeV d+Au collisions,
the $R_{dAu}$ of protons rise faster than $R_{dAu}$ of pions and
kaons. The $R_{dAu}$ of proton is larger than 1 at intermediate
$p_T$ while the proton production follows binary scaling at the
same $p_T$ range in 200 GeV Au+Au collisions. These results
further prove that the suppression observed in Au+Au collisions at
intermediate and high $p_T$ is due to final state interactions in
a dense and dissipative medium produced during the collision and
not due to the initial state wave function of the Au nucleus.
Additionally, the particle-species dependence of the Cronin effect
is found to be significantly smaller than that from lower energy
p+A collisions. In $\sqrt{s_{_{NN}}} = 200$ GeV d+Au collisions,
the ratio of the nuclear modification factor $R_{dAu}$ between
$(p+\bar{p})$ and charged hadrons ($h$) in the $p_{T}$ range $1.2<
p_{T}<3.0$ GeV/c was measured to be
$1.19\pm0.05$(stat)$\pm0.03$(syst) in minimum-bias collisions.
Both the $R_{dAu}$ values and $(p+\bar{p})/h$ ratios show little
centrality dependence, in contrast to previous measurements in
Au+Au collisions at $\sqrt{s_{NN}}$ = 130 and 200 GeV. The ratios
of protons over charged hadrons in d+Au and p+p collisions are
found to be about a factor of 2 lower than that from Au+Au
collisions, indicating that the relative baryon enhancement
observed in heavy ion collisions at RHIC is due to the final state
effects in Au+Au collisions.

The identified particle spectra in d+Au and p+p collisions not
only provide the reference for those in Au+Au collisions, but also
provide a chance to see the mechanism of the Cronin effect itself
clearly. Usually the Cronin effect has been explained to be the
initial state effect only since 1970s~\cite{accardi}. However, we
compare our pion and proton spectra in minimum-bias and
centrality-selected d+Au collisions with the recombination
model~\cite{hwayang}. The recombination model can reproduce both
the pion spectra and proton spectra. This recombination model is
built on the hadronization process, which is a final-state effect,
while the initial multiple parton scattering model~\cite{accardi}
can't reproduce the difference of the Cronin effect between pions
and protons. From these comparisons, we conclude that the Cronin
effect in $\sqrt{s_{_{NN}}} = 200$ GeV d+Au collisions is not the
initial state effect only, and that final state effect plays an
important role.

The integral yield $dN/dy$ and $\langle p_T \rangle$ in p+p and
d+Au collisions were estimated from the power law fit and thermal
model fit. The integral yield $R_{dAu}$ of $\pi^{\pm}$, $K^{\pm}$,
$p$ and $\bar{p}$ are observed to be smaller than 1 while those of
$p$ and $\bar{p}$ are close to 1. The $\pi^{-}/\pi^{+}$,
$K^{-}/K^{+}$ and $\bar{p}/p$ ratios as a function of $p_{T}$ are
observed to be flat with $p_T$ within the errors in d+Au and p+p
minimum-bias collisions and show little centrality dependence in
d+Au collisions. The integral yield ratios of $K^{-}/\pi^{-}$ and
$\bar{p}/\pi^{-}$ as a function of $dN/d\eta$ were also presented
in p+p and d+Au collisions.

\section{Outlook}
For the outlook, I will discuss whether the Cronin effect is mass
dependent or baryon/meson dependent at 200 GeV. What other physics
topic have we done from MRPC-TOFr in d+Au and p+p collisions in
2003 run? If we have the full time-of-flight (Full-TOF) coverage,
what can we do? Also I will discuss a little bit about the low
energy 63 GeV Au+Au run.

\subsection{Cronin effect at 200 GeV: Mass dependent or baryon/meson dependent?}
We know that recombination model can reproduce the spectra of
pions and protons in d+Au collisions. Also the $R_{CP}$ of
identified particles in Au+Au collisions suggest that the degree
of suppression depends on particle species(baryon/meson) at
intermediate $p_T$. Does the Cronin effect in 200 GeV d+Au
collisions depend on the particle species (baryon/meson) or depend
on the particle mass? From our data, it shows the Cronin effect
for proton is bigger than those for pion and kaon. And the Cronin
effect of pion shows little difference from that of kaon at
$p_T<1.5$ GeV/c. In order to see the Cronin effect is baryon/meson
dependent or mass dependent, we can compare the Cronin effect of
proton with those of $K^*$ and $\phi$ since the mass of $K^*$ and
$\phi$ are close to that of proton while $K^*$ and $\phi$ are
mesons and proton is a baryon. The preliminary results show that
the Cronin effect of $K^*$ and $\phi$~\cite{kstarphi} are similar
to that of pion and different from that of proton. However, the
final results from $K^*$ and $\phi$ are needed to confirm this
issue.

\subsection{Electron PID from MRPC-TOFr}
The production and spectra of hadrons with heavy flavor are
sensitive to initial conditions and the later stage dynamical
evolution in high energy nuclear collisions, and may be less
affected by the non-perturbative complication in theoretical
calculations~\cite{charm1}. Charm production has been proposed as
a sensitive measurement of parton distribution function in nucleon
and the nuclear shadowing effect by systematically studying p+p,
and p+A collisions~\cite{lin96}. The relatively reduced energy
loss of heavy quark traversing a quark-gluon plasma will help us
distinguish the medium in which the jet loses its
energy~\cite{dokshitzer01}. A possible enhancement of charmonia
($J/\Psi$) production can be present at RHIC energies~\cite{jpsi}
due to the coalescence of the copiously produced charm
quarks~\cite{opencharm}.

\begin{figure}[h]
\centering
\includegraphics[height=20pc,width=24pc]{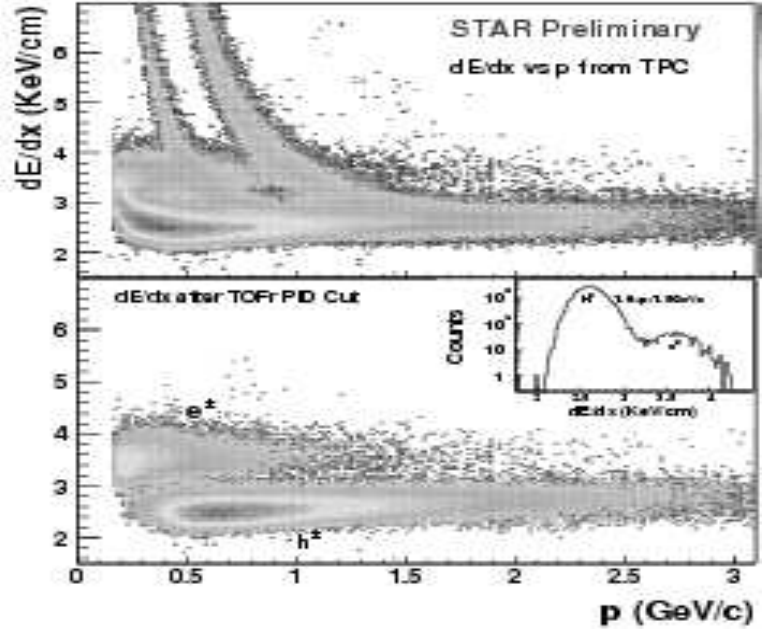}
\caption{$dE/dx$ in TPC versus $p$ without(the upper panel) or
with (the lower pannel) the TOFr velocity cut $|1/\beta-
1|\le0.03$. The insert shows $dE/dx$ distribution for 1 $\le p
\le$ 1.5 GeV/c. } \label{electron}
\end{figure}

The recent STAR results on the absolute open charm cross section
measurements from direct charmed hadron $D^0$
reconstruction~\cite{Haibin:03} in d+Au collisions and electrons
from charm semileptonic decay in both p+p and d+Au collisions at
200 GeV were presented~\cite{opencharm}. Based on the capability
of hadron identification~\cite{startof1} from the MRPC-TOFr tray
in 2003, electrons could be identified at low momentum
($p_{T}\le3$ GeV/c) by the combination of velocity ($\beta$) from
TOFr~\cite{startof} and the particle ionization energy loss
($dE/dx$) from TPC~\cite{tpc}. Figure~\ref{electron} shows that
the electrons are clearly identified as a separate band in the
$dE/dx$ versus momentum ($p$) with a selection on $\beta$ at
$|1/\beta-1|\le0.03$ in d+Au collisions. At higher $p_{T}$ (2--4
GeV/c), negative electrons were also identified directly by TPC
since hadrons have lower $dE/dx$ due to the relativisitic rise of
electron $dE/dx$. Based on the clear electron identification, the
open-charm-decayed electron spectra was derived~\cite{opencharm}.
Combined with $D^0$ measurement from TPC, the total charm cross
section was obtained~\cite{opencharm} in d+Au collisions.

\subsection{Full-TOF Physics}
Based on the hadron PID and electron PID of MRPC-TOFr in 2003, we
can imagine how many physics we can do if we have full
time-of-flight coverage based on MRPC technology. The
proposal~\cite{tofproposal} for large area time-of-flight system
for STAR has been proposed. Since the pion/kaon can be separated
up to transverse momentum 1.6 GeV/c and proton can be identified
up to 3.0 GeV/c from time-of-flight system. The resonance spectra
measured from hadronic decay will be extended to much higher
$p_T$. The direct open charm spectra from its hadronic decay
channel will reach higher $p_T$ with much more precise
measurement. Since the electron can be clearly identified up to
transverse momentum 3$\sim$4 GeV/c by the combination of velocity
($\beta$) from TOFr ~\cite{startof} and the particle ionization
energy loss ($dE/dx$) from TPC ~\cite{tpc}, the electron spectra
from charm-semi-leptonic decay will be measured precisely. As we
know that the measurement of the di-leptonic decays of vector
mesons are very difficult since the branch ratios are too small
and it's really hard to subtract the background. But with TOF
upgrading together with the SVT and micro-vertex detector
upgrading, the di-leptonic decays of vector mesons will be
measured much more easily, which will bring the direct information
of QGP since the electron is a lepton and the cross section of
interaction between electrons and hadrons is little. Thus we can
see directly the property of quark-gluon plasma such as the
temperature and the chiral symmetry restoration. This will be the
most interesting and meaningful thing for the QGP
search~\cite{xzb}. Besides, there are many other physics
topics~\cite{tofproposal} such as identified particle correlation
and fluctuation, particle composition of jet fragmentation, and
anti-nuclei etc.

\subsection{63 GeV Au+Au collisions at RHIC}
The bulk properties such as elliptic flow $v_2$ and particle
production show smooth trend from AGS, SPS to RHIC energy. One
energy point $\sqrt{s_{_{NN}}} = 63$ GeV, which is between SPS and
full RHIC energy, was selected since high quality charged-particle
and $\pi^0$ inclusive spectra have been measured in p+p collisions
at 63 GeV at Intersecting Storage Rings (ISR) and will serve as
the reference spectra for computing the nuclear modification
factor for Au+Au collisions measured at the same energy. The $v_2$
and particle production at 63 GeV will be studied at RHIC.
Besides, the nuclear modification factor as a function of $p_T$
will also be studied in this collision system in which the hard
scattering component has been significantly reduced. The results
from 63 GeV Au+Au collisions will be helpful for us to understand
the property of dense medium created in 200 GeV Au+Au collisions.

{
\appendix

\chapter{{\hspace{3.5cm}Tables of the Invariant Yields}}

\begin{table*}
\begin{scriptsize}
  \centering
 \begin{tabular} {|c|c|c|c|} \hline

        $p_{T}$  & $p_{T}$ width  & M.B. & 0\%-20\%  \\ \hline
3.50e-01& 1.00e-01 & $3.14e+00\pm3.65e-02\pm2.51e-01$ &
$6.70e+00\pm9.69e-02\pm5.36e-01$ \\ \hline 4.50e-01& 1.00e-01 &
$1.82e+00\pm2.45e-02\pm1.45e-01$ &
$3.88e+00\pm6.51e-02\pm3.10e-01$ \\ \hline 5.50e-01& 1.00e-01 &
$1.08e+00\pm1.73e-02\pm8.67e-02$ &
$2.38e+00\pm4.68e-02\pm1.90e-01$ \\ \hline 6.50e-01& 1.00e-01 &
$6.18e-01\pm1.22e-02\pm4.94e-02$ &
$1.38e+00\pm3.28e-02\pm1.10e-01$ \\ \hline 7.50e-01& 1.00e-01 &
$4.13e-01\pm9.15e-03\pm3.31e-02$ &
$9.53e-01\pm2.56e-02\pm7.63e-02$ \\ \hline 8.50e-01& 1.00e-01 &
$2.66e-01\pm6.98e-03\pm2.13e-02$ &
$6.11e-01\pm1.93e-02\pm4.88e-02$ \\ \hline 9.50e-01& 1.00e-01 &
$1.70e-01\pm5.36e-03\pm1.36e-02$ &
$3.79e-01\pm1.44e-02\pm3.03e-02$ \\ \hline 1.05e+00& 1.00e-01 &
$1.14e-01\pm4.07e-03\pm9.09e-03$ &
$2.68e-01\pm1.15e-02\pm2.14e-02$ \\ \hline 1.15e+00& 1.00e-01 &
$7.70e-02\pm3.27e-03\pm6.16e-03$ &
$1.71e-01\pm8.82e-03\pm1.37e-02$ \\ \hline 1.30e+00& 2.00e-01 &
$4.43e-02\pm1.66e-03\pm3.55e-03$ &
$1.06e-01\pm4.71e-03\pm8.45e-03$ \\ \hline 1.50e+00& 2.00e-01 &
$2.10e-02\pm1.07e-03\pm1.68e-03$ &
$4.65e-02\pm2.96e-03\pm3.72e-03$ \\ \hline 1.70e+00& 2.00e-01 &
$1.05e-02\pm7.19e-04\pm8.40e-04$ &
$2.29e-02\pm1.92e-03\pm1.83e-03$ \\ \hline
 \end{tabular}
\caption{$\pi^{+}$ spectra in minimum-bias and 0-20\% d+Au
collisions. The unit of $p_{T}$ and $p_{T}$ width is $GeV/c$. }
\label{pionplusspectratable1}
\end{scriptsize}
\end{table*}

\begin{table*}
\begin{scriptsize}
  \centering
  \begin{tabular}{|c|c|c|c|} \hline
        $p_{T}$ & $p_{T}$ width  & 20\%-40\% & 40\%-100\%  \\ \hline
3.50e-01& 1.00e-01 & $4.73e+00\pm6.94e-02\pm3.78e-01$ &
$1.93e+00\pm2.85e-02\pm1.55e-01$ \\ \hline 4.50e-01& 1.00e-01 &
$2.80e+00\pm4.73e-02\pm2.24e-01$ &
$1.10e+00\pm1.89e-02\pm8.81e-02$ \\ \hline 5.50e-01& 1.00e-01 &
$1.68e+00\pm3.36e-02\pm1.34e-01$ &
$6.43e-01\pm1.31e-02\pm5.14e-02$ \\ \hline 6.50e-01& 1.00e-01 &
$9.50e-01\pm2.30e-02\pm7.60e-02$ &
$3.68e-01\pm9.03e-03\pm2.94e-02$ \\ \hline 7.50e-01& 1.00e-01 &
$6.41e-01\pm1.76e-02\pm5.13e-02$ &
$2.40e-01\pm6.80e-03\pm1.92e-02$ \\ \hline 8.50e-01& 1.00e-01 &
$4.20e-01\pm1.36e-02\pm3.36e-02$ &
$1.52e-01\pm5.09e-03\pm1.22e-02$ \\ \hline 9.50e-01& 1.00e-01 &
$2.58e-01\pm1.00e-02\pm2.07e-02$ &
$1.01e-01\pm3.97e-03\pm8.05e-03$ \\ \hline 1.05e+00& 1.00e-01 &
$1.80e-01\pm7.92e-03\pm1.44e-02$ &
$6.10e-02\pm2.84e-03\pm4.88e-03$ \\ \hline 1.15e+00& 1.00e-01 &
$1.21e-01\pm6.35e-03\pm9.72e-03$ &
$4.38e-02\pm2.37e-03\pm3.50e-03$ \\ \hline 1.30e+00& 2.00e-01 &
$6.49e-02\pm3.02e-03\pm5.19e-03$ &
$2.40e-02\pm1.16e-03\pm1.92e-03$ \\ \hline 1.50e+00& 2.00e-01 &
$3.28e-02\pm2.05e-03\pm2.63e-03$ &
$1.03e-02\pm7.12e-04\pm8.27e-04$ \\ \hline 1.70e+00& 2.00e-01 &
$1.57e-02\pm1.31e-03\pm1.26e-03$ &
$4.92e-03\pm4.49e-04\pm3.94e-04$ \\ \hline
 \end{tabular}
\caption{$\pi^{+}$ spectra in 20-40\% and 40-100\% d+Au
collisions. The unit of $p_{T}$ and $p_{T}$ width is $GeV/c$.}
\label{pionplusspectratable2}
\end{scriptsize}
\end{table*}

\begin{table}[h]
\begin{scriptsize}
  \centering
  \begin{tabular}{|c|c|c|} \hline
        $p_{T}$  & $p_{T}$ width  & p+p \\ \hline
3.50e-01& 1.00e-01 & $9.71e-01\pm1.21e-02\pm7.76e-02$ \\ \hline
4.50e-01& 1.00e-01 & $5.32e-01\pm7.84e-03\pm4.26e-02$ \\ \hline
5.50e-01& 1.00e-01 & $3.14e-01\pm5.46e-03\pm2.51e-02$ \\ \hline
6.50e-01& 1.00e-01 & $1.74e-01\pm3.72e-03\pm1.40e-02$ \\ \hline
7.50e-01& 1.00e-01 & $1.08e-01\pm2.65e-03\pm8.64e-03$ \\ \hline
8.50e-01& 1.00e-01 & $6.42e-02\pm1.89e-03\pm5.14e-03$ \\ \hline
9.50e-01& 1.00e-01 & $4.03e-02\pm1.42e-03\pm3.22e-03$ \\ \hline
1.05e+00& 1.00e-01 & $2.40e-02\pm9.94e-04\pm1.92e-03$ \\ \hline
1.15e+00& 1.00e-01 & $1.55e-02\pm7.71e-04\pm1.24e-03$ \\ \hline
1.30e+00& 2.00e-01 & $8.19e-03\pm3.86e-04\pm6.55e-04$ \\ \hline
1.50e+00& 2.00e-01 & $3.77e-03\pm2.46e-04\pm3.02e-04$ \\ \hline
1.70e+00& 2.00e-01 & $1.84e-03\pm1.96e-04\pm1.47e-04$ \\ \hline
 \end{tabular}
\caption{$\pi^{+}$ spectra in p+p collisions. The unit of $p_{T}$
and $p_{T}$ width is $GeV/c$.} \label{pionplusspectratable3}
\end{scriptsize}
\end{table}

\begin{table}[h]
\begin{scriptsize}
  \centering
  \begin{tabular}{|c|c|c|c|} \hline
        $p_{T}$  & $p_{T}$ width  & M.B. & 0\%-20\%  \\ \hline
3.50e-01& 1.00e-01 & $3.20e+00\pm3.73e-02\pm2.56e-01$ &
$6.77e+00\pm9.88e-02\pm5.42e-01$ \\ \hline 4.50e-01& 1.00e-01 &
$1.82e+00\pm2.46e-02\pm1.45e-01$ &
$3.92e+00\pm6.60e-02\pm3.13e-01$ \\ \hline 5.50e-01& 1.00e-01 &
$1.08e+00\pm1.72e-02\pm8.60e-02$ &
$2.42e+00\pm4.71e-02\pm1.94e-01$ \\ \hline 6.50e-01& 1.00e-01 &
$6.70e-01\pm1.30e-02\pm5.36e-02$ &
$1.49e+00\pm3.51e-02\pm1.20e-01$ \\ \hline 7.50e-01& 1.00e-01 &
$4.05e-01\pm9.06e-03\pm3.24e-02$ &
$9.22e-01\pm2.50e-02\pm7.38e-02$ \\ \hline 8.50e-01& 1.00e-01 &
$2.59e-01\pm6.89e-03\pm2.07e-02$ &
$6.13e-01\pm1.96e-02\pm4.91e-02$ \\ \hline 9.50e-01& 1.00e-01 &
$1.68e-01\pm5.38e-03\pm1.34e-02$ &
$3.85e-01\pm1.48e-02\pm3.08e-02$ \\ \hline 1.05e+00& 1.00e-01 &
$1.16e-01\pm4.23e-03\pm9.29e-03$ &
$2.70e-01\pm1.18e-02\pm2.16e-02$ \\ \hline 1.15e+00& 1.00e-01 &
$7.63e-02\pm3.28e-03\pm6.11e-03$ &
$1.71e-01\pm8.83e-03\pm1.37e-02$ \\ \hline 1.30e+00& 2.00e-01 &
$4.36e-02\pm1.66e-03\pm3.49e-03$ &
$9.67e-02\pm4.45e-03\pm7.73e-03$ \\ \hline 1.50e+00& 2.00e-01 &
$2.04e-02\pm1.07e-03\pm1.63e-03$ &
$4.94e-02\pm3.06e-03\pm3.96e-03$ \\ \hline 1.70e+00& 2.00e-01 &
$1.03e-02\pm7.23e-04\pm8.26e-04$ &
$2.54e-02\pm2.07e-03\pm2.03e-03$ \\ \hline
 \end{tabular}
\caption{$\pi^{-}$ spectra in minimum-bias and 0-20\% d+Au
collisions. The unit of $p_{T}$ and $p_{T}$ width is $GeV/c$. }
\label{pionminusspectratable1}
\end{scriptsize}
\end{table}

\begin{table}[h]
\begin{scriptsize}
  \centering
  \begin{tabular}{|c|c|c|c|} \hline
        $p_{T}$ & $p_{T}$ width  & 20\%-40\% & 40\%-100\%  \\ \hline
3.50e-01& 1.00e-01 & $4.81e+00\pm7.11e-02\pm3.85e-01$ &
$2.00e+00\pm2.95e-02\pm1.60e-01$ \\ \hline 4.50e-01& 1.00e-01 &
$2.76e+00\pm4.73e-02\pm2.21e-01$ &
$1.12e+00\pm1.93e-02\pm8.95e-02$ \\ \hline 5.50e-01& 1.00e-01 &
$1.65e+00\pm3.28e-02\pm1.32e-01$ &
$6.43e-01\pm1.31e-02\pm5.14e-02$ \\ \hline 6.50e-01& 1.00e-01 &
$1.03e+00\pm2.47e-02\pm8.23e-02$ &
$4.04e-01\pm9.83e-03\pm3.23e-02$ \\ \hline 7.50e-01& 1.00e-01 &
$6.15e-01\pm1.71e-02\pm4.92e-02$ &
$2.40e-01\pm6.80e-03\pm1.92e-02$ \\ \hline 8.50e-01& 1.00e-01 &
$3.92e-01\pm1.30e-02\pm3.13e-02$ &
$1.51e-01\pm5.10e-03\pm1.21e-02$ \\ \hline 9.50e-01& 1.00e-01 &
$2.61e-01\pm1.03e-02\pm2.09e-02$ &
$9.62e-02\pm3.91e-03\pm7.70e-03$ \\ \hline 1.05e+00& 1.00e-01 &
$1.81e-01\pm8.11e-03\pm1.45e-02$ &
$6.65e-02\pm3.07e-03\pm5.32e-03$ \\ \hline 1.15e+00& 1.00e-01 &
$1.20e-01\pm6.26e-03\pm9.58e-03$ &
$4.23e-02\pm2.31e-03\pm3.38e-03$ \\ \hline 1.30e+00& 2.00e-01 &
$6.82e-02\pm3.18e-03\pm5.46e-03$ &
$2.42e-02\pm1.18e-03\pm1.93e-03$ \\ \hline 1.50e+00& 2.00e-01 &
$3.24e-02\pm2.13e-03\pm2.59e-03$ &
$1.05e-02\pm7.69e-04\pm8.42e-04$ \\ \hline 1.70e+00& 2.00e-01 &
$1.55e-02\pm1.34e-03\pm1.24e-03$ &
$5.02e-03\pm1.06e-03\pm4.01e-04$ \\ \hline
 \end{tabular}
\caption{$\pi^{-}$ spectra in 20-40\% and 40-100\% d+Au
collisions. The unit of $p_{T}$ and $p_{T}$ width is $GeV/c$.}
\label{pionminusspectratable2}
\end{scriptsize}
\end{table}

\begin{table}[h]
\begin{scriptsize}
  \centering
  \begin{tabular}{|c|c|c|} \hline
        $p_{T}$  & $p_{T}$ width  & p+p \\ \hline
3.50e-01& 1.00e-01 & $9.71e-01\pm1.22e-02\pm7.76e-02$ \\ \hline
4.50e-01& 1.00e-01 & $5.47e-01\pm8.02e-03\pm4.37e-02$ \\ \hline
5.50e-01& 1.00e-01 & $3.09e-01\pm5.38e-03\pm2.47e-02$ \\ \hline
6.50e-01& 1.00e-01 & $1.84e-01\pm3.89e-03\pm1.47e-02$ \\ \hline
7.50e-01& 1.00e-01 & $1.00e-01\pm2.50e-03\pm8.01e-03$ \\ \hline
8.50e-01& 1.00e-01 & $6.36e-02\pm1.89e-03\pm5.09e-03$ \\ \hline
9.50e-01& 1.00e-01 & $3.80e-02\pm1.38e-03\pm3.04e-03$ \\ \hline
1.05e+00& 1.00e-01 & $2.44e-02\pm1.03e-03\pm1.95e-03$ \\ \hline
1.15e+00& 1.00e-01 & $1.57e-02\pm7.87e-04\pm1.25e-03$ \\ \hline
1.30e+00& 2.00e-01 & $8.70e-03\pm4.02e-04\pm6.96e-04$ \\ \hline
1.50e+00& 2.00e-01 & $3.62e-03\pm2.45e-04\pm2.90e-04$ \\ \hline
1.70e+00& 2.00e-01 & $1.69e-03\pm1.76e-04\pm1.35e-04$ \\ \hline
 \end{tabular}
\caption{$\pi^{-}$ spectra in p+p collisions. The unit of $p_{T}$
and $p_{T}$ width is $GeV/c$.} \label{pionminusspectratable3}
\end{scriptsize}
\end{table}

\begin{table}[h]
\begin{scriptsize}
  \centering
  \begin{tabular}{|c|c|c|c|} \hline
        $p_{T}$  & $p_{T}$ width  & M.B. & 0\%-20\%  \\ \hline
4.57e-01& 1.00e-01 & $2.41e-01\pm9.47e-03\pm1.93e-02$ &
$4.83e-01\pm2.75e-02\pm3.86e-02$ \\ \hline 5.56e-01& 1.00e-01 &
$1.87e-01\pm7.53e-03\pm1.49e-02$ &
$3.87e-01\pm2.14e-02\pm3.10e-02$ \\ \hline 6.55e-01& 1.00e-01 &
$1.33e-01\pm5.14e-03\pm1.07e-02$ &
$2.81e-01\pm1.53e-02\pm2.25e-02$ \\ \hline 7.54e-01& 1.00e-01 &
$1.01e-01\pm3.18e-03\pm8.07e-03$ &
$2.14e-01\pm1.09e-02\pm1.72e-02$ \\ \hline 8.54e-01& 1.00e-01 &
$6.97e-02\pm2.48e-03\pm5.57e-03$ &
$1.58e-01\pm8.74e-03\pm1.27e-02$ \\ \hline 9.54e-01& 1.00e-01 &
$5.01e-02\pm1.95e-03\pm4.01e-03$ &
$9.95e-02\pm6.30e-03\pm7.96e-03$ \\ \hline 1.05e+00& 1.00e-01 &
$3.77e-02\pm1.65e-03\pm3.02e-03$ &
$8.92e-02\pm5.83e-03\pm7.14e-03$ \\ \hline 1.15e+00& 1.00e-01 &
$2.73e-02\pm1.31e-03\pm2.18e-03$ &
$5.61e-02\pm4.30e-03\pm4.49e-03$ \\ \hline 1.30e+00& 2.00e-01 &
$1.78e-02\pm7.25e-04\pm1.42e-03$ &
$4.00e-02\pm2.49e-03\pm3.20e-03$ \\ \hline 1.50e+00& 2.00e-01 &
$9.12e-03\pm5.25e-04\pm7.30e-04$ &
$1.95e-02\pm1.89e-03\pm1.56e-03$ \\ \hline 1.70e+00& 2.00e-01 &
$4.68e-03\pm3.96e-04\pm3.75e-04$ & $---$ \\ \hline

 \end{tabular}
\caption{$K^{+}$ spectra in minimum-bias and 0-20\% d+Au
collisions. The unit of $p_{T}$ and $p_{T}$ width is $GeV/c$. }
\label{kaonplusspectratable1}
\end{scriptsize}
\end{table}

\begin{table}[h]
\begin{scriptsize}
  \centering
  \begin{tabular}{|c|c|c|c|} \hline
        $p_{T}$ & $p_{T}$ width  & 20\%-40\% & 40\%-100\%  \\ \hline
4.57e-01& 1.00e-01 & $3.63e-01\pm2.06e-02\pm2.91e-02$ &
$1.27e-01\pm7.70e-03\pm1.02e-02$ \\ \hline 5.56e-01& 1.00e-01 &
$2.74e-01\pm1.54e-02\pm2.19e-02$ &
$1.07e-01\pm6.16e-03\pm8.60e-03$ \\ \hline 6.55e-01& 1.00e-01 &
$2.01e-01\pm1.11e-02\pm1.60e-02$ &
$7.37e-02\pm4.26e-03\pm5.89e-03$ \\ \hline 7.54e-01& 1.00e-01 &
$1.58e-01\pm8.18e-03\pm1.27e-02$ &
$5.25e-02\pm2.99e-03\pm4.20e-03$ \\ \hline 8.54e-01& 1.00e-01 &
$1.01e-01\pm6.16e-03\pm8.06e-03$ &
$3.73e-02\pm2.35e-03\pm2.98e-03$ \\ \hline 9.54e-01& 1.00e-01 &
$7.97e-02\pm4.89e-03\pm6.37e-03$ &
$2.80e-02\pm1.86e-03\pm2.24e-03$ \\ \hline 1.05e+00& 1.00e-01 &
$5.34e-02\pm3.83e-03\pm4.27e-03$ &
$1.91e-02\pm1.47e-03\pm1.52e-03$ \\ \hline 1.15e+00& 1.00e-01 &
$3.91e-02\pm3.12e-03\pm3.13e-03$ &
$1.36e-02\pm1.18e-03\pm1.09e-03$ \\ \hline 1.30e+00& 2.00e-01 &
$2.51e-02\pm1.67e-03\pm2.01e-03$ &
$8.65e-03\pm6.83e-04\pm6.92e-04$ \\ \hline 1.50e+00& 2.00e-01 &
$1.32e-02\pm1.19e-03\pm1.05e-03$ &
$4.69e-03\pm5.07e-04\pm3.75e-04$ \\ \hline

 \end{tabular}
\caption{$K^{+}$ spectra in 20-40\% and 40-100\% d+Au collisions.
The unit of $p_{T}$ and $p_{T}$ width is $GeV/c$.}
\label{kaonplusspectratable2}
\end{scriptsize}
\end{table}

\begin{table}[h]
\begin{scriptsize}
  \centering
  \begin{tabular}{|c|c|c|} \hline
        $p_{T}$  & $p_{T}$ width  & p+p \\ \hline
4.57e-01& 1.00e-01 & $6.47e-02\pm3.01e-03\pm5.18e-03$ \\ \hline
5.56e-01& 1.00e-01 & $4.32e-02\pm2.10e-03\pm3.45e-03$ \\ \hline
6.55e-01& 1.00e-01 & $3.18e-02\pm1.51e-03\pm2.54e-03$ \\ \hline
7.54e-01& 1.00e-01 & $2.18e-02\pm9.70e-04\pm1.74e-03$ \\ \hline
8.54e-01& 1.00e-01 & $1.57e-02\pm7.69e-04\pm1.25e-03$ \\ \hline
9.54e-01& 1.00e-01 & $9.92e-03\pm5.60e-04\pm7.93e-04$ \\ \hline
1.05e+00& 1.00e-01 & $7.17e-03\pm4.74e-04\pm5.74e-04$ \\ \hline
1.15e+00& 1.00e-01 & $5.60e-03\pm4.18e-04\pm4.48e-04$ \\ \hline
1.30e+00& 2.00e-01 & $3.73e-03\pm2.72e-04\pm2.98e-04$ \\ \hline
1.50e+00& 2.00e-01 & $1.81e-03\pm1.18e-04\pm1.45e-04$ \\ \hline
 \end{tabular}
\caption{$K^{+}$ spectra in p+p collisions. The unit of $p_{T}$
and $p_{T}$ width is $GeV/c$.} \label{kaonplusspectratable3}
\end{scriptsize}
\end{table}

\begin{table}[h]
\begin{scriptsize}
  \centering
  \begin{tabular}{|c|c|c|c|} \hline
        $p_{T}$  & $p_{T}$ width  & M.B. & 0\%-20\%  \\ \hline
4.57e-01& 1.00e-01 & $2.34e-01\pm9.47e-03\pm1.87e-02$ &
$4.87e-01\pm2.81e-02\pm3.90e-02$ \\ \hline 5.56e-01& 1.00e-01 &
$1.92e-01\pm7.98e-03\pm1.54e-02$ &
$4.01e-01\pm2.28e-02\pm3.21e-02$ \\ \hline 6.55e-01& 1.00e-01 &
$1.39e-01\pm5.81e-03\pm1.12e-02$ &
$2.91e-01\pm1.69e-02\pm2.33e-02$ \\ \hline 7.54e-01& 1.00e-01 &
$9.21e-02\pm3.10e-03\pm7.37e-03$ &
$1.90e-01\pm1.04e-02\pm1.52e-02$ \\ \hline 8.54e-01& 1.00e-01 &
$7.11e-02\pm2.61e-03\pm5.69e-03$ &
$1.45e-01\pm8.76e-03\pm1.16e-02$ \\ \hline 9.54e-01& 1.00e-01 &
$4.70e-02\pm1.94e-03\pm3.76e-03$ &
$9.49e-02\pm6.34e-03\pm7.59e-03$ \\ \hline 1.05e+00& 1.00e-01 &
$3.11e-02\pm1.49e-03\pm2.48e-03$ &
$6.56e-02\pm5.07e-03\pm5.25e-03$ \\ \hline 1.15e+00& 1.00e-01 &
$2.28e-02\pm1.21e-03\pm1.82e-03$ &
$4.74e-02\pm4.00e-03\pm3.79e-03$ \\ \hline 1.30e+00& 2.00e-01 &
$1.61e-02\pm7.10e-04\pm1.29e-03$ &
$3.70e-02\pm2.41e-03\pm2.96e-03$ \\ \hline 1.50e+00& 2.00e-01 &
$9.47e-03\pm5.54e-04\pm7.58e-04$ &
$2.01e-02\pm1.78e-03\pm1.61e-03$ \\ \hline 1.70e+00& 2.00e-01 &
$4.44e-03\pm4.19e-04\pm3.55e-04$ & $---$ \\ \hline

 \end{tabular}
\caption{$K^{-}$ spectra in minimum-bias and 0-20\% d+Au
collisions. The unit of $p_{T}$ and $p_{T}$ width is $GeV/c$. }
\label{kaonminusspectratable1}
\end{scriptsize}
\end{table}

\begin{table}[h]
\begin{scriptsize}
  \centering
  \begin{tabular}{|c|c|c|c|} \hline
        $p_{T}$ & $p_{T}$ width  & 20\%-40\% & 40\%-100\%  \\ \hline
4.57e-01& 1.00e-01 & $3.24e-01\pm1.95e-02\pm2.59e-02$ &
$1.39e-01\pm8.25e-03\pm1.11e-02$ \\ \hline 5.56e-01& 1.00e-01 &
$2.66e-01\pm1.58e-02\pm2.13e-02$ &
$1.13e-01\pm6.63e-03\pm9.03e-03$ \\ \hline 6.55e-01& 1.00e-01 &
$1.99e-01\pm1.18e-02\pm1.59e-02$ &
$7.50e-02\pm4.61e-03\pm6.00e-03$ \\ \hline 7.54e-01& 1.00e-01 &
$1.36e-01\pm7.60e-03\pm1.08e-02$ &
$5.62e-02\pm3.16e-03\pm4.50e-03$ \\ \hline 8.54e-01& 1.00e-01 &
$1.13e-01\pm6.64e-03\pm9.04e-03$ &
$3.78e-02\pm2.47e-03\pm3.03e-03$ \\ \hline 9.54e-01& 1.00e-01 &
$7.35e-02\pm4.83e-03\pm5.88e-03$ &
$2.55e-02\pm1.82e-03\pm2.04e-03$ \\ \hline 1.05e+00& 1.00e-01 &
$4.51e-02\pm3.57e-03\pm3.61e-03$ &
$1.93e-02\pm1.51e-03\pm1.54e-03$ \\ \hline 1.15e+00& 1.00e-01 &
$3.03e-02\pm2.74e-03\pm2.43e-03$ &
$1.36e-02\pm1.20e-03\pm1.09e-03$ \\ \hline 1.30e+00& 2.00e-01 &
$2.35e-02\pm1.66e-03\pm1.88e-03$ &
$7.84e-03\pm6.41e-04\pm6.27e-04$ \\ \hline 1.50e+00& 2.00e-01 &
$1.05e-02\pm1.20e-03\pm8.36e-04$ &
$5.10e-03\pm5.72e-04\pm4.08e-04$ \\ \hline
 \end{tabular}
\caption{$K^{-}$ spectra in 20-40\% and 40-100\% d+Au collisions.
The unit of $p_{T}$ and $p_{T}$ width is $GeV/c$.}
\label{kaonminusspectratable2}
\end{scriptsize}
\end{table}

\begin{table}[h]
\begin{scriptsize}
  \centering
  \begin{tabular}{|c|c|c|} \hline
        $p_{T}$  & $p_{T}$ width  & p+p \\ \hline
4.57e-01& 1.00e-01 & $5.99e-02\pm2.91e-03\pm4.79e-03$ \\ \hline
5.56e-01& 1.00e-01 & $4.79e-02\pm2.35e-03\pm3.83e-03$ \\ \hline
6.55e-01& 1.00e-01 & $3.06e-02\pm1.58e-03\pm2.45e-03$ \\ \hline
7.54e-01& 1.00e-01 & $2.15e-02\pm9.81e-04\pm1.72e-03$ \\ \hline
8.54e-01& 1.00e-01 & $1.46e-02\pm7.67e-04\pm1.17e-03$ \\ \hline
9.54e-01& 1.00e-01 & $1.01e-02\pm5.86e-04\pm8.09e-04$ \\ \hline
1.05e+00& 1.00e-01 & $7.87e-03\pm5.22e-04\pm6.30e-04$ \\ \hline
1.15e+00& 1.00e-01 & $5.25e-03\pm4.27e-04\pm4.20e-04$ \\ \hline
1.30e+00& 2.00e-01 & $3.42e-03\pm2.57e-04\pm2.74e-04$ \\ \hline
1.50e+00& 2.00e-01 & $1.75e-03\pm1.61e-04\pm1.40e-04$ \\ \hline
 \end{tabular}
\caption{$K^{-}$ spectra in p+p collisions. The unit of $p_{T}$
and $p_{T}$ width is $GeV/c$.} \label{kaonminusspectratable3}
\end{scriptsize}
\end{table}

\begin{table}[h]
\begin{scriptsize}
  \centering
  \begin{tabular}{|c|c|c|c|}
    \hline
        $p_{T}$  & $p_{T}$ width  & M.B. & 0\%-20\%  \\ \hline
4.68e-01& 1.00e-01 & $1.88e-01\pm1.87e-02\pm2.44e-02$ &
$3.96e-01\pm4.51e-02\pm5.15e-02$ \\ \hline 5.63e-01& 1.00e-01 &
$1.35e-01\pm1.11e-02\pm1.75e-02$ &
$2.80e-01\pm2.67e-02\pm3.64e-02$ \\ \hline 6.61e-01& 1.00e-01 &
$1.07e-01\pm8.89e-03\pm1.39e-02$ &
$2.10e-01\pm2.01e-02\pm2.73e-02$ \\ \hline 7.59e-01& 1.00e-01 &
$8.02e-02\pm6.78e-03\pm1.04e-02$ &
$1.74e-01\pm1.67e-02\pm2.27e-02$ \\ \hline 8.58e-01& 1.00e-01 &
$5.82e-02\pm5.20e-03\pm7.57e-03$ &
$1.25e-01\pm1.27e-02\pm1.63e-02$ \\ \hline 9.57e-01& 1.00e-01 &
$4.45e-02\pm4.85e-03\pm5.78e-03$ &
$1.05e-01\pm1.25e-02\pm1.37e-02$ \\ \hline 1.06e+00& 1.00e-01 &
$3.63e-02\pm4.11e-03\pm2.90e-03$ &
$8.16e-02\pm1.02e-02\pm6.53e-03$ \\ \hline 1.16e+00& 1.00e-01 &
$2.82e-02\pm2.11e-03\pm2.26e-03$ &
$5.08e-02\pm5.11e-03\pm4.06e-03$ \\ \hline 1.31e+00& 2.00e-01 &
$1.86e-02\pm1.14e-03\pm1.49e-03$ &
$4.14e-02\pm3.21e-03\pm3.31e-03$ \\ \hline 1.50e+00& 2.00e-01 &
$1.02e-02\pm7.85e-04\pm8.14e-04$ &
$2.46e-02\pm2.34e-03\pm1.97e-03$ \\ \hline 1.70e+00& 2.00e-01 &
$5.64e-03\pm3.24e-04\pm4.51e-04$ &
$1.21e-02\pm1.03e-03\pm9.66e-04$ \\ \hline 1.90e+00& 2.00e-01 &
$3.14e-03\pm2.33e-04\pm2.51e-04$ &
$8.08e-03\pm8.38e-04\pm6.46e-04$ \\ \hline 2.25e+00& 5.00e-01 &
$1.39e-03\pm9.47e-05\pm1.12e-04$ &
$3.37e-03\pm3.21e-04\pm2.70e-04$ \\ \hline 2.75e+00& 5.00e-01 &
$2.75e-04\pm3.88e-05\pm2.20e-05$ &
$6.19e-04\pm1.26e-04\pm4.95e-05$ \\ \hline 3.50e+00& 1.00e+00 &
$8.13e-05\pm1.37e-05\pm6.50e-06$ & $---$ \\ \hline

 \end{tabular}
\caption{$p$ spectra in minimum-bias and 0-20\% d+Au collisions.
The unit of $p_{T}$ and $p_{T}$ width is $GeV/c$. }
\label{protonspectratable1}
\end{scriptsize}
\end{table}

\begin{table}[h]
\begin{scriptsize}
  \centering
  \begin{tabular}{|c|c|c|c|}
    \hline
        $p_{T}$ & $p_{T}$ width  & 20\%-40\% & 40\%-100\%  \\ \hline
4.68e-01& 1.00e-01 & $2.78e-01\pm3.24e-02\pm3.61e-02$ &
$1.21e-01\pm1.40e-02\pm1.57e-02$ \\ \hline 5.63e-01& 1.00e-01 &
$1.90e-01\pm1.84e-02\pm2.48e-02$ &
$7.84e-02\pm7.61e-03\pm1.02e-02$ \\ \hline 6.61e-01& 1.00e-01 &
$1.56e-01\pm1.49e-02\pm2.03e-02$ &
$6.18e-02\pm5.96e-03\pm8.03e-03$ \\ \hline 7.59e-01& 1.00e-01 &
$1.18e-01\pm1.14e-02\pm1.53e-02$ &
$4.84e-02\pm4.71e-03\pm6.29e-03$ \\ \hline 8.58e-01& 1.00e-01 &
$8.54e-02\pm8.68e-03\pm1.11e-02$ &
$2.98e-02\pm3.12e-03\pm3.87e-03$ \\ \hline 9.57e-01& 1.00e-01 &
$6.45e-02\pm7.77e-03\pm8.38e-03$ &
$2.39e-02\pm2.93e-03\pm3.11e-03$ \\ \hline 1.06e+00& 1.00e-01 &
$5.48e-02\pm6.88e-03\pm4.38e-03$ &
$2.12e-02\pm2.69e-03\pm1.69e-03$ \\ \hline 1.16e+00& 1.00e-01 &
$4.21e-02\pm3.97e-03\pm3.37e-03$ &
$1.52e-02\pm1.49e-03\pm1.22e-03$ \\ \hline 1.31e+00& 2.00e-01 &
$2.68e-02\pm2.09e-03\pm2.15e-03$ &
$9.80e-03\pm7.91e-04\pm7.84e-04$ \\ \hline 1.50e+00& 2.00e-01 &
$1.41e-02\pm1.38e-03\pm1.13e-03$ &
$5.05e-03\pm5.17e-04\pm4.04e-04$ \\ \hline 1.70e+00& 2.00e-01 &
$7.69e-03\pm7.02e-04\pm6.15e-04$ &
$2.54e-03\pm2.55e-04\pm2.03e-04$ \\ \hline 1.90e+00& 2.00e-01 &
$4.12e-03\pm5.31e-04\pm3.30e-04$ &
$1.43e-03\pm2.03e-04\pm1.15e-04$ \\ \hline 2.25e+00& 5.00e-01 &
$2.30e-03\pm2.28e-04\pm1.84e-04$ &
$6.31e-04\pm7.57e-05\pm5.05e-05$ \\ \hline 2.75e+00& 5.00e-01 &
$4.36e-04\pm1.13e-04\pm3.49e-05$ &
$1.62e-04\pm4.14e-05\pm1.29e-05$ \\ \hline

 \end{tabular}
\caption{$p$ spectra in 20-40\% and 40-100\% d+Au collisions. The
unit of $p_{T}$ and $p_{T}$ width is $GeV/c$.}
\label{protonspectratable2}
\end{scriptsize}
\end{table}

\begin{table}[h]
\begin{scriptsize}
  \centering
  \begin{tabular}{|c|c|c|}
    \hline
        $p_{T}$  & $p_{T}$ width  & p+p \\ \hline
4.68e-01& 1.00e-01 & $4.51e-02\pm5.24e-03\pm5.86e-03$ \\ \hline
5.63e-01& 1.00e-01 & $3.33e-02\pm3.29e-03\pm4.33e-03$ \\ \hline
6.61e-01& 1.00e-01 & $2.60e-02\pm2.57e-03\pm3.38e-03$ \\ \hline
7.59e-01& 1.00e-01 & $2.03e-02\pm2.12e-03\pm2.64e-03$ \\ \hline
8.58e-01& 1.00e-01 & $1.14e-02\pm1.30e-03\pm1.49e-03$ \\ \hline
9.57e-01& 1.00e-01 & $8.93e-03\pm1.20e-03\pm1.16e-03$ \\ \hline
1.06e+00& 1.00e-01 & $6.98e-03\pm1.00e-03\pm5.59e-04$ \\ \hline
1.16e+00& 1.00e-01 & $4.68e-03\pm5.95e-04\pm3.75e-04$ \\ \hline
1.31e+00& 2.00e-01 & $2.90e-03\pm3.14e-04\pm2.32e-04$ \\ \hline
1.50e+00& 2.00e-01 & $1.34e-03\pm2.04e-04\pm1.07e-04$ \\ \hline
1.70e+00& 2.00e-01 & $6.61e-04\pm6.44e-05\pm5.29e-05$ \\ \hline
1.90e+00& 2.00e-01 & $4.73e-04\pm5.58e-05\pm3.78e-05$ \\ \hline
2.25e+00& 5.00e-01 & $1.84e-04\pm2.23e-05\pm1.47e-05$ \\ \hline
2.75e+00& 5.00e-01 & $3.06e-05\pm7.26e-06\pm2.45e-06$ \\ \hline

 \end{tabular}
\caption{$p$ spectra in p+p collisions. The unit of $p_{T}$ and
$p_{T}$ width is $GeV/c$.} \label{protonspectratable3}
\end{scriptsize}
\end{table}

\begin{table}[h]
\begin{scriptsize}
  \centering
  \begin{tabular}{|c|c|c|c|}
    \hline
        $p_{T}$  & $p_{T}$ width  & M.B. & 0\%-20\%  \\ \hline
4.68e-01& 1.00e-01 & $1.22e-01\pm7.62e-03\pm1.59e-02$ &
$2.57e-01\pm2.17e-02\pm3.34e-02$ \\ \hline 5.63e-01& 1.00e-01 &
$1.08e-01\pm5.87e-03\pm1.40e-02$ &
$2.26e-01\pm1.62e-02\pm2.94e-02$ \\ \hline 6.61e-01& 1.00e-01 &
$8.86e-02\pm4.69e-03\pm1.15e-02$ &
$1.73e-01\pm1.23e-02\pm2.25e-02$ \\ \hline 7.59e-01& 1.00e-01 &
$6.40e-02\pm3.47e-03\pm8.32e-03$ &
$1.39e-01\pm9.76e-03\pm1.81e-02$ \\ \hline 8.58e-01& 1.00e-01 &
$5.51e-02\pm3.04e-03\pm7.16e-03$ &
$1.19e-01\pm8.58e-03\pm1.54e-02$ \\ \hline 9.57e-01& 1.00e-01 &
$3.80e-02\pm2.30e-03\pm4.94e-03$ &
$8.99e-02\pm6.92e-03\pm1.17e-02$ \\ \hline 1.06e+00& 1.00e-01 &
$3.27e-02\pm1.41e-03\pm2.62e-03$ &
$7.36e-02\pm5.01e-03\pm5.89e-03$ \\ \hline 1.16e+00& 1.00e-01 &
$2.36e-02\pm1.12e-03\pm1.89e-03$ &
$4.24e-02\pm3.50e-03\pm3.39e-03$ \\ \hline 1.31e+00& 2.00e-01 &
$1.55e-02\pm6.16e-04\pm1.24e-03$ &
$3.48e-02\pm2.14e-03\pm2.78e-03$ \\ \hline 1.50e+00& 2.00e-01 &
$8.00e-03\pm4.04e-04\pm6.40e-04$ &
$1.93e-02\pm1.46e-03\pm1.54e-03$ \\ \hline 1.70e+00& 2.00e-01 &
$4.78e-03\pm2.97e-04\pm3.82e-04$ &
$1.01e-02\pm9.74e-04\pm8.11e-04$ \\ \hline 1.90e+00& 2.00e-01 &
$2.56e-03\pm2.04e-04\pm2.05e-04$ &
$4.99e-03\pm6.54e-04\pm3.99e-04$ \\ \hline 2.25e+00& 5.00e-01 &
$1.06e-03\pm7.97e-05\pm8.49e-05$ &
$2.69e-03\pm2.85e-04\pm2.15e-04$ \\ \hline 2.75e+00& 5.00e-01 &
$3.32e-04\pm4.61e-05\pm2.66e-05$ &
$5.86e-04\pm1.35e-04\pm4.69e-05$ \\ \hline 3.50e+00& 1.00e+00 &
$7.89e-05\pm1.53e-05\pm6.31e-06$ & $---$ \\ \hline
 \end{tabular}
\caption{$\bar{p}$ spectra in minimum-bias and 0-20\% d+Au
collisions. The unit of $p_{T}$ and $p_{T}$ width is $GeV/c$.}
\label{pbarspectratable1}
\end{scriptsize}
\end{table}

\begin{table}[h]
\begin{scriptsize}
  \centering
  \begin{tabular}{|c|c|c|c|}
    \hline
        $p_{T}$ & $p_{T}$ width  & 20\%-40\% & 40\%-100\%  \\ \hline
4.68e-01& 1.00e-01 & $1.63e-01\pm1.45e-02\pm2.12e-02$ &
$7.50e-02\pm6.44e-03\pm9.75e-03$ \\ \hline 5.63e-01& 1.00e-01 &
$1.65e-01\pm1.20e-02\pm2.14e-02$ &
$6.04e-02\pm4.55e-03\pm7.85e-03$ \\ \hline 6.61e-01& 1.00e-01 &
$1.24e-01\pm8.94e-03\pm1.62e-02$ &
$5.51e-02\pm3.89e-03\pm7.16e-03$ \\ \hline 7.59e-01& 1.00e-01 &
$9.35e-02\pm6.78e-03\pm1.22e-02$ &
$3.61e-02\pm2.68e-03\pm4.69e-03$ \\ \hline 8.58e-01& 1.00e-01 &
$7.98e-02\pm5.96e-03\pm1.04e-02$ &
$2.85e-02\pm2.22e-03\pm3.70e-03$ \\ \hline 9.57e-01& 1.00e-01 &
$5.22e-02\pm4.33e-03\pm6.78e-03$ &
$1.95e-02\pm1.68e-03\pm2.54e-03$ \\ \hline 1.06e+00& 1.00e-01 &
$4.14e-02\pm3.20e-03\pm3.31e-03$ &
$1.82e-02\pm1.37e-03\pm1.45e-03$ \\ \hline 1.16e+00& 1.00e-01 &
$3.61e-02\pm2.82e-03\pm2.89e-03$ &
$1.22e-02\pm1.04e-03\pm9.73e-04$ \\ \hline 1.31e+00& 2.00e-01 &
$2.32e-02\pm1.50e-03\pm1.85e-03$ &
$7.64e-03\pm5.47e-04\pm6.11e-04$ \\ \hline 1.50e+00& 2.00e-01 &
$1.13e-02\pm9.51e-04\pm9.04e-04$ &
$3.51e-03\pm3.35e-04\pm2.81e-04$ \\ \hline 1.70e+00& 2.00e-01 &
$6.41e-03\pm6.63e-04\pm5.13e-04$ &
$2.21e-03\pm2.48e-04\pm1.77e-04$ \\ \hline 1.90e+00& 2.00e-01 &
$4.64e-03\pm5.37e-04\pm3.71e-04$ &
$9.80e-04\pm1.70e-04\pm7.84e-05$ \\ \hline 2.25e+00& 5.00e-01 &
$1.46e-03\pm1.78e-04\pm1.17e-04$ &
$5.66e-04\pm7.10e-05\pm4.53e-05$ \\ \hline 2.75e+00& 5.00e-01 &
$4.43e-04\pm1.38e-04\pm3.54e-05$ &
$1.29e-04\pm4.62e-05\pm1.03e-05$ \\ \hline

 \end{tabular}
\caption{$\bar{p}$ spectra in 20-40\% and 40-100\% d+Au
collisions. The unit of $p_{T}$ and $p_{T}$ width is $GeV/c$.}
\label{pbarspectratable2}
\end{scriptsize}
\end{table}

\begin{table}[h]
\begin{scriptsize}
  \centering
  \begin{tabular}{|c|c|c|}
    \hline
        $p_{T}$  & $p_{T}$ width  & p+p \\ \hline
4.68e-01& 1.00e-01 & $3.74e-02\pm2.60e-03\pm4.86e-03$ \\ \hline
5.63e-01& 1.00e-01 & $2.84e-02\pm1.77e-03\pm3.70e-03$ \\ \hline
6.61e-01& 1.00e-01 & $2.27e-02\pm1.38e-03\pm2.96e-03$ \\ \hline
7.59e-01& 1.00e-01 & $1.32e-02\pm8.65e-04\pm1.71e-03$ \\ \hline
8.58e-01& 1.00e-01 & $1.05e-02\pm7.20e-04\pm1.37e-03$ \\ \hline
9.57e-01& 1.00e-01 & $7.17e-03\pm5.40e-04\pm9.32e-04$ \\ \hline
1.06e+00& 1.00e-01 & $5.60e-03\pm3.77e-04\pm4.48e-04$ \\ \hline
1.16e+00& 1.00e-01 & $3.74e-03\pm2.86e-04\pm2.99e-04$ \\ \hline
1.31e+00& 2.00e-01 & $2.31e-03\pm1.49e-04\pm1.84e-04$ \\ \hline
1.50e+00& 2.00e-01 & $9.69e-04\pm8.84e-05\pm7.75e-05$ \\ \hline
1.70e+00& 2.00e-01 & $5.95e-04\pm6.65e-05\pm4.76e-05$ \\ \hline
1.90e+00& 2.00e-01 & $3.57e-04\pm4.79e-05\pm2.86e-05$ \\ \hline
2.25e+00& 5.00e-01 & $1.12e-04\pm1.90e-05\pm8.93e-06$ \\ \hline
2.75e+00& 5.00e-01 & $3.87e-05\pm1.07e-05\pm3.10e-06$ \\ \hline

 \end{tabular}
\caption{$\bar{p}$ spectra in p+p collisions. The unit of $p_{T}$
and $p_{T}$ width is $GeV/c$.} \label{pbarspectratable3}
\end{scriptsize}
\end{table}


\chapter{{\hspace{3.5cm}How to make MRPC}}
This appendix is based on the procedure of the MRPC production in
USTC. I will introduce the material preparations and then the
chamber installation.
\section{Preparations}
\subsection{Glass} (1) Check the glass very carefully by eye. The
glass with scrapes is not accepted. (2) Measure the size of the
glass with the digital vernier caliper. The errors of the length
and width are required to be within 0.1 mm.  Measure the thickness
in several different places.  The precision of the thickness is
required to be 0.01 mm for each glass. (3) Use the micrometer to
measure the flatness, which is required to be less than 0.01 mm.
Use the mirror and observe the stripes of interference. (4) Grind
the edge and the corner of the glass, and clean it. The size of
outer glass is $78(width)\times 206(length) \times 1.1(height)$
$mm^3$ and the size of inner glass is $61\times 200\times 0.54$
$mm^3$.

\subsection{Graphite Layer} (1) Stick the layer in the middle of the outer
glass. Squeeze the air out. (2) Stick a small copper tape, which
is for high voltage (HV) applying, on to the graphite layer, which
is in the middle of the long side, and 0$\sim$0.5 mm away from the
edge of the glass. The size of the graphite layer is $74\times202$
$mm^2$. The size of the copper tape (the rectangle with the round
angle) is $6\times10$ $mm^2$.

\subsection{Mylar layer} Cut the mylar layer, and see if there is any tiny holes or
scrapes. If yes, don't use it. The size of mylar is
$84\times212\times0.35$ $mm^3$.

\subsection{Honeycomb board} Measure the size and flatness. The
error of the length and width is required to be within 0.2 mm, the
error of the thickness is required to be within 0.05 mm. The
flatness is required to be within 0.1 mm. The size of honeycomb
board is $84\times 208\times 4$ $mm^3$.

\subsection{The printed circuit board (PCB)} (1) Check the surface of the metal
which is used as read-out strips carefully, and see the position
of the HV-holes is right or not. Check the size of the metal
holes, whose diameters are required to be larger than 0.9 mm. (2)
Use double side tape to stick the PCB board with the Mylar. The
size of the double side tape is the same as the PCB board. The
length of the mylar is 1 mm longer than that of the PCB board. (3)
Use sealing ion to open a $\phi$ 3 mm hole, the center of the hole
is in the middle of the HV holes. The size of PCB is $94\times
210\times 1.5$ $mm^3$. The size of metal holes are $\phi$ 1 mm.

\subsection{Lucite cylinder}
Use digital vernier caliper to measure the length of the Lucite
cylinder. Clean it and stick a double side tape on one side. The
size is $\phi$ 3 mm and $3.87<length<3.93$ mm.

\subsection{Other stuff}
Besides, we also need pins, fish line and little plastic cannula.
The size of the pin is 2-2.1 cm long. The fish line is $\phi$ 0.22
mm. The plastic cannula is $\phi$ 1 mm. One kind of the cannula is
7 mm long, and the other is 5.6 mm long. Table~\ref{mrpcmaterial}
lists the main materials for 1 MRPC.

\section{Installation}
\subsection{The outer glass and mylar and PCB}
(1) stick the outer glass on to the center of the mylar. (2) Use
the sealing ion to connect the HV conductive line with the copper
tape. Apply the HV to measure the noise rate and dark current. (3)
Between the mylar and outer glass edge, on each side, use silica
gel to seal. Attention: keep the surface clean and smooth.
Attention: If one side is done, wait till the silica gel becomes
solid. (4) Stick pins. Seal the pins which are used for the fish
line coiling, into the metal holes of the PCB board. (5) Use the
inner glass to fix on the position of Lucite cylinders, and keep
them away from the pins for fish line. Then stick the 8$\sim$10
Lucite cylinders onto the outer glass.

\subsection{Inner glass and fish-line coiling} (1) Pre-install.
Don't use fish-line. Pay attention to adjust the position of the
pins. (2) This is now the real installation and fish line coiling.
Clean the outer glass and inner glass carefully, coil a loop of
fish line, add a piece of glass, then coil another loop of fish
line, add another piece of glass, and so on and so forth.
Attention: clean the fish line before it coils, and blow the
surface of glass to protect it from the dirt with nitrogen jet.
(3) Another time for pre-installation. Pay attention to the
position of the upper and lower electrodes and adjust the position
of pins. (4) Paste 3140 RTV coating onto the surface of Lucite
cylinders. (5) Connect the two electrodes. Make sure all the pins
connect right into the metal holes. Then lay the whole flat, and
put on a block which is 4 kilogram weight. (6) After 2 hours,
stick the honeycomb. (7) Measure the thickness of the whole. Make
sure the precision is within 0.05 mm. (8) Connect the
conductive-line for the read-out strips, and then put the whole
into a bag. Attention: the conductive-line should not be broken.

\begin{table*}
\begin{scriptsize}
  \centering
 \begin{tabular}{|c|c|c|c|c|} \hline

 material& type  &  character & number  & source  \\ \hline
outer glass& window glass & 1.1 mm thick, VR: $8.7\times 10^{12}$
ohm.cm & 2 & Shanghai \\ \hline inner glass&window glass & 0.54 mm
thick, VR: $8.5\times 10^{12}$ ohm.cm & 5 & USA \\ \hline graphite
layer& T9149& 0.13 mm thick, SR : 2M ohm/square & 2 & Japan \\
\hline Mylar& M0 & 0.35 mm thick& 2& Dupont Corp.
\\ \hline
honeycomb & & 4 mm thick & 2& Shanghai \\ \hline PCB & gold & 1.58
 mm thick& 2 & Shenzhen \\ \hline copper tape & & 0.08 mm thick & 2
& 3M Comp. \\ \hline LC & Lucite& $\phi$ 3 mm,3.9 mm long & 8-10 &
processing \\ \hline pins (single) &metal pin&21.5 mm long& 14 &
\\ \hline pins (pair) &metal pin&21.5 mm long& 12 &
\\ \hline cannula & F-plastic & $\phi$ 1.4 mm & 38 & \\ \hline
fish line & top line & $\phi$ 0.22 mm & &Switzerland \\
\hline DST & 9690& 0.13 mm thick && 3M Comp. \\ \hline silica gel
& CAF4 & high-voltage insulation & &Switzerland \\ \hline
 \end{tabular}
\caption{The material for 1 MRPC model. VR is the volume
resistivity and SR is the surface resistivity. LC is the Lucite
cylinder. DST is the double side tape.} \label{mrpcmaterial}
\end{scriptsize}
\end{table*}

\chapter{{\hspace{3.5cm}List of Publications}}
\hspace{0.7cm}1.  \emph{Pion, kaon, proton and anti-proton
transverse momentum distributions from p+p and d+Au collisions at
$\sqrt{s_{NN}}$ = 200 GeV,} STAR Collaboration, e-Print Archives
(nu-ex/0309012), submitted.

2.  \emph{Open Charm Yields in 200 GeV p+p and d+Au Collisions at
RHIC,} Lijuan Ruan (for the STAR Collaboration), Journal of
Physics G, 30 (2004) S1197-S1200, contributed to 17th
International Conference on Ultra Relativistic Nucleus-Nucleus
Collisions (Quark Matter 2004).

3.  \emph{A Monte Carlo Simulation of Multi-gap Resistive Plate
Chamber and comparision with Experimental Results,} RUAN Li-Juan,
SHAO Ming, CHEN Hong-Fang, {\it et al.}, HEP and NP, Vol. 27, No.
8 (2003) 712-715.

4.  \emph{ Monte Carlo Study of the Property of Multi-gap
Resistive Plate Chambers,} Shao Ming, Ruan Lijuan, Chen Hongfang,
{\it et al.}, HEP and NP, Vol. 27, No. 1 (2003) 67-71, (in
Chinese).

5.  \emph{Study on Light Collection and its Uniformity of Long
Lead Tungstate crystal by Monte Carlo Method,} Ruan Lijuan, Shao
Ming, Xu Tong, {\it et al.}, Chinese Journal of Computational
Physics, Vol. 19, No. 5 (2002) 453-458, (in Chinese).

6.  \emph{Beam test results of two kinds of multi-gap resistive
plate chambers,}  M. Shao, L. J. Ruan, H. F. Chen, J. Wu, , C. Li,
Z. Z. Xu, X. L. Wang, S.L. Huang, Z. M. Wang and Z. P. Zhang,
Nucl. Instri. and Meth. A 492 (2002) 344-350.

7.  \emph{The Study of the Resistive Property of the Electrode
Material of MRPC,} Ruan Lijuan, Wang Xiaolian, Li Cheng , {\it et
al.}, to be published in Journal of University of Science and
Technology of China (in Chinese).

8.  \emph{The Calibration Method of TOFr in the STAR Experiment,}
RUAN Lijuan, WU Jian, DONG Xin , {\it et al.}, to be published in
HEP and NP (in Chinese).

9.  \emph{Spectra of $\pi$ K p $K^{*}$ $\phi$ from Au+Au
Collisions at 62.4 GeV,} Lijuan Ruan (for the STAR Collaboration),
to be published in Journal of Physics G, Contributed to 8th
International Conference on Strangeness in Quark Matter (SQM
2004).

10. \emph{Pseudorapidity Asymmetry and Centrality Dependence of
Charged Hadron Spectra in d+Au Collisions at $\sqrt{s_{NN}}$ = 200
GeV,} STAR Collaboration: e-Print Archives (nu-ex/0408016),
submitted.

11. \emph{Transverse momentum correlations and minijet dissipation
in Au-Au collisions at $\sqrt{s_{NN}}$ = 130 GeV,} STAR
Collaboration, e-Print Archives (nu-ex/0408012), submitted.

12. \emph{Azimuthal anisotropy and correlations at large
transverse momenta in p+p and Au+Au collisions at $\sqrt{s_{NN}}$=
200 GeV,} STAR Collaboration, e-Print Archives (nu-ex/0407007),
submitted.

13. \emph{Open charm yields in d+Au collisions at $\sqrt{s_{NN}}$
= 200 GeV,} STAR Collaboration, e-Print Archives (nu-ex/0407006),
submitted.

14. \emph{Measurements of transverse energy distributions in Au+Au
collisions at $\sqrt{s_{NN}}$ = 200 GeV,} STAR Collaboration,
e-Print Archives (nu-ex/0407003), submitted.

15. \emph{Transverse-momentum dependent modification of dynamic
texture in central Au+Au collisions at $\sqrt{s_{NN}}$ = 200 GeV,}
STAR Collaboration, e-Print Archives (nu-ex/0407001), submitted.

16. \emph{Hadronization geometry and charge-dependent number
autocorrelations on axial momentum space in Au-Au collisions at
$\sqrt{s_{NN}}$ = 130 GeV,} STAR Collaboration, e-Print
Archives(nu-ex/0406035), submitted.

17. \emph{Phi meson production in Au+Au and p+p collisions at
sqrt(s)=200 GeV,} STAR Collaboration, e-Print Archives
(nu-ex/0406003), submitted.

18. \emph{Centrality and pseudorapidity dependence of charged
hadron production at intermediate pT in Au+Au collisions at
$\sqrt{s_{NN}}$ = 130 GeV,} STAR Collaboration, e-Print Archives
(nu-ex/0404020), to be published in Physical Review C.

19. \emph{Production of e$+$e$-$ Pairs Accompanied by Nuclear
Dissociation in Ultra-Peripheral Heavy Ion Collision,} STAR
Collaboration, e-Print Archives (nu-ex/0404012), to be published
in Physical Review C.

20.  \emph{Photon and neutral pion production in Au+Au collisions
at $\sqrt{s_{NN}}$ = 130 GeV,} STAR Collaboration, e-Print
Archives (nu-ex/0401008), to be published in Physical Review C.

21. \emph{Azimuthally sensitive HBT in Au+Au collisions at
$\sqrt{s_{NN}}$ = 200 GeV,} STAR Collaboration, Phys. Rev. Lett.
93, 012301 (2004).

22. \emph{Production of Charged Pions and Hadrons in Au+Au
Collisions at $\sqrt{s_{NN}}$=130 GeV,} STAR Collaboration,
e-Print Archives (nu-ex/0311017), submitted.

23. \emph{Azimuthal anisotropy at RHIC: the first and fourth
harmonics,} STAR Collaboration, Phys. Rev. Lett. 92, 062301
(2004).

24.  \emph{Cross Sections and Transverse Single-Spin Asymmetries
in Forward Neutral Pion Production from Proton Collisions at
sqrt(s) = 200 GeV,}
 STAR Collaboration, Phys. Rev. Lett. 92, 171801 (2004).

25.  \emph{Identified particle distributions in pp and Au+Au
collisions at sqrt{snn}=200 GeV,} STAR Collaboration, Phys. Rev.
Lett. 92, 112301 (2004).

26.  \emph{Event-by-Event (pt) fluctuations in Au-Au collisions at
$\sqrt{s_{NN}}$ = 130 GeV,} STAR Collaboration, e-Print Archives
(nu-ex/0308033), submitted.

27.  \emph{Multi-strange baryon production in Au-Au collisions at
$\sqrt{s_{NN}}$ = 130 GeV,} STAR Collaboration, Phys. Rev. Lett.
92, 182301 (2004).

28.  \emph{Pion-Kaon Correlations in Central Au+Au Collisions at
$\sqrt{s_{NN}}$ = 130 GeV,} STAR Collaboration, Phys. Rev. Lett.
91, 262302 (2003).

29.  \emph{rho-0 Production and Possible Modification in Au+Au and
p+p Collisions at $\sqrt{s_{NN}}$ = 200 GeV,} STAR Collaboration,
Phys. Rev. Lett. 92, 092301 (2004).

30.  \emph{Net charge fluctuations in Au+Au collisions at
$\sqrt{s_{NN}}$ = 130 GeV,} STAR Collaboration, Phys. Rev. C 68,
044905 (2003).

31.  \emph{Rapidity and Centrality Dependence of Proton and
Anti-proton Production from Au+Au Collisions at $\sqrt{s_{NN}}$ =
130 GeV,} STAR Collaboration, e-Print Archives (nu-ex/0306029),
submitted.

32.  \emph{Three-Pion Hanbury Brown-Twiss Correlations in
Relativistic Heavy-Ion Collisions from the STAR Experiment,} STAR
Collaboration, Phys. Rev. Lett. 91, 262301 (2003).

33.  \emph{Evidence from d+Au measurements for final-state
suppression of high pT hadrons in Au+Au collisions at RHIC,} STAR
Collaboration, Phys. Rev. Lett. 91, 072304 (2003).

34.  \emph{Particle-type dependence of azimuthal anisotropy and
nuclear modification of particle production in Au+Au collisions at
$\sqrt{s_{NN}}$ = 200 GeV,} STAR Collaboration, Phys. Rev. Lett.
92, 052302 (2004).

35.  \emph{Transverse momentum and collision energy dependence of
high pT hadron suppression in Au+Au collisions at
ultrarelativistic energies,} STAR Collaboration, Phys. Rev. Lett.
91, 172302 (2003). 

\chapter{{\hspace{3.5cm}STAR Collaboration}}
\begin{figure}[htb]
\hspace{-6pc}
\includegraphics[width=45pc]{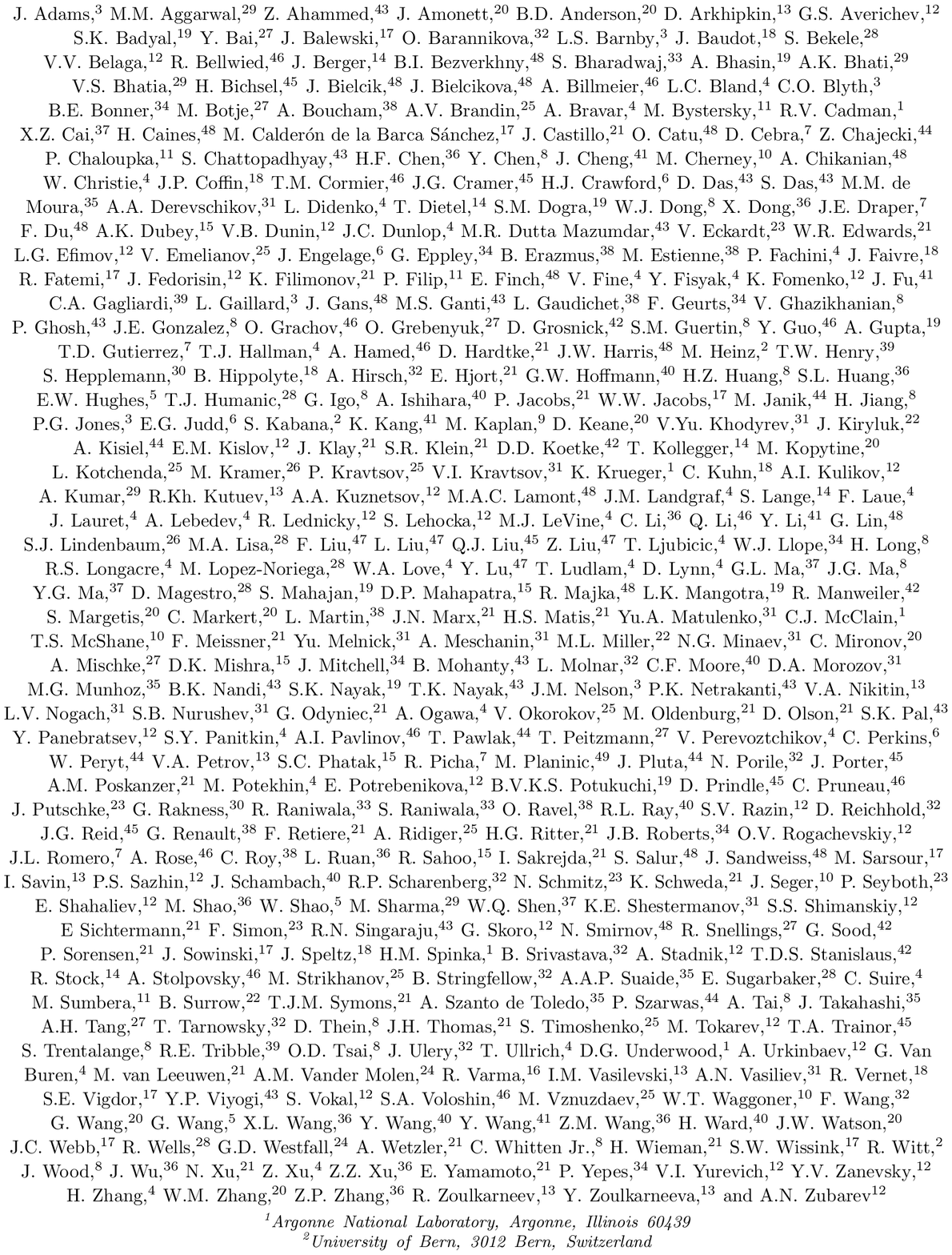}
\end{figure}

\begin{figure}[htb]
\hspace{-6pc}
\includegraphics[width=45pc]{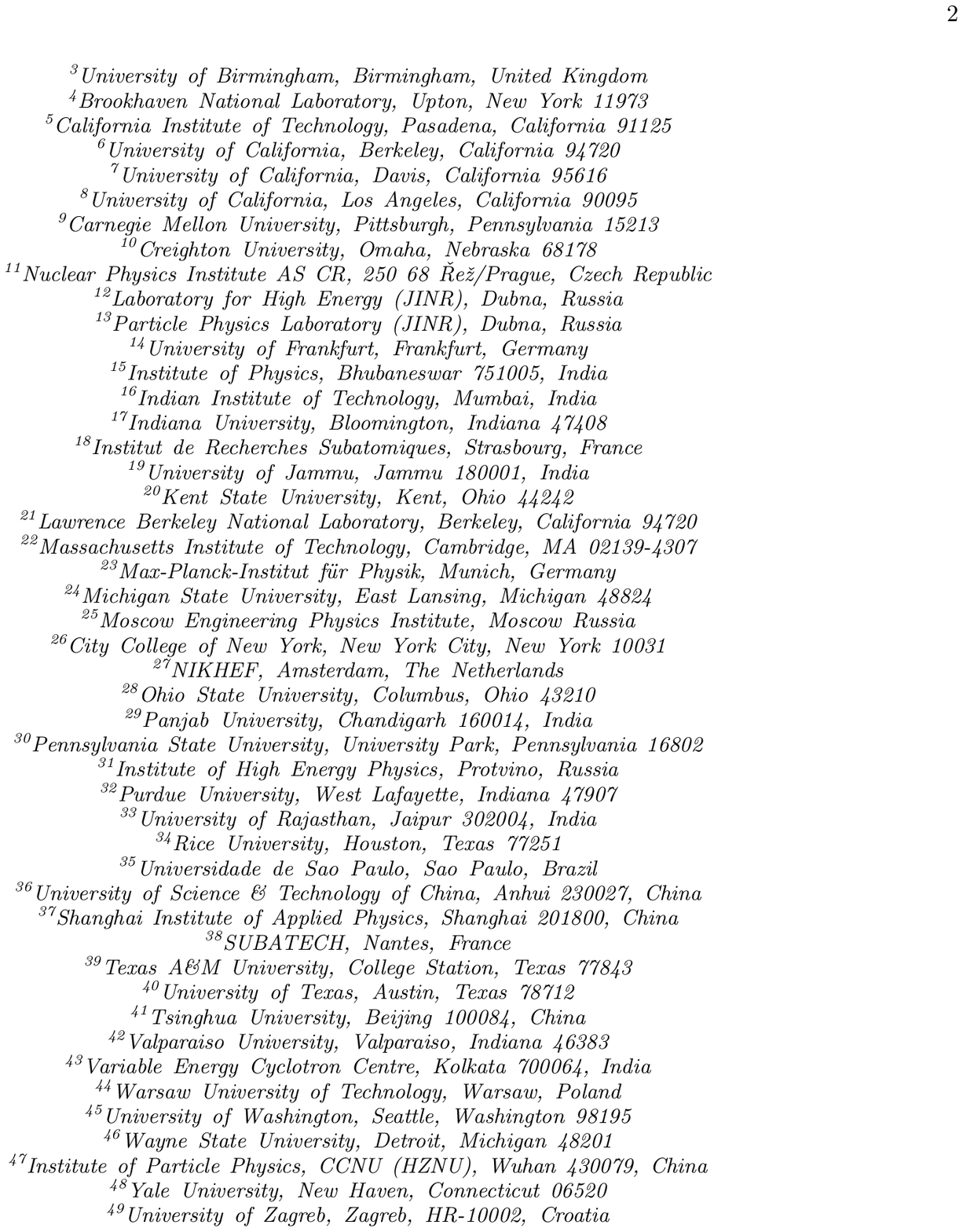}
\end{figure}
} \pagebreak
\bibliographystyle{prsty}
\addcontentsline{toc}{chapter}{Bibliography}
\bibliography{thesis}
\end{document}